\documentclass[review]{elsarticle}
\usepackage{lineno,hyperref}

\let\today\relax
\makeatletter
\def\ps@pprintTitle{%
    \let\@oddhead\@empty
    \let\@evenhead\@empty
    \def\@oddfoot{\footnotesize\itshape
         {} \hfill\today}%
    \let\@evenfoot\@oddfoot
    }
\makeatother

\usepackage{supertabular}
\usepackage{rotating}
\usepackage{subcaption}
\usepackage{float}
\newsavebox{\largestimage}
\usepackage{siunitx}
\usepackage{ulem}
\usepackage{array}
\newcolumntype{C}[1]{>{\centering\arraybackslash}p{#1}}

\usepackage{graphicx}
\usepackage{mwe}
\usepackage{subfiles}

\usepackage[utf8]{inputenc}
\usepackage{newunicodechar,graphicx}

\DeclareRobustCommand{\okina}{%
  \raisebox{\dimexpr\fontcharht\font`A-\height}{%
    \scalebox{0.8}{`}%
  }%
}
\newunicodechar{ʻ}{\okina}

\usepackage[autostyle=false, style=english]{csquotes}
\MakeOuterQuote{"}

\usepackage{filecontents}
\usepackage{amsmath}
\usepackage{textcomp}
\biboptions{authoryear}

\begin{document}
\begin{frontmatter}
\pagenumbering{arabic}

\title{Possible Interstellar meteoroids detected by the Canadian Meteor Orbit Radar}

\author[1]{Mark Froncisz }
\author[1,2]{Peter Brown \fnref{myfootnote}}
\author[3]{Robert J. Weryk}
\fntext[myfootnote]{Correspondence to: Peter Brown (pbrown@uwo.ca)}
\address[1]{Dept. of Physics and Astronomy, University of Western Ontario, London, Ontario, Canada N6A 3K7}
\address[2]{Centre for Planetary Science and Exploration, University of Western Ontario, London, Ontario, Canada N6A 5B7}
\address[3]{Institute for Astronomy, University of Hawaii, Honolulu, HI, 96822, USA}

\begin{abstract}

We examine meteoroid orbits recorded by the Canadian Meteor Orbit Radar (CMOR) from 2012-2019, consisting of just over 11 million orbits in a search for potential interstellar meteoroids. Our 7.5 year survey consists of an integrated time-area product of $\sim$ 7$\times$ 10$^6$ km$^2$ hours. Selecting just over 160000 six station meteor echoes having the highest measured velocity accuracy from within our sample, we found five candidate interstellar events. These five potential interstellar meteoroids were found to be hyperbolic at the 2$\sigma$-level using only their raw measured speed. Applying a new atmospheric deceleration correction algorithm developed for CMOR, we show that all five candidate events were likely hyperbolic at better than 3$\sigma$, the most significant being a 3.7$\sigma$ detection. Assuming all five detections are true interstellar meteoroids, we estimate the interstellar meteoroid flux at Earth to be at least 6.6 $\times$ 10$^{-7}$ meteoroids/km$^{2}$/hr appropriate to a mass of 2 $\times$ 10$^{-7}$kg.

Using estimated measurement uncertainties directly extracted from CMOR data, we simulated CMOR's ability to detect a hypothetical ʻOumuamua - associated hyperbolic meteoroid stream. Such a stream was found to be significant at the 1.8$\sigma$ level, suggesting that CMOR would likely detect such a stream of meteoroids as hyperbolic. We also show that CMOR's sensitivity to interstellar meteoroid detection is directionally dependent.

\end{abstract}


\end{frontmatter}

\section{Introduction}\label{Introduction}
Interstellar meteoroids offer a source of direct sampling of material that originated from beyond our solar system. Presolar grains embedded within meteorites already provide samples of small solids formed in other star systems \citep{zinner2014}, but the particular system where such grains are produced is unknown. Observations having trajectory information of large (\textgreater10 $\mu$m) interstellar meteoroids may retain information on their source region as well as providing clues as to the environment through which they traversed before being detected, as such large particles are not coupled to the local gas flow. Even larger interstellar particles (ISPs) on the order \textgreater 100 $\mu$m in radius \citep{Murray2004} may travel through the interstellar medium (ISM) for great distances with little perturbation by Lorentz forces. Integrating the motion of these meteoroids may allow for their original sources to be determined. Unique identification of the stellar system from which a particular interstellar meteoroid originates can help constrain the planet formation process, provide limits of the spatial density of larger grains in the interstellar medium, and probe debris disks. 

Small interstellar particles have been directly detected in-situ in the solar system by various spacecraft impact ionization sensors, including the Ulysses \citep{Grn1993}, Galileo \citep{Baguhl1995} and Cassini spacecraft \citep{Altobelli2003}, though the particle masses derived from impact sensors are very uncertain. The impact-detected particles were on the order of 1 $\mu$m in radius (10$^{-14}$ kg) and smaller, assuming an average density of 3000 kg/m$^{3}$.  Plasma producing EM emission from impacts of micron-sized and smaller interstellar particles have been detected by antennas on the STEREO \citep{Zaslavsky2012} and WIND \citep{Wood2015} spacecraft. These measurements showed that sub-micron interstellar dust shows a strong flux variation with time and orbital location. Such small particles are coupled to the local gas flow in the solar neighborhood, which has an upstream flow direction in galactic coordinates ofl$_{\textit{gal}}$ = \ang{3}, b$_{\textit{gal}}$ = \ang{16} \citep{Frisch1999} with a speed of 25 km/s. A good review summarizing our understanding of the small interstellar dust detected in the solar system is given by \citet{Sterken2019}.

The radar detection of interstellar meteoroids was reported by \citet{Baggaley2000} using the Advanced Meteor Orbit Radar (AMOR) in New Zealand and by \citet{Meisel2002} using the Arecibo radio telescope in Puerto Rico. Masses and sizes of these detections were 9 $\times$ 10$^{-9}$ kg (89 $\mu$m) and 7 $\times$ 10$^{-11}$ kg (18 $\mu$m), respectively, again assuming a meteoroid density of 3000 kg/m$^{3}$. However, the veracity of these detections as real interstellar meteoroids has been questioned by other authors (eg. \citep{Hajduk2001,Murray2004, Musci2012}).

\citet{Hajdukov2013} examined a catalogue comprised of 64650 meteors observed by a multi-station video meteor network in Japan between 2007 and 2008 \citep{SonotaCo2009}. Of these detections, 7489 appeared to have hyperbolic orbits. After filtering for meteors with low error in measured velocity and rejecting meteors associated with showers, 238 retained hyperbolic orbits and showed no prior  significant gravitational perturbations from close encounters with planets within our solar system. Their main conclusion, that most apparently hyperbolic optically measured meteoroid orbits are due to measurement error, is similar to the main conclusion from other similar studies [eg. \cite{Musci2012, hajdukova2012}].

\citet{Murray2004} provided estimates for the flux of interstellar particles and examined potential ISP-producing sources. They suggested the most prolific source would be dust grains produced as condensates in the atmospheres of asymptotic giant branch stars, which are blown into the ISM by stellar winds. Other sources they considered included dust grains from young main-sequence stars ejected by dynamical interactions with planets, and dust ejected due to radiation pressure from a host star. Additionally, they identified dust grains emitted from high-speed narrow jets formed during the accretion phase of young stellar objects to be a possible source of interstellar meteoroids. 
\citet{Weryk2005} analyzed data collected by the Canadian Meteor Orbit Radar (CMOR) between May 2002 and September 2004, where more than 1.5 million meteor orbits were computed. From this study, 40 meteoroids were found to have heliocentric speeds 2$\sigma$ above the hyperbolic limit, and 12 of these meteoroids had heliocentric speeds 3$\sigma$ above the hyperbolic limit.  

The present study may be considered an extension to the work of \citet{Weryk2005}. Here we expand on that study, in particular, by examining CMOR echoes collected on six receiver stations (five remote, as compared to only two available remote stations for the previous study). These additional stations significantly increase the confidence in time of flight velocity solutions, compared to the minimum three required for a unique time of flight solution as was the case for the earlier \citet{Weryk2005} study. We also develop an improved velocity correction for atmospheric deceleration for CMOR echoes based on examination of shower-associated meteors. This improves our confidence in derived out-of-atmosphere velocities and therefore calculated heliocentric orbits.  

The discovery of 1I/ ʻOumuamua, in 2017 \citep{Meech2017} was the first definitive observation of a large interstellar object transiting through our solar system. A two day observation arc established that the orbit of ʻOumuamua was hyperbolic. Over 200 subsequent observations over 34 days provided a more accurate assessment of its orbit and detailed lightcurve measurements provide information on its size, morphology \citep{Trilling2018}, and clues to its composition \citep{Jewitt2017}. The present study uses the orbital elements of  ʻOumuamua, as a test case to model whether CMOR could detect a hypothetical hyperbolic meteoroid stream of ʻOumuamua-associated meteoroids.  

\section{Instrumentation and Initial Data Processing}

The Canadian Meteor Orbit Radar (CMOR) is a multi-frequency HF/VHF radar array located in Tavistock, Ontario, Canada (43.26\textdegree N, 80.77\textdegree W). It consists of a main site three-element Yagi-Uda vertically directed antenna as a transmitter,  five co-located receiving antennas, and five additional remote station receivers. All receiver antennas are two-element vertically directed Yagi-Uda antennas. The five receiving antennas at the main site are arranged as an interferometer, allowing for positional measurements of meteor echoes \citep{Jones2005} with accuracy of order 1$^\circ$. CMOR operates at 17.45, 29.85 and 38.15 MHz; however for this study only data collected from the 29.85 MHz radar was used, as that frequency alone has orbital capability. More details of the operation and hardware of the CMOR system can be found in \citet{Webster2004, Jones2005, Brown2008}.

CMOR transmits Gaussianly-tapered radar pulses which are reflected off the electrons left behind by the meteoroid ablating in the atmosphere. Only the portion of these meteor trails which are oriented orthogonal to the receiver station (ie: their specular points) are detectable at a particular receiver through transverse scattering (see eg. \citet{Ceplecha1998}). With three or more receiver stations and interferometric (echo direction location in the sky) capability, complete trajectory and time-of-flight velocity measurements can be obtained \citep{Jones2005}. Confidence in time-of-flight velocity measurements increases with the number of remote receiver stations, as the velocity solution  becomes over constrained beyond the minimum of three stations required for a velocity solution. 

For CMOR, meteor echoes detected at multiple receiver stations are automatically correlated and trajectories automatically computed \citep{Weryk2012a}. In addition, meteor echoes are automatically filtered to remove dubious or poor-quality echoes. The Fresnel phase-time method (hereafter referred to as "pre-t0") is also employed as a validation check on meteor speed. This method is described in detail in \citet{Ceplecha1998}. Pre-t0 velocities are automatically computed from echoes detected at the main receiver station (T0) as described in \citet{mazur2019}.

The 29.85 MHz system transmits at 15kW peak power using a pulse repetition frequency of 532 Hz, and has an effective collecting area of between approximately 100 and 400 km$^2$ (Table \ref{tab:cmorparam}), dependent on the radiant declination \citep{CampbellBrown2006}. Using the equations from \citet{Verniani1973} we estimate that the minimum detectable meteor magnitude for orbit measurements under the requirement of detection at all six stations is +6 mag, corresponding to a meteoroid diameter of approximately 400 $\mu$m at a velocity of 45 km/s with an effective limiting mass of 1.8 $\times$ $10^{-7}$ kg. CMOR has been in near continuous operation since 2002 with various upgrades to transmitter power and the addition (and repositioning) of remote receiver stations. It has remained in its present configuration since late 2011. Data collected for this study spans January 2012 through June 2019 when all transmit and receive locations and parameters were constant. In total, during this time period, 11073016 radar meteor orbits were recorded.

\begin{table}[H]
\centering

\begin{tabular}{ll}
\hline
\textbf{Transmitter Location}                         & 43.26\textdegree N, 80.77\textdegree W              \\
\textbf{Frequency}                &  29.85 MHz       \\
\textbf{Pulse Repetition Frequency} & 532 pps                       \\
\textbf{Sample Rate}                & 50 ksps                       \\
\textbf{Range Sampling}            & 3 km                          \\
\textbf{Peak Transmit Power}                 & 15 kW                         \\
\textbf{Minimum detectable echo Power}                 & 10$^{-13}$ W                         \\
\textbf{Collecting Area}            & 100-400 km$^2$\\
\textbf{Magnitude Limit}            & +8 (+6 for six station orbits)                           \\
\textbf{Height Range}               & 70-120 km                      \\
\textbf{No. of Receiver Stations}   & 6                             \\
\textbf{Data Collection Dates}      & January 2012 - June 2019      \\
\textbf{Total orbits measured}      & 11073016                    \\
\textbf{Number of 6 Station orbits Recorded}  & 395973                 
\\
\hline    
\end{tabular}
\caption[CMOR parameters]{Parameters of the 29.85 MHz CMOR orbital system and the experimental configuration for this study.}
\label{tab:cmorparam}
\end{table}

\section{Searching for Interstellar Meteoroids in CMOR Data}\label{IScmor}
\subsection{Filtering}

CMOR multi-station echoes collected between January 2012 and June 2019 were examined to find evidence of interstellar meteoroids. This data set comprised over 11 million individual multi-station echoes for which time of flight velocities could be determined. We restricted our search  to those echoes recorded on all six receiver stations - this yielded 395973 echoes. This was done  as using all receiver stations gives the most accurate velocities. 

Among this dataset, we further selected only events which appeared to have hyperbolic heliocentric trajectories as computed from the raw (uncorrected) measured time of flight velocities ($V_m$). This produced  7282 candidate events for initial examination. Finally, we limited our search to events with estimated velocity errors of \textless10\%, based on a Monte Carlo routine simulating each echo and its geometry together with the CMOR detection algorithm (see \citet{Weryk2012a} for details). 

From previous experience examining six station echoes from meteor showers with CMOR, we have found that solutions begin to scatter significantly once the average time of flight residuals exceeds 2.0. Based on this experience, we also required events to have an average time of flight fit residuals (which we term sdel) of \textless2.0 samples. This gives a global metric of the goodness of fit between the measured inflection time picks at each site and the equivalent time picks from the best fit velocity vector for echoes having four or more station detections. The value of sdel of 2.0 was found through manual examination of the goodness of fit, and represents a qualitative cut for six station orbit detections which retains only fits where all inflection points are self-consistent between the measurement and solution to within better than 3-4 pulses at all stations. 

Once our final automated list of six station events was assembled, we manually examined each remaining echo (a total of 461) to ensure reliable time picks on all stations. As well we manually ensured the echoes had correct interferometry (meaning the directional solutions for all pairwise antenna combinations converge to one and not multiple solutions \citep{Jones1998}) and ``smooth'' amplitude profiles such that the profiles are consistent with the expected signal affected by ambipolar diffusion, which should produce a smooth exponential amplitude decay after the peak \citep{Ceplecha1998} . While echoes are automatically correlated across receiver sites and trajectories computed without manual intervention for these selected echoes, a detailed manual analysis is essential to remove bad measurements. 

We performed detailed inspection of the amplitude vs. time profile of echoes at all six receiver stations, compared automatic and manual selection of time picks for time-of-flight velocities, and examined interferometry solutions. We also compared rise time speed, Fresnel amplitude oscillation speed and Fresnel phase slope (pre-t0) speeds to the time of flight velocity for consistency. In addition to these quality controls,  to check the overall robustness of each trajectory solution we removed individual receiver station time picks to verify that the computed  velocity was minimally affected by removal of any one station, typically showing solution differences in speed of less than 5\% . 

\subsection{Results: Possible Interstellar Candidates based on Raw Measured Velocity}

After applying all filters and completing manual examination, a total of five apparently hyperbolic events were identified. These events had time of flight velocities consistent with  calculated pre-t0 velocities, calculated rise time velocities and, when visible, calculated Fresnel amplitude-time velocities (method described in \citet{Ceplecha1998}). The observed in-atmosphere radiant and speeds (without deceleration correction) in all cases produced hyperbolic orbits. Potentially erroneous interstation echo matches were checked by removing individual receiver stations from the calculated solution and verifying that the computed orbits remained consistently hyperbolic.  

\begin{figure}[H]
        \vspace*{-4cm}
        \advance\leftskip-2cm
        \advance\rightskip-2cm

        \begin{subfigure}{0.7\textwidth}

            \includegraphics[trim=0.5cm 0.8cm 0.5cm 1.5cm, clip,width=\textwidth]{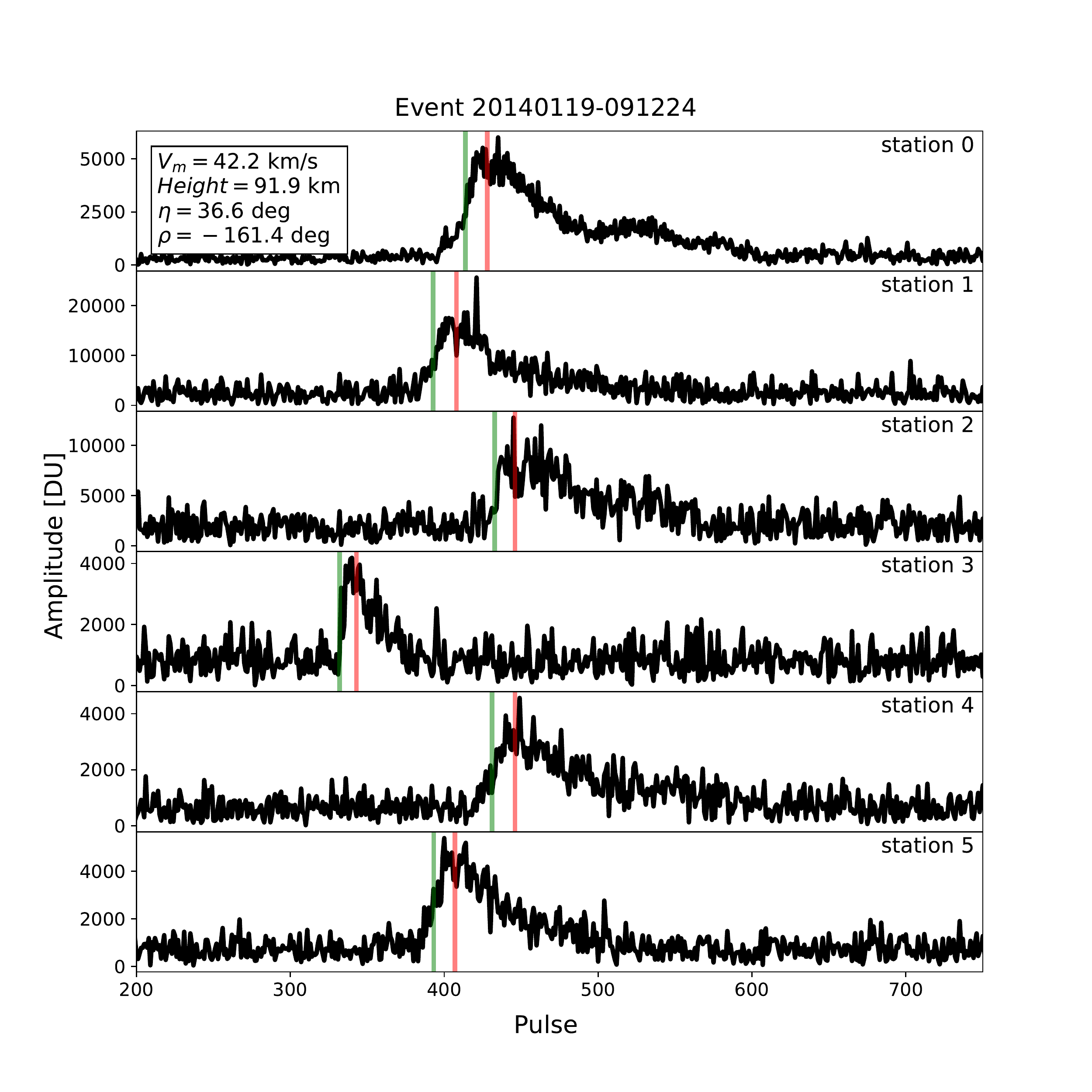}

        \end{subfigure}
        \smallskip
        \vspace*{-0.25cm}
        \begin{subfigure}{0.7\textwidth}  

            \includegraphics[trim=0.5cm 0.8cm 0.5cm 1.5cm, clip,width=\textwidth]{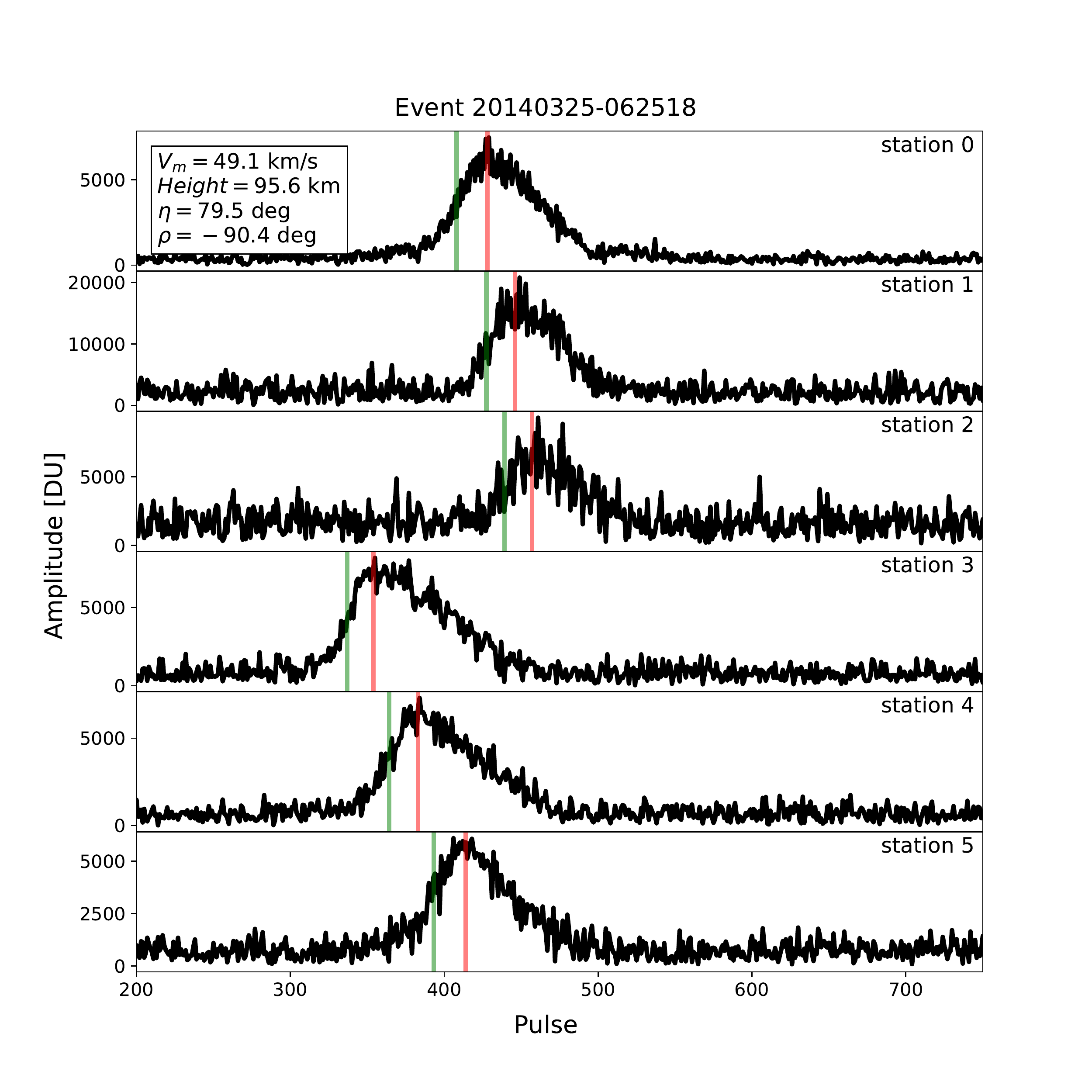}

        \end{subfigure}

        \begin{subfigure}{0.7\textwidth}   

            \includegraphics[trim=0.5cm 0.8cm 0.5cm 1.5cm, clip,width=\textwidth]{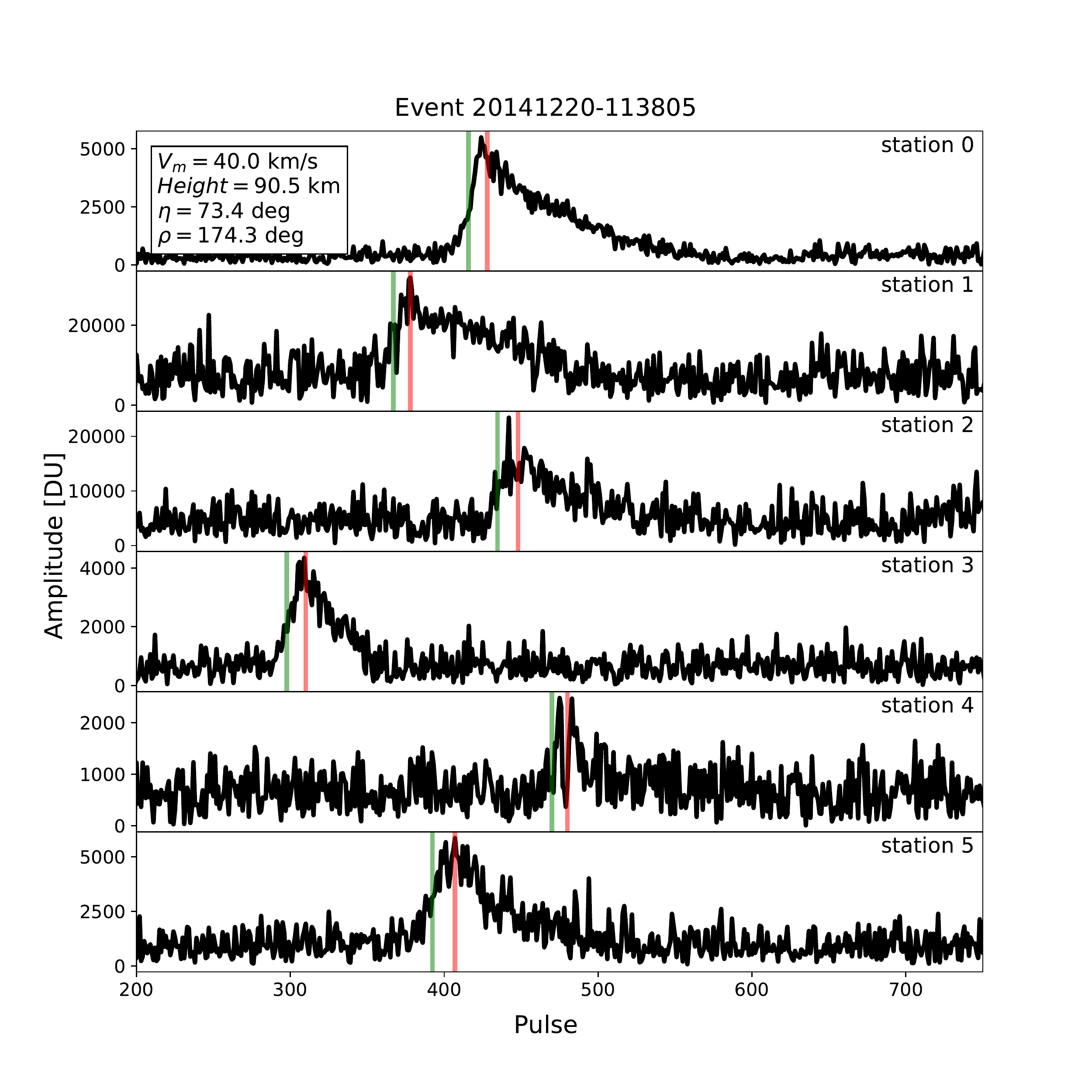}

        \end{subfigure}
           \smallskip
            \vspace*{-0.25cm}
        \begin{subfigure}{0.7\textwidth}   

            \includegraphics[trim=0.5cm 0.8cm 0.5cm 1.5cm, clip,width=\textwidth]{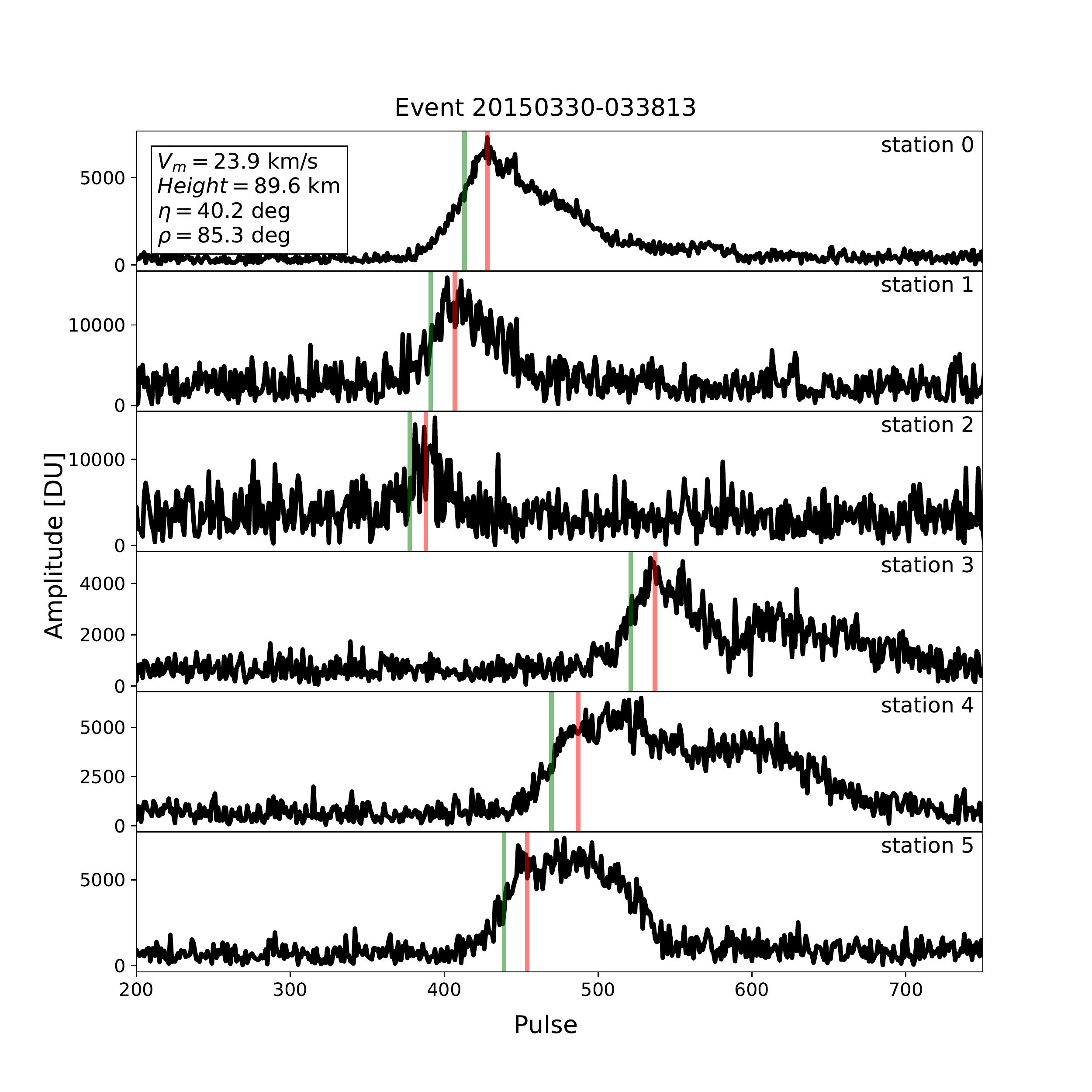}

        \end{subfigure}

        \begin{subfigure}[b]{0.7\textwidth}   
      
            \includegraphics[trim=0.5cm 0.8cm 0.5cm 1.5cm, clip,width=\textwidth]{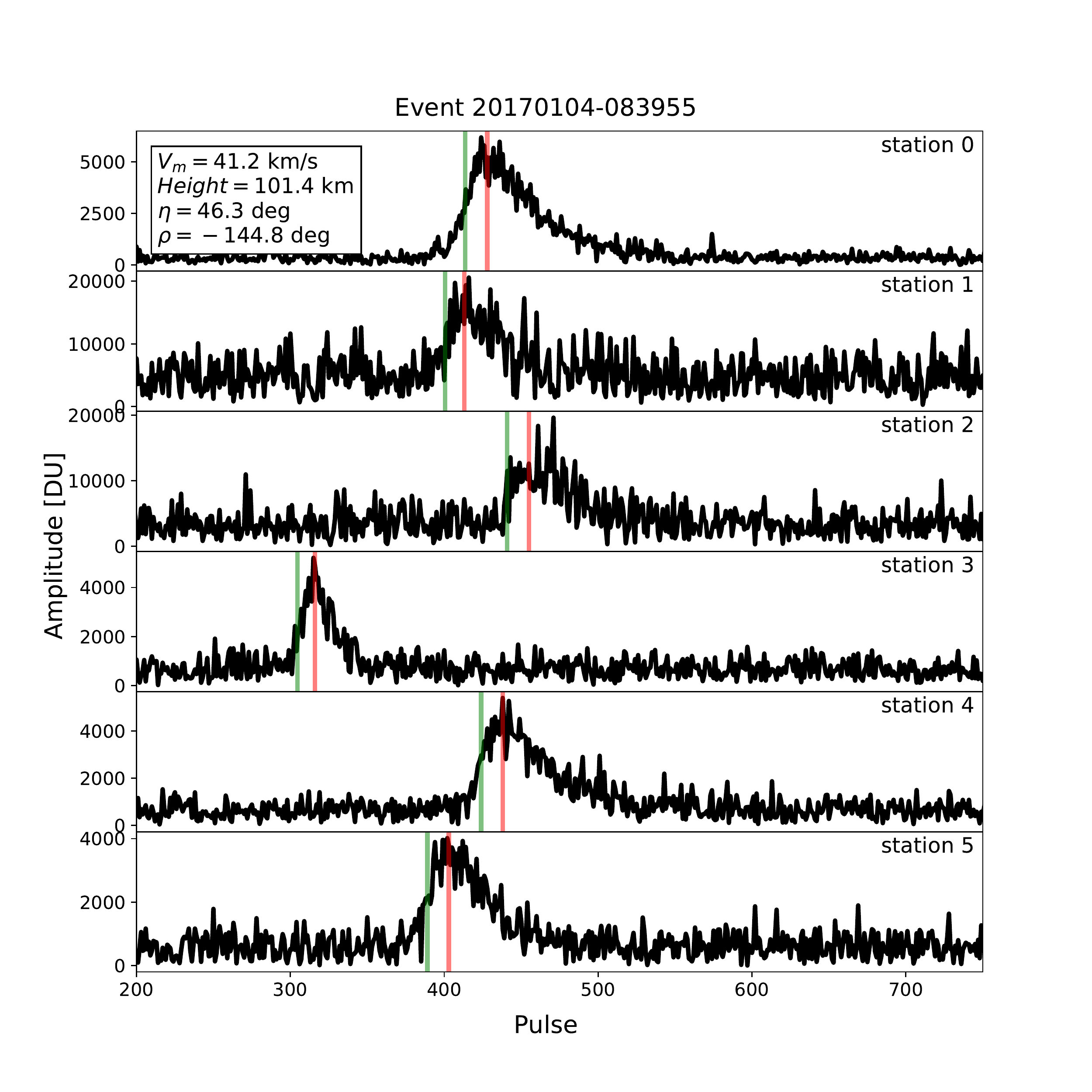}

        \end{subfigure}
        \smallskip
        \begin{minipage}[b]{0.7\textwidth}
        
        \caption[Amplitude versus time plots of candidate events]{Amplitude versus pulses (time) plots at each receiver station of the candidate interstellar events. Green lines represent inflection points, red lines represent peak points.
        \vspace{3cm}}
        \label{candipulse}
        \end{minipage}
\end{figure}

These five events are summarized in Table \ref{tab:candivelparams} and the amplitude profile for each station given in Figure \ref{candipulse}.  We emphasize that these events are hyperbolic as measured with time of flight in-atmosphere measured speeds. These time of flight speeds represent the average over the height range of specular points across all six stations. As these speeds are lower than top of the atmosphere speed due to atmospheric deceleration in reality these should be more hyperbolic than observed, as described later. We examine the uncertainties in these quantities in the next sections.

\begin{table}[H]
\advance\leftskip-1.5cm

\begin{tabular}{cccccccp{3cm}}

\hline
\textbf{Event} & \multicolumn{1}{C{1cm}}{\textbf{$\eta$ [deg]}} & \multicolumn{1}{C{1cm}}{\textbf{$\rho$ [deg]}} &  \multicolumn{1}{C{1cm}}{\textbf{$V_c$ [km/s]}} & \multicolumn{1}{C{1cm}}{\textbf{$V_m$ [km/s]}} & \multicolumn{1}{C{1cm}}{\textbf{$V_f$ [km/s]}} &
\multicolumn{1}{C{1cm}}{\textbf{$V_{\textit{t0}}$ [km/s]}} &
\multicolumn{1}{C{3cm}}{\textbf{$H_{\textit{0,1,2,3,4,5}}$ \newline [km]}} \\
\hline
2014-268-1026        & 73.37        & 174.34       & 41.16         & 40.00         & 38.53         & 37.2            & 90.5,\phantom{X}89.0,\phantom{X}91.1, 86.8,\phantom{X}92.1,\phantom{X}89.7       \\
2015-008-1D0C       & 40.21        & 85.30        & 24.31         & 23.90         & N/A           & 22.96           & 89.6,\phantom{X}88.6,\phantom{X}91.2, 94.6,\phantom{X}92.1,\phantom{X}90.9       \\
2014-004-0805       & 79.50        & -90.45       & 50.15         & 49.10         & N/A           & 39.85*          & 95.6,\phantom{X}96.6,\phantom{X}97.3, 91.2,\phantom{X}92.8,\phantom{X}94.6       \\
2017-283-2484       & 46.31        & -144.81      & 43.24         & 41.20         & N/A           & 42.24           & 101.4,\phantom{X}101.8,\phantom{X}100.9, 103.9,\phantom{X}101.2,\phantom{X}102.0 \\
2014-299-0152        & 36.55        & -161.39      & 43.18         & 42.20         & N/A           & 35.57**           & 91.9,\phantom{X}93.7,\phantom{X}90.6, 98.2,\phantom{X}90.7,\phantom{X}93.6         
\\
\hline
\end{tabular}
\caption[Trajectories of candidate interstellar meteors]{Trajectories of candidate interstellar meteors as measured by CMOR. The event name tag is year-Solar-longitude followed by a unique internal file name for a particular echo. Here $\eta$ is the local zenith angular distance of the apparent radiant, $\rho$ is the local azimuth of the apparent radiant, measured counter-clockwise as seen from above from due East, $V_c$ is the deceleration corrected velocity at the top of the atmosphere (used to compute the orbit - see \ref{decelCor}), $V_m$ is the measured time of flight speed in the atmosphere, $V_f$ is the fresnel amplitude velocity (if available), $V_{\textit{t0}}$ is the pre-t0 velocity and $H_{\textit{0,1,2,3,4,5}}$ is the height of the specular point from each receiver station, where the main transmit/receiver station is station 0. *Note: $V_{\textit{t0}}$ automatically calculated from the method described in \citet{mazur2019} for event 2014-004-0805 was 39.85 km/s far below $V_m$ of 49.10 km/s. This appears to be an edge case scenario where the automatic pre-t0 velocity calculation fails. Manual inspection of the pre-t0 velocity shows that it is 49.87 km/s, consistent with $V_m$.** Similarly, manual inspection of the pre-t0 for this echo shows a speed closer to 39 km/s.}
\label{tab:candivelparams}
\end{table}

\subsection{Monte Carlo Simulation}\label{ISMonteCarlo}
To establish the significance of the hyperbolic excess speeds as measured, we need to estimate uncertainties for each echo. To do this, we performed Monte Carlo simulations of each meteor echo by randomly varying parameters drawn from empirically derived error distributions estimated directly from CMOR data. Many of these empirical estimates for error were based on six-station CMOR detected shower echoes (see Section \ref{decelCor}). This simulation uses the observed geometry, range, interferometry, and speed of each echo and generates a synthetic echo based on the ideal transverse scattering amplitude vs. time profile produced by solving the Fresnel integrals (see \citet{Weryk2012}). 

\begin{figure}[H]
\vspace*{2mm}
\begin{center}
\includegraphics[width=9cm]{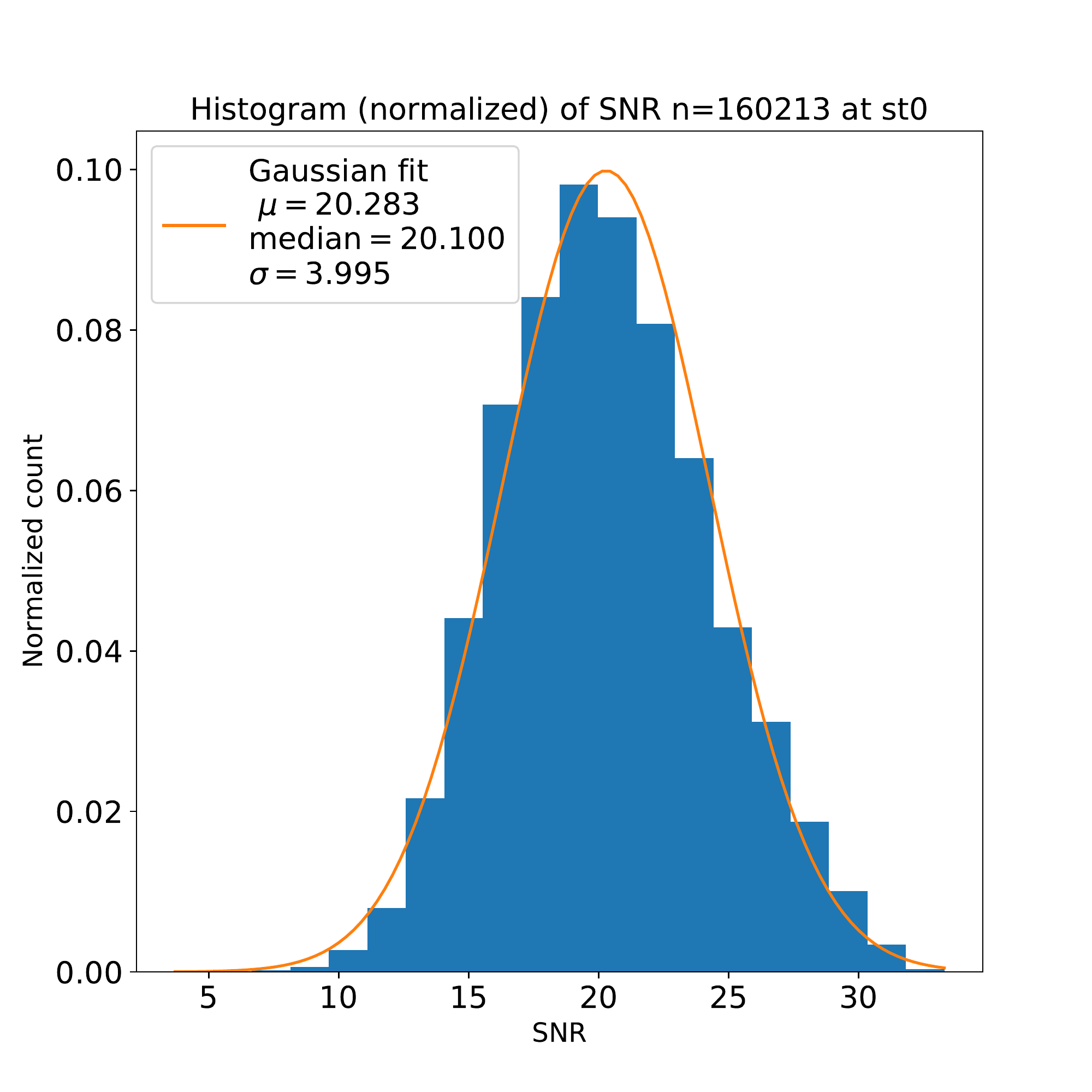}
\caption[SNR for Station 0]{Signal-to-noise ratio at receiver station 0 (main site) for all six station meteor echoes with orbital measurements. The plot is normalized to our bin sizes such that the total area under the curve is unity.}
\label{snr0}
\end{center}
\end{figure}

\begin{figure}[H]
\vspace*{2mm}
\begin{center}
\includegraphics[trim=0.5cm 0.5cm 0.5cm 1.5cm, clip,width=10cm]{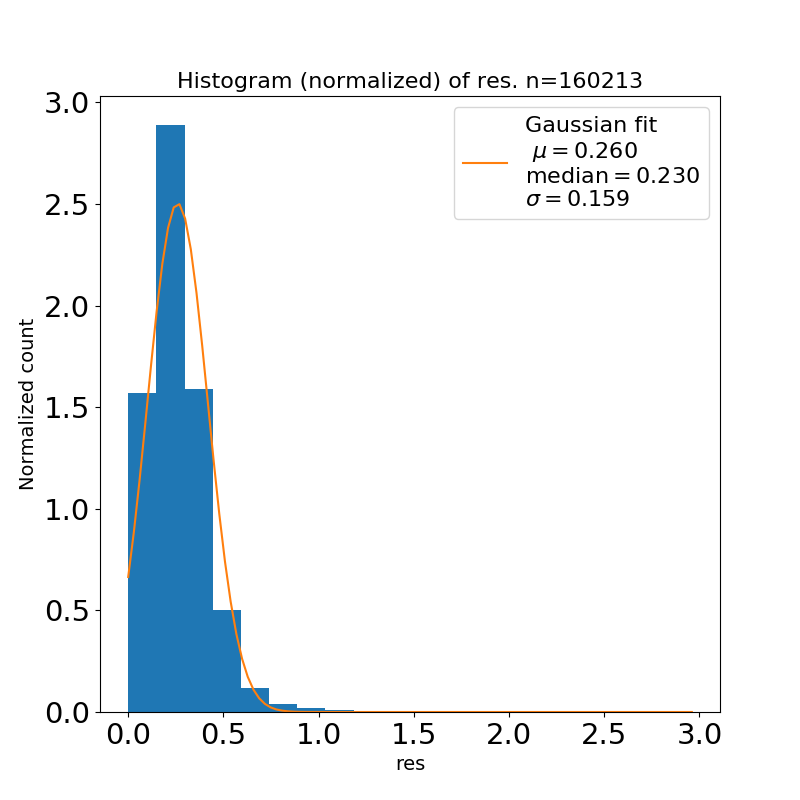}
\caption[Interferometry solution residuals for shower meteors]{Histogram showing the mean interferometry solution residuals (in degrees) for station 0 from examination of 160213 meteors detected on all 6-stations. More details are given in Section \ref{decelCor}.}
\label{res}
\end{center}
\end{figure}

The CMOR time inflection pick algorithm and interferometry algorithm is then applied to estimate the model speed. Uncertainties in range were simulated by varying the observed range $\pm$ 1.5 km in a uniform distribution, representing a precision of one full range gate (3 km) \citep{Brown2008}. The mean signal-to-noise ratio (SNR) at receiver station 0 (main site) for all six station echoes was found to be 20.3 (Figure \ref{snr0}). Mean SNRs for stations 1,2,3,4 and 5 were 11.8, 12.6, 14.5, 16.4 and 17.9, respectively. Standard deviation of errors in the echo direction (interferometry) were found to be on average 0.159 degrees (representing a positional uncertainty of 300m at 100 km range) from examination of 160213  meteors (Figure \ref{res}) which were selected based on the methodology described in Section \ref{decelCor}. For the simulations, a more conservative value for interferometry uncertainty of 1 degree was used, consistent with differences between optical and interferometric specular points measures for CMOR reported in past studies \citep{Weryk2012a}. 

\begin{figure}[H]

        \begin{subfigure}{0.55\textwidth}

            \includegraphics[trim=0.5cm 0.8cm 0.5cm 1.5cm, clip,width=\textwidth]{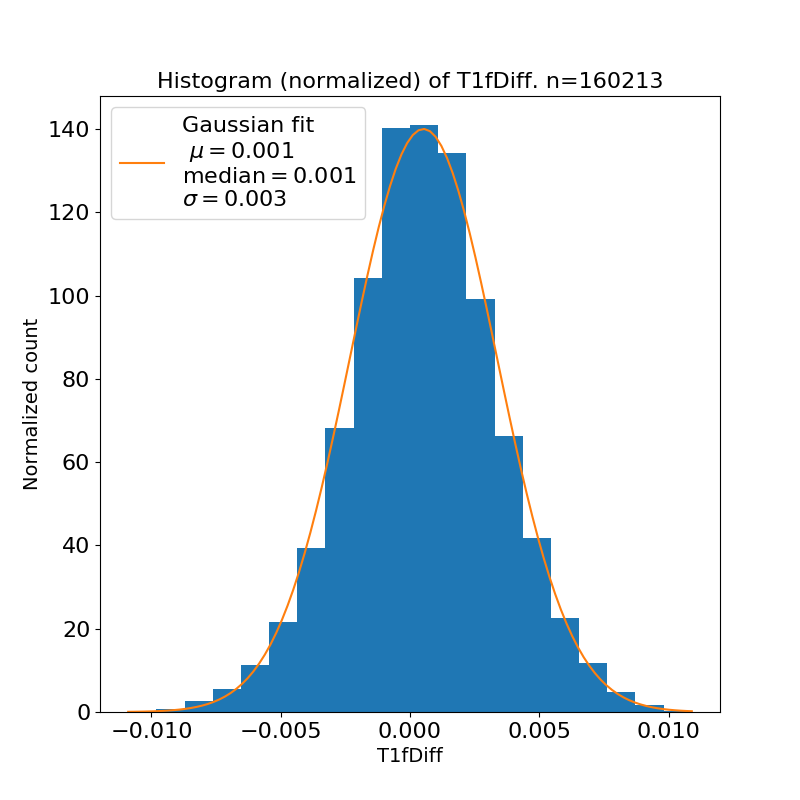}

        \end{subfigure}
        \medskip
        \begin{subfigure}{0.55\textwidth}  

            \includegraphics[trim=0.5cm 0.8cm 0.5cm 1.5cm, clip,width=\textwidth]{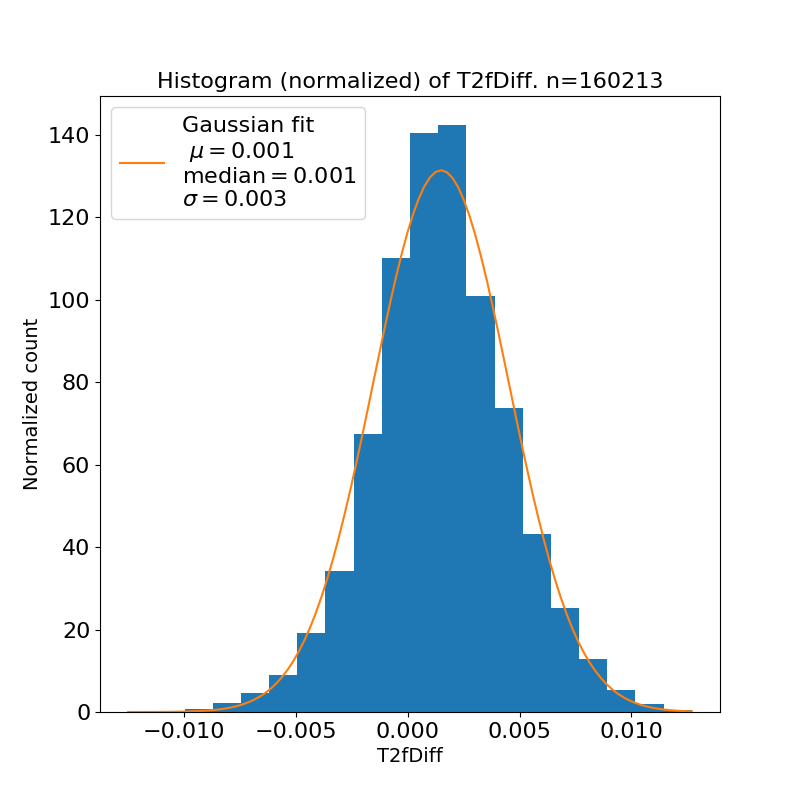}

        \end{subfigure}

        \begin{subfigure}{0.55\textwidth}   

            \includegraphics[trim=0.5cm 0.8cm 0.5cm 1.5cm, clip,width=\textwidth]{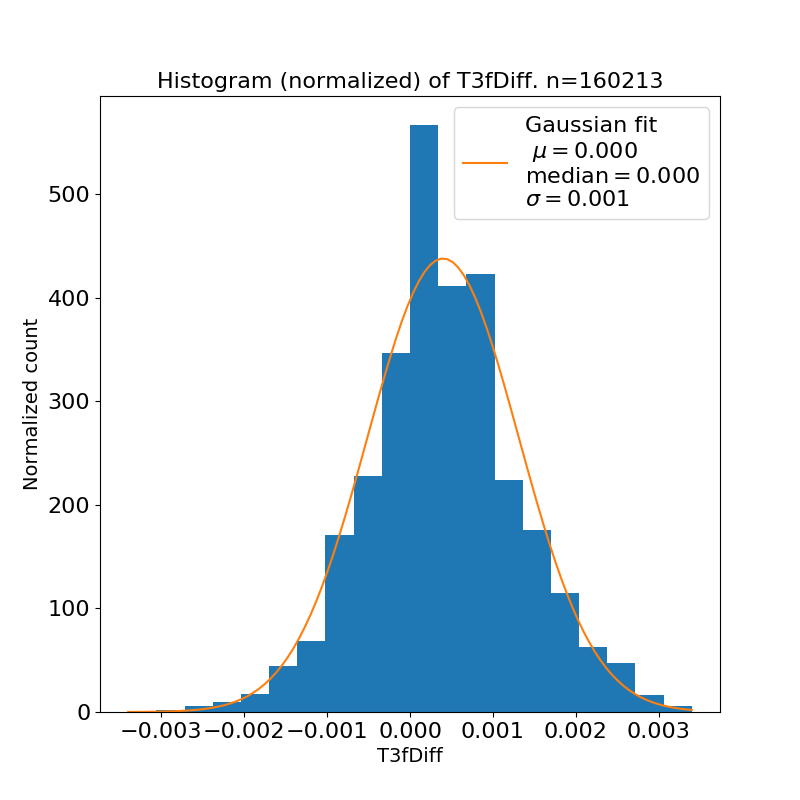}

        \end{subfigure}
           \medskip
        \begin{subfigure}{0.55\textwidth}   

            \includegraphics[trim=0.5cm 0.8cm 0.5cm 1.5cm, clip,width=\textwidth]{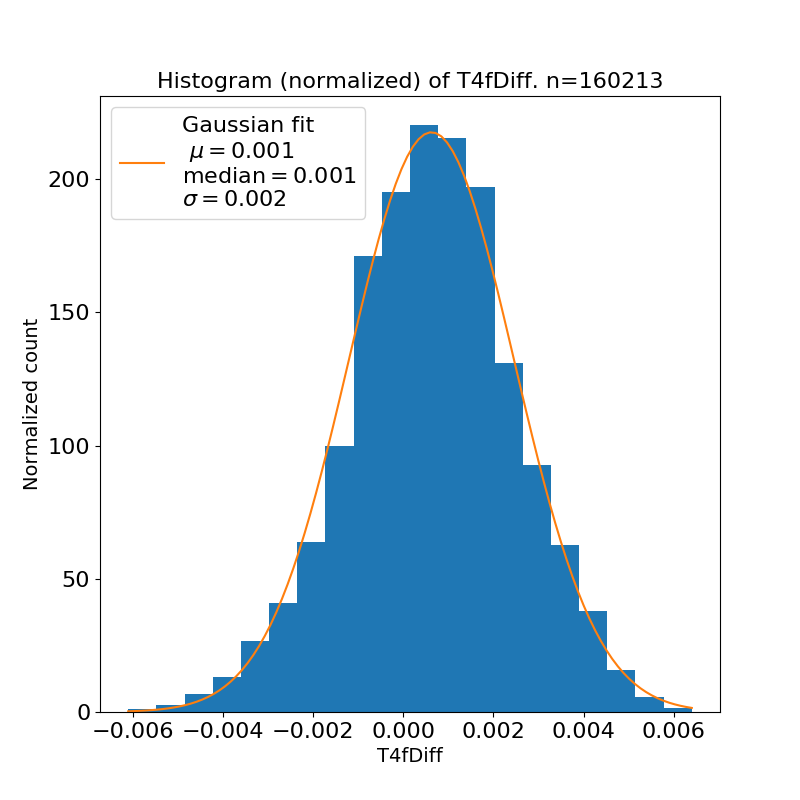}

        \end{subfigure}

        \begin{subfigure}[b]{0.55\textwidth}   
      
            \includegraphics[trim=0.5cm 0.8cm 0.5cm 1.5cm, clip,width=\textwidth]{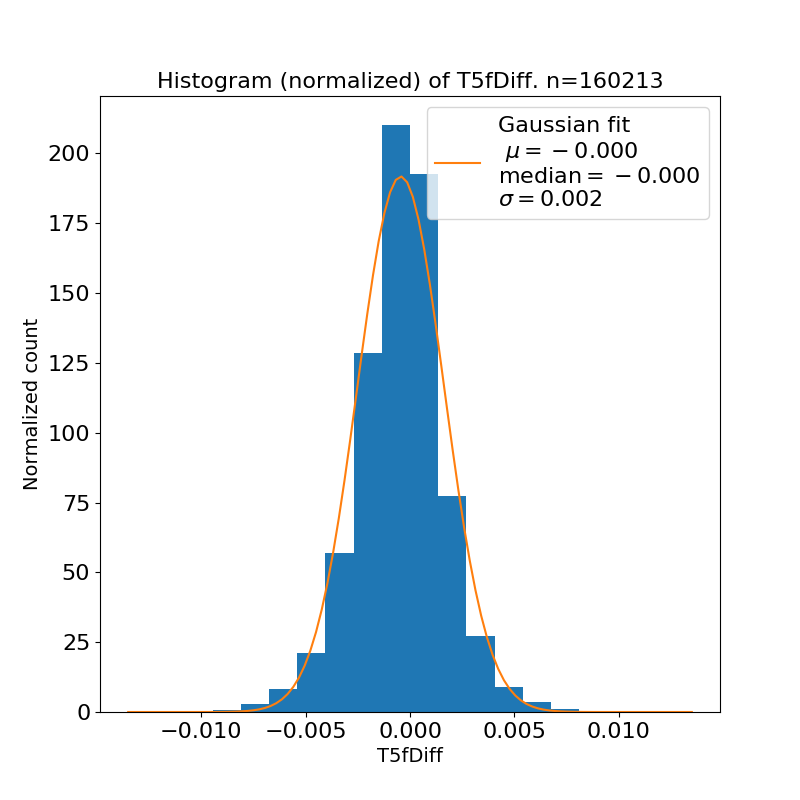}

        \end{subfigure}
        \medskip
        \begin{minipage}[b]{0.55\textwidth}
        
        \caption[Time pick differences  for stations T1-T5]{Histograms of mean time pick differences (in seconds) for stations 1-5 between observed and computed solutions from examination of 160213 meteors meteors recorded on all 6-stations. This provides an empirical estimate of the time pick uncertainty which is then used in the echo Monte Carlo simulations. Note that the timing is relative to T0 (the main site) and that 1 pulse represents a timing difference of 1.8$\times 10^{-3}$ sec.\vspace{1cm}}
        \label{Tdiff}
        \end{minipage}
\end{figure}

Finally, uncertainty in time picks for inflection points of echoes is varied over a random (assumed to be) gaussian distribution for time picks which were found to have a standard deviation of approximately 1 pulse or 1/532 second (Figure \ref{Tdiff}) based on examination of all six station meteors (Table \ref{tab:mcparams}). Here the time pick uncertainty is estimated by noting the difference between the observed time pick and the overall trajectory best-fit time pick per echo and per station - it represents the observed empirical spread in time picks across all six station trajectories. 

\begin{table}[h]
\centering

\begin{tabular}{lll}
\hline
\textbf{Error Paramater} & \textbf{Error Value}             & \textbf{Distribution} \\ \hline
Range                    & 1.5 km {[}$\pm${]}                 & random, uniform       \\
Interferometry           & 1 degree {[}SD{]}                & random, gaussian      \\
Time picks               & 1 pulse (1/532 seconds) {[}SD{]} & random, gaussian     
\\
\hline
\end{tabular}
\caption[Monte Carlo error parameters]{Error parameters used as input to Monte Carlo simulations for interstellar candidate meteors detected by CMOR.}
\label{tab:mcparams}
\end{table}

This Monte Carlo procedure can be applied to either the directly measured velocities $V_m$ or to the deceleration corrected velocity $V_c$ providing distributions of significance in the individual hyperbolic measurements. To estimate the deceleration correction which should be applied to the observed speed to recover the initial pre-atmosphere speed, we would need to know in detail the ablation behaviour of a particular echo, information which is not available. To approximate this correction, we instead use shower meteor echoes identified in the CMOR data set and bootstrap our observed speed to the "reference" shower speed. This provides a correction as a function of speed and height. The resulting average correction is applicable to CMOR-sized meteoroids and cannot be easily applied to other systems. 

A similar approach using much less data and less secure meteor shower speeds was attempted by \citet{Brown2005}. We also explored the deceleration dependence on entry angle and found it to be much less significant than the height and speed dependence and hence omit entry angle from the correction. We develop this deceleration correction in the next section. 

\section{Deceleration Correction} \label{decelCor}

A more accurate deceleration correction for meteor echoes detected by CMOR allows for improved pre-atmosphere velocity determination leading to more accurate orbit calculations. In particular, this correction  provides our best statistical estimate to the true top of atmosphere velocity, recognizing that all our candidate events have measured in-atmosphere speeds which already produce hyperbolic orbits.

Shower meteors, being on common orbits (by definition) should have similar entry velocities. There is, however, no accepted quantitative metric for shower association, but we will attempt to define some criteria to use for our study in Sections \ref{sec:radsel} and \ref{sec:radradius}. However, there may be some spread in velocities based on time of occurrence in the shower (ie. velocity changes with solar longitude) or with meteoroid mass. Several surveys have previously estimated initial shower velocities. In the past these have applied a deceleration model \citep{Jacchia1961} to estimate the pre-atmosphere speed. Ideally, measuring shower meteoroid velocities at very high heights, before deceleration becomes significant, is better. A dataset of shower meteors which uses the latter approach for the first time was recently collected by the Middle Atmosphere Alomar Radar System from head echo shower measurements \citep{Schult2018}, which are detected at very high (typically above 100 km) height and have very precisely measured doppler speeds. We adopt MAARSY shower speeds when available as the most probable "reference" pre-atmosphere speed. 

Since the total deceleration depends on the amount of atmosphere encountered for a fixed mass meteoroid (eg. \citet{Ceplecha1998}), in general, we expect the measured speed of a particular shower echo to depend most significantly on height, though entry angle and physical structure may also play a secondary role. However, the height at which initial deceleration occurs for similar masses will be velocity dependent as the starting height for ablation is speed dependent \citep{Koten2004, Hawkes1975} as is the intercepted momentum per unit atmospheric mass. 

We examined the average measured time-of-flight velocity \citep{Jones2005} from CMOR shower echoes within specific height bins to find the average shower velocity for typical CMOR echoes as a function of height.  The basic approach follows the procedure outlined in \citet{Browna}. However, this study uses a much larger sample and more recently measured values for initial ("reference") shower velocities.  

We used the literature value of reference speed to estimate the height at which shower echoes show no noticeable deceleration relative to the top of atmosphere, which we term $H_0$. For CMOR, this varies from roughly 95 km at slow speeds to 105 km for higher speed meteoroids.  The associated slopes (change in speed per km below the $H_0$ height) and $H_0$ are determined for several showers spanning a spread of velocities. We found that linear fits reproduced the observed speed vs. height behaviour for most showers. Thus, linear fits for slope and $H_0$ were determined for each shower. These fits as a function of speed are then combined into a single correction term as a function of specular height and time-of-flight speed which can then be applied to any CMOR echo to give a best-estimate for its pre-atmosphere speed. 

The resulting fit confidence bounds in slope and $H_0$ as a function of velocity place limits on the total correction as well as a best estimate for the nominal correction. Data collected from CMOR between January 2012 and December 2018 was used to estimate this velocity correction.  

We use as ground-truth top of atmosphere reference velocities for known showers reported by MAARSY \citep{Schult2018} where possible and otherwise  literature sources (Table \ref{tab:showervel}) summarized on the IAU Meteor Data Center established shower list \footnote{\url{https://www.ta3.sk/IAUC22DB/MDC2007/}}. The literature sources vary in their estimates of velocity. Therefore, velocities which are closest to the observed CMOR values where noticeable deceleration begins are chosen. For meteors from a particular shower, we take CMOR measured velocities that fall between $\pm$ 20\%  of the expected top of atmosphere velocity (based on the literature values) for a meteor shower for expected velocities of \textless 40 km/s and $\pm$ 30\% for velocities \textgreater 40 km/s. This includes extremely decelerated meteors ($\approx$5 km/s) at heights as low as $\approx$80 km. As lower velocities can be measured with greater precision with CMOR (due to the larger time offsets between stations), we use a smaller spread in velocities (20\%) for these velocities compared to 30\% used for faster velocities which are measured less accurately. The expected $V_{\textit{inf}}$ (expected velocity at top of atmosphere) were taken from literature values (Table \ref{tab:showervel}) using the expected geocentric velocity ($V_{g}$) of a shower and adding the acceleration due to gravity of the Earth taken from infinity (Equation \ref{Vinf}). $V_{\textit{esc}}$ is taken to be the escape velocity of the Earth at 100 km above the surface, 11.1 km/s. 

\begin{equation}
\label{Vinf}
V_{\textit{inf}} = \sqrt{{V_g}^2 + {V_{\textit{esc}}}^2}
\end{equation}

Only meteor echoes with individually estimated velocity ($V_{m}$) errors of \textless 10\% of measured velocity were included in the selection of shower meteors. Errors in $V_{m}$ were computed following the approach outlined in \citet{Weryk2012}, which we briefly describe. 

In this approach we extract the observed speed, radiant, azimuth, elevation and each station's observed echo signal-to-noise ratio. This input information is then used to create a synthetic amplitude - time profile based on the standard Fresnel integrals for meteor backscattering \citep{Ceplecha1998} with Gaussian noise added to each amplitude profile per station according to the observed SNR. From these noise-generated synthetic profiles, the same algorithms used for time-of-flight measurements which estimate interferometry and inflection points per station are applied and the resulting speed found per synthetic echo. The error is taken to be the standard deviation in the speed about the mean from all Monte Carlo runs (10000 in our case).

Additionally, only shower echoes with pre-t0 velocity errors \citep{mazur2019} of \textless 10\% were included.

\begin{table}[H]
\centering
\advance\leftskip-2cm
\advance\rightskip-2cm

\begin{tabular}{llccl}
\hline
\multicolumn{1}{c}{\textbf{Shower}} &
\multicolumn{1}{C{1.25cm}}{\textbf{IAU code}} &\multicolumn{1}{C{1.25cm}}{\textbf{$V_{g}$ {[}km/s{]}}} & \multicolumn{1}{C{1.25cm}}{\textbf{$V_{\textit{inf}}$ {[}km/s{]}}} & \multicolumn{1}{c}{\textbf{Source}} \\ \hline
Daytime Arietids & ARI                     & 38.6                     & 40.2                     & \citet{Schult2018}                 \\
Daytime Sextantids & DSX                   & 31.2                     & 33.1                     & \citet{Galligan2002}         \\
Draconids & DRA                           & 20.7                     & 23.5                     & \citet{Jenniskens2016}             \\
Geminids & GEM                            & 33.1                     & 34.9                     & \citet{Schult2018}                 \\
January Leonids & JLE                      & 51.4                     & 52.6                     & \citet{Jenniskens2016}             \\
Leonids & LEO                             & 69.3                     & 70.2                     & \citet{Schult2018}                \\
November Omega Orionids & NOO               & 42.7                     & 44.1                     & \citet{Schult2018}                 \\
Orionids & ORI                            & 66.3                     & 67.2                     & \citet{Jenniskens2016}              \\
Perseids & PER                            & 57.9                     & 59.0                       & \citet{Molau2007}                          \\
Quadrantids & QUA                         & 40.0                     & 41.5                     & \citet{Schult2018}                 \\
Southern Delta Aquarids & SDA               & 39.4                     & 41.0                       & \citet{Molau2007}                         \\
Xi Coronae Borealids & XCB                  & 44.9                     & 46.3                     & \citet{Schult2018}    
\\
\hline            
\end{tabular}
\caption[Sources for $V_{g}$ and $V_{\textit{inf}}$ velocities for shower meteors]{Literature sources used for reference $V_{g}$ and $V_{\textit{inf}}$ velocities for shower meteors.}
\label{tab:showervel}
\end{table}

\begin{table}[H]
\advance\leftskip-0.5cm
\advance\rightskip-0.5cm

\begin{tabular}{lcccccc}
\hline
\multicolumn{1}{c}{\textbf{Shower}} & \multicolumn{1}{C{1cm}}{\textbf{$\lambda -\lambda_\odot$ {[}deg{]}}} & \multicolumn{1}{C{1cm}}{\textbf{$\beta$ {[}deg{]}}} & \multicolumn{1}{C{1cm}}{\textbf{Peak [$\lambda_\odot$]}} & \multicolumn{1}{C{1.5cm}}{\textbf{Peak [date]}} & \multicolumn{1}{C{1cm}}{\textbf{Spread {[}days{]}}} & \multicolumn{1}{C{1cm}}{\textbf{Radius {[}deg{]}}} \\ \hline
Daytime Arietids                     & 331.0                                        & 8.0                                           & 77                                                         & 8-Jun                                              & 5                                              & 4.4                                                  \\
Daytime Sextantids                   & 329.6                                      & -11.5                                       & 189                                                        & 2-Oct                                              & 4                                              & 3.0                                                    \\
Draconids                           & 51.7                                     & 77.6                                       & 195                                                        & 8-Oct                                              & 1                                              & 3.8                                                  \\
Geminids                            & 208.0                                      & 10.5                                        & 261                                                        & 13-Dec                                             & 10                                             & 4.2                                                  \\
January Leonids                      & 219.0                                        & 10.2                                        & 282                                                        & 3-Jan                                              & 3                                              & 3.5                                                  \\
Leonids                             & 272.0                                      & 10.0                                        & 237                                                        & 20-Nov                                             & 5                                              & 5.0                                                    \\
November Omega Orionids               & 203.5                                      & -7.8                                        & 247                                                        & 29-Nov                                             & 3                                              & 2.8                                                  \\
Orionids                            & 247.0                                      & -7.8                                        & 208                                                        & 22-Oct                                             & 5                                              & 3.5                                                  \\
Perseids                            & 282.0                                      & 38.5                                        & 140                                                        & 13-Aug                                             & 5                                              & 3.0                                                    \\
Quadrantids                         & 277.0                                      & 63.2                                        & 283                                                        & 4-Jan                                              & 2                                              & 5.5                                                  \\
Southern Delta Aquarids               & 210.2                                      & -7.3                                        & 124                                                        & 27-Jul                                             & 3                                              & 3.5                                                  \\
Xi Coronae Borealids                  & 301.5                                      & 51.5                                        & 296                                                        & 16-Jan                                             & 2                                              & 5.0         
\\
\hline                                          
\end{tabular}
\caption[Shower parameters from wavelet analysis of CMOR data]{Shower parameters used to associate individual echoes with a particular shower. This is based on the 3D-wavelet analysis methodology applied to CMOR velocities \citep{Brown2010} using all radiants recorded between 2002-2016. Here the spread refers to the number of days around the maximum the shower was detectable and/or showed radiant motion in sun-centred coordinates was less than one degree. Radius refers to the size of the radiant area about the point of maximum where radiant density is above the background. The ecliptic longitude and latitude of the radiant is given by $\lambda$ and $\beta$ respectively,  while the solar longitude is $\lambda_\odot$.}
\label{tab:showerparam}
\end{table}

\subsection{Radiant Selection} \label{sec:radsel}

The radiant of a particular shower in sun-centred ecliptic coordinates were taken from the $\lambda -\lambda_0$, $\beta$ of the shower during the peak solar longitude bin, as measured by CMOR using the 3D wavelet procedure described in \citet{Brown2010} as shown in Table \ref{tab:showerparam}. Here all CMOR radiants from 2002 - 2016 were combined to better estimate the shower radiant and its drift. $\lambda -\lambda_0$ and $\beta$ were used to select shower echoes as radiant drift is minimal in sun-centred coordinates for any given shower. 

The solar longitude ($\lambda_\odot$) of peak activity is defined as the solar longitude (in the J2000.0 equinox) with highest wavelet coefficient excursion over the annual background at the same sun-centred ecliptic coordinates averaged over the entire year (excluding the shower activity interval) as described in \citet{Brown2010}. The range of solar longitudes used to identify individual CMOR shower echoes with a specific shower is based on the time of peak activity and the wavelet determined observed radiant ($\lambda -\lambda_0$ and $\beta$) at the peak. From this starting point, we included only the days before and after the peak where the sun-centred radiant as measured by the wavelet procedure described in \citet{Brown2010} remains within 1 degree of $\lambda -\lambda_0$ and $\beta$ values at the peak, a value much smaller than our radiant selection radius.

\subsection{Radiant Selection Radius} \label{sec:radradius}

The radiant selection radius is the maximum angular separation in the sky we adopted between the CMOR wavelet measured radiant at peak activity and the radiant of an individual CMOR echo used to associate a meteor radiant as being part of the shower. To establish this radius for each shower we first estimated the background density of sporadic radiants at the same sun-centred ecliptic coordinates. Starting from the nominal sun-centred shower radiant we expect the radiant density to fall as the radius is increased; once the density reaches the background, we declare this the effective shower selection radius. This follows a similar procedure described in \citet{Ye2013}. 

This involves first using an initial radius of 8 degrees from the shower radiant and counting all meteor echo radiants which fit the above criteria. A background count of sporadic meteors with the same velocity restrictions is also taken at the same sun-centred location starting with an 8 degree radiant radius but separated in time by $\pm$ (2 $\times$ day spread + 5 degrees of solar longitude) beyond the shower activity interval (defined from the wavelet duration of the shower) plus the day spread. This is done in order to ensure that equal time windows (and hence collecting area - time products) are used for both the intervals where shower meteors are counted and background meteors counted. If there is a difference between the total number of days used for shower meteor counts and the total number of days used in background meteor counts (i.e. in some cases there were no days with recorded data due to radar downtime), then the count of background meteors is normalized to the number of collection days used to identify the shower meteor echoes.   

\begin{figure}[H]
\vspace*{2mm}
\begin{center}
\includegraphics[width=10cm]{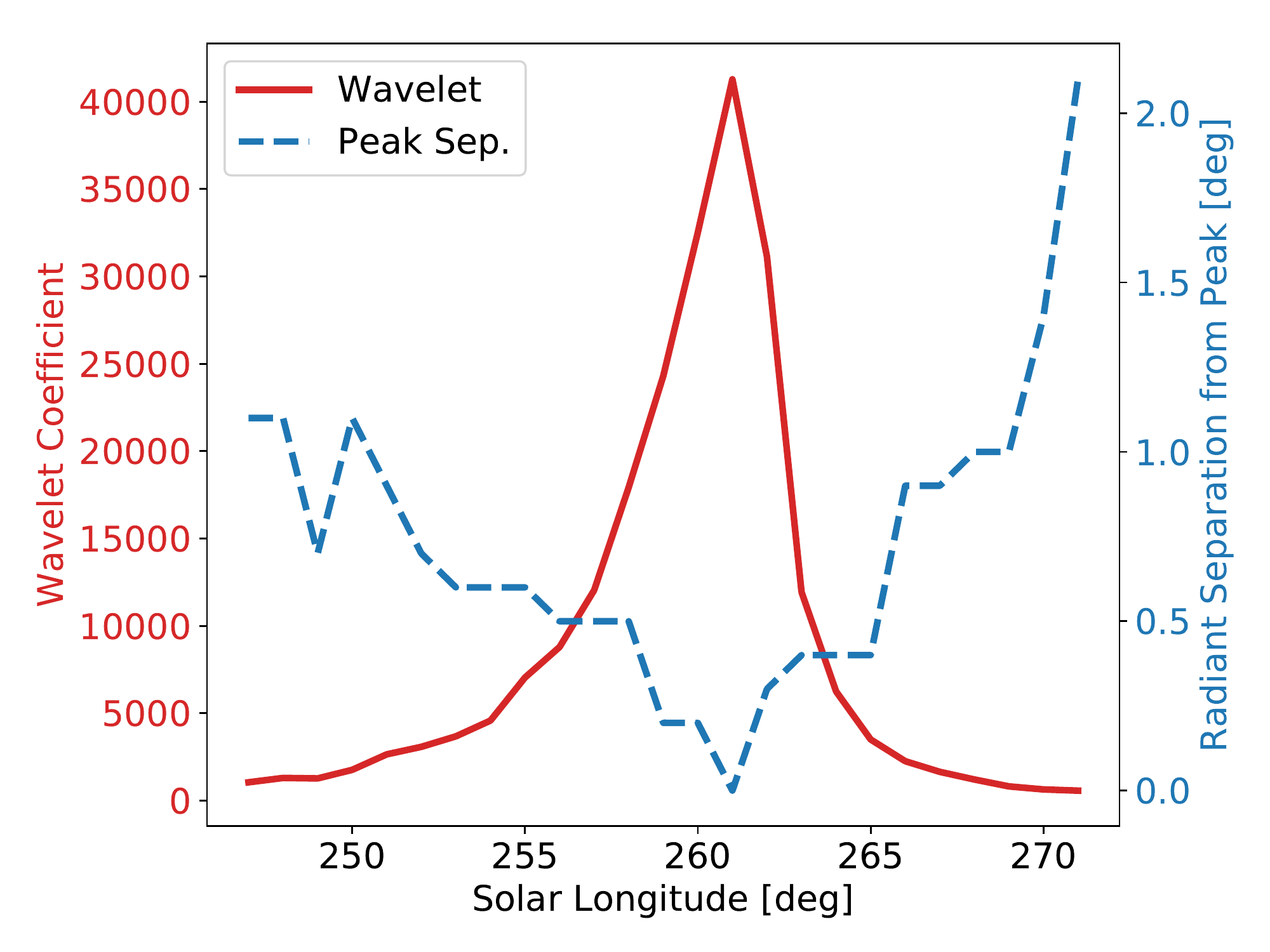}
\centering
\caption[Geminids shower activity]{Geminid meteor shower activity from stacked wavelet radiant measurements collected between 2002 and 2016. The total day spread for an angular separation of wavelet-determined observed radiant ($\lambda -\lambda_0$ and $\beta$) from peak activity $\approx$  \textless 1 $\deg$ is from $\lambda_\odot$ 251 to 269, or 19 days total. Therefore 19/2  $\approx$ 10 days (before and after the peak activity) of spread is what is used to identify shower echoes. Here the radiant separation relates the radiant location at the peak based on the wavelet analysis to the radiant drift on days on either side of the peak. The wavelet coefficient is a proxy for relative shower activity.}
\label{Gemsactivity}
\end{center}
\end{figure}

As an example, for the 2002-2016 Geminid shower (Figure \ref{Gemsactivity}), the total shower duration is 2 $\times$ 10 days, and 20 days of background meteors are taken outside this interval starting 25 days before and 25 days after the shower activity period. But there are 3 days of radar dropouts, so the background meteor counts would be multiplied by 20/17 to correct for this. 

\begin{figure}[H]
\vspace*{2mm}
\begin{center}
\includegraphics[trim = 1cm 1cm 1cm 2.5cm, clip, width=12cm]{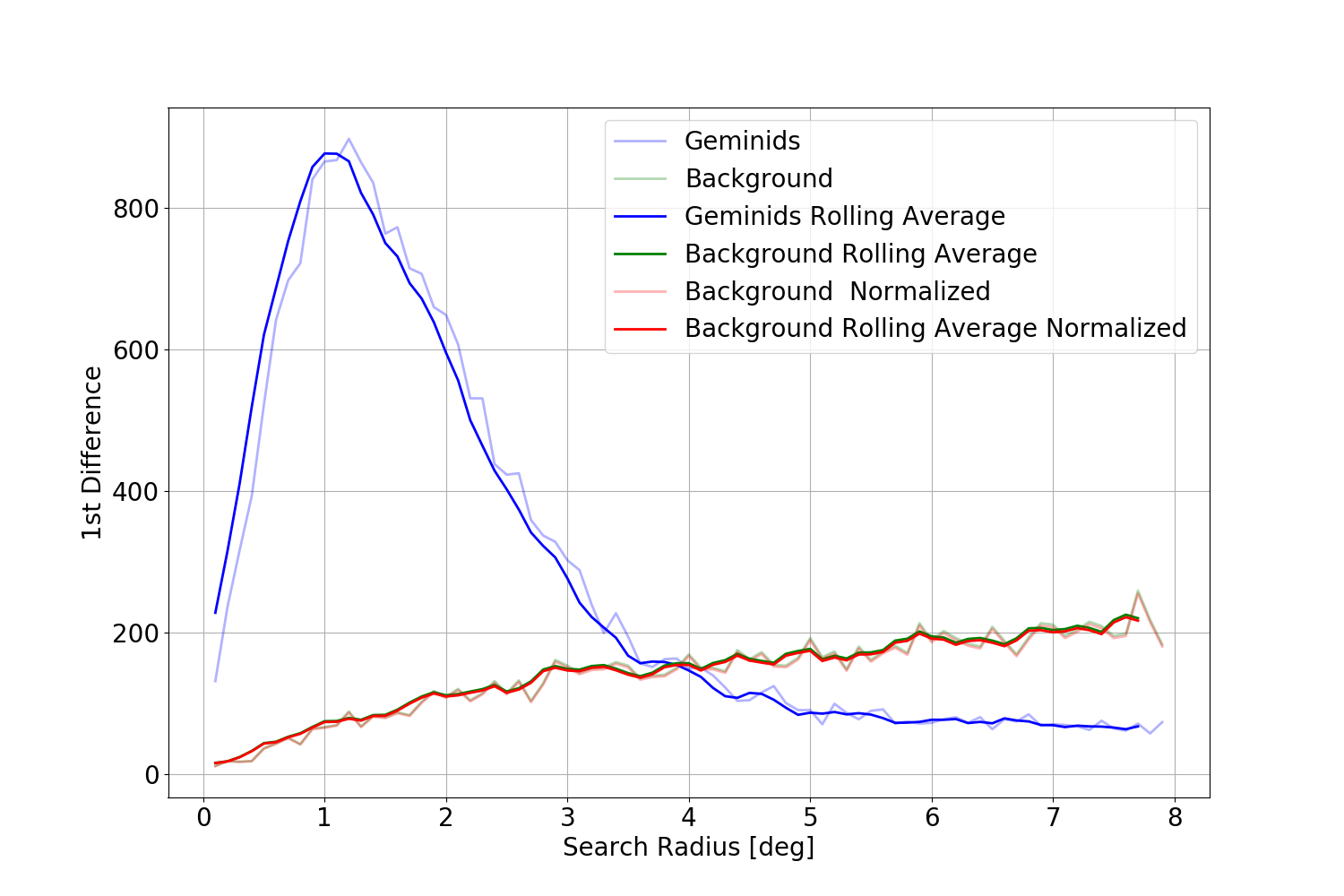}
\caption[1st-difference of meteor count for Geminids] {A 1st-difference of meteor radiant counts as a function of search radius about the Geminid radiant during its activity period and the same for sporadic meteor radiants at the same sun-centred location 25 days after the peak.}
\label{1stdiffGems}
\end{center}
\end{figure}

\begin{figure}[H]
\vspace*{2mm}
\begin{center}
\includegraphics[trim = 2cm 1cm 1cm 1cm, clip,width=15cm]{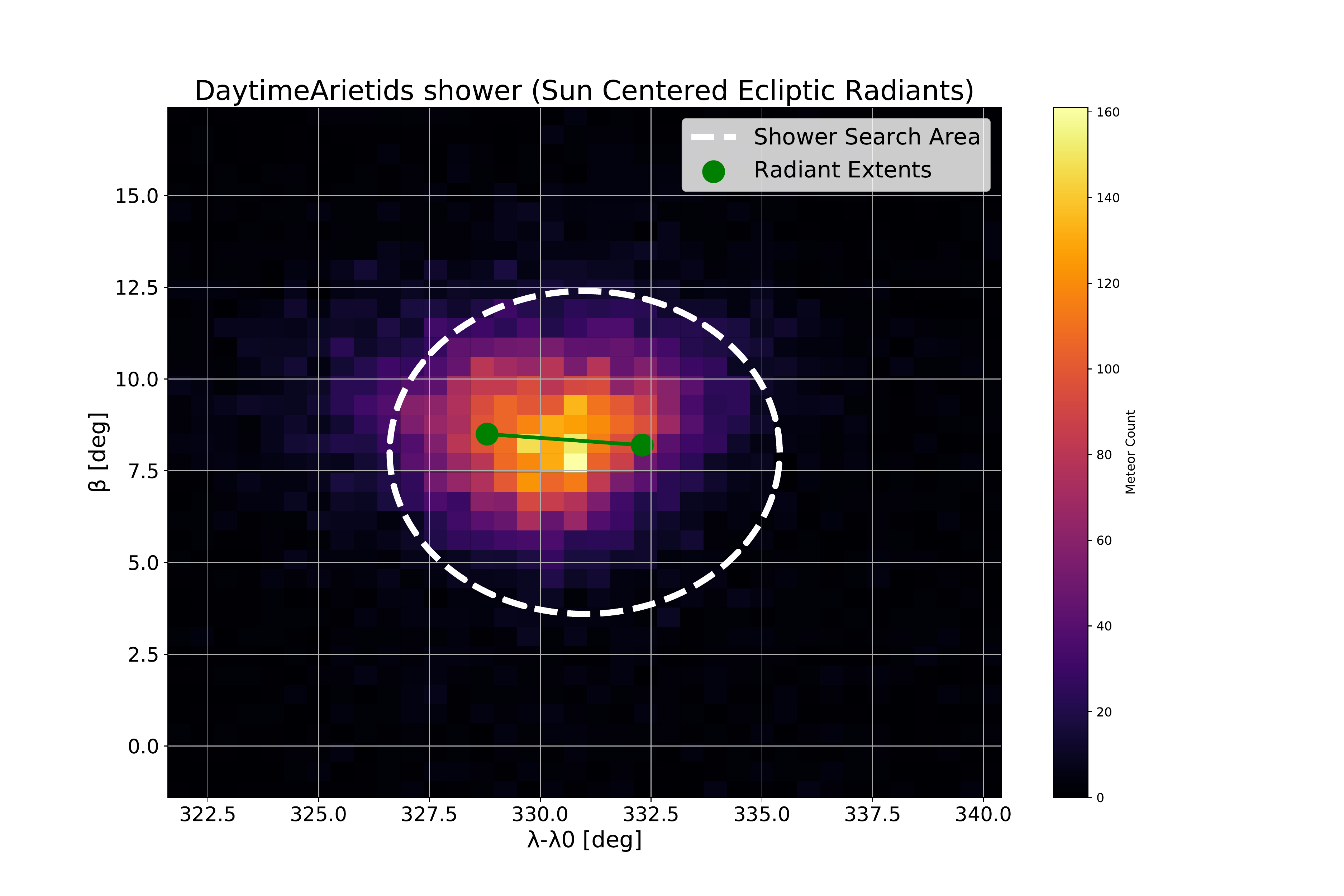}
\caption[Density plot of count of Daytime Arietids meteor shower]{Density plot of radiants for the Daytime Arietids with the shower search area (white dashed circle) overlaid for comparison. Green dots and line represent the extent of the radiant drift between the start and end of the search period based on the stacked wavelet radiant location as described in the text.}
\label{meteordens}
\end{center}
\end{figure}

Starting at this 8 degree initial value, the radiant acceptance radius is decreased in 0.1 degree steps for both the shower meteor radiant search area and the background sporadic meteor radiant search area, within their respective time windows. A first difference in count numbers between each successive decrease in search radius is calculated and a running average of 0.3 degrees of the count is then computed (Figure \ref{1stdiffGems}). The radius at which the rolling average of the first difference in the count for the shower radiants crosses the first difference in the count of the background sporadic meteor radiants is taken to be the radius over which meteor radiants can be reliably associated with the shower without significant background contamination. This becomes the search radius used for that particular shower (Figure \ref{meteordens}) in our deceleration analysis. Only meteor echoes with individual radiant uncertainties, as determined from the Monte Carlo procedure described in \citet{Weryk2012}, of less than 1 degree are included. 

\begin{figure}[H]
\centering
        \advance\leftskip-3cm
        \advance\rightskip-3cm
\begin{subfigure}[b]{0.7\textwidth}
   \includegraphics[trim=1cm 1cm 2cm 2cm,clip,width=\textwidth]{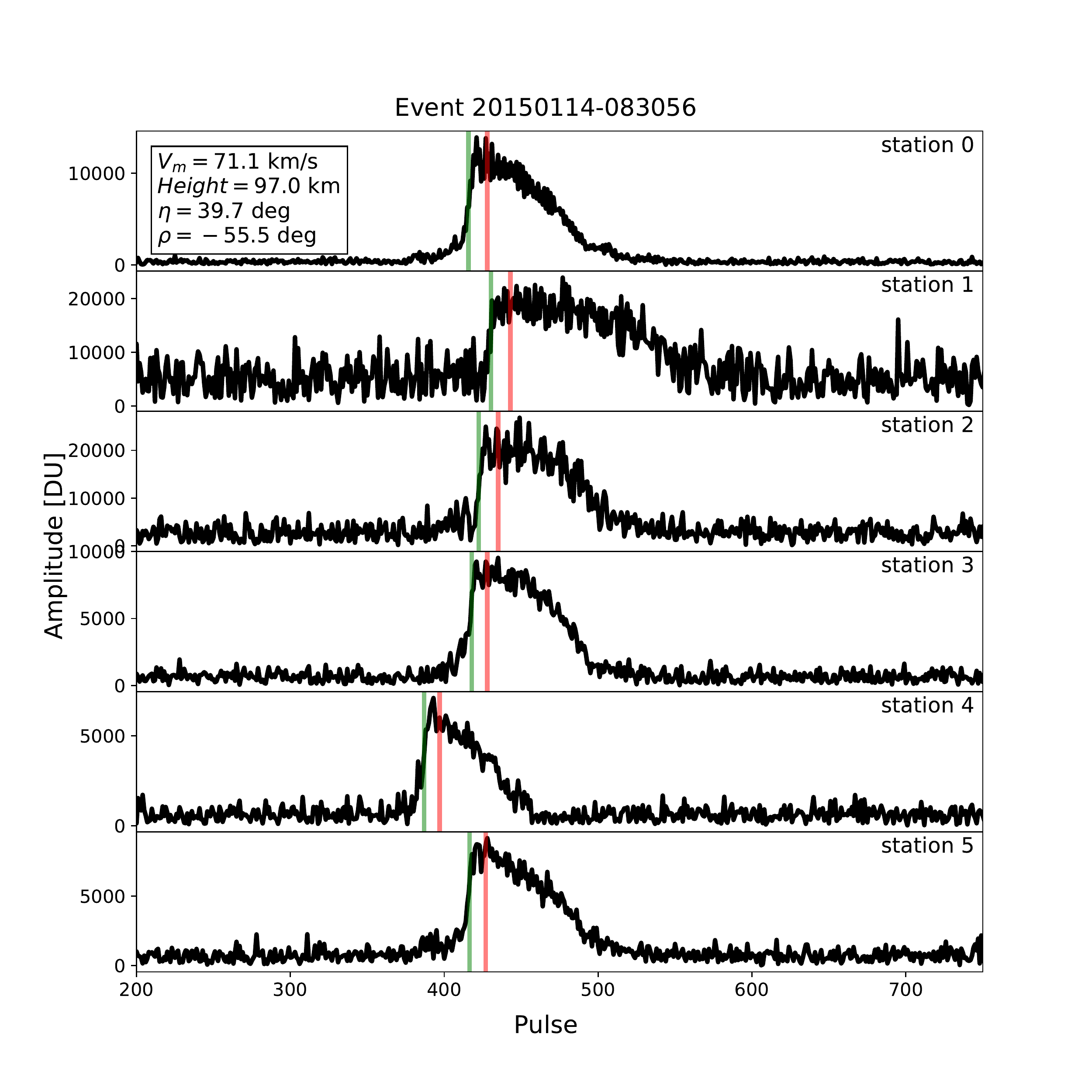}

\end{subfigure}
\begin{subfigure}[b]{0.7\textwidth}
   \includegraphics[trim=1cm 1cm 2cm 2cm,clip,width=\textwidth]{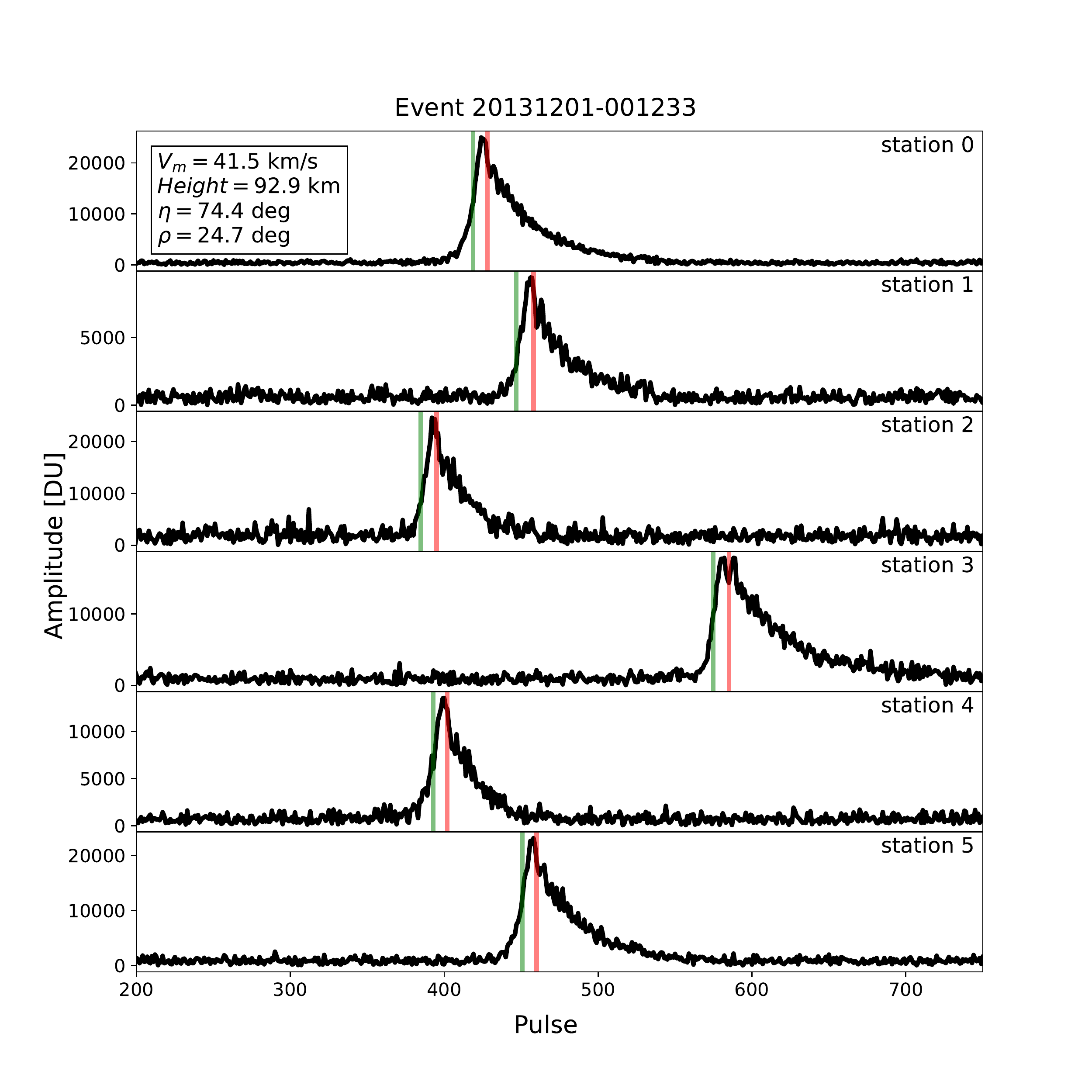}

\end{subfigure}

\caption[6-station amplitude vs time plots of good and bad timing spread]{An echo detected at 6-stations showing amplitude versus time for each site. Event 20150114-083056 shows bad timing spread where inflection points appear in near vertical line and all coincide in time. The error in TOF speed for this event is 135.93 km/s. Event 20131201-001233 shows good timing spread where inflection points are spread out. The error in TOF speed for this event is 0.12 km/s. Green lines represent inflection points, red lines represent peak points. The time of flight speed and specular height (as measured by interferometry from station 0) are given in the upper left of each sub-caption. The local radiant zenith angle ($\eta$) and azimuth ($\rho$) are also shown.}
\label{LODex}
\end{figure}

Large uncertainties in radiant solutions are often caused by meteor echoes which have specular points from multiple stations very near each other along the trail. Such echoes have very small difference in timing across different receiver stations. This decreases the accuracy of the time of flight velocity measurement. When the echo amplitudes as a function of time are plotted per station, the inflection points for these echoes appear to occur in or near a straight vertical line as shown in Figure \ref{LODex}. We wish to remove these meteors from further analysis as they have large uncertainties. From empirical tests where we compared the Monte Carlo error in V$_m$ with the time lags, we arrived at a criteria for rejecting meteor echoes not having sufficient time delays for good velocity solutions as being when the average time offset between echoes measured relative to the main (T0) receiver/transmitter station and other receiver stations as described in Equation \ref{LOD}.

\begin{equation}
\label{LOD}
\text{Offset} = \frac{V_{g}}{V_{\text{Geminids}}} \times 0.1 s
\end{equation}

The Geminid meteor shower was used for calibration of this selection filter. For the Geminids, we removed meteor echoes with average time offsets below 0.1 seconds (roughly 5 pulses) for the Geminids. This minimum offset value was then scaled proportional to the shower velocity relative to the Geminid velocity. Two showers were not filtered for time offsets: the Draconids, where  including this offset filter left too few echoes for analysis (less than five useable height bins), and the Perseids, whose average path lengths ($\approx$7 km) are significantly shorter than the path lengths of other showers ($\approx$10 km) and whose speed is very high (60 km/s) resulting in necessarily small offsets.

\subsection{Estimation of Deceleration Slope and $H_0$} 

\begin{figure}[H]
\begin{center}
\includegraphics[width=11cm]{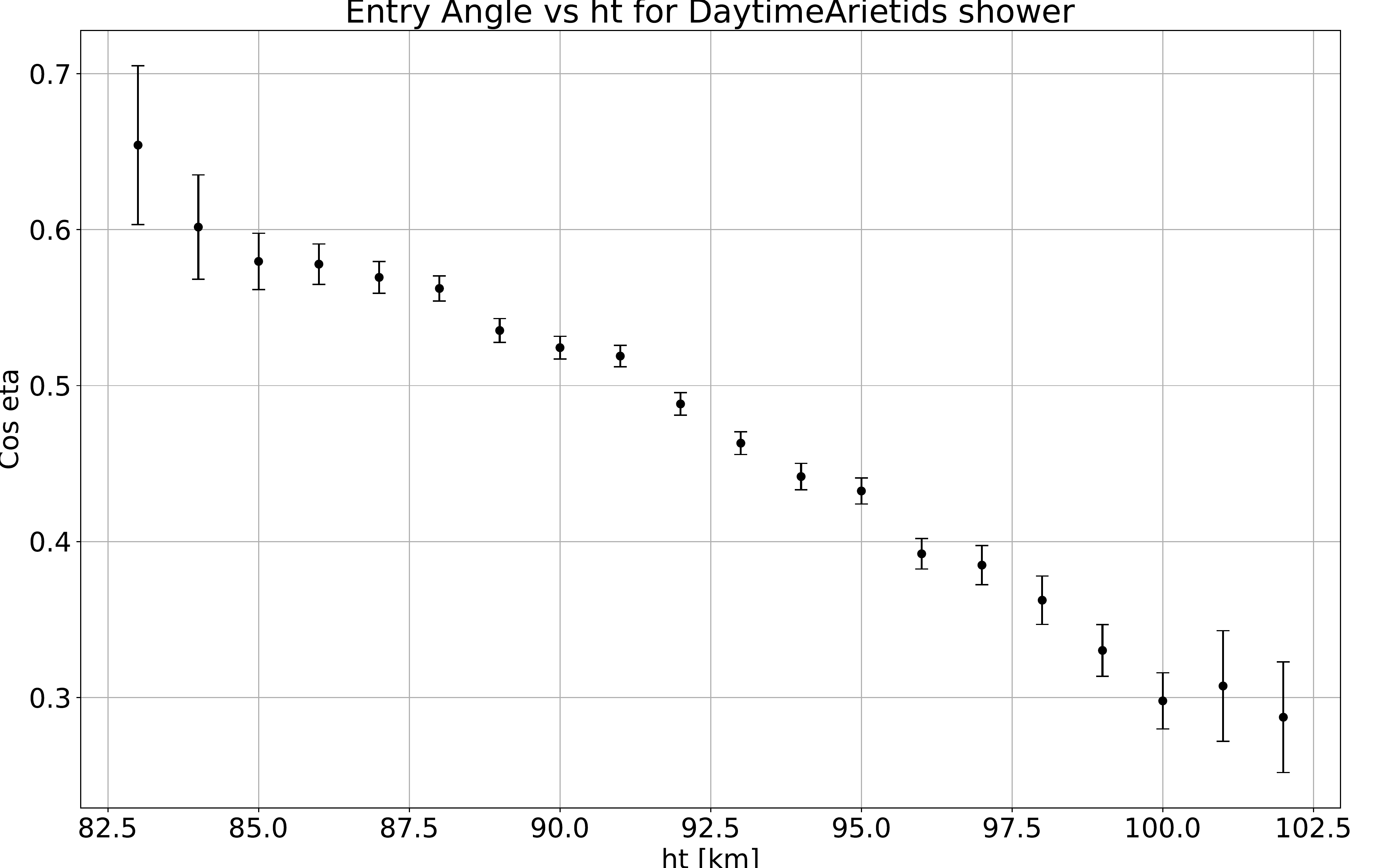}
\caption[Cosine $\eta$ (radiant zenith distance) vs height for the Daytime Arietids]{Mean cosine of local zenith distance of the apparent radiant ($\eta$) versus height for the Daytime Arietids. The uncertainty bounds per point represent the standard error of the mean per bin. Here shallower entry angles are seen at preferentially higher heights, as expected.}
\label{DAcoseta}
\end{center}
\end{figure}

\begin{figure}[H]
\begin{center}
\includegraphics[width=11.5cm]{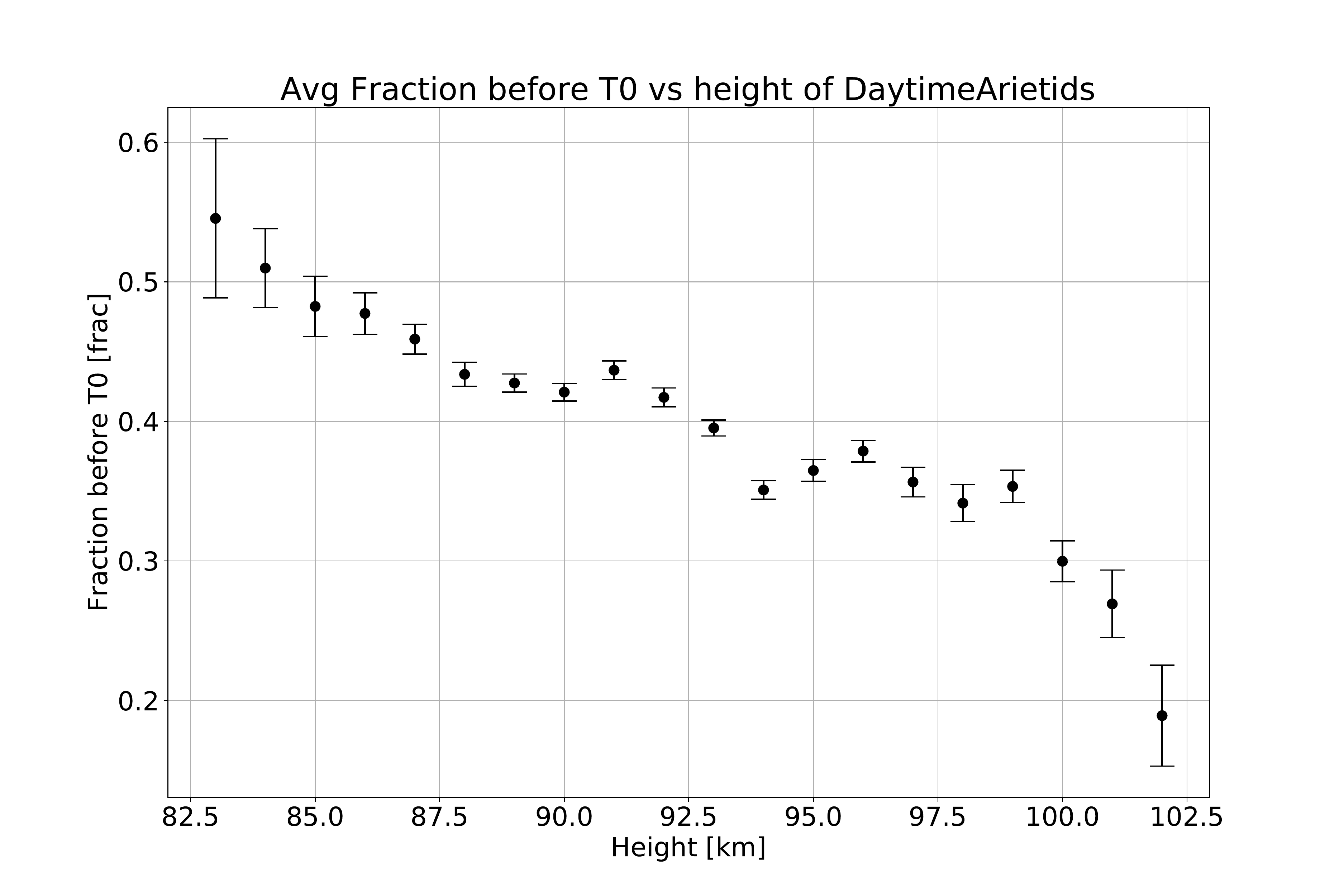}
\caption[Fraction of meteor trail path length before T0 vs height for Daytime Arietids]{The height here is the specular echo height as measured from station 0. For all echoes in each height bin we calculate the mean fraction of the total observed meteor trail length which occurs before the specular point, T0 as well as the standard deviation (uncertainty bounds per bin). In this example we are using only echoes associated with the Daytime Arietid meteor shower. The trend here is as expected, namely that the lower the specular height as seen  from the main station, the larger is the fraction of the trail which occurs above this specular point - ie. the specular point is occurring near the end of the measured trail at lower heights.}
\label{DAT1frac}
\end{center}
\end{figure}

\begin{figure}[H]
\begin{center}
\includegraphics[width=11.5cm]{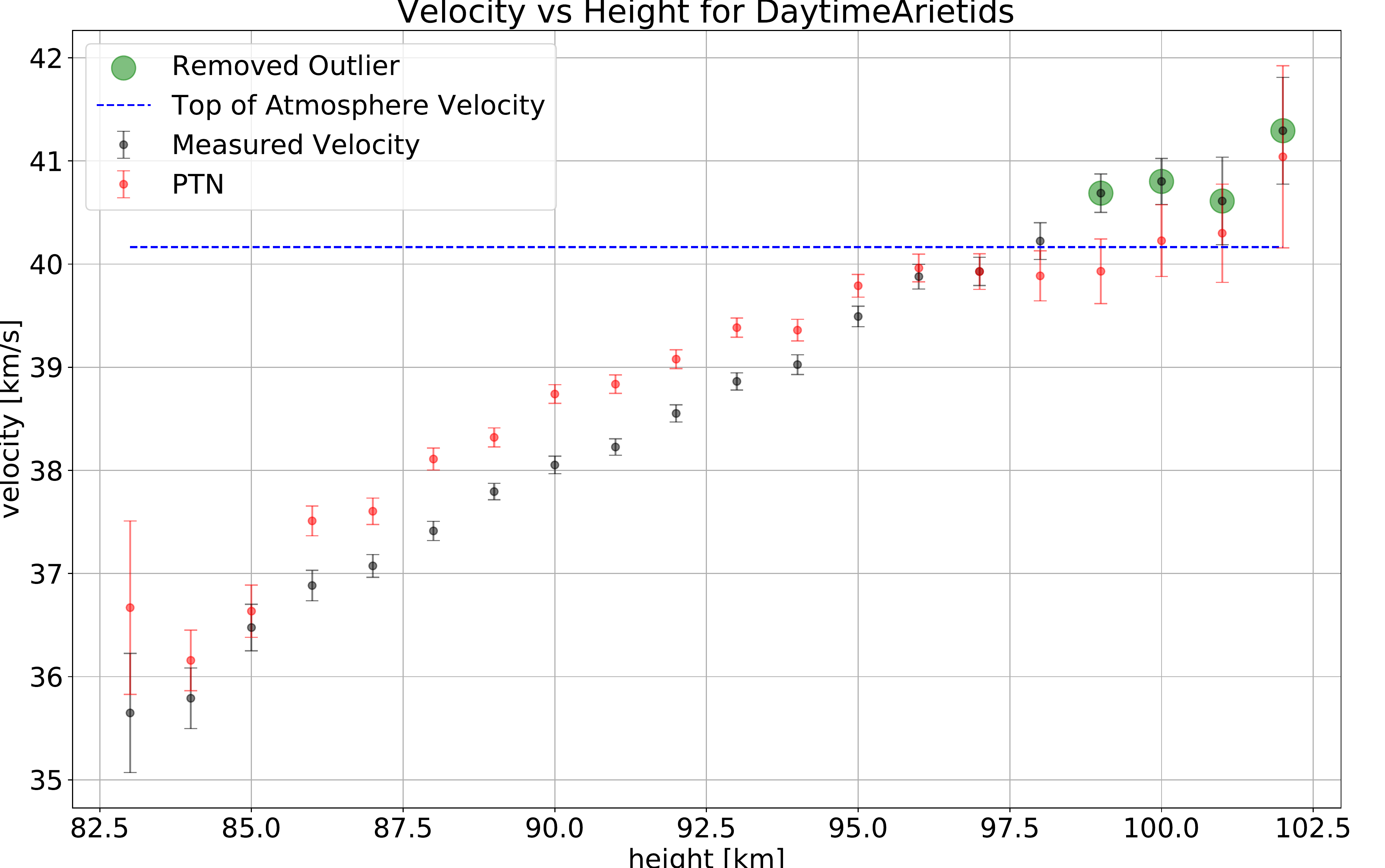}
\caption[Time of flight and pre-t0 velocities vs height for Daytime Arietids]{The average and standard error of the mean of the measured time of flight and pre-t0 (PTN) velocities per 1 km height bin for the Daytime Arietids. A general plateau in the speed vs. height is evident above 96 km, particularly in the pre-t0 speeds which are a better point estimate of speed.}
\label{DAvels}
\end{center}
\end{figure}

After filtering using the foregoing criteria, average measured time of flight velocities, pre-t0 velocities, the cosine of local zenith angular distance of apparent radiant, and echo ranges (among other parameters) were computed in 1 km height bins between 80 km up to 120 km for each shower, and the standard error of these averages also calculated. Bins which contain less than 10 meteors are ignored. Plots were created of average measured velocity, pre-t0 velocity, cosine of the local zenith angular distancge of apparent radiant ($\eta$), path length (calculated as the time of flight $\times$ difference between minimum and maximum time offsets of receiver stations) as a function of height. These plots are shown in the appendix. Additionally, the location of the main site specular point (which determines the height) as a fraction of the total trail length was computed. These plots were generated for each shower and were manually inspected to ensure that they followed expected trends, which include: 

\begin{itemize}
    \item Cosine of $\eta$ (radiant zenith distance) should remain approximately constant or vary monotonically with respect to height (Figure \ref{DAcoseta}), with lower height bins on average accessible for more steep (lower $\eta$) entries. Note that the entry angle  is also correlated with the speed as a function of height and is related to the local radiant height through the specular reflection condition; ie. low entry angle trajectories are always associated with high local echo elevations.  
    \item The fraction of the echo path length observed before the station 0 specular point should decrease with height (Figure \ref{DAT1frac}). That is we expect the main site specular point to lie near the end of the trail as height decreases.  
    \item The measured time of flight-based speed and pre-t0 speed should increase with height and asymptotically approach $V_{\textit{inf}}$ quite sharply. We reject as outliers time of flight velocity points above the crossing of $V_{\textit{inf}}$ / start of plateau (Figure \ref{DAvels}). The measured time of flight and pre-t0 velocities per bin are expected to differ because the former is a measure of the average speed over a larger segment of the trail, but the slope/trends should be comparable.  
\end{itemize}

Height bins which fall substantially outside of these expected trends from visual inspection are added to an outlier list for each shower and are not included in the final analysis.

Once outliers have been removed, a second linear fit to the measured velocity ($V_{m}$) versus height ($H$) is generated (Equation \ref{Vmslope}), weighted to the reciprocal of the standard error of each average measured velocity. The slope coefficient ($S$ in km/s/km) of this fit and its intersection height ($H_0$ in km) with the literature $V_{\textit{inf}}$ are found for each shower (Equation \ref{H0}) as:

\begin{equation}
\label{Vmslope}
V_m = S_{\textit{coeff}} H + S_{\textit{const}}
\end{equation}

\begin{equation}
\label{H0}
H_0 = (V_{\textit{inf}} - S_{\textit{const}}) / S_{\textit{coeff}}
\end{equation}

\subsection{Combining All Showers} 

The individual velocity vs. height slopes obtained for each shower as a function of $V_{\textit{inf}}$ are used to construct an overall linear fit. The deceleration slope across all showers is generated from a linear fit weighted by the standard error of each shower slope. If all meteoroids from different streams had similar physical properties/masses and were observed under similar geometries we would expect this slope to have a single value. This simply reflects the fact that the deceleration is driven by the mass of atmosphere intercepted, which is independent of speed.  

\begin{figure}[H]
\begin{center}
\includegraphics[width=12cm]{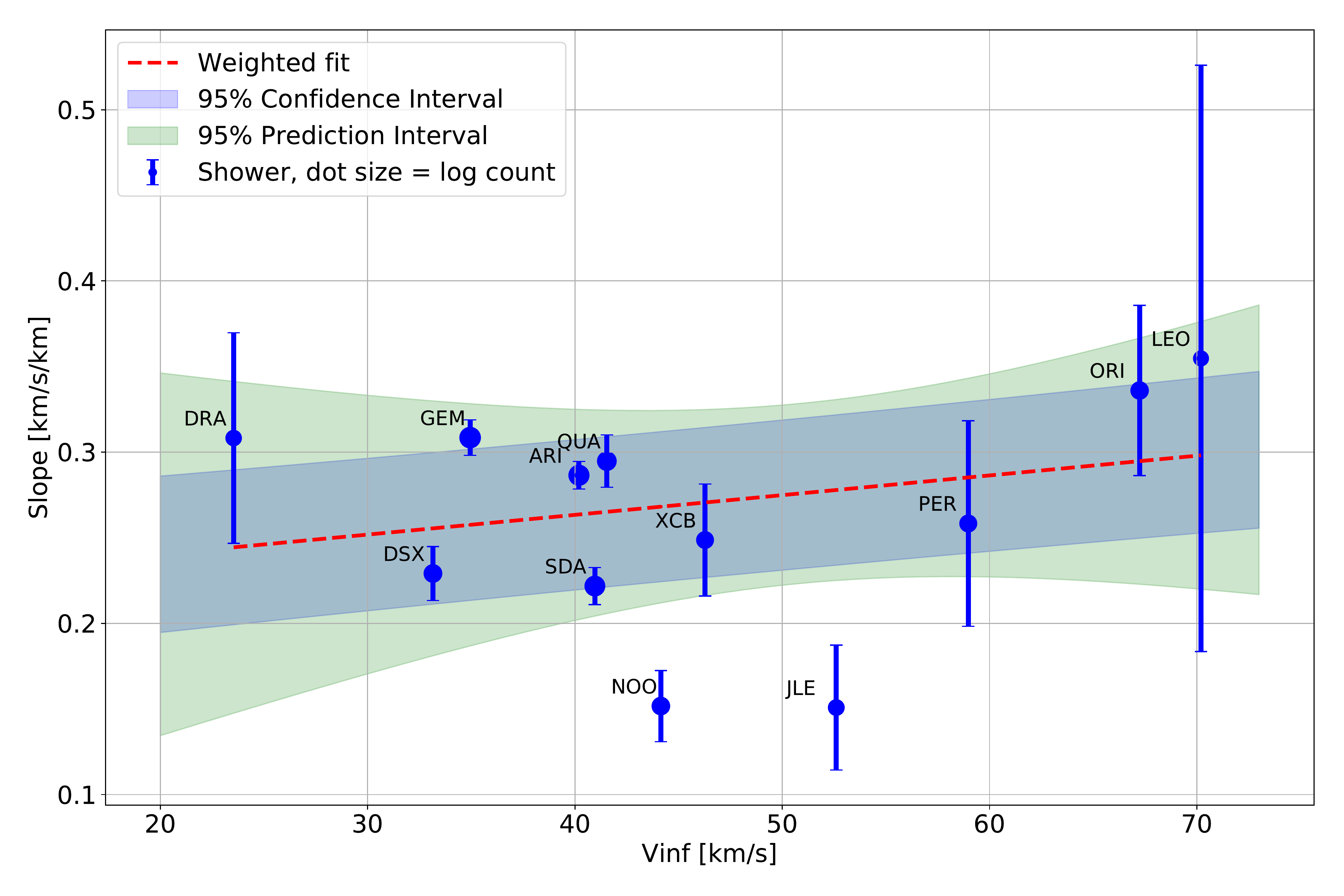}
\caption[Slope vs $V_{\textit{inf}}$ fit for showers]{Slope of the change in speed with height as a function of $V_{\textit{inf}}$ for all twelve of the showers (identified by their three letter IAU designation) used for calibration of the deceleration correction.}
\label{slopeVinf}
\end{center}
\end{figure}

The $S$ versus $V_{\textit{inf}}$ , with a linear regression weighted by the count per shower and 95\% Confidence and Prediction Intervals shaded are shown in Figure \ref{slopeVinf}. While there is scatter, reflecting both physical differences between meteoroids,  the way they decelerate as well as variations in CMOR observing geometry between showers, as shown in Table \ref{tab:paramval} the slope coefficient is within one standard error of zero,  much as expected from theory.  

\begin{figure}[H]
\begin{center}
\includegraphics[width=12cm]{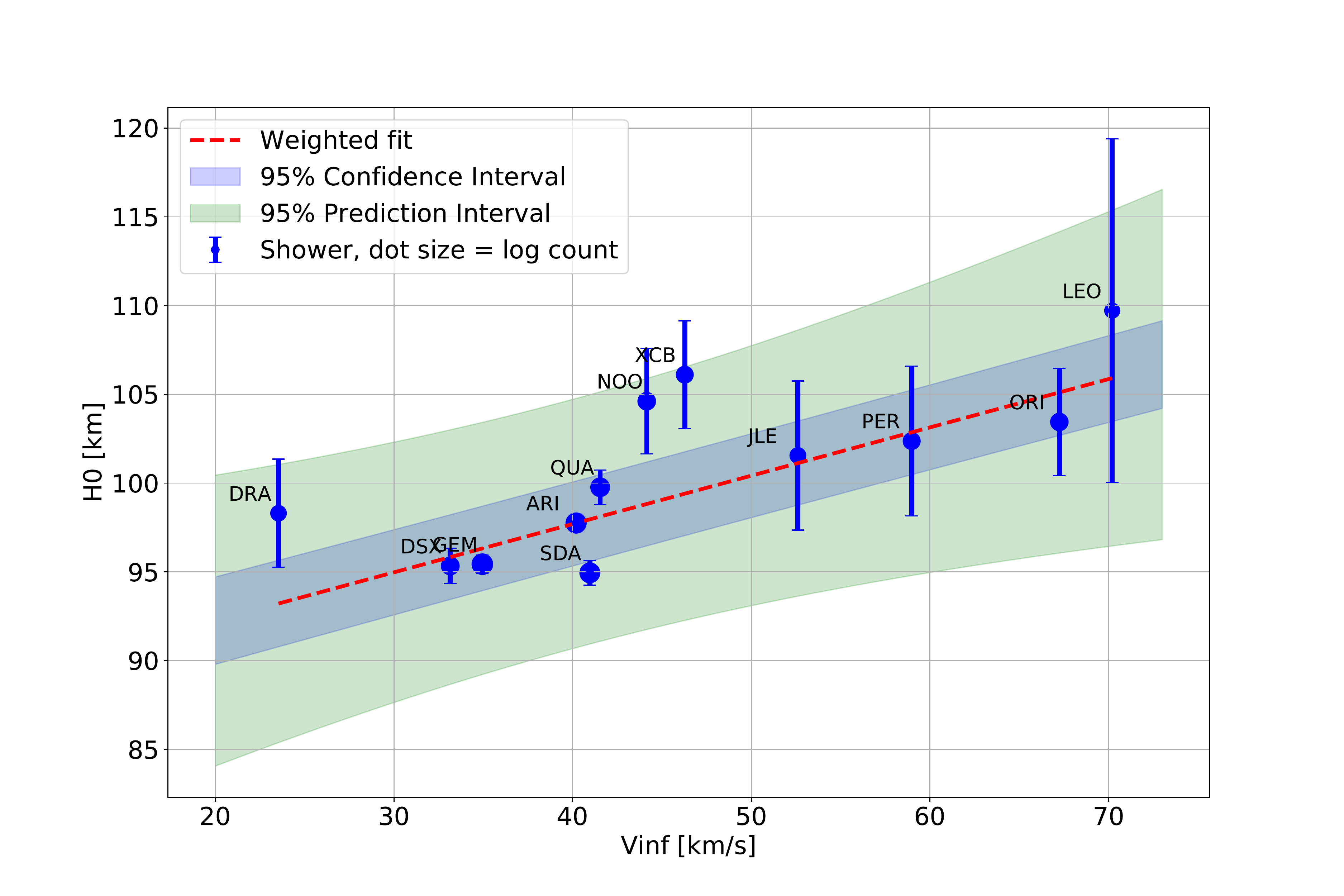}
\caption[$H_0$ vs $V_{\textit{inf}}$ fit for showers]{The height at which negligible deceleration in shower meteors is measured by CMOR, $H_0$, as a function of  $V_{\textit{inf}}$ for all twelve measured showers.}
\label{H0Vinf}
\end{center}
\end{figure}

The $H_0$ intercept as a function of $V_{\textit{inf}}$  across all showers is similarly obtained by extracting the linear fit for each shower, which is weighted by the error in $H_0$ for each shower (Figure \ref{H0Vinf}). This error is calculated by taking the extremes of the $H_0$ intercept for each shower from the velocity versus height slope $\pm$ the standard error of the slope, beginning at the calculated slope’s intercept at 80 km height. A height of 80 km was chosen as very few meteor echoes were observed below this height and most meteors echoes observed between 80 km and the $H_0$ intercept height would fall within the region contained between the height slope $\pm$ standard error extremes. Equations \ref{H0min} and \ref{H0max} summarize the resulting $H_0$ limits.

\begin{equation}
\label{H0min}
H0_{\textit{min}} = \frac{V_{\textit{inf}}-(80 \textit{km} \times S_{\textit{coeff}}-80 \textit{km} \times(S_{\textit{coeff}}  + S_{\textit{error}})) + S_{\textit{const}}}{S_{\textit{coeff}}  + S_{\textit{error}}}
\end{equation}

\begin{equation}
\label{H0max}
H0_{\textit{max}} = \frac{V_{\textit{inf}}-(80 \textit{km} \times S_{\textit{coeff}}-80 \textit{km} \times(S_{\textit{coeff}}  - S_{\textit{error}})) + S_{\textit{const}}}{S_{\textit{coeff}}  - S_{\textit{error}}}
\end{equation}

As expected, the $H_0$ value generally increases with speed, reflecting the higher beginning heights of ablation for faster meteoroids (which therefore receive more energy per atmospheric molecule encountered), paralleling the increase in beginning height with speed observed by optical cameras (eg. \citet{ceplecha68}). However, for heights $\approx$105 km, the echo height ceiling becomes significant at CMOR's 29.85 MHz frequency, so the trend at higher speeds also reflects the lack of detectable echoes above this height as opposed to real differences in beginning ablation heights. 

To independently assess how reasonable is our empirical estimate that $H_0$ represents the height above which negligible deceleration occurs, we applied the ablation modelling approach of \citet{Vida2018}. We assumed that the meteoroids were cometary and used a peak magnitude of +7 (appropriate to the median size of CMOR detected echoes as described in \citet{Brown2008}) and examined the expected deceleration as a function of height. For the range of values of $H_0$ in Fig \ref{H0Vinf} we find that the \citet{Vida2018} formalism predicts decelerations of order 0.5 km/s at the highest speeds (heights of 105 km) to almost 1 km/s at speeds of 20 km/s for heights of 92 km.  

It is notable that the Draconids in the lowest speed portion of the distribution is anomalous in that radar Draconids tend to begin deceleration at higher heights than a simple extrapolation would predict from the other showers. This reflects the well known fragility of the Draconids and their higher starting heights \citep{Koten2007} compared to other meteors of similar initial speed.

\subsection{Final Velocity Correction} 

The $H_0$ and $S$ fit coefficients and constants (Table \ref{tab:paramval}) are combined to obtain a final measured velocity correction ($V_c$) (Equation \ref{Vc}).

\begin{equation}
\label{Vc}
V_c = V_m - [(H - (H0^{\textit{fit}}_{\textit{coeff}}V_m + H0^{\textit{fit}}_{\textit{const}})) \times (S^{\textit{fit}}_{\textit{coeff}}V_m + S^{\textit{fit}}_{\textit{const}})]
\end{equation}

This correction uses the observed echo height from the main site and the time of flight speed to then estimate the equivalent average amount of velocity correction needed to bring the meteoroid to the top of the atmosphere. This general correction is applicable to all CMOR-detected meteor echoes, but should not be applied to other systems. 

\begin{figure}
\begin{center}
\includegraphics[scale=0.9]{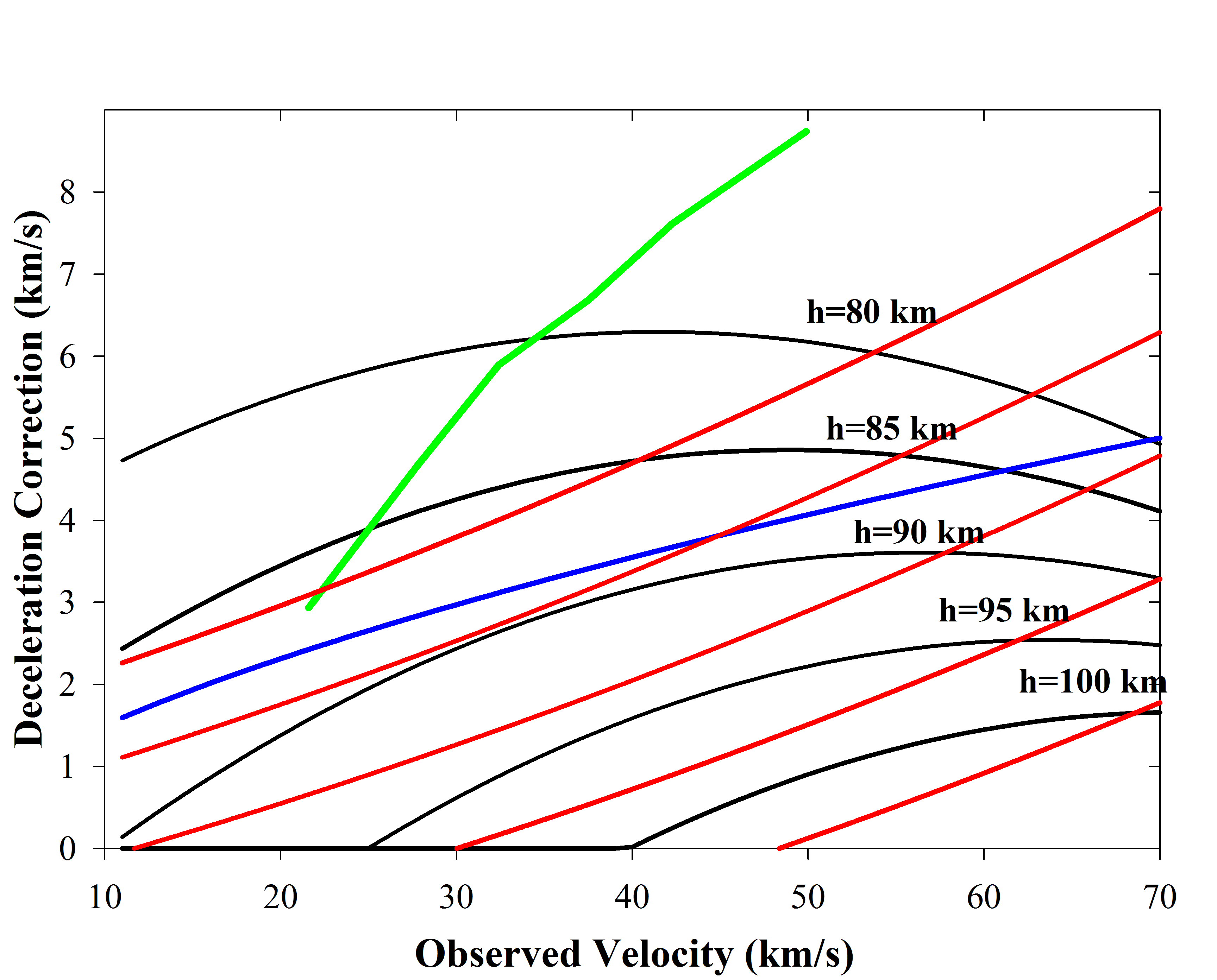}
\caption[New-velocity-correction]{New velocity correction for CMOR (red lines) as a function of height compared to original correction (black lines) from \citet{Brown2005}. Also shown are the deceleration correction used for AMOR (green line) \citep{Baggaley1994} and for the Harvard Radio Meteor Project (blue line) \citep{Verniani1973}. Note that both AMOR and HRMP used average corrections with speed without an explicit height dependence.} 
\label{new-vel-cor}
\end{center}
\end{figure}

Figure \ref{new-vel-cor} shows the magnitude of the correction as a function of velocity for various specular heights. Also shown are the corrections used by the Advanced Meteor Orbit Radar (AMOR) \citep{Baggaley1994} and the Harvard Radio Meteor Project \citep{Verniani1973} (HRMP) together with the original deceleration correction for CMOR given by \citet{Brown2005}. 

The largest difference between the old and new correction are for low velocities and low heights, where extreme differences of 2-3 km/s are found. However, unlike the old correction which rolled over at high speeds for a given height (an unphysical result), the new correction becomes linearly larger for both lower heights and higher speeds. The new correction most closely resembles the mean HRMP correction and is less than the AMOR correction. Since HRMP sampled masses similar to CMOR while AMOR masses were much smaller this is both physically consistent and more realistic than the earlier CMOR deceleration correction. Note that for most heights and speeds the new and old corrections differ by of order 1 km/s on average.   

\begin{figure}
\begin{center}
\includegraphics[trim = 2cm 1cm 1cm 1cm,clip,width=12cm]{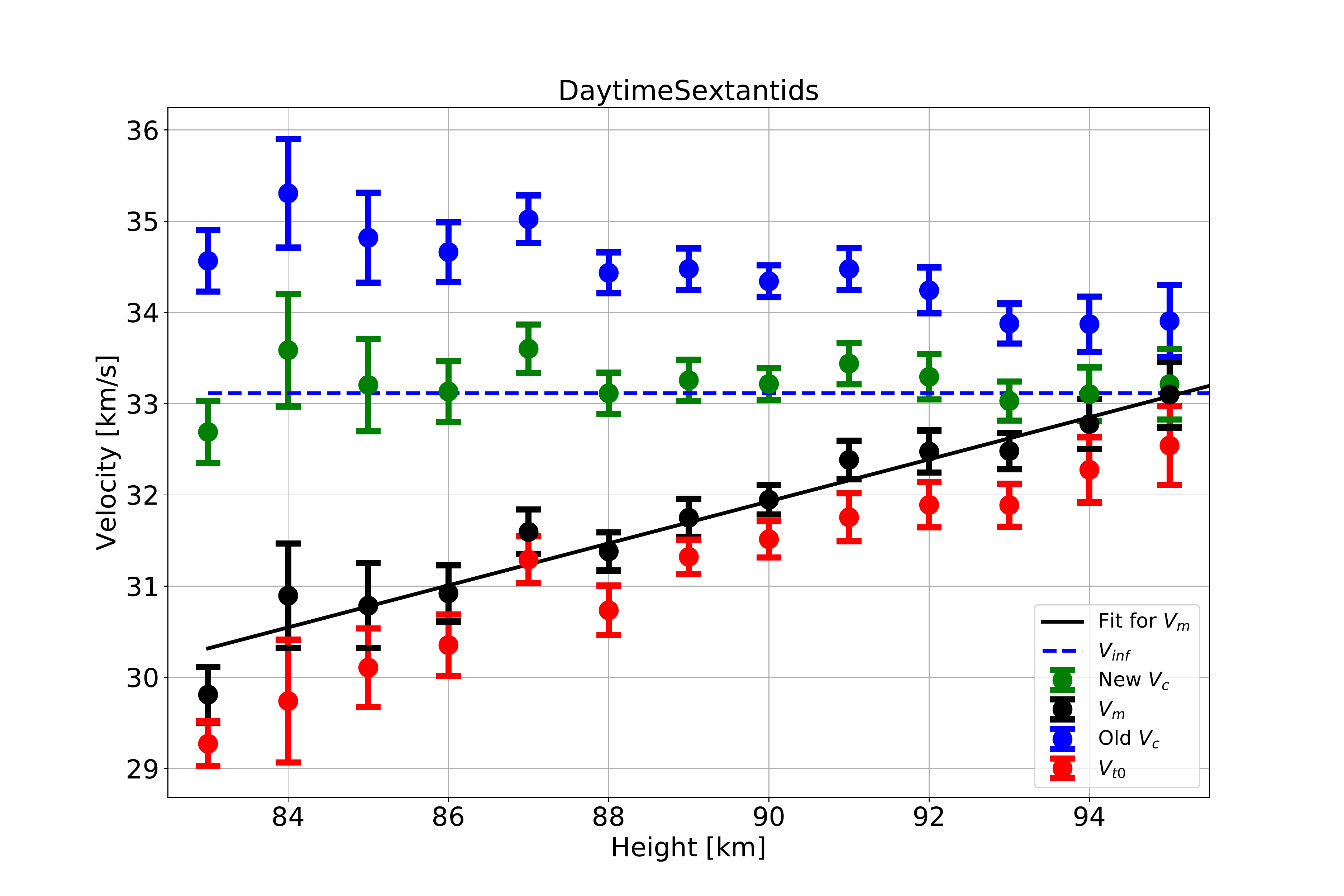}
\caption[Velocities vs height for the Daytime Sextandids]{Velocities versus height in 1 km height bins for the Daytime Sextandids.}
\label{DSXvels}
\end{center}
\end{figure}

As an example application of the new correction,  in Figure \ref{DSXvels}, we apply the correction to the Daytime Sextantid shower.  The resulting corrected speeds better reproduce the expected top of atmosphere speed, showing a near constant corrected speed as a function of echo height. In contrast, the earlier correction from \citet{Brown2005} (shown as blue symbols) tends to over correct the speed, particularly at lower heights for this shower.

We obtain lower and upper limits encompassing the majority of the expected distribution to the 1$\sigma$ level for the velocity correction (Equations \ref{Vclower} and \ref{Vcupper}) by applying the standard errors to the slope and $H_0$ values (Table \ref{tab:paramval}) for Slope (Equations \ref{Slower} and \ref{Supper} and $H_0$ (Equations \ref{H0lower} and \ref{H0upper} based on these fits. If the lower limit falls below the measured velocity, then we use the measured velocity as the lower limit.

\begin{equation}
\label{H0lower}
H0_{\textit{lower}} = V_m(H0^{\textit{fit}}_{\textit{coeff}} - \sigma_{\bar{x}} H0^{\textit{fit}}_{\textit{coeff}}) + H0^{\textit{fit}}_{\textit{const}} - \sigma_{\bar{x}} H0^{\textit{fit}}_{\textit{const}}
\end{equation}

\begin{equation}
\label{Slower}
S_{\textit{lower}} = V_m(S^{\textit{fit}}_{\textit{coeff}} + \sigma_{\bar{x}}S^{\textit{fit}}_{\textit{coeff}}) + S^{\textit{fit}}_{\textit{const}} + \sigma_{\bar{x}}S^{\textit{fit}}_{\textit{const}}
\end{equation}
\begin{equation}
\label{Vclower}
V^{\textit{lower}}_c = V_m-[(H-H0_{\textit{lower}})\times S_{\textit{lower}})]
\end{equation}
\begin{equation}
\label{H0upper}
H0_{\textit{upper}} = V_m(H0^{\textit{fit}}_{\textit{coeff}} + \sigma_{\bar{x}} H0^{\textit{fit}}_{\textit{coeff}}) + H0^{\textit{fit}}_{\textit{const}} + \sigma_{\bar{x}} H0^{\textit{fit}}_{\textit{const}}
\end{equation}
\begin{equation}
\label{Supper}
S_{\textit{upper}} = V_m(S^{\textit{fit}}_{\textit{coeff}} + \sigma_{\bar{x}}S^{\textit{fit}}_{\textit{coeff}}) + S^{\textit{fit}}_{\textit{const}} + \sigma_{\bar{x}}S^{\textit{fit}}_{\textit{const}}
\end{equation}
\begin{equation}
\label{Vcupper}
V^{\textit{upper}}_c = V_m-[(H-H0_{\textit{upper}})\times S_{\textit{upper}})]
\end{equation}

\begin{table}[H]
\centering

\begin{tabular}{lc}
\hline
\multicolumn{1}{c}{\textbf{Parameter}} & \textbf{Value} \\ \hline
$H0^{\textit{fit}}_{\textit{coeff}}$                         & 0.2726   \\
$H0^{\textit{fit}}_{\textit{const}}$                           & 86.8152  \\
$S^{\textit{fit}}_{\textit{coeff}}$                        & 0.0012   \\
$S^{\textit{fit}}_{\textit{const}}$                           & 0.2171   \\
$H0^{\textit{fit}}_{\textit{coeff}}$  standard error          & 0.0950   \\
$H0^{\textit{fit}}_{\textit{const}}$ standard error          & 3.8797   \\
$S^{\textit{fit}}_{\textit{coeff}}$ standard error       & 0.0013   \\
$S^{\textit{fit}}_{\textit{const}}$ standard error          & 0.0720
\\
\hline
\end{tabular}
\caption[Parameter values used for velocity correction]{Parameter values used for velocity correction}
\label{tab:paramval}
\end{table}

Of the 12 showers used in calibrating the new velocity correction, ten showed as good or was an improvement in the average velocity correction with height, producing better agreement with literature values of top-of-atmosphere velocity (Table \ref{tab:velcompare}) as compared to the earlier correction from \citet{Brown2005}. 

\begin{table}[H]
\centering

\begin{tabular}{lcccc}
\hline
\multicolumn{1}{c}{\textbf{Shower}} & \multicolumn{1}{C{1.5cm}}{\textbf{no. of Meteors}} & \multicolumn{1}{C{1cm}}{\textbf{$V_{\textit{inf}}$ {[}km/s{]}}} & \multicolumn{1}{C{2cm}}{\textbf{Mean Old $V_c$ - $V_{\textit{inf}}$ {[}km/s{]}}}  & \multicolumn{1}{C{2cm}}{\textbf{Mean New $V_c$ - $V_{\textit{inf}}$ {[}km/s{]}}} \\ \hline
Draconids                           & 159                                       & 23.5                     & -0.9                                         & -1.6                                                    \\
\textit{Daytime Sextantids}         & \textit{756}                      & \textit{33.1}            & \textit{1.3}                                & \textit{0.1}                                            \\
\textit{Geminids}                   & \textit{7672}                     & \textit{34.9}            & \textit{0.9}                                 & \textit{-0.4}                                           \\
\textit{Daytime Arietids}           & \textit{6023}                     & \textit{40.2}            & \textit{0.6}                                 & \textit{-0.5}                                           \\
\textit{Southern Delta Aquariids}   & \textit{4051}                   & \textit{40.9}            & \textit{2.1}                                 & \textit{0.9}                                            \\
Quadrantids                         & 1629                                      & 41.5                     & 0.0                                          & -1.0                                                    \\
\textit{November Omega Orionids}    & \textit{685}                      & \textit{44.1}            & \textit{0.6}                                 & \textit{-0.3}                                           \\
\textit{Xi Coronae Borealids}                & \textit{397}                                       & \textit{46.3}                     & \textit{-1.0}                                         & \textit{-1.9}\\
\textit{January Leonids}            & \textit{190}                     & \textit{52.6}            & 
\textit{1.3}                                 & \textit{0.7}                                            \\
\textit{Perseids}                   & \textit{427}                    & \textit{59}              & 
\textit{0.3}                                 & \textit{0.0}                                            \\
\textit{Orionids}                            & \textit{624}                                      & \textit{67.2}                     & \textit{-0.3}                                         & \textit{-0.2}                                                   \\
\textit{Leonids}                    & \textit{112}                      & \textit{70.2}            & \textit{-2.7}                                & \textit{-2.5}                                           \\
\hline                            
\end{tabular}
\caption[Comparison of old velocity correction to new velocity correction results]{Comparison of old velocity correction to new velocity correction results ($V_c$). Italicized showers indicate smaller (absolute value) residuals between new $V_c$ - $V_{\textit{inf}}$ versus old $V_c$ - $V_{\textit{inf}}$, where $V_{\textit{inf}}$ is the literature value of top-of-atmosphere speed.}
\label{tab:velcompare}
\end{table}

\subsection{CMOR detected Interstellar Candidates with Deceleration Correction}

We now apply our new deceleration correction  to the measured time of flight-derived velocities of all CMOR-detected candidate interstellar meteoroids following the procedure outlined in Section  \ref{decelCor}. This provides a more realistic estimate of the top of atmosphere velocity, though we again emphasize that these candidate events have measured nominal in atmosphere speeds which already place them in hyperbolic orbits.

A Monte Carlo simulation (10000 runs) was performed for each of the five interstellar meteoroid candidates as described in Section \ref{ISMonteCarlo}. Based on their observed in-atmosphere speeds ($V_m$), all five candidates showed eccentricities \textgreater 2$\sigma$ above the hyperbolic limit with one event yielding \textgreater 3$\sigma$ for e \textgreater 1 as summarized in Table \ref{tab:corvelIS}. Upon applying the deceleration correction to estimate the "true" speed, $V_c$, all five events showed \textgreater 3$\sigma$ for e \textgreater 1. 

\begin{figure}[H]
        \advance\leftskip-3cm
        \advance\rightskip-3cm

        \begin{subfigure}{0.7\textwidth}

            \includegraphics[trim=1cm 0cm 2cm 2cm,clip,width=\textwidth]{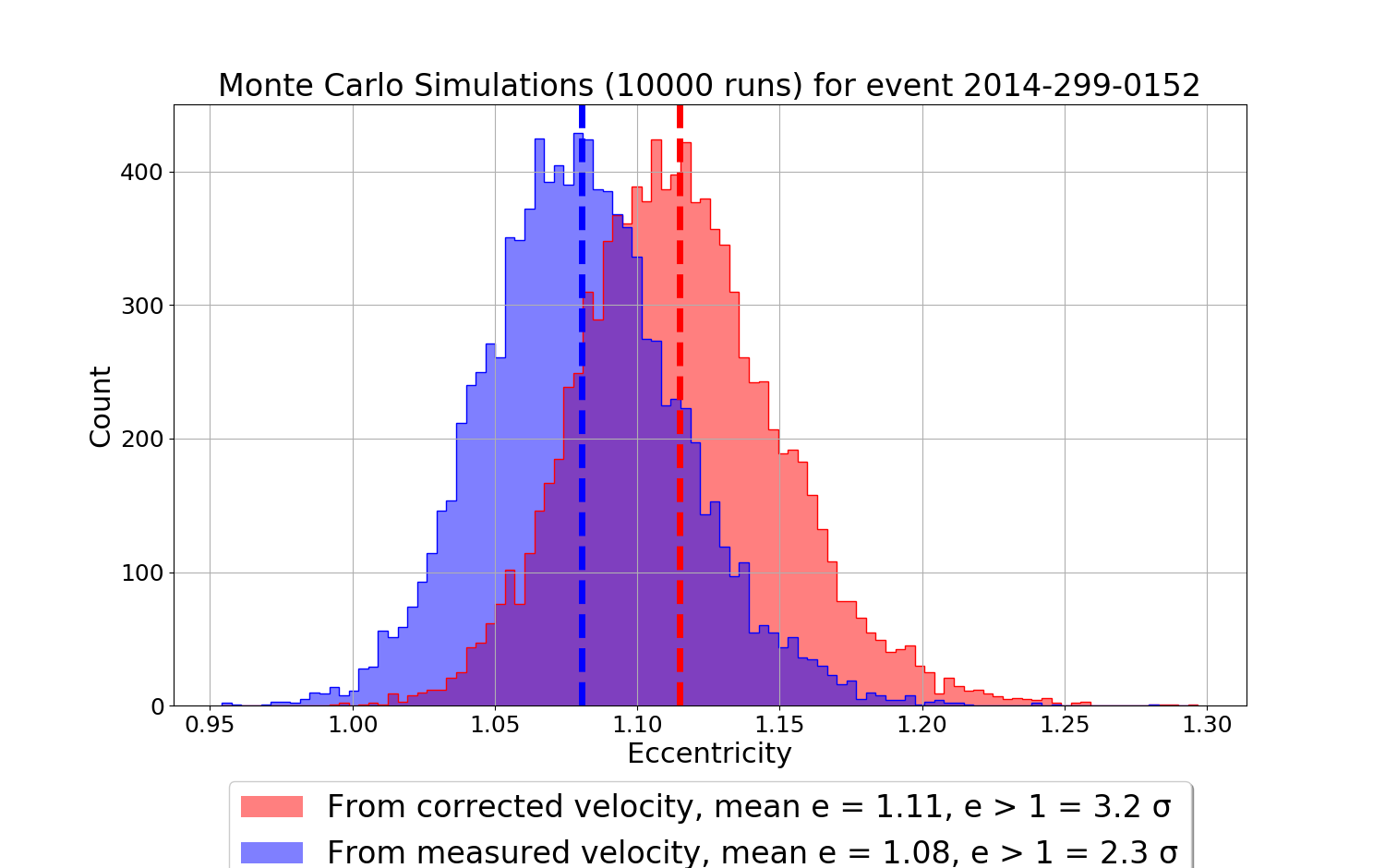}

        \end{subfigure}
        \medskip
        \begin{subfigure}{0.7\textwidth}  

            \includegraphics[trim=1cm 0cm 2cm 2cm,clip, width=\textwidth]{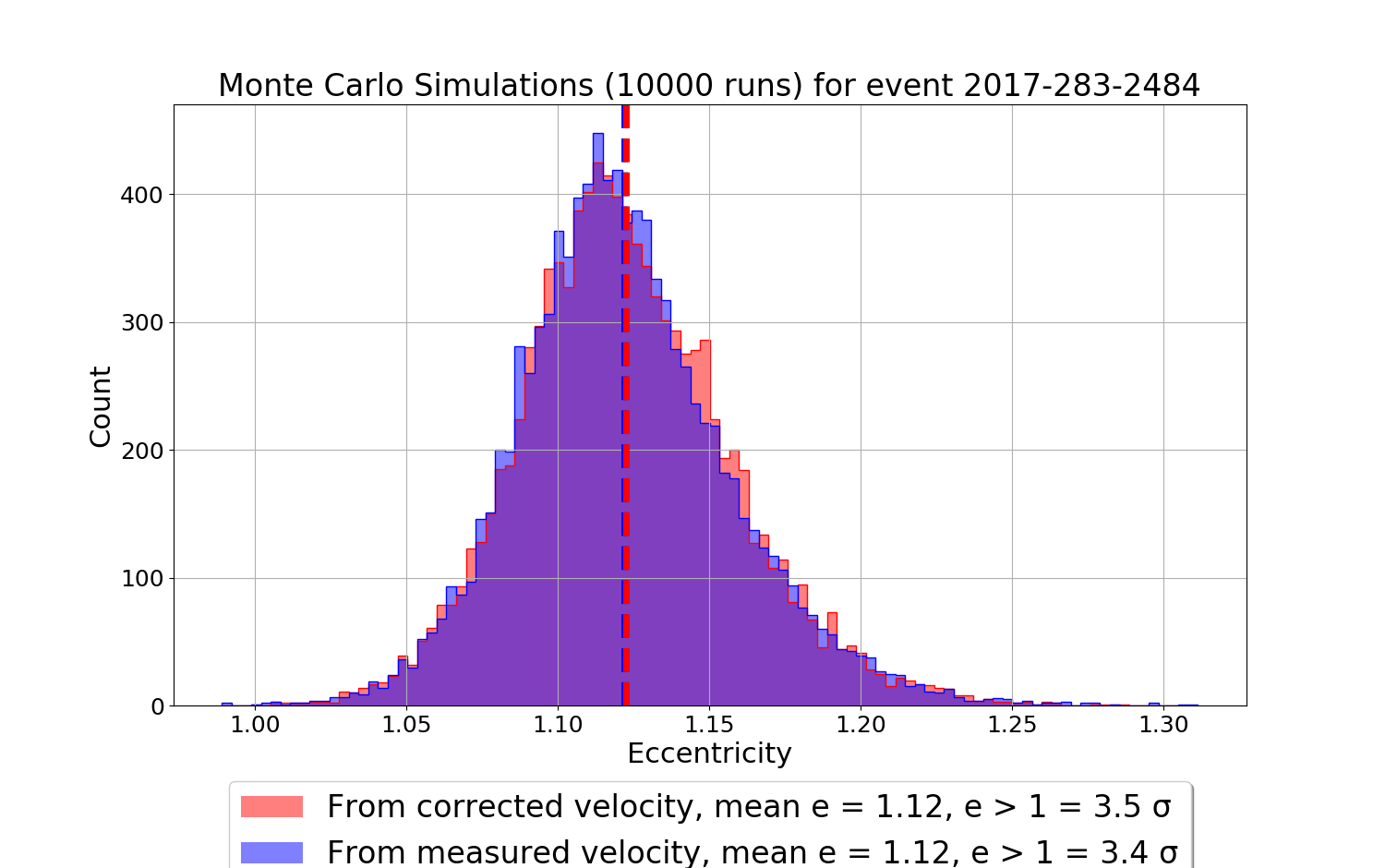}

        \end{subfigure}

        \begin{subfigure}{0.7\textwidth}   

            \includegraphics[trim=1cm 0cm 2cm 2cm,clip,width=\textwidth]{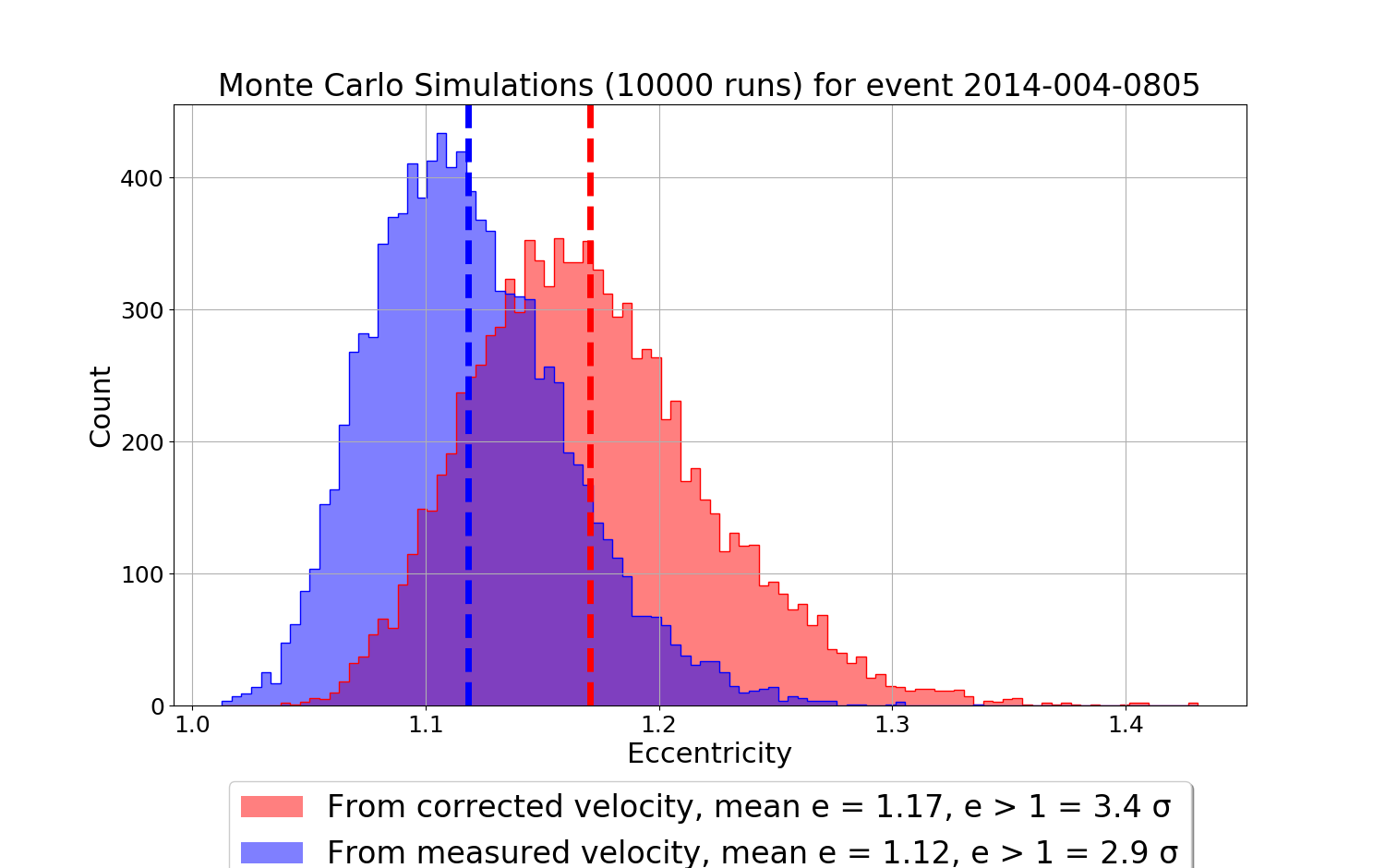}

        \end{subfigure}
           \medskip
        \begin{subfigure}{0.7\textwidth}   

            \includegraphics[trim=1cm 0cm 2cm 2cm,clip,width=\textwidth]{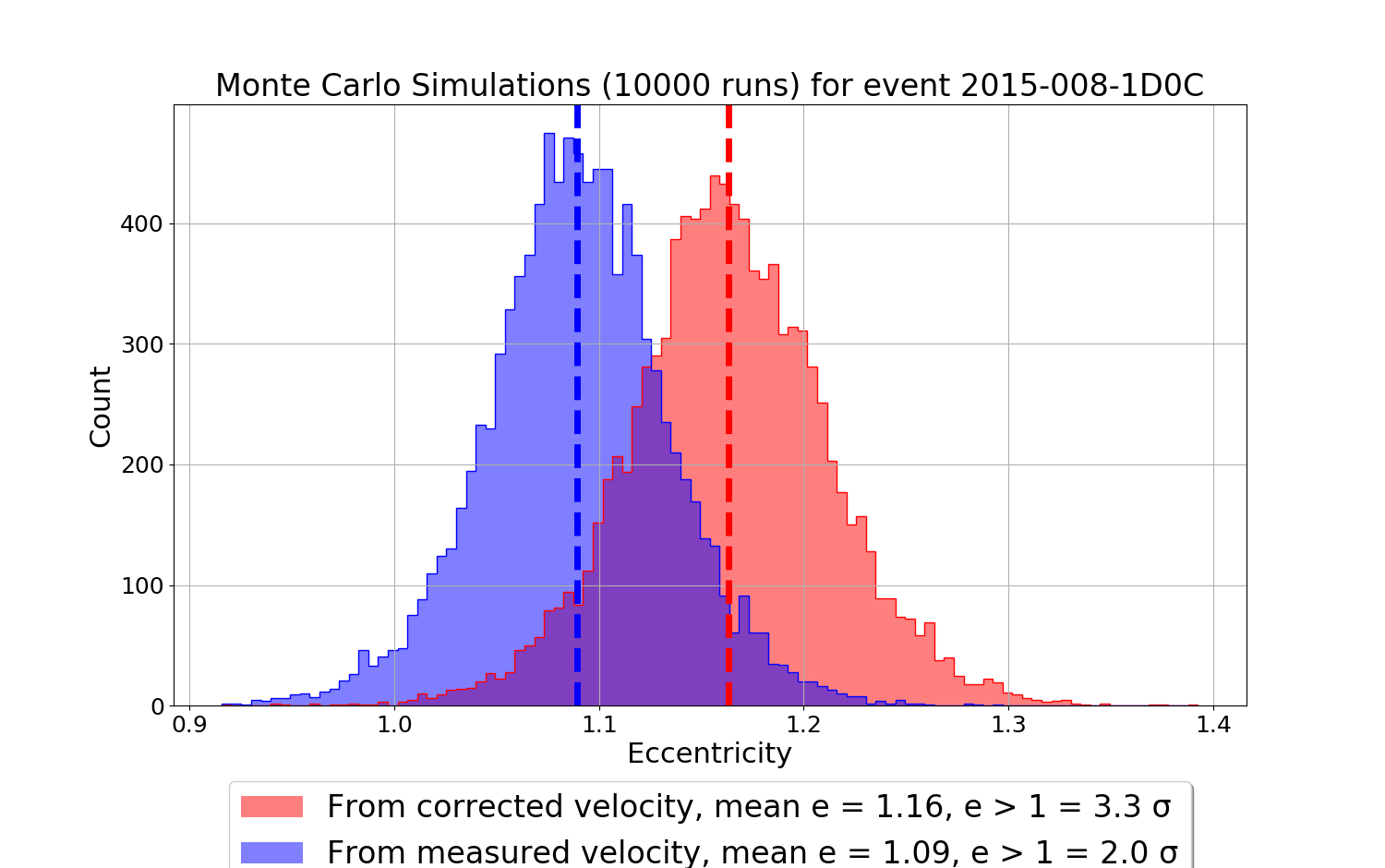}

        \end{subfigure}
        \caption[Eccentricities for simulations of interstellar meteoroid candidates]{Histogram of eccentricities based on 10000 Monte Carlo simulations of four interstellar candidate CMOR echoes showing eccentricities derived from raw measured velocities $V_m$ (blue histogram) and corrected velocities $V_c$ (red histogram). The eccentricity based on the original measured speed, $V_m$ is shown as a vertical dashed blue line while the nominal eccentricity found after correcting for atmosphere deceleration is the vertical dashed red line.}
        \label{IScandis}
    \end{figure}

\begin{figure}[H]
\vspace*{2mm}
\begin{center}
\includegraphics[width=11cm]{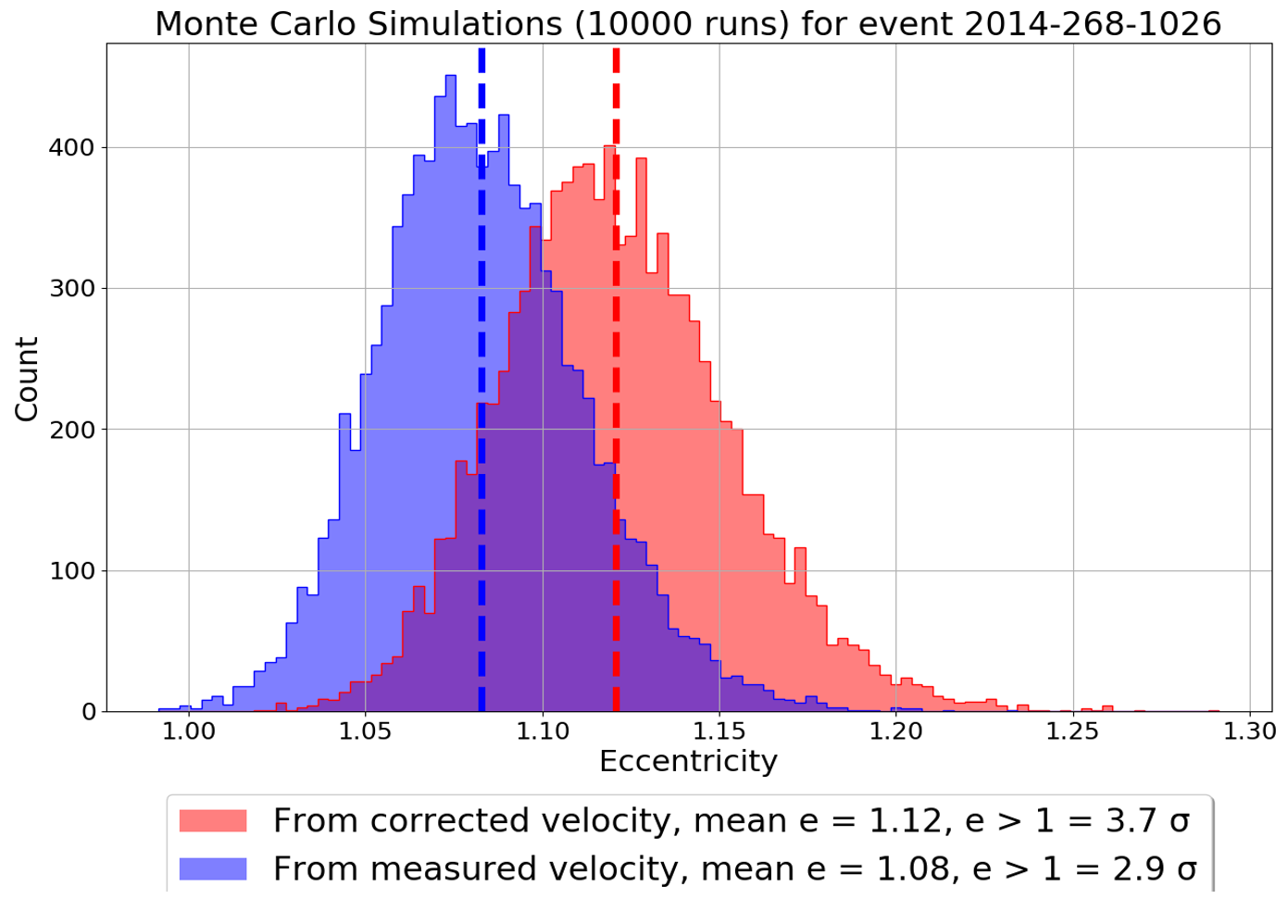}
\caption[Eccentricities of simulations of Alpha candidate event]{Histogram of eccentricities from 10000-run Monte Carlo simulations of the best CMOR IS candidate event based on raw measured velocity $V_m$ and corrected velocity $V_c$.}
\label{alpha_candi}
\end{center}
\end{figure}

Figure \ref{IScandis} shows the resulting Monte Carlo distributions of eccentricity for four of these events based on both the measured time of flight speed (blue histogram) and our best estimate of the deceleration corrected top of atmosphere speed (red histogram). The most promising interstellar (IS) candidate  which showed \textgreater 3.7$\sigma$ for e \textgreater 1 was event 2014-268-1026. This is shown in Figure \ref{alpha_candi}.

\begin{table}[H]
\centering

\begin{tabular}{ccccc}
\hline
\textbf{Event} & \multicolumn{1}{C{1.5cm}}{\textbf{Mean e from $V_m$}} & \multicolumn{1}{C{1.25cm}}{\textbf{$\sigma$ e \textgreater{}1 for $V_m$}} & \multicolumn{1}{C{1.5cm}}{\textbf{Mean e from $V_c$}} & \multicolumn{1}{C{1.25cm}}{\textbf{$\sigma$ e\textgreater{}1 for $V_c$}} \\ \hline
2017-283-2438  & 1.12                     & 3.4$\sigma$                                      & 1.12                     & 3.5$\sigma$                                      \\
2014-004-0805  & 1.12                     & 2.9$\sigma$                                      & 1.17                     & 3.4$\sigma$                                      \\
2014-299-0152  & 1.08                     & 2.3$\sigma$                                      & 1.11                     & 3.2$\sigma$                                      \\
2015-008-1D0C  & 1.09                     & 2.0$\sigma$                                      & 1.16                     & 3.3$\sigma$                                      \\
2014-268-1026  & 1.08                     & 2.9$\sigma$                                      & 1.12                     & 3.7$\sigma$                                     
\\
\hline
\end{tabular}
\caption[Eccentricities from simulations for interstellar meteoroid candidates]{Eccentricities from Monte Carlo simulations for all five interstellar meteoroid candidates calculated using raw measured velocities ($V_m$) and atmospheric deceleration corrected velocities ($V_c$).}
\label{tab:corvelIS}
\end{table}

We summarize our best estimate for the orbital and radiant parameters for each of our five candidate IS events in Table \ref{tab:IScandiOrbs}.

\subsection{Estimated Interstellar Meteoroid Flux}\label{physicprop}

To estimate the equivalent IS flux from our five possible CMOR echo detections, we need to determine the limiting mass and integrated collecting area-time product of our survey.

\begin{figure}[H]
\vspace*{2mm}
\begin{center}
\includegraphics[trim = 0cm 0.5cm 0cm 1cm, clip,width=9.5cm]{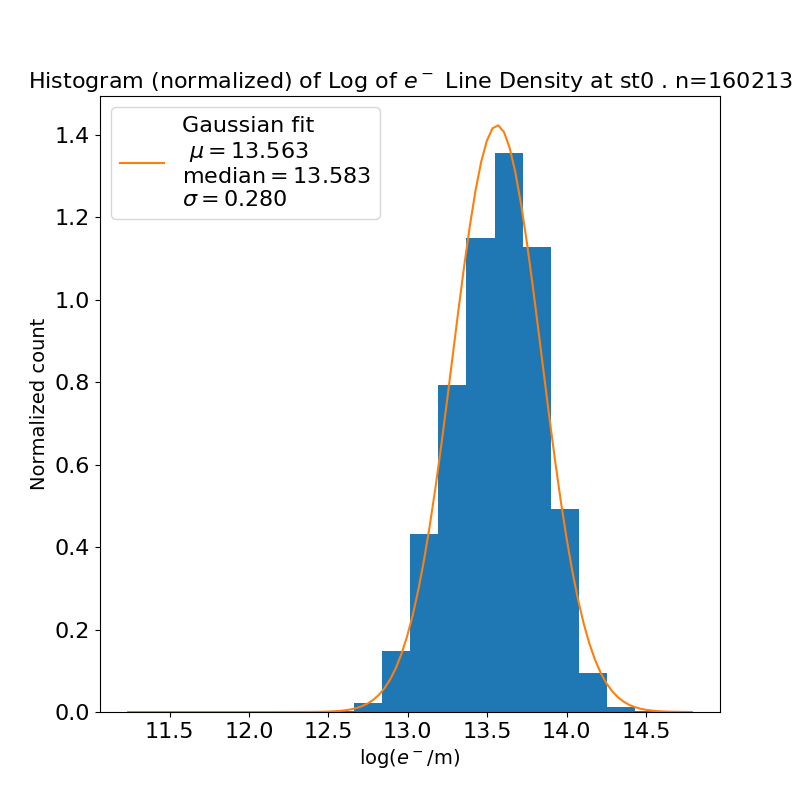}
\caption[Electron Line Density Histogram]{Log of electron line density histogram for all six station events}
\label{logqud0}
\end{center}
\end{figure}

The weakest echoes CMOR is sensitive to  approaches +8.5 radio magnitude \citep{Brown2008} (equivalent to an estimated limiting electron line density of $\approx$ 2$\times 10^{12}$ $e^-$/m). By limiting our survey to echoes which appear only on all six receiver stations, we expect the limiting line density to be larger than this value. Figure \ref{logqud0} shows the distribution of electron line densities for all six station events in our data set showing that the effective completeness limit is near 2$\times 10^{13}$ $e^-$/m or close to radio magnitude of +6. Using the mass-magnitude-velocity relation from \citet{Verniani1973}, this corresponds to meteoroid masses on the order of 10$^{-7}$ kg for events with in-atmosphere velocities of 45 km/s. These are equivalent to diameters ranging from 400 $\mu$m for meteoroids with bulk densities similar to asteroidal and chondritic meteoroids (4200 kg/m$^{3}$), to 800 $\mu$m for meteoroids with similar bulk densities to meteoroids found in Halley type orbits (360 kg/m$^{3}$) \citep{Kikwaya2011}.

\begin{figure}[H]
\vspace*{2mm}
\advance\leftskip-1.5cm
\includegraphics[width=15cm]{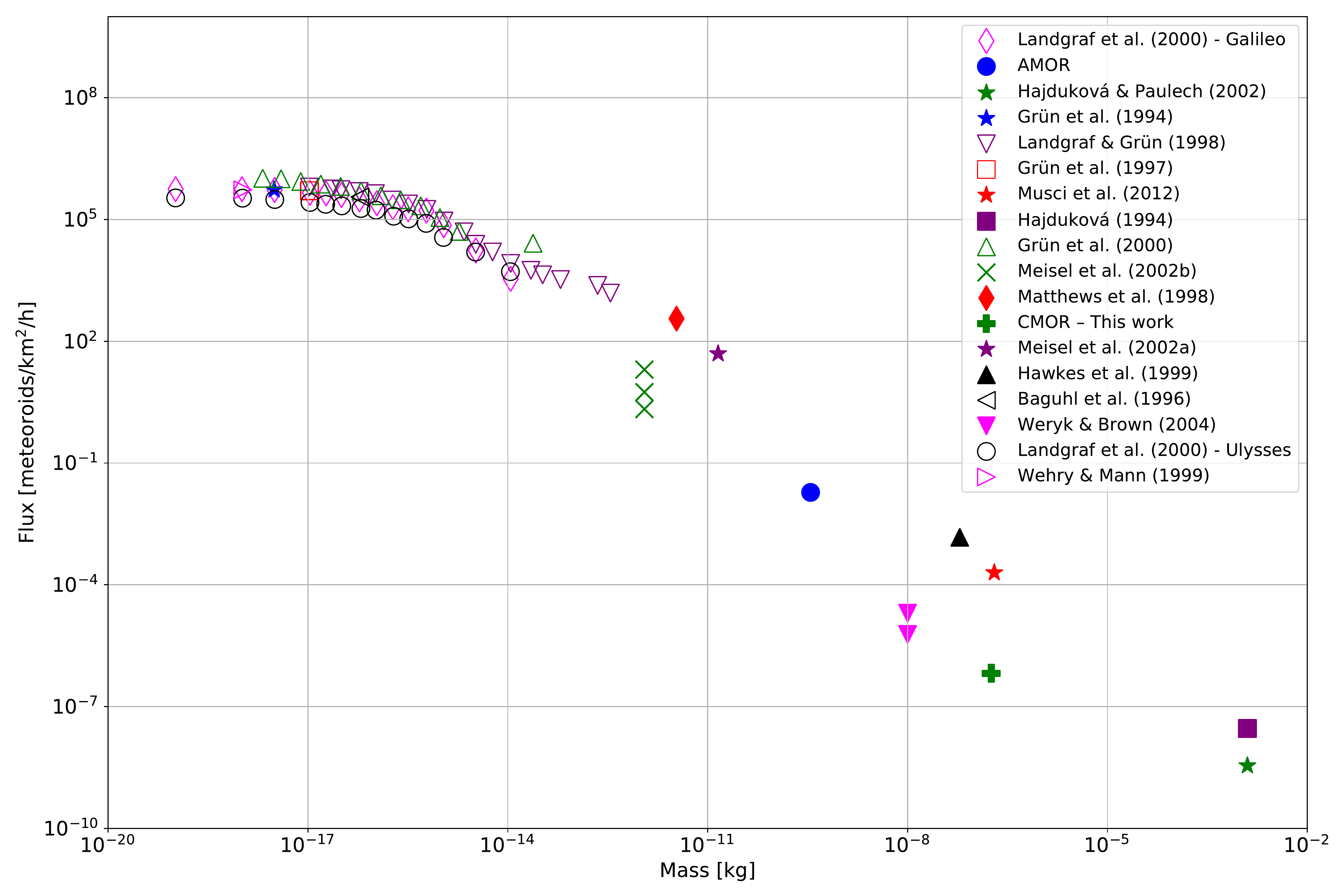}
\caption[Interstellar meteoroid flux estimates]{Summary of interstellar meteoroid flux estimates. Table modified from \citet{Musci2012}. AMOR results from \citet{Baggaley2000}. Note that interstellar dust with masses below $\approx 10^{-13}$ kg are coupled to the local interstellar wind flow, while larger particles, in the range of our measurements, are not \citep{Morfill1979}. Hence these two mass regimes are expected to have different dependencies and a single power law is unlikely to match across the full range shown.}
\label{isflux}
\end{figure}

The effective integrated collecting area of CMOR is dependent on the declination of the radiant. From \citet{CampbellBrown2006}, we use the effective average daily integrated collecting area based on the radiant declination of each of the five candidate meteoroids to estimate a lower bound flux for interstellar meteoroids based on our 7.5 years of observations. The resulting estimated flux is found to be 6.6 $\times$ 10$^{-7}$ meteoroids/km$^{2}$/hr (Figure \ref{isflux}). We note that given our coarse measurement precision it is entirely possible there are more hyperbolic meteoroids in our survey at lower velocities; in this sense our flux value may be interpreted as a lower bound. This estimate is an order of magnitude less than that of \citet{Weryk2005} and their 3$\sigma$ flux estimate. The limiting ISP mass in that study was an order of magnitude lower than this work, as the current study was restricted to larger meteoroids (having lower radio magnitude) as described above.

\nocite{Landgraf2000}\nocite{Landgraf}\nocite{Grn2000}\nocite{Grn1997}\nocite{Grn1994}\nocite{Wehry1999}\nocite{Baguhl1996}\nocite{Meisel2002}\nocite{Meisel2002b}\nocite{Hawkes1999}\nocite{Hajdukova1994}\nocite{Hajdukova2002}\nocite{Mathews1998}

\subsection{Timing and Geometric Similarity Among Candidate Interstellar Meteor Echoes}\label{simcandi}

\begin{figure}[H]
        \advance\leftskip-3cm
        \advance\rightskip-3cm

        \begin{subfigure}{0.65\textwidth}

            \includegraphics[trim=1cm 1cm 2cm 2cm,clip,width=\textwidth]{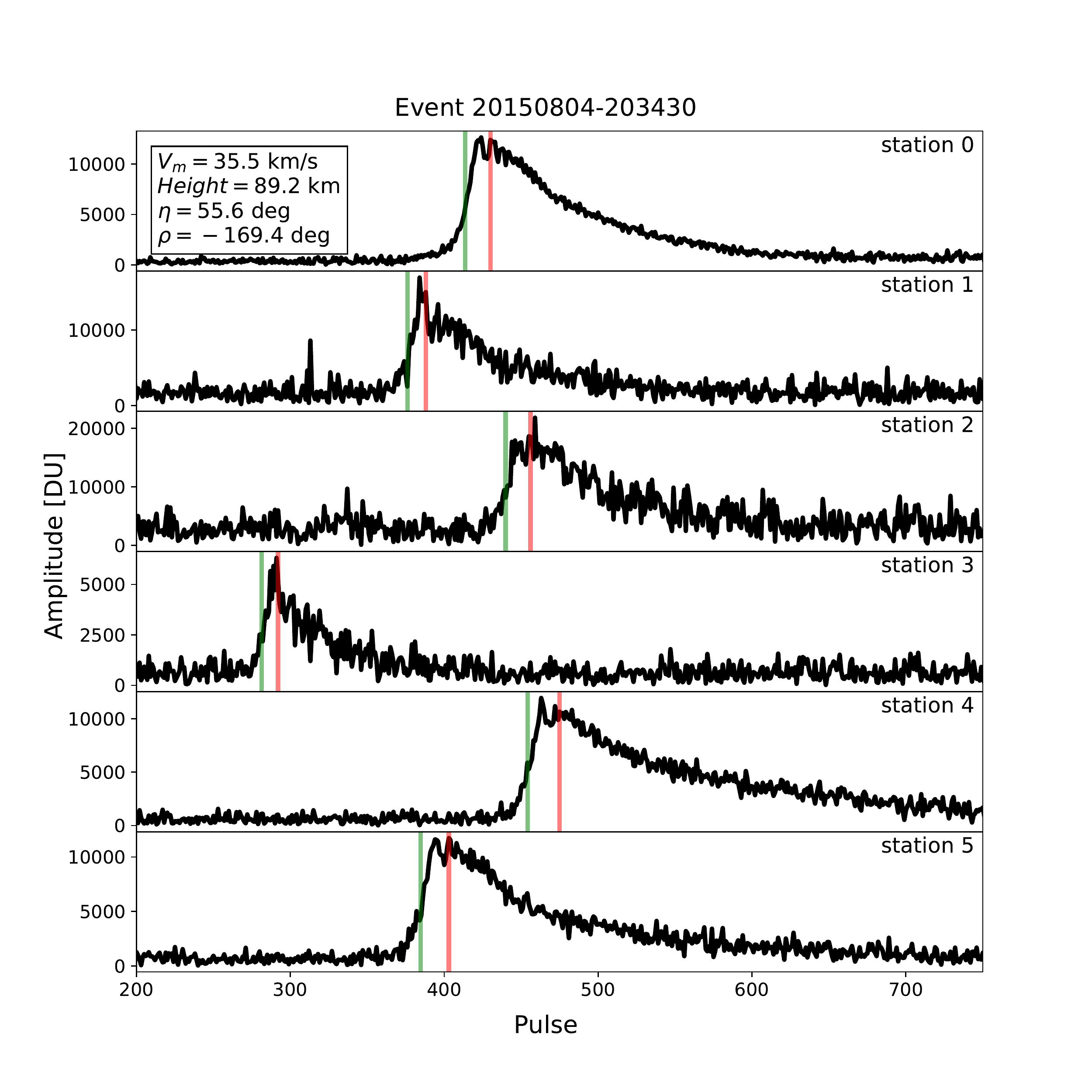}

        \end{subfigure}
        \medskip
        \begin{subfigure}{0.65\textwidth}  

            \includegraphics[trim=1cm 1cm 2cm 2cm,clip, width=\textwidth]{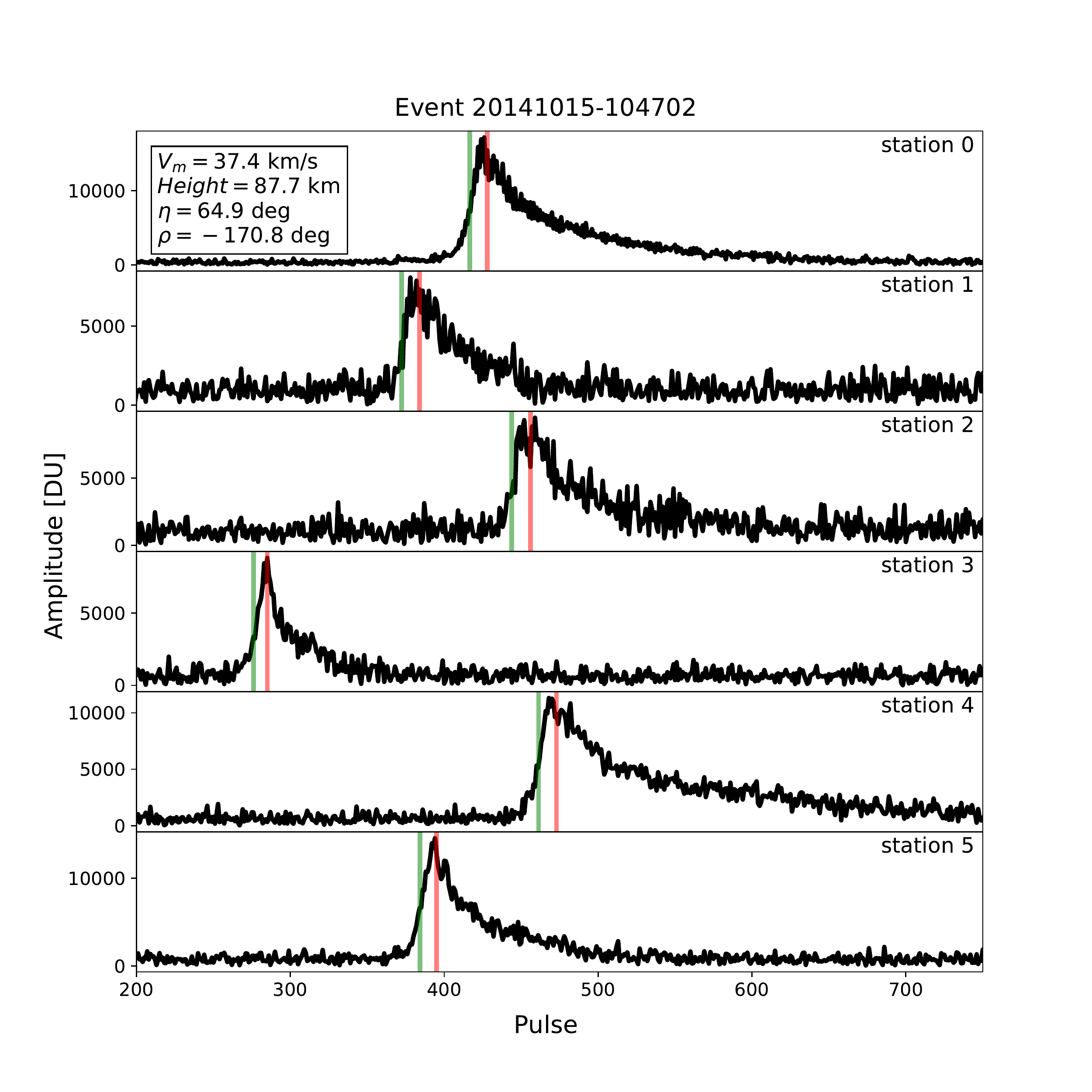}

        \end{subfigure}

        \begin{subfigure}{0.65\textwidth}   

            \includegraphics[trim=1cm 1cm 2cm 2cm,clip,width=\textwidth]{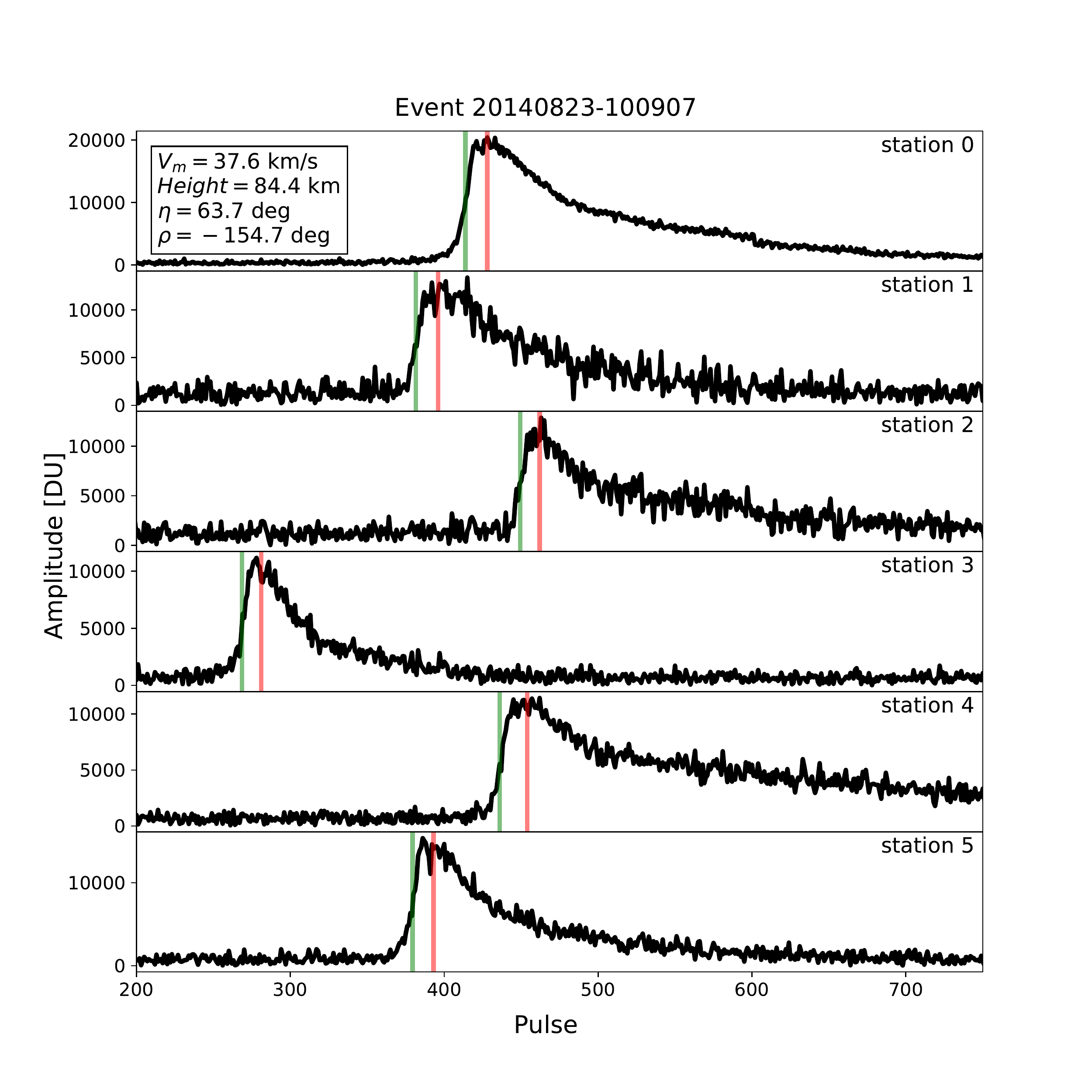}

        \end{subfigure}
           \medskip
        \begin{subfigure}{0.65\textwidth}   

            \includegraphics[trim=1cm 1cm 2cm 2cm,clip,width=\textwidth]{Amp_pulses-20131201-001233.pdf}

        \end{subfigure}
        \caption[Amplitude versus time plots]{Amplitude versus pulses (time) plots at each receiver station of 3 different nominally hyperbolic events with similar profile geometries (events 2015-804-203430, 20140823-100907, 20141015-104702) and 1 event with apparent mirrored profile geometry (event 20131201-001233). Green lines represent inflection points, red lines represent peak points. These events failed other selection criteria and hence are not included in our final set of five events.}
        \label{syncprofiles}
    \end{figure}

In manually reviewing all possible candidates, we observed that a number of candidate interstellar echoes, as well as several that appeared at or close to the hyperbolic limit but did not strictly meet our criteria, exhibited remarkably similar echo timing offsets (Figure \ref{syncprofiles}). Additionally, by mirroring the echo timings (essentially treating the meteor as if it were travelling in the opposite azimuthal direction), several other near hyperbolic events were noted. Most of these echoes have radiants of similar zenith distance and azimuths (or approximately 180 degrees opposite azimuths) as well as comparable speeds. We note that all of these meteors had very different radiants (in both galactic and ecliptic coordinates) and occurred at different solar longitude, implying that they do not have a common astronomical source origin. 

This raises the possibility that the similarity is due to a geometrical bias (or sensitivity) in the layout of CMOR remote stations together with our filtering criteria which preferentially selects apparent interstellar-appearing meteors having radiants in particular local directions. While this does not rule out our events being potentially hyperbolic within the stated statistical limits, such a geometrical trend suggests instrumental effects are present among our detected population warranting caution in interpretation of the results.  

\begin{table}[H]
\centering
\advance\leftskip-4.5cm

{\footnotesize
{\linespread{1.2}

\begin{tabular}{lp{1.5cm}p{1cm}p{1cm}p{1cm}p{1cm}p{0.8cm}p{1cm}p{1cm}p{1cm}p{1cm}p{0.9cm}p{0.9cm}p{0.9cm}}

\hline
\multicolumn{1}{c}{\textbf{Event}} & \multicolumn{1}{C{1.5cm}}{\textbf{epoch [UTC]}}&
\multicolumn{1}{c}{\textbf{e}} &
\multicolumn{1}{C{1cm}}{\textbf{i \newline [deg]}} & \multicolumn{1}{C{1cm}}{\textbf{$\omega$ [deg]}} & \multicolumn{1}{C{1cm}}{\textbf{$\Omega$ [deg]}} & \multicolumn{1}{C{0.8cm}}{\textbf{q [au]}} &
\multicolumn{1}{C{1cm}}{\textbf{$V_g$ [km/s]}} & 
\multicolumn{1}{C{1cm}}{\textbf{$V_H$ [km/s]}} & 
\multicolumn{1}{C{1cm}}{\textbf{$\alpha_g$ [deg]}} & 
\multicolumn{1}{C{1cm}}{\textbf{$\delta_g$ \newline [deg]}} &
\multicolumn{1}{C{0.9cm}}{\textbf{l$_{\textit{gal}}$ [deg]}} &
\multicolumn{1}{C{0.9cm}}{\textbf{b$_{\textit{gal}}$ [deg]}} &
\multicolumn{1}{C{0.9cm}}{\textbf{$\psi$ [deg]}} \\ \hline

2014-268-1026 & 2014/12/20 11:38:05 & 1.121 {[}0.032{]} & 16.572 {[}1.207{]} & 112.684 {[}3.457{]} & 88.295 {[}0.0{]}  & 0.264 {[}0.021{]} & 42.388 {[}0.770{]} & 46.946 {[}0.910{]} & 100.325 {[}1.395{]} & 14.302 {[}0.702{]}  & 174.25  & -40.58 & 154.8 \\
2015-008-1D0C & 2015/03/30 3:38:13  & 1.163 {[}0.049{]} & 28.246 {[}0.846{]} & 179.944 {[}0.982{]} & 8.916 {[}0.0{]}   & 0.998 {[}0.000{]} & 22.543 {[}0.776{]} & 43.836 {[}0.497{]} & 192.356 {[}9.423{]} & 85.227 {[}1.002{]}  & 140.40  & -33.82 & 138.0 \\
2014-004-0805 & 2014/03/25 6:25:18  & 1.170 {[}0.051{]} & 67.025 {[}2.607{]} & 102.770 {[}5.196{]} & 184.329 {[}0.0{]} & 0.342 {[}0.038{]} & 50.509 {[}1.272{]} & 47.064 {[}1.080{]} & 196.792 {[}0.758{]} & -37.596 {[}2.112{]} & 262.64  & +16.96 & 94.9 \\
2017-283-2484 & 2017/01/04 8:39:55  & 1.123 {[}0.035{]} & 21.842 {[}2.491{]} & 109.681 {[}1.809{]} & 103.955 {[}0.0{]} & 0.289 {[}0.010{]} & 42.021 {[}1.345{]} & 46.691 {[}1.153{]} & 115.374 {[}0.604{]} & 9.108 {[}1.257{]}   & 186.49 & -30.62 & 165.7 \\
2014-299-0152 & 2014/01/19 9:12:25  & 1.115 {[}0.036{]} & 14.842 {[}1.289{]} & 297.365 {[}2.367{]} & 299.018 {[}0.0{]} & 0.227 {[}0.012{]} & 44.139 {[}1.356{]} & 47.411 {[}1.389{]} & 136.795 {[}1.193{]} & 23.848 {[}0.605{]}  & 175.63 & -1.16 & 162.8
\\
\hline

\end{tabular}
}
}
\caption[Orbital elements and velocities of interstellar meteoroid candidate simulations]{Orbital elements and velocities of interstellar meteoroid candidates, taken from mean values of Monte Carlo simulations using $V_c$, our best estimate of the initial speed of the meteoroid prior to atmospheric deceleration.   The two quantities $\alpha_g$ and $\delta_g$ represent the geocentric radiant (in J2000). $V_H$ represents heliocentric speed, and $V_g$ represents the geocentric speed. l$_{\textit{gal}}$ and b$_{\textit{gal}}$ are the asymptotic galactic radiant coordinates. Values in brackets are standard deviations from Monte Carlo simulations.For comparison the galactic coordinates for the upstream direction of the local interstellar gas inflow is l$_{\textit{gal}}$ = \ang{3}, b$_{\textit{gal}}$ = \ang{16} \citep{Frisch1999}. All of our candidate asymptotic radiants are tens of degrees or more from the the direction of the local interstellar dust inflow direction, specified in the $\psi$ column.}
\label{tab:IScandiOrbs}
\end{table}

\section{CMOR Detection Sensitivity to Hyperbolic Meteors - Case Study of a Hypothetical ʻOumuamuids shower}\label{Oumuamuids}

The foregoing analysis suggests that some already detected CMOR meteors may be hyperbolic, but more generally can CMOR detect interstellar (hyperbolic) meteoroids and with what degree of confidence? Without a clear positive detection of many hyperbolic events, answering this question requires simulating CMOR's response to meteoroids moving on hyperbolic orbits.

1I/ ʻOumuamua was the first large interstellar objected detected transiting through our solar system. It was discovered by the Panoramic Survey Telescope And Rapid Response System (Pan-STARRS) telescope in October 2017, one month after it had passed its perihelion \citep{Meech2017}. Subsequent observations were made including pre-discovery observations, over a 34-day observation arc by various telescopes before ʻOumuamua became too faint to be detected. Its orbital eccentricity was found to be 1.20 with a perihelion distance of 0.26 au and inclination of 122.74 deg (Table \ref{tab:Oumparamval}). The estimated size based on light-curve analysis and modelling from data obtained by the Spitzer Space Telescope ranged from  240 by 40 m, assuming a high albedo of 0.2 to 1080 by 180 m for a more comet-like albedo of 0.01 \citep{Trilling2018} and an elongation ratio of 6:1 \citep{McNeill2018}. 

Additionally, it was observed that ʻOumuamua experiences non-gravitational acceleration, for which comet-like outgassing has been proposed as its source \citep{Micheli2018}. Photometric observations from the NOT and WYNN telescopes reveal that ʻOumuamua is similar in colour to that of D-type asteroids \citep{Jewitt2017} found within our solar system. 

We use the observed orbit of 1I/ʻOumuamua as a real-world proxy for the type of orbit probable for larger interstellar meteoroids.  This addresses the question of whether CMOR could detect a hypothetical ʻOumuamua-associated meteoroid stream and in particular at what significance level would such a hyperbolic stream be detectable? 

\subsection{Theoretical Radiant}

We determine the theoretical radiant of an ʻOumuamuid-type shower (Table \ref{tab:Oumuarad}) from 1I/ʻOumuamua's orbital elements (Table \ref{tab:Oumparamval}). It is instructive to note that 1I is hyperbolic with an observational significance of 10000$\sigma$. Using the method described in \citet{Nesl1998} we computed the expected radiant for 1I-related meteoroids.  The B+ method, which  adjusts the perihelion distance and eccentricity, as described by \citet{Svoren1993}, on the inbound leg of the the orbit, yielded the lowest D discriminant. Thus, the radiant parameters which were computed by this method were adopted as the theoretical radiant for the stream.
\begin{table}[H]
\centering

\begin{tabular}{cccc}
\hline
\multicolumn{4}{c}{\textbf{\begin{tabular}[c]{@{}c@{}}Orbital Elements at Epoch 2458080.5 (2017-Nov-23.0) TDB\\ Reference: JPL16 (heliocentric ecliptic J2000)\\\footnotesize{JPL Small-Body Database - https://ssd.jpl.nasa.gov/sbdb.cgi?sstr=3788040}\\ \end{tabular}}} \\ \hline
\textbf{Element}                      & \textbf{Value}                            & \textbf{Uncertainty (1$\sigma$)}                      & \textbf{Units}                      \\ \hline
e                                     & 1.201133796102373                         & 2.1064e-05                                          & none                                    \\
a                                     & -1.27234500742808                         & 0.00010015                                          & au                                  \\
q                                     & 0.2559115812959116                        & 6.6635e-06                                          & au                                  \\
i                                     & 122.7417062847286                         & 0.00028826                                          & deg                                 \\
node                                  & 24.59690955523242                         & 0.00025422                                          & deg                                 \\
peri                                  & 241.8105360304898                         & 0.0012495                                           & deg                                 \\
M                                     & 51.1576197938249                          & 0.0061155                                           & deg                                 \\
tp                                    & 2458006.007321375231                      & 0.00026424                                          & TDB                                 \\
n                                     & 0.6867469493413392                        & 8.1084e-05                                          & deg/d
\\
\hline
\end{tabular}
\caption[Orbital elements of ʻOumuamua]{Orbital Elements of ʻOumuamua (1I/2017 U1).}
\label{tab:Oumparamval}
\end{table}

\begin{table}[H]
\centering

\begin{tabular}{cccccccc}
\hline
\multicolumn{8}{c}{\textbf{Equinox: 2000.0, Data for Year: 2017}}                                                                                                              \\ \hline
\textbf{Method} & \multicolumn{1}{C{1cm}}{\textbf{$\alpha$ [deg]}} & \multicolumn{1}{C{1cm}}{\textbf{$\delta$ [deg]}} & \multicolumn{1}{C{1cm}}{\textbf{$V_g$ [km/s]}} & \multicolumn{1}{C{1cm}}{\textbf{$V_H$ [km/s]}} & \multicolumn{1}{C{1cm}}{\textbf{L [deg]}} & \multicolumn{1}{C{1cm}}{\textbf{Peak [Date]}} & \multicolumn{1}{C{1cm}}{\textbf{D-disc.}} \\ \hline
-Q    & 267.7 & 5.80   & 64.190 & 44.93 & 24.6  & APR. 14.6 & 0.458   \\
-B    & 267.1 & 5.80   & 64.740 & 45.60 & 24.6  & APR. 14.6 & 0.456   \\
-W    & 248.5 & -4.80  & 64.080 & 49.66 & 24.6  & APR. 14.6 & 1.008   \\
-A    & 323.6 & 2.50   & 63.840 & 49.42 & 102.9 & JULY  5.0 & 1.038   \\
-H    & 282.3 & -9.00  & 65.330 & 49.51 & 57.3  & MAY  18.2 & 0.770   \\
-P    & 269.6 & 1.60   & 64.200 & 49.57 & 42.9  & MAY   3.4 & 0.558   \\
Q+    & 160.3 & -7.50  & 65.320 & 51.88 & 204.6 & OCT. 18.0 & 0.060   \\
B+    & 159.5 & -7.50  & 64.600 & 50.93 & 204.6 & OCT. 18.0 & 0.057   \\
W+    & 156.5 & -8.50  & 64.340 & 49.78 & 204.6 & OCT. 18.0 & 0.146   \\
A+    & 166.9 & -11.70 & 65.140 & 49.84 & 216.9 & OCT. 30.4 & 0.193   \\
H+    & 159.8 & -9.90  & 64.410 & 49.80 & 208.4 & OCT. 21.9 & 0.118   \\
P+    & 159.2 & -7.50  & 65.240 & 49.80 & 207.9 & OCT. 21.3 & 0.150  
\\
\hline
\end{tabular}
\caption[Theoretical radiant of ʻOumuamuids]{Theoretical Radiant of ʻOumuamuids using the orbit from Table \ref{tab:Oumparamval} and the technique of \cite{Nesl1998}. We used the B+ method.}
\label{tab:Oumuarad}
\end{table}

\subsection{Monte Carlo Simulation Results}

\begin{figure}[H]
\begin{center}
\includegraphics[width=11cm]{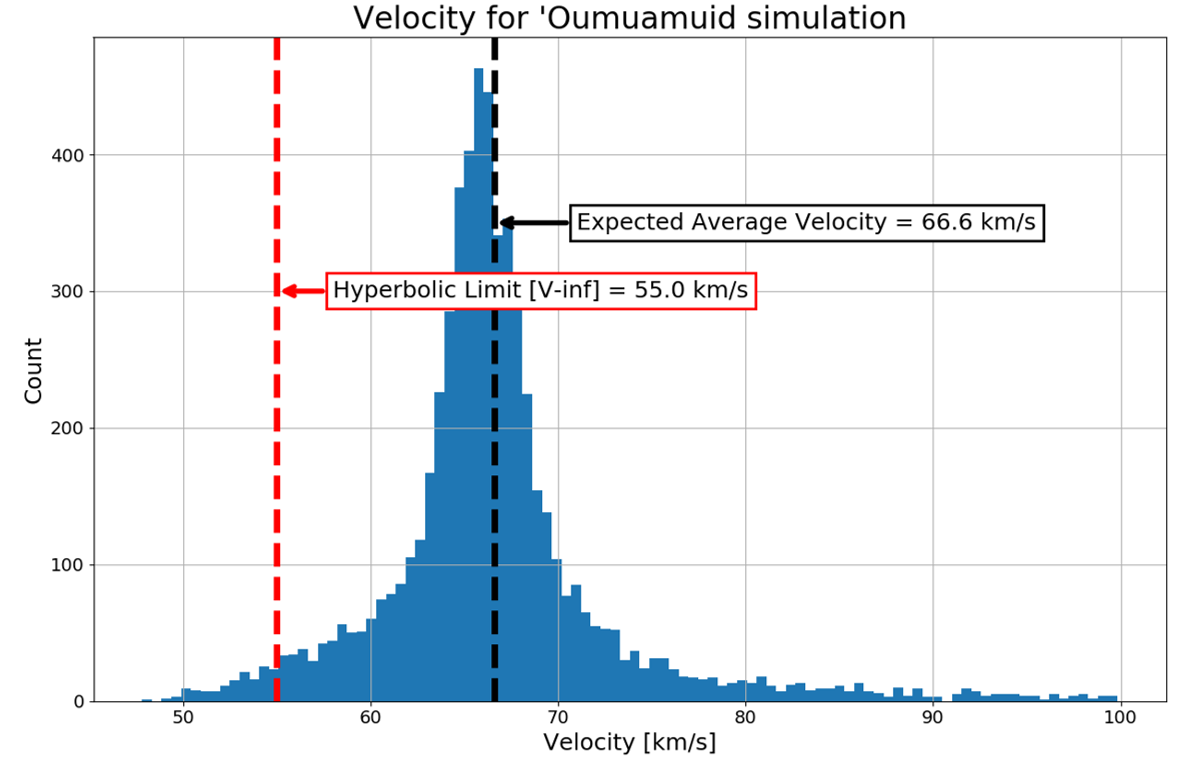}
\caption[Eccentricities of simulations of ʻOumuamuid shower]{Histogram of synthetically measured eccentricities based on /$\approx$ 6000 simulated hypothetical ʻOumuamuid shower meteor echoes.}
\label{OumuamuaVel}
\end{center}
\end{figure}

To test CMOR detectability for 1I-related meteoroids, we used the theoretical radiant and a Monte Carlo simulation of $\approx$6000 meteors from this radiant at the expected peak date (Oct 18) to generate synthetic echoes. The time pick uncertainty and interferometry errors were chosen in the same manner as described in Section \ref{ISMonteCarlo}. We found that the expected average top of atmosphere velocity of the ʻOumuamuids to be 11.6 km/s above the hyperbolic limit (Figure \ref{OumuamuaVel}) which produced a 1.8$\sigma$ average detection given CMORs speed uncertainty for this radiant. That is, among all the many echoes detected by CMOR from a 1I-type stream, on the average they would be 1.8$\sigma$ hyperbolic. Such a hyperbolic radiant would easily be detectable by CMOR. Only 211 of the simulated meteors (3.6\% of the retained meteors) produced orbits which appeared to have eccentricities of \textless 1. 

\begin{figure}[H]
\begin{center}
\includegraphics[width=11cm]{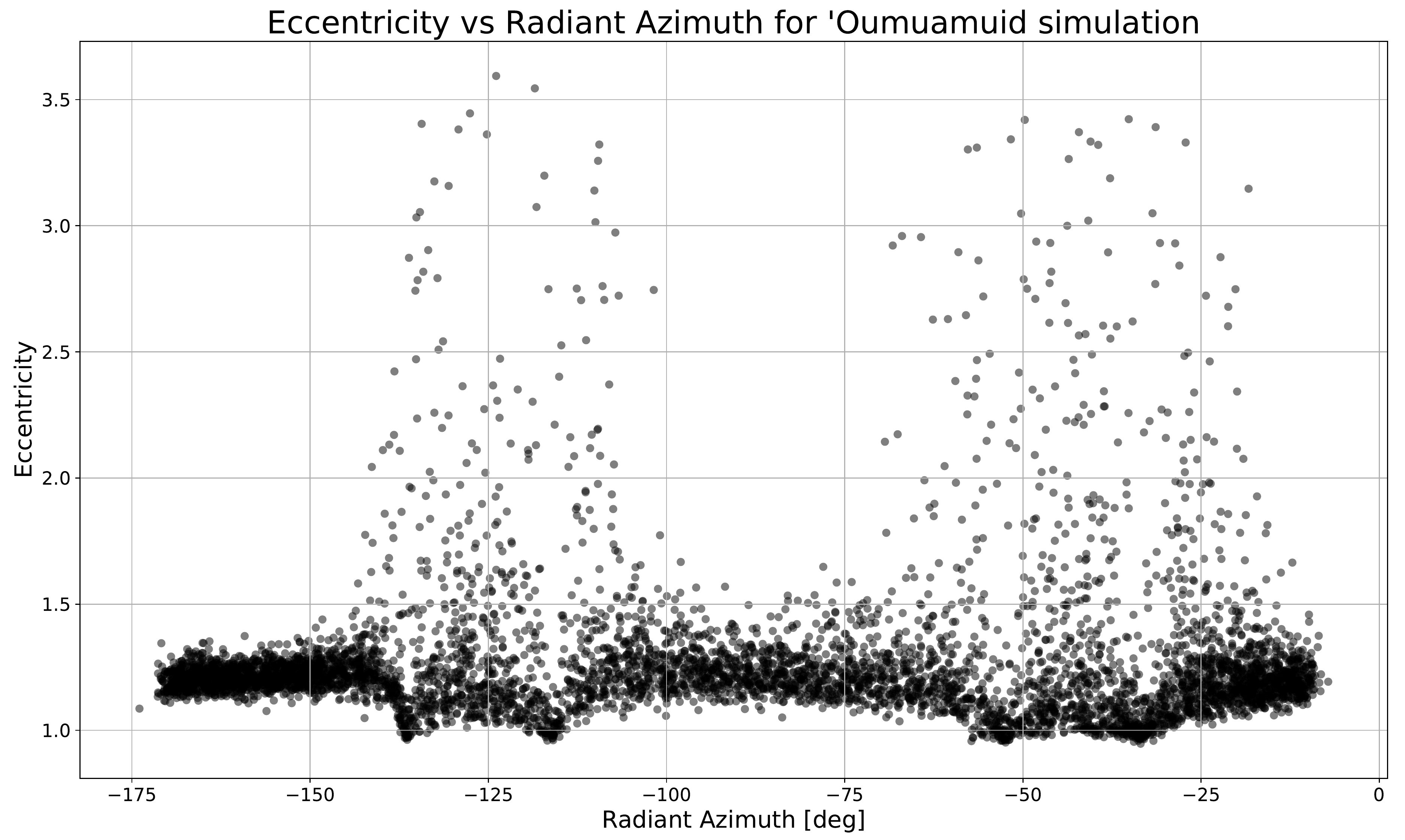}
\caption[Eccentricity vs $\eta$ of ʻOumuamuid shower]{Eccentricity versus Radiant Azimuth based on $\approx$ 6000 simulated hypothetical ʻOumuamuid shower meteor echoes.}
\label{OumuamuaAzi}
\end{center}
\end{figure}

\begin{figure}[H]
\begin{center}
\includegraphics[width=11cm]{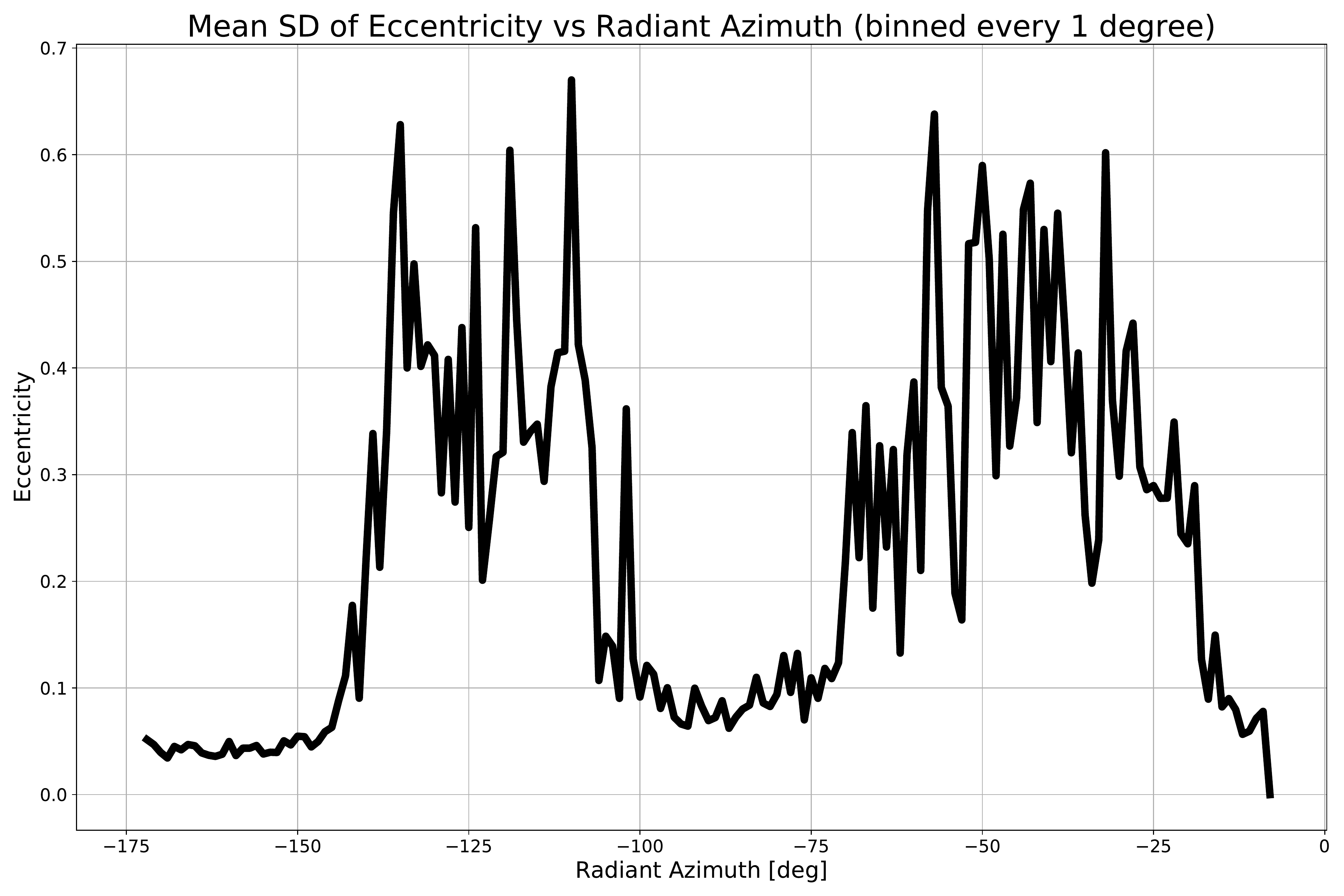}
\caption[Standard deviation of eccentricity vs $\eta$ of simulated ʻOumuamuid orbits]{Mean standard deviation of eccentricity versus radiant azimuth based on $\approx$ 6000 simulated hypothetical ʻOumuamuid shower meteor echoes.}
\label{OumuamuaEstd}
\end{center}
\end{figure}

As the radiant azimuth affects the geometry of detected echoes, we expect the precision for the velocity to vary with azimuth. We see that for most radiant azimuth's the eccentricity is largely consistent with the expected theoretical result $\approx$1.2 (Figure \ref{OumuamuaAzi}) which is the approximate orbital eccentricity of ʻOumuamua. However, at certain radiant azimuths, the eccentricity values have a large variance, reflecting poor relative geometry for CMOR detection. Examination of the standard deviation of the simulated eccentricities within 1 degree bins for each radiant azimuth shows low standard deviations (approximately \textless 0.1) occurs between radiant azimuths of 0 to -17, -74 to -103 and -141 to -180 degrees (Figure \ref{OumuamuaEstd}). High deviations in eccentricity occur at the remaining radiant azimuths where particular geometric configurations for which CMOR is less precise in time of flight speed estimation. 

Overall, a 1I-like stream would be resolvable and detectable on average at the 1.8$\sigma$ level, but it is important to recognize that a short lived outburst could have better (or worse) velocity precision depending on radiant geometry. This demonstrates that knowing the instrumental sensitivity for CMOR as a function of local radiant location, one can fine-tune such a search and focus on directions for which CMOR has maximum velocity precision, as any real interstellar meteoroid stream shows a wide variation in local radiant direction \citep{Baggaley2002}. An examination of CMOR data collected on and about the days of expected peak activity was performed to search for any evidence of ʻOumuamua associated meteors - none were observed

It should be noted that as this current study was being finalised, the Minor Planet Center\footnote{https://minorplanetcenter.net/mpec/K19/K19S72.html} assigned the numbered desigation 2I/Borisov to a second interstellar object.  While the perihelion distance of 2I is much larger than ʻOumuamua at 2.0 au, it has a very cometary apperance, suggesting that meteoroid streams may indeed exist from interstellar comets, though 2I/Borisov meteoroids would not be detectable at Earth.

\section{Conclusion}

We used the Canadian Meteor Orbit Radar to detect in excess of 160000 six station meteor echoes over a 7.5 year timespan in a search for potential interstellar meteoroids. Our integrated collecting area time product for the survey was $\approx$ 7$\times$10$^6$ km$^2$ hours. 

Among this sample, we found five candidate interstellar meteoroids with eccentricities $\textgreater$2$\sigma$ above the hyperbolic limit based purely on their in-atmosphere speeds, by restricting our search to high-quality, 6-station events with low velocity errors. We calculated orbits using raw, measured time of flight velocities to isolate potential interstellar candidates. We also performed Monte Carlo simulations to establish likely uncertainty distributions in the orbits using average errors in time picks and interferometry based on measured CMOR data. These events had raw velocities which produced equivalent eccentricities more than 2$\sigma$ above the hyperbolic limit. 

An improved velocity correction for atmospheric deceleration for CMOR was developed and applied to these candidate interstellar meteoroids, increasing the likelihood of e $\textgreater$ 1 to $\textgreater$ 3$\sigma$ for all five meteors,  the highest value being 3.7$\sigma$. Assuming all five hyperbolic events were true interstellar meteoroids, we derive an equivalent lower bound to the flux of 6.6$\times 10^{-7}$ meteoroids km$^{-2}$ hour$^{-1}$ to a limiting mass of 10$^{-7}$ kg.

These five events have nominal hyperbolic excess speeds relative to the solar system of $\approx$ 20km/s, which is in the range expected for ISPs arriving from local stellar sources at Earth \citep{Murray2004}. We note that the theoretical analysis by \citet{Murray2004} also suggests there may be many more interstellar meteoroids with lower speeds, which our analysis may not identify. However, the five detections are all in the 3-4$\sigma$ range after applying atmospheric deceleration correction. This significance is similar to the best candidates found by \citet{Musci2012} and \citet{Weryk2005}. Given the large number of events examined, this suggests to us the most likely explanation in these cases is simple measurement error, though we cannot rule out true interstellar origins for these events at the significance levels quoted.

A simulation of hypothetical meteoroids associated with 1I/ʻOumuamua confirmed that CMOR would detect such a stream with an average significance of 1.8$\sigma$ above the hyperbolic limit, but with noticeable variations in velocity precision as a function of radiant direction. Examination of CMOR data on expected days of ʻOumuamuid’ peak activity did not yield any detectable hyperbolic meteoroids. 

Further examination of specific radar-meteor radiant geometries (as detailed in Section \ref{simcandi}) which may be favourable to detecting interstellar meteoroids may be useful in adding confidence to future candidate detections. 

\section*{Acknowledgements}

This work was supported in part by the NASA Meteoroid Environment Office under cooperative agreement 80NSSC18M0046. PGB also acknowledges funding support from the Natural Sciences and Engineering Research council of Canada and the Canada Research Chairs program. We thank Denis Vida for providing model estimates for the deceleration of CMOR-sized meteoroids for comparison to our empirical values. An earlier version of this manuscript was improved considerably through comments of two anonymous referees.

Datasets used for production of this work are available at: 
\newline
http://dx.doi.org/10.17632/77ff732y5m.1

\clearpage
\newpage

\bibliography{references.bib}

\begin{thebibliography}{64}
\expandafter\ifx\csname natexlab\endcsname\relax\def\natexlab#1{#1}\fi
\providecommand{\bibinfo}[2]{#2}
\ifx\xfnm\relax \def\xfnm[#1]{\unskip,\space#1}\fi
\bibitem[{Altobelli(2003)}]{Altobelli2003}
\bibinfo{author}{Altobelli, N.} (\bibinfo{year}{2003}).
\newblock \bibinfo{title}{{Cassini between Venus and Earth: Detection of
  interstellar dust}}.
\newblock {\it \bibinfo{journal}{Journal of Geophysical Research}\/},  {\it
  \bibinfo{volume}{108}\/}, \bibinfo{pages}{LIS 7--1--LIS 7--9}.
\bibitem[{Baggaley(1994)}]{Baggaley1994}
\bibinfo{author}{Baggaley, W.} (\bibinfo{year}{1994}).
\newblock \bibinfo{title}{{The Advanced Meteor Orbit Radar Facility: AMOR}}.
\newblock {\it \bibinfo{journal}{Quarterly Journal of the Royal Astronomical
  Society}\/},  {\it \bibinfo{volume}{35}\/}, \bibinfo{pages}{293--320}.
\bibitem[{Baggaley \& Nesluan(2002)}]{Baggaley2002}
\bibinfo{author}{Baggaley, W.}, \& \bibinfo{author}{Nesluan, L.}
  (\bibinfo{year}{2002}).
\newblock \bibinfo{title}{{A model of the heliocentric orbits of a stream of
  Earth-impacting interstellar meteoroids}}.
\newblock {\it \bibinfo{journal}{Astronomy and Astrophysics}\/},  {\it
  \bibinfo{volume}{382}\/}, \bibinfo{pages}{1118--1124}.
\bibitem[{Baggaley(2000)}]{Baggaley2000}
\bibinfo{author}{Baggaley, W.~J.} (\bibinfo{year}{2000}).
\newblock \bibinfo{title}{Advanced meteor orbit radar observations of
  interstellar meteoroids}.
\newblock {\it \bibinfo{journal}{Journal of Geophysical Research: Space
  Physics}\/},  {\it \bibinfo{volume}{105}\/}, \bibinfo{pages}{10353--10361}.
\bibitem[{Baguhl et~al.(1995)Baguhl, Gr\"{u}n, Hamilton, Linkert, Riemann,
  Staubach \& Zook}]{Baguhl1995}
\bibinfo{author}{Baguhl, M.}, \bibinfo{author}{Gr\"{u}n, E.},
  \bibinfo{author}{Hamilton, D.~P.}, \bibinfo{author}{Linkert, G.},
  \bibinfo{author}{Riemann, R.}, \bibinfo{author}{Staubach, P.}, \&
  \bibinfo{author}{Zook, H.~A.} (\bibinfo{year}{1995}).
\newblock \bibinfo{title}{The flux of interstellar dust observed by ulysses and
  galileo}.
\newblock {\it \bibinfo{journal}{Space Science Reviews}\/},  {\it
  \bibinfo{volume}{72}\/}, \bibinfo{pages}{471--476}.
\bibitem[{Baguhl et~al.(1996)Baguhl, Gr\"{u}n \& Landgraf}]{Baguhl1996}
\bibinfo{author}{Baguhl, M.}, \bibinfo{author}{Gr\"{u}n, E.}, \&
  \bibinfo{author}{Landgraf, M.} (\bibinfo{year}{1996}).
\newblock \bibinfo{title}{In situ measurements of interstellar dust with the
  ulysses and galileo spaceprobes}.
\newblock {\it \bibinfo{journal}{Space Science Reviews}\/},  {\it
  \bibinfo{volume}{78}\/}, \bibinfo{pages}{165--172}.
\bibitem[{Brown et~al.(2005{\natexlab{a}})Brown, Jones, Weryk \&
  Campbell-Brown}]{Brown2005}
\bibinfo{author}{Brown, P.}, \bibinfo{author}{Jones, J.},
  \bibinfo{author}{Weryk, R.}, \& \bibinfo{author}{Campbell-Brown, M.~D.}
  (\bibinfo{year}{2005}{\natexlab{a}}).
\newblock \bibinfo{title}{{The velocity distribution of meteoroids at the Earth
  as measured by the Canadian Meteor Orbit Radar (CMOR)}}.
\newblock {\it \bibinfo{journal}{Earth, Moon and Planets}\/},  (pp.
  \bibinfo{pages}{617--626}).
\bibitem[{Brown et~al.(2005{\natexlab{b}})Brown, Jones, Weryk \&
  Campbell-Brown}]{Browna}
\bibinfo{author}{Brown, P.}, \bibinfo{author}{Jones, J.},
  \bibinfo{author}{Weryk, R.~J.}, \& \bibinfo{author}{Campbell-Brown, M.~D.}
  (\bibinfo{year}{2005}{\natexlab{b}}).
\newblock \bibinfo{title}{The velocity distribution of meteoroids at the earth
  as measured by the canadian meteor orbit radar ({CMOR})}.
\newblock In {\it \bibinfo{booktitle}{Modern Meteor Science An
  Interdisciplinary View}\/} (pp. \bibinfo{pages}{617--626}).
\newblock \bibinfo{publisher}{Springer-Verlag}.
\bibitem[{Brown et~al.(2008)Brown, Weryk, Wong \& Jones}]{Brown2008}
\bibinfo{author}{Brown, P.}, \bibinfo{author}{Weryk, R.},
  \bibinfo{author}{Wong, D.}, \& \bibinfo{author}{Jones, J.}
  (\bibinfo{year}{2008}).
\newblock \bibinfo{title}{A meteoroid stream survey using the canadian meteor
  orbit radar}.
\newblock {\it \bibinfo{journal}{Icarus}\/},  {\it \bibinfo{volume}{195}\/},
  \bibinfo{pages}{317--339}.
\bibitem[{Brown et~al.(2010)Brown, Wong, Weryk \& Wiegert}]{Brown2010}
\bibinfo{author}{Brown, P.}, \bibinfo{author}{Wong, D.},
  \bibinfo{author}{Weryk, R.}, \& \bibinfo{author}{Wiegert, P.}
  (\bibinfo{year}{2010}).
\newblock \bibinfo{title}{A meteoroid stream survey using the canadian meteor
  orbit radar}.
\newblock {\it \bibinfo{journal}{Icarus}\/},  {\it \bibinfo{volume}{207}\/},
  \bibinfo{pages}{66--81}.
\bibitem[{Campbell-Brown \& Jones(2006)}]{CampbellBrown2006}
\bibinfo{author}{Campbell-Brown, M.~D.}, \& \bibinfo{author}{Jones, J.}
  (\bibinfo{year}{2006}).
\newblock \bibinfo{title}{Annual variation of sporadic radar meteor rates}.
\newblock {\it \bibinfo{journal}{Monthly Notices of the Royal Astronomical
  Society}\/},  {\it \bibinfo{volume}{367}\/}, \bibinfo{pages}{709--716}.
\bibitem[{Ceplecha(1968)}]{ceplecha68}
\bibinfo{author}{Ceplecha, Z.} (\bibinfo{year}{1968}).
\newblock \bibinfo{title}{{Discrete levels of meteor beginning height}}.
\newblock {\it \bibinfo{journal}{SAO Special Report {\#}279}\/},  {\it
  \bibinfo{volume}{279}\/}, \bibinfo{pages}{1--52}.
\bibitem[{Ceplecha et~al.(1998)Ceplecha, Borovi{\v{c}}ka, Elford, ReVelle,
  Hawkes, Porub{\v{c}}an \& {\v{S}}imek}]{Ceplecha1998}
\bibinfo{author}{Ceplecha, Z.}, \bibinfo{author}{Borovi{\v{c}}ka, J.},
  \bibinfo{author}{Elford, W.~G.}, \bibinfo{author}{ReVelle, D.~O.},
  \bibinfo{author}{Hawkes, R.~L.}, \bibinfo{author}{Porub{\v{c}}an, V.}, \&
  \bibinfo{author}{{\v{S}}imek, M.} (\bibinfo{year}{1998}).
\newblock \bibinfo{title}{Meteor phenomena and bodies}.
\newblock {\it \bibinfo{journal}{Space Science Reviews}\/},  {\it
  \bibinfo{volume}{84}\/}, \bibinfo{pages}{327--471}.
\bibitem[{Frisch et~al.(1999)Frisch, Dorschner, Landgraf, Hoppe, Jones, Slavin,
  Witt \& Zank}]{Frisch1999}
\bibinfo{author}{Frisch, P.~C.}, \bibinfo{author}{Dorschner, J.~M.},
  \bibinfo{author}{Landgraf, M.}, \bibinfo{author}{Hoppe, P.},
  \bibinfo{author}{Jones, A.~P.}, \bibinfo{author}{Slavin, J.~D.},
  \bibinfo{author}{Witt, A.~N.}, \& \bibinfo{author}{Zank, G.~P.}
  (\bibinfo{year}{1999}).
\newblock \bibinfo{title}{{Dust in the local interstellar wind}}.
\newblock {\it \bibinfo{journal}{Astrophysical Journal}\/},  {\it
  \bibinfo{volume}{525}\/}, \bibinfo{pages}{492--516}.
\bibitem[{Galligan \& Baggaley(2002)}]{Galligan2002}
\bibinfo{author}{Galligan, D.}, \& \bibinfo{author}{Baggaley, W.}
  (\bibinfo{year}{2002}).
\newblock \bibinfo{title}{Wavelet enhancement for detecting shower structure in
  radar meteoroid orbit data. {II}. application to the {AMOR} data set}.
\newblock In {\it \bibinfo{booktitle}{Dust in the Solar System and other
  Planetary Systems, Proceedings of the {IA} U Colloquium 181 held at the
  University of Kent}\/} (pp. \bibinfo{pages}{48--60}).
\newblock \bibinfo{publisher}{Elsevier}.
\bibitem[{{Gr\"{u}n} et~al.(1994){Gr\"{u}n}, {Gustafson}, {Mann}, {Baguhl},
  {Morfill}, {Staubach}, {Taylor} \& {Zook}}]{Grn1994}
\bibinfo{author}{{Gr\"{u}n}, E.}, \bibinfo{author}{{Gustafson}, B.},
  \bibinfo{author}{{Mann}, I.}, \bibinfo{author}{{Baguhl}, M.},
  \bibinfo{author}{{Morfill}, G.~E.}, \bibinfo{author}{{Staubach}, P.},
  \bibinfo{author}{{Taylor}, A.}, \& \bibinfo{author}{{Zook}, H.~A.}
  (\bibinfo{year}{1994}).
\newblock \bibinfo{title}{{Interstellar dust in the heliosphere}}.
\newblock {\it \bibinfo{journal}{Astronomy and Astrophysics}\/},  {\it
  \bibinfo{volume}{286}\/}, \bibinfo{pages}{915--924}.
\bibitem[{Gr\"{u}n et~al.(2000)Gr\"{u}n, Landgraf, Hor{\'{a}}nyi, Kissel,
  Kr\"{u}ger, Srama, Svedhem \& Withnell}]{Grn2000}
\bibinfo{author}{Gr\"{u}n, E.}, \bibinfo{author}{Landgraf, M.},
  \bibinfo{author}{Hor{\'{a}}nyi, M.}, \bibinfo{author}{Kissel, J.},
  \bibinfo{author}{Kr\"{u}ger, H.}, \bibinfo{author}{Srama, R.},
  \bibinfo{author}{Svedhem, H.}, \& \bibinfo{author}{Withnell, P.}
  (\bibinfo{year}{2000}).
\newblock \bibinfo{title}{Techniques for galactic dust measurements in the
  heliosphere}.
\newblock {\it \bibinfo{journal}{Journal of Geophysical Research: Space
  Physics}\/},  {\it \bibinfo{volume}{105}\/}, \bibinfo{pages}{10403--10410}.
\bibitem[{Gr\"{u}n et~al.(1997)Gr\"{u}n, Staubach, Baguhl, Hamilton, Zook,
  Dermott, Gustafson, Fechtig, Kissel, Linkert, Linkert, Srama, Hanner,
  Polanskey, Horanyi, Lindblad, Mann, McDonnell, Morfill \& Schwehm}]{Grn1997}
\bibinfo{author}{Gr\"{u}n, E.}, \bibinfo{author}{Staubach, P.},
  \bibinfo{author}{Baguhl, M.}, \bibinfo{author}{Hamilton, D.},
  \bibinfo{author}{Zook, H.}, \bibinfo{author}{Dermott, S.},
  \bibinfo{author}{Gustafson, B.}, \bibinfo{author}{Fechtig, H.},
  \bibinfo{author}{Kissel, J.}, \bibinfo{author}{Linkert, D.},
  \bibinfo{author}{Linkert, G.}, \bibinfo{author}{Srama, R.},
  \bibinfo{author}{Hanner, M.}, \bibinfo{author}{Polanskey, C.},
  \bibinfo{author}{Horanyi, M.}, \bibinfo{author}{Lindblad, B.},
  \bibinfo{author}{Mann, I.}, \bibinfo{author}{McDonnell, J.~A.},
  \bibinfo{author}{Morfill, G.}, \& \bibinfo{author}{Schwehm, G.}
  (\bibinfo{year}{1997}).
\newblock \bibinfo{title}{South{\textendash}north and radial traverses through
  the interplanetary dust cloud}.
\newblock {\it \bibinfo{journal}{Icarus}\/},  {\it \bibinfo{volume}{129}\/},
  \bibinfo{pages}{270--288}.
\bibitem[{Gr\"{u}n et~al.(1993)Gr\"{u}n, Zook, Baguhl, Balogh, Bame, Fechtig,
  Forsyth, Manner, Horanyi, Kissel, Lindblad, Linkert, Linkert, Mann,
  McDonnell, Morfill, Phillips, Polanskey, Schwehm, Siddique, Staubach, Svestka
  \& Taylor}]{Grn1993}
\bibinfo{author}{Gr\"{u}n, E.}, \bibinfo{author}{Zook, H.~A.},
  \bibinfo{author}{Baguhl, M.}, \bibinfo{author}{Balogh, A.},
  \bibinfo{author}{Bame, S.~J.}, \bibinfo{author}{Fechtig, H.},
  \bibinfo{author}{Forsyth, R.}, \bibinfo{author}{Manner, M.~S.},
  \bibinfo{author}{Horanyi, M.}, \bibinfo{author}{Kissel, J.},
  \bibinfo{author}{Lindblad, B.-A.}, \bibinfo{author}{Linkert, D.},
  \bibinfo{author}{Linkert, G.}, \bibinfo{author}{Mann, I.},
  \bibinfo{author}{McDonnell, J. A.~M.}, \bibinfo{author}{Morfill, G.~E.},
  \bibinfo{author}{Phillips, J.~L.}, \bibinfo{author}{Polanskey, C.},
  \bibinfo{author}{Schwehm, G.}, \bibinfo{author}{Siddique, N.},
  \bibinfo{author}{Staubach, P.}, \bibinfo{author}{Svestka, J.}, \&
  \bibinfo{author}{Taylor, A.} (\bibinfo{year}{1993}).
\newblock \bibinfo{title}{Discovery of jovian dust streams and interstellar
  grains by the ulysses spacecraft}.
\newblock {\it \bibinfo{journal}{Nature}\/},  {\it \bibinfo{volume}{362}\/},
  \bibinfo{pages}{428--430}.
\bibitem[{Hajduk(2001)}]{Hajduk2001}
\bibinfo{author}{Hajduk, A.} (\bibinfo{year}{2001}).
\newblock \bibinfo{title}{{On the very high velocity meteors}}.
\newblock In \bibinfo{editor}{B.~Warmbein.} (Ed.), {\it
  \bibinfo{booktitle}{Proceedings of the Meteoroids 2001 Conference, 6 - 10
  August 2001, Kiruna, Sweden}\/} (pp. \bibinfo{pages}{557 -- 559}).
\newblock \bibinfo{publisher}{ESA Sp-495}.
\bibitem[{Hajdukov{\'{a}}(2012)}]{hajdukova2012}
\bibinfo{author}{Hajdukov{\'{a}}, M.} (\bibinfo{year}{2012}).
\newblock \bibinfo{title}{{Population of hyperbolic meteoroids}}.
\newblock In {\it \bibinfo{booktitle}{Proceedings of the IMC}\/}
  \bibinfo{number}{1993} (pp. \bibinfo{pages}{98--104}).
\bibitem[{Hajdukov{\'{a}} et~al.(2013)Hajdukov{\'{a}}, Korno{\v{s}} \&
  T{\'{o}}th}]{Hajdukov2013}
\bibinfo{author}{Hajdukov{\'{a}}, M.}, \bibinfo{author}{Korno{\v{s}}, L.}, \&
  \bibinfo{author}{T{\'{o}}th, J.} (\bibinfo{year}{2013}).
\newblock \bibinfo{title}{Frequency of hyperbolic and interstellar meteoroids}.
\newblock {\it \bibinfo{journal}{Meteoritics {\&} Planetary Science}\/},  {\it
  \bibinfo{volume}{49}\/}, \bibinfo{pages}{63--68}.
\bibitem[{{Hajdukov{\'a}}(1994)}]{Hajdukova1994}
\bibinfo{author}{{Hajdukov{\'a}}, M., Jr.} (\bibinfo{year}{1994}).
\newblock \bibinfo{title}{{On the frequency of interstellar meteoroids}}.
\newblock {\it \bibinfo{journal}{Astronomy and Astrophysics}\/},  {\it
  \bibinfo{volume}{288}\/}, \bibinfo{pages}{330--334}.
\bibitem[{Hajdukov{\'a} \& Paulech(2002)}]{Hajdukova2002}
\bibinfo{author}{Hajdukov{\'a}, M., Jr.}, \& \bibinfo{author}{Paulech, T.}
  (\bibinfo{year}{2002}).
\newblock \bibinfo{title}{Interstellar and interplanetary meteoroid flux from
  updated iau mdc data}.
\newblock In \bibinfo{editor}{B.~{Warmbein}} (Ed.), {\it
  \bibinfo{booktitle}{Asteroids, Comets, and Meteors: ACM 2002}\/} (pp.
  \bibinfo{pages}{173--176}).
\newblock volume \bibinfo{volume}{500} of {\it \bibinfo{series}{ESA Special
  Publication}\/}.
\bibitem[{Hawkes et~al.(1999)Hawkes, Close \& S.~Woodworth}]{Hawkes1999}
\bibinfo{author}{Hawkes, L.}, \bibinfo{author}{Close, T.}, \&
  \bibinfo{author}{S.~Woodworth, S.} (\bibinfo{year}{1999}).
\newblock \bibinfo{title}{Proc. int. conf., meteoroids 1998}.
\newblock Bratislava: Astronomical Institute of the Slovak Academy of Sciences
  (p. \bibinfo{pages}{257}).
\bibitem[{Hawkes \& Jones(1975)}]{Hawkes1975}
\bibinfo{author}{Hawkes, R.~L.}, \& \bibinfo{author}{Jones, J.}
  (\bibinfo{year}{1975}).
\newblock \bibinfo{title}{A quantitative model for the ablation of dustball
  meteors}.
\newblock {\it \bibinfo{journal}{Monthly Notices of the Royal Astronomical
  Society}\/},  {\it \bibinfo{volume}{173}\/}, \bibinfo{pages}{339--356}.
\bibitem[{Jacchia \& Whipple(1961)}]{Jacchia1961}
\bibinfo{author}{Jacchia, L.~G.}, \& \bibinfo{author}{Whipple, F.~L.}
  (\bibinfo{year}{1961}).
\newblock \bibinfo{title}{Precision orbits of 413 photographic meteors}.
\newblock {\it \bibinfo{journal}{Smithsonian Contributions to Astrophysics}\/},
   {\it \bibinfo{volume}{4}\/}, \bibinfo{pages}{97--129}.
\bibitem[{Jenniskens et~al.(2016)Jenniskens, N{\'{e}}non, Albers, Gural,
  Haberman, Holman, Morales, Grigsby, Samuels \& Johannink}]{Jenniskens2016}
\bibinfo{author}{Jenniskens, P.}, \bibinfo{author}{N{\'{e}}non, Q.},
  \bibinfo{author}{Albers, J.}, \bibinfo{author}{Gural, P.},
  \bibinfo{author}{Haberman, B.}, \bibinfo{author}{Holman, D.},
  \bibinfo{author}{Morales, R.}, \bibinfo{author}{Grigsby, B.},
  \bibinfo{author}{Samuels, D.}, \& \bibinfo{author}{Johannink, C.}
  (\bibinfo{year}{2016}).
\newblock \bibinfo{title}{The established meteor showers as observed by
  {CAMS}}.
\newblock {\it \bibinfo{journal}{Icarus}\/},  {\it \bibinfo{volume}{266}\/},
  \bibinfo{pages}{331--354}.
\bibitem[{Jewitt et~al.(2017)Jewitt, Luu, Rajagopal, Kotulla, Ridgway, Liu \&
  Augusteijn}]{Jewitt2017}
\bibinfo{author}{Jewitt, D.}, \bibinfo{author}{Luu, J.},
  \bibinfo{author}{Rajagopal, J.}, \bibinfo{author}{Kotulla, R.},
  \bibinfo{author}{Ridgway, S.}, \bibinfo{author}{Liu, W.}, \&
  \bibinfo{author}{Augusteijn, T.} (\bibinfo{year}{2017}).
\newblock \bibinfo{title}{Interstellar interloper 1i/2017 u1: Observations from
  the {NOT} and {WIYN} telescopes}.
\newblock {\it \bibinfo{journal}{The Astrophysical Journal}\/},  {\it
  \bibinfo{volume}{850}\/}, \bibinfo{pages}{L36}.
\bibitem[{Jones et~al.(2005)Jones, Brown, Ellis, Webster, Campbell-Brown,
  Krzemenski \& Weryk}]{Jones2005}
\bibinfo{author}{Jones, J.}, \bibinfo{author}{Brown, P.},
  \bibinfo{author}{Ellis, K.}, \bibinfo{author}{Webster, A.},
  \bibinfo{author}{Campbell-Brown, M.}, \bibinfo{author}{Krzemenski, Z.}, \&
  \bibinfo{author}{Weryk, R.} (\bibinfo{year}{2005}).
\newblock \bibinfo{title}{The canadian meteor orbit radar: system overview and
  preliminary results}.
\newblock {\it \bibinfo{journal}{Planetary and Space Science}\/},  {\it
  \bibinfo{volume}{53}\/}, \bibinfo{pages}{413--421}.
\bibitem[{Jones et~al.(1998)Jones, Webster \& Hocking}]{Jones1998}
\bibinfo{author}{Jones, J.}, \bibinfo{author}{Webster, A.~R.}, \&
  \bibinfo{author}{Hocking, W.~K.} (\bibinfo{year}{1998}).
\newblock \bibinfo{title}{An improved interferometer design for use with meteor
  radars}.
\newblock {\it \bibinfo{journal}{Radio Science}\/},  {\it
  \bibinfo{volume}{33}\/}, \bibinfo{pages}{55--65}.
\bibitem[{Kikwaya et~al.(2011)Kikwaya, Campbell-Brown \& Brown}]{Kikwaya2011}
\bibinfo{author}{Kikwaya, J.-B.}, \bibinfo{author}{Campbell-Brown, M.}, \&
  \bibinfo{author}{Brown, P.~G.} (\bibinfo{year}{2011}).
\newblock \bibinfo{title}{Bulk density of small meteoroids}.
\newblock {\it \bibinfo{journal}{Astronomy {\&} Astrophysics}\/},  {\it
  \bibinfo{volume}{530}\/}, \bibinfo{pages}{A113}.
\bibitem[{Koten et~al.(2004)Koten, Borovi{\v{c}}ka, Spurn{\'{y}}, Betlem \&
  Evans}]{Koten2004}
\bibinfo{author}{Koten, P.}, \bibinfo{author}{Borovi{\v{c}}ka, J.},
  \bibinfo{author}{Spurn{\'{y}}, P.}, \bibinfo{author}{Betlem, H.}, \&
  \bibinfo{author}{Evans, S.} (\bibinfo{year}{2004}).
\newblock \bibinfo{title}{Atmospheric trajectories and light curves of shower
  meteors}.
\newblock {\it \bibinfo{journal}{Astronomy {\&} Astrophysics}\/},  {\it
  \bibinfo{volume}{428}\/}, \bibinfo{pages}{683--690}.
\bibitem[{Koten et~al.(2007)Koten, Borovi{\v{c}}ka, Spurn{\'{y}} \&
  Stork}]{Koten2007}
\bibinfo{author}{Koten, P.}, \bibinfo{author}{Borovi{\v{c}}ka, J.},
  \bibinfo{author}{Spurn{\'{y}}, P.}, \& \bibinfo{author}{Stork, R.}
  (\bibinfo{year}{2007}).
\newblock \bibinfo{title}{{Optical observations of enhanced activity of the
  2005 Draconid meteor shower}}.
\newblock {\it \bibinfo{journal}{Astronomy and Astrophysics}\/},  {\it
  \bibinfo{volume}{466}\/}, \bibinfo{pages}{729--735}.
\bibitem[{Landgraf et~al.(2000)Landgraf, Baggaley, Gr\"{u}n, Kr\"{u}ger \&
  Linkert}]{Landgraf2000}
\bibinfo{author}{Landgraf, M.}, \bibinfo{author}{Baggaley, W.~J.},
  \bibinfo{author}{Gr\"{u}n, E.}, \bibinfo{author}{Kr\"{u}ger, H.}, \&
  \bibinfo{author}{Linkert, G.} (\bibinfo{year}{2000}).
\newblock \bibinfo{title}{Aspects of the mass distribution of interstellar dust
  grains in the solar system from in situ measurements}.
\newblock {\it \bibinfo{journal}{Journal of Geophysical Research: Space
  Physics}\/},  {\it \bibinfo{volume}{105}\/}, \bibinfo{pages}{10343--10352}.
\bibitem[{Landgraf \& Gr\"{u}n(1998)}]{Landgraf}
\bibinfo{author}{Landgraf, M.}, \& \bibinfo{author}{Gr\"{u}n, E.}
  (\bibinfo{year}{1998}).
\newblock \bibinfo{title}{In situ measurements of interstellar dust}.
\newblock In {\it \bibinfo{booktitle}{The Local Bubble and Beyond
  Lyman-Spitzer-Colloquium}\/} (pp. \bibinfo{pages}{381--384}).
\newblock \bibinfo{publisher}{Springer Berlin Heidelberg}.
\bibitem[{Mathews et~al.(1999)Mathews, Meisel, Janches, Getman \&
  Zhou}]{Mathews1998}
\bibinfo{author}{Mathews, D.~J.}, \bibinfo{author}{Meisel, D.~D.},
  \bibinfo{author}{Janches, D.}, \bibinfo{author}{Getman, S.~V.}, \&
  \bibinfo{author}{Zhou, Q.-H.} (\bibinfo{year}{1999}).
\newblock \bibinfo{title}{Proc. int. conf., meteoroids 1998}.
\newblock Bratislava: Astronomical Institute of the Slovak Academy of Sciences
  (p.~\bibinfo{pages}{79}).
\bibitem[{Mazur et~al.(2019)Mazur, Pokorn\'{y}, Weryk, Brown, Vida, Schult,
  Stober \& Agrawal}]{mazur2019}
\bibinfo{author}{Mazur, M.}, \bibinfo{author}{Pokorn\'{y}, P.},
  \bibinfo{author}{Weryk, R.~J.}, \bibinfo{author}{Brown, P.},
  \bibinfo{author}{Vida, D.}, \bibinfo{author}{Schult, C.},
  \bibinfo{author}{Stober, G.}, \& \bibinfo{author}{Agrawal, A.}
  (\bibinfo{year}{2019}).
\newblock \bibinfo{title}{An algorithm for measurement of radar transverse
  scattering meteor echo speeds using pre-$t_{0}$ phases}.
\newblock {\it \bibinfo{journal}{Radio Science}\/},  {\it
  \bibinfo{volume}{submitted}\/}.
\bibitem[{McNeill et~al.(2018)McNeill, Trilling \& Mommert}]{McNeill2018}
\bibinfo{author}{McNeill, A.}, \bibinfo{author}{Trilling, D.~E.}, \&
  \bibinfo{author}{Mommert, M.} (\bibinfo{year}{2018}).
\newblock \bibinfo{title}{Constraints on the density and internal strength of
  1i/'oumuamua}.
\newblock {\it \bibinfo{journal}{The Astrophysical Journal}\/},  {\it
  \bibinfo{volume}{857}\/}, \bibinfo{pages}{L1}.
\bibitem[{Meech et~al.(2017)Meech, Weryk, Micheli, Kleyna, Hainaut, Jedicke,
  Wainscoat, Chambers, Keane, Petric, Denneau, Magnier, Berger, Huber,
  Flewelling, Waters, Schunova-Lilly \& Chastel}]{Meech2017}
\bibinfo{author}{Meech, K.~J.}, \bibinfo{author}{Weryk, R.},
  \bibinfo{author}{Micheli, M.}, \bibinfo{author}{Kleyna, J.~T.},
  \bibinfo{author}{Hainaut, O.~R.}, \bibinfo{author}{Jedicke, R.},
  \bibinfo{author}{Wainscoat, R.~J.}, \bibinfo{author}{Chambers, K.~C.},
  \bibinfo{author}{Keane, J.~V.}, \bibinfo{author}{Petric, A.},
  \bibinfo{author}{Denneau, L.}, \bibinfo{author}{Magnier, E.},
  \bibinfo{author}{Berger, T.}, \bibinfo{author}{Huber, M.~E.},
  \bibinfo{author}{Flewelling, H.}, \bibinfo{author}{Waters, C.},
  \bibinfo{author}{Schunova-Lilly, E.}, \& \bibinfo{author}{Chastel, S.}
  (\bibinfo{year}{2017}).
\newblock \bibinfo{title}{A brief visit from a red and extremely elongated
  interstellar asteroid}.
\newblock {\it \bibinfo{journal}{Nature}\/},  {\it \bibinfo{volume}{552}\/},
  \bibinfo{pages}{378--381}.
\bibitem[{Meisel et~al.(2002{\natexlab{a}})Meisel, Janches \&
  Mathews}]{Meisel2002}
\bibinfo{author}{Meisel, D.~D.}, \bibinfo{author}{Janches, D.}, \&
  \bibinfo{author}{Mathews, J.~D.} (\bibinfo{year}{2002}{\natexlab{a}}).
\newblock \bibinfo{title}{Extrasolar micrometeors radiating from the vicinity
  of the local interstellar bubble}.
\newblock {\it \bibinfo{journal}{The Astrophysical Journal}\/},  {\it
  \bibinfo{volume}{567}\/}, \bibinfo{pages}{323--341}.
\bibitem[{Meisel et~al.(2002{\natexlab{b}})Meisel, Janches \&
  Mathews}]{Meisel2002b}
\bibinfo{author}{Meisel, D.~D.}, \bibinfo{author}{Janches, D.}, \&
  \bibinfo{author}{Mathews, J.~D.} (\bibinfo{year}{2002}{\natexlab{b}}).
\newblock \bibinfo{title}{The size distribution of arecibo interstellar
  particles and its implications}.
\newblock {\it \bibinfo{journal}{The Astrophysical Journal}\/},  {\it
  \bibinfo{volume}{579}\/}, \bibinfo{pages}{895--904}.
\bibitem[{Micheli et~al.(2018)Micheli, Farnocchia, Meech, Buie, Hainaut,
  Prialnik, Sch\"{o}rghofer, Weaver, Chodas, Kleyna, Weryk, Wainscoat, Ebeling,
  Keane, Chambers, Koschny \& Petropoulos}]{Micheli2018}
\bibinfo{author}{Micheli, M.}, \bibinfo{author}{Farnocchia, D.},
  \bibinfo{author}{Meech, K.~J.}, \bibinfo{author}{Buie, M.~W.},
  \bibinfo{author}{Hainaut, O.~R.}, \bibinfo{author}{Prialnik, D.},
  \bibinfo{author}{Sch\"{o}rghofer, N.}, \bibinfo{author}{Weaver, H.~A.},
  \bibinfo{author}{Chodas, P.~W.}, \bibinfo{author}{Kleyna, J.~T.},
  \bibinfo{author}{Weryk, R.}, \bibinfo{author}{Wainscoat, R.~J.},
  \bibinfo{author}{Ebeling, H.}, \bibinfo{author}{Keane, J.~V.},
  \bibinfo{author}{Chambers, K.~C.}, \bibinfo{author}{Koschny, D.}, \&
  \bibinfo{author}{Petropoulos, A.~E.} (\bibinfo{year}{2018}).
\newblock \bibinfo{title}{Non-gravitational acceleration in the trajectory of
  1i/2017 u1 (`oumuamua)}.
\newblock {\it \bibinfo{journal}{Nature}\/},  {\it \bibinfo{volume}{559}\/},
  \bibinfo{pages}{223--226}.
\bibitem[{Molau(2007)}]{Molau2007}
\bibinfo{author}{Molau, S.} (\bibinfo{year}{2007}).
\newblock \bibinfo{title}{How good is the imo working list of meteor showers? a
  complete analysis of the imo video meteor database.}
\newblock In \bibinfo{editor}{F.~{Bettonvil}}, \& \bibinfo{editor}{J.~{Kac}}
  (Eds.), {\it \bibinfo{booktitle}{Proceedings of the International Meteor
  Conference, 25th IMC, Roden, Netherlands, 2006}\/} (pp.
  \bibinfo{pages}{38--55}).
\bibitem[{Morfill \& Gr{\"{u}}n(1979)}]{Morfill1979}
\bibinfo{author}{Morfill, G.}, \& \bibinfo{author}{Gr{\"{u}}n, E.}
  (\bibinfo{year}{1979}).
\newblock \bibinfo{title}{{The motion of charged dust particles in
  interplanetary space—II. Interstellar grains}}.
\newblock {\it \bibinfo{journal}{Planetary and Space Science}\/},  {\it
  \bibinfo{volume}{27}\/}, \bibinfo{pages}{1283--1292}.
\bibitem[{Murray et~al.(2004)Murray, Weingartner \& Capobianco}]{Murray2004}
\bibinfo{author}{Murray, N.}, \bibinfo{author}{Weingartner, J.~C.}, \&
  \bibinfo{author}{Capobianco, C.} (\bibinfo{year}{2004}).
\newblock \bibinfo{title}{On the flux of extrasolar dust in earth's
  atmosphere}.
\newblock {\it \bibinfo{journal}{The Astrophysical Journal}\/},  {\it
  \bibinfo{volume}{600}\/}, \bibinfo{pages}{804--827}.
\bibitem[{Musci et~al.(2012)Musci, Weryk, Brown, Campbell-Brown \&
  Wiegert}]{Musci2012}
\bibinfo{author}{Musci, R.}, \bibinfo{author}{Weryk, R.~J.},
  \bibinfo{author}{Brown, P.}, \bibinfo{author}{Campbell-Brown, M.~D.}, \&
  \bibinfo{author}{Wiegert, P.~A.} (\bibinfo{year}{2012}).
\newblock \bibinfo{title}{An optical survey for millimeter-sized interstellar
  meteoroids.}
\newblock {\it \bibinfo{journal}{The Astrophysical Journal}\/},  {\it
  \bibinfo{volume}{745}\/}, \bibinfo{pages}{161}.
\bibitem[{Neslušan et~al.(1998)Neslušan, Svoren \& Porubcan}]{Nesl1998}
\bibinfo{author}{Neslušan, L.}, \bibinfo{author}{Svoren, J.}, \&
  \bibinfo{author}{Porubcan, V.} (\bibinfo{year}{1998}).
\newblock \bibinfo{title}{A computer program for calculation of a theoretical
  meteor-stream radiant}.
\newblock {\it \bibinfo{journal}{Astronomy and Astrophysics}\/},  {\it
  \bibinfo{volume}{331}\/}, \bibinfo{pages}{411--413}.
\bibitem[{Schult et~al.(2018)Schult, Brown, Pokorn{\'{y}}, Stober \&
  Chau}]{Schult2018}
\bibinfo{author}{Schult, C.}, \bibinfo{author}{Brown, P.},
  \bibinfo{author}{Pokorn{\'{y}}, P.}, \bibinfo{author}{Stober, G.}, \&
  \bibinfo{author}{Chau, J.~L.} (\bibinfo{year}{2018}).
\newblock \bibinfo{title}{A meteoroid stream survey using meteor head echo
  observations from the middle atmosphere {ALOMAR} radar system ({MAARSY})}.
\newblock {\it \bibinfo{journal}{Icarus}\/},  {\it \bibinfo{volume}{309}\/},
  \bibinfo{pages}{177--186}.
\bibitem[{SonotaCo(2009)}]{SonotaCo2009}
\bibinfo{author}{SonotaCo} (\bibinfo{year}{2009}).
\newblock \bibinfo{title}{A meteor shower catalog based on video observations
  in 2007-2008}.
\newblock {\it \bibinfo{journal}{WGN, Journal of the International Meteor
  Organization}\/},  {\it \bibinfo{volume}{37}\/}, \bibinfo{pages}{55--62}.
\bibitem[{Sterken et~al.(2019)Sterken, Westphal, Altobelli, Malaspina \&
  Postberg}]{Sterken2019}
\bibinfo{author}{Sterken, V.~J.}, \bibinfo{author}{Westphal, A.~J.},
  \bibinfo{author}{Altobelli, N.}, \bibinfo{author}{Malaspina, D.}, \&
  \bibinfo{author}{Postberg, F.} (\bibinfo{year}{2019}).
\newblock \bibinfo{title}{{Interstellar Dust in the Solar System}}.
\newblock {\it \bibinfo{journal}{Space Science Reviews}\/},  {\it
  \bibinfo{volume}{215}\/}.
\bibitem[{Svoren et~al.(1993)Svoren, Neslušan \& Porubcan}]{Svoren1993}
\bibinfo{author}{Svoren, J.}, \bibinfo{author}{Neslušan, L.}, \&
  \bibinfo{author}{Porubcan, V.} (\bibinfo{year}{1993}).
\newblock \bibinfo{title}{Applicability of meteor radiant determination methods
  depending on orbit type. i. high-eccentric orbits}.
\newblock {\it \bibinfo{journal}{Contributions of the Astronomical Observatory
  Skalnate Pleso}\/},  {\it \bibinfo{volume}{23}\/}, \bibinfo{pages}{23--44}.
\bibitem[{Trilling et~al.(2018)Trilling, Mommert, Hora, Farnocchia, Chodas,
  Giorgini, Smith, Carey, Lisse, Werner, McNeill, Chesley, Emery, Fazio,
  Fernandez, Harris, Marengo, Mueller, Roegge, Smith, Weaver, Meech \&
  Micheli}]{Trilling2018}
\bibinfo{author}{Trilling, D.~E.}, \bibinfo{author}{Mommert, M.},
  \bibinfo{author}{Hora, J.~L.}, \bibinfo{author}{Farnocchia, D.},
  \bibinfo{author}{Chodas, P.}, \bibinfo{author}{Giorgini, J.},
  \bibinfo{author}{Smith, H.~A.}, \bibinfo{author}{Carey, S.},
  \bibinfo{author}{Lisse, C.~M.}, \bibinfo{author}{Werner, M.},
  \bibinfo{author}{McNeill, A.}, \bibinfo{author}{Chesley, S.~R.},
  \bibinfo{author}{Emery, J.~P.}, \bibinfo{author}{Fazio, G.},
  \bibinfo{author}{Fernandez, Y.~R.}, \bibinfo{author}{Harris, A.},
  \bibinfo{author}{Marengo, M.}, \bibinfo{author}{Mueller, M.},
  \bibinfo{author}{Roegge, A.}, \bibinfo{author}{Smith, N.},
  \bibinfo{author}{Weaver, H.~A.}, \bibinfo{author}{Meech, K.}, \&
  \bibinfo{author}{Micheli, M.} (\bibinfo{year}{2018}).
\newblock \bibinfo{title}{Spitzer observations of interstellar object
  1i/`oumuamua}.
\newblock {\it \bibinfo{journal}{The Astronomical Journal}\/},  {\it
  \bibinfo{volume}{156}\/}, \bibinfo{pages}{261}.
\bibitem[{Verniani(1973)}]{Verniani1973}
\bibinfo{author}{Verniani, F.} (\bibinfo{year}{1973}).
\newblock \bibinfo{title}{An analysis of the physical parameters of 5759 faint
  radio meteors}.
\newblock {\it \bibinfo{journal}{Journal of Geophysical Research}\/},  {\it
  \bibinfo{volume}{78}\/}, \bibinfo{pages}{8429--8462}.
\bibitem[{Vida et~al.(2018)Vida, Brown \& Campbell-Brown}]{Vida2018}
\bibinfo{author}{Vida, D.}, \bibinfo{author}{Brown, P.~G.}, \&
  \bibinfo{author}{Campbell-Brown, M.~D.} (\bibinfo{year}{2018}).
\newblock \bibinfo{title}{{Modelling the measurement accuracy of pre-atmosphere
  velocities of meteoroids}}.
\newblock {\it \bibinfo{journal}{Monthly Notices of the Royal Astronomical
  Society}\/},  {\it \bibinfo{volume}{479}\/}, \bibinfo{pages}{4307--4319}.
\bibitem[{Webster et~al.(2004)Webster, Brown, Jones, Ellis \&
  Campbell-Brown}]{Webster2004}
\bibinfo{author}{Webster, A.}, \bibinfo{author}{Brown, P.},
  \bibinfo{author}{Jones, J.}, \bibinfo{author}{Ellis, K.~J.}, \&
  \bibinfo{author}{Campbell-Brown, M.~D.} (\bibinfo{year}{2004}).
\newblock \bibinfo{title}{{Canadian Meteor Orbit Radar (CMOR)}}.
\newblock {\it \bibinfo{journal}{Atmospheric Chemistry and Physics}\/},  {\it
  \bibinfo{volume}{4}\/}, \bibinfo{pages}{679--684}.
\bibitem[{{Wehry} \& {Mann}(1999)}]{Wehry1999}
\bibinfo{author}{{Wehry}, A.}, \& \bibinfo{author}{{Mann}, I.}
  (\bibinfo{year}{1999}).
\newblock \bibinfo{title}{{Identification of beta -meteoroids from measurements
  of the dust detector onboard the ULYSSES spacecraft}}.
\newblock {\it \bibinfo{journal}{Astronomy and Astrophysics}\/},  {\it
  \bibinfo{volume}{341}\/}, \bibinfo{pages}{296--303}.
\bibitem[{Weryk \& Brown(2012{\natexlab{a}})}]{Weryk2012a}
\bibinfo{author}{Weryk, R.}, \& \bibinfo{author}{Brown, P.}
  (\bibinfo{year}{2012}{\natexlab{a}}).
\newblock \bibinfo{title}{{Simultaneous radar and video meteors—I: Metric
  comparisons}}.
\newblock {\it \bibinfo{journal}{Planetary and Space Science}\/},  {\it
  \bibinfo{volume}{62}\/}, \bibinfo{pages}{132--152}.
\bibitem[{Weryk \& Brown(2005)}]{Weryk2005}
\bibinfo{author}{Weryk, R.~J.}, \& \bibinfo{author}{Brown, P.}
  (\bibinfo{year}{2005}).
\newblock \bibinfo{title}{A search for interstellar meteoroids using the
  canadian meteor orbit radar ({CMOR})}.
\newblock {\it \bibinfo{journal}{Earth, Moon, and Planets}\/},  {\it
  \bibinfo{volume}{95}\/}, \bibinfo{pages}{221--227}.
\bibitem[{Weryk \& Brown(2012{\natexlab{b}})}]{Weryk2012}
\bibinfo{author}{Weryk, R.~J.}, \& \bibinfo{author}{Brown, P.~G.}
  (\bibinfo{year}{2012}{\natexlab{b}}).
\newblock \bibinfo{title}{Simultaneous radar and video meteors{\textemdash}i:
  Metric comparisons}.
\newblock {\it \bibinfo{journal}{Planetary and Space Science}\/},  {\it
  \bibinfo{volume}{62}\/}, \bibinfo{pages}{132--152}.
\bibitem[{Wood et~al.(2015)Wood, Malaspina, Andersson \& Horanyi}]{Wood2015}
\bibinfo{author}{Wood, S.~R.}, \bibinfo{author}{Malaspina, D.~M.},
  \bibinfo{author}{Andersson, L.}, \& \bibinfo{author}{Horanyi, M.}
  (\bibinfo{year}{2015}).
\newblock \bibinfo{title}{{Hypervelocity dust impacts on the Wind spacecraft:
  Correlations between Ulysses and Wind interstellar dust detections}}.
\newblock {\it \bibinfo{journal}{Journal of Geophysical Research: Space
  Physics}\/},  {\it \bibinfo{volume}{120}\/}, \bibinfo{pages}{7121--7129}.
\bibitem[{Ye et~al.(2013)Ye, Wiegert, Brown, Campbell-Brown \& Weryk}]{Ye2013}
\bibinfo{author}{Ye, Q.}, \bibinfo{author}{Wiegert, P.},
  \bibinfo{author}{Brown, P.}, \bibinfo{author}{Campbell-Brown, M.~D.}, \&
  \bibinfo{author}{Weryk, R.} (\bibinfo{year}{2013}).
\newblock \bibinfo{title}{{The unexpected 2012 Draconid meteor storm}}.
\newblock {\it \bibinfo{journal}{Monthly Notices of the Royal Astronomical
  Society}\/},  {\it \bibinfo{volume}{437}\/}, \bibinfo{pages}{3812--3823}.
\bibitem[{Zaslavsky et~al.(2012)Zaslavsky, Meyer-Vernet, Mann, Czechowski,
  Issautier, {Le Chat}, Pantellini, Goetz, Maksimovic, Bale \&
  Kasper}]{Zaslavsky2012}
\bibinfo{author}{Zaslavsky, A.}, \bibinfo{author}{Meyer-Vernet, N.},
  \bibinfo{author}{Mann, I.}, \bibinfo{author}{Czechowski, A.},
  \bibinfo{author}{Issautier, K.}, \bibinfo{author}{{Le Chat}, G.},
  \bibinfo{author}{Pantellini, F.}, \bibinfo{author}{Goetz, K.},
  \bibinfo{author}{Maksimovic, M.}, \bibinfo{author}{Bale, S.~D.}, \&
  \bibinfo{author}{Kasper, J.~C.} (\bibinfo{year}{2012}).
\newblock \bibinfo{title}{{Interplanetary dust detection by radio antennas:
  Mass calibration and fluxes measured by STEREO/WAVES}}.
\newblock {\it \bibinfo{journal}{Journal of Geophysical Research: Space
  Physics}\/},  {\it \bibinfo{volume}{117}\/}, \bibinfo{pages}{1--13}.
\bibitem[{{Zinner}(2014)}]{zinner2014}
\bibinfo{author}{{Zinner}, E.} (\bibinfo{year}{2014}).
\newblock \bibinfo{title}{{Presolar Grains}}.
\newblock In \bibinfo{editor}{A.~M. {Davis}} (Ed.), {\it
  \bibinfo{booktitle}{Meteorites and Cosmochemical Processes}\/} (pp.
  \bibinfo{pages}{181--213}).

\end{thebibliography}


\clearpage
\newpage

\newenvironment{changemargin}[3]{%
\begin{list}{}{%
\setlength{\topsep}{0pt}%
\setlength{\leftmargin}{#1}%
\setlength{\rightmargin}{#2}
\setlength{\topmargin}{#3}
\setlength{\footskip}{4cm}
\setlength{\listparindent}{\parindent}%
\setlength{\itemindent}{\parindent}%
\setlength{\parsep}{\parskip}%
}%
\item[]}{\end{list}}


\section*{Appendix: Deceleration Correction Data in support of Section \ref{decelCor}}\label{sec:appendix}
\addcontentsline{toc}{section}{\nameref{sec:appendix}}
\begin{changemargin}{0cm}{0cm}{-2cm}

\setcounter{figure}{0}
\setcounter{table}{0}

\renewcommand{\thetable}{A\arabic{table}}  
\renewcommand{\thefigure}{A\arabic{figure}}

\begin{figure}[h!]
	\advance\leftskip-2cm
	\advance\rightskip-2cm

	\begin{subfigure}{0.6\textwidth}

		\includegraphics[trim = 2cm 1cm 3.5cm 1.5cm, clip, width=1\linewidth]{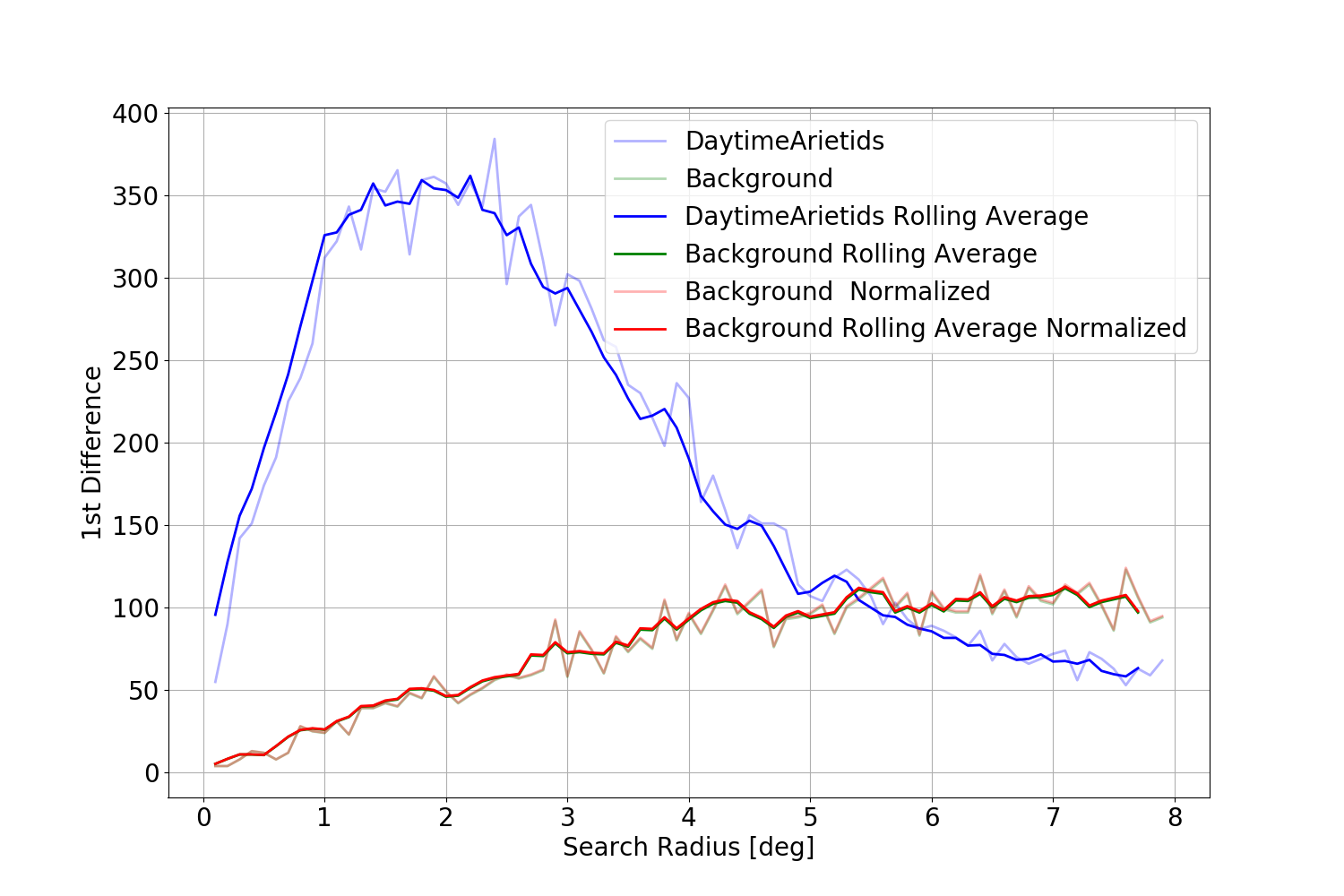}
	
	\end{subfigure}
	\hspace*{\fill}
	\begin{subfigure}{0.6\textwidth}
	
		\includegraphics[trim = 2cm 1cm 3.5cm 1.5cm, clip, width=1\linewidth]{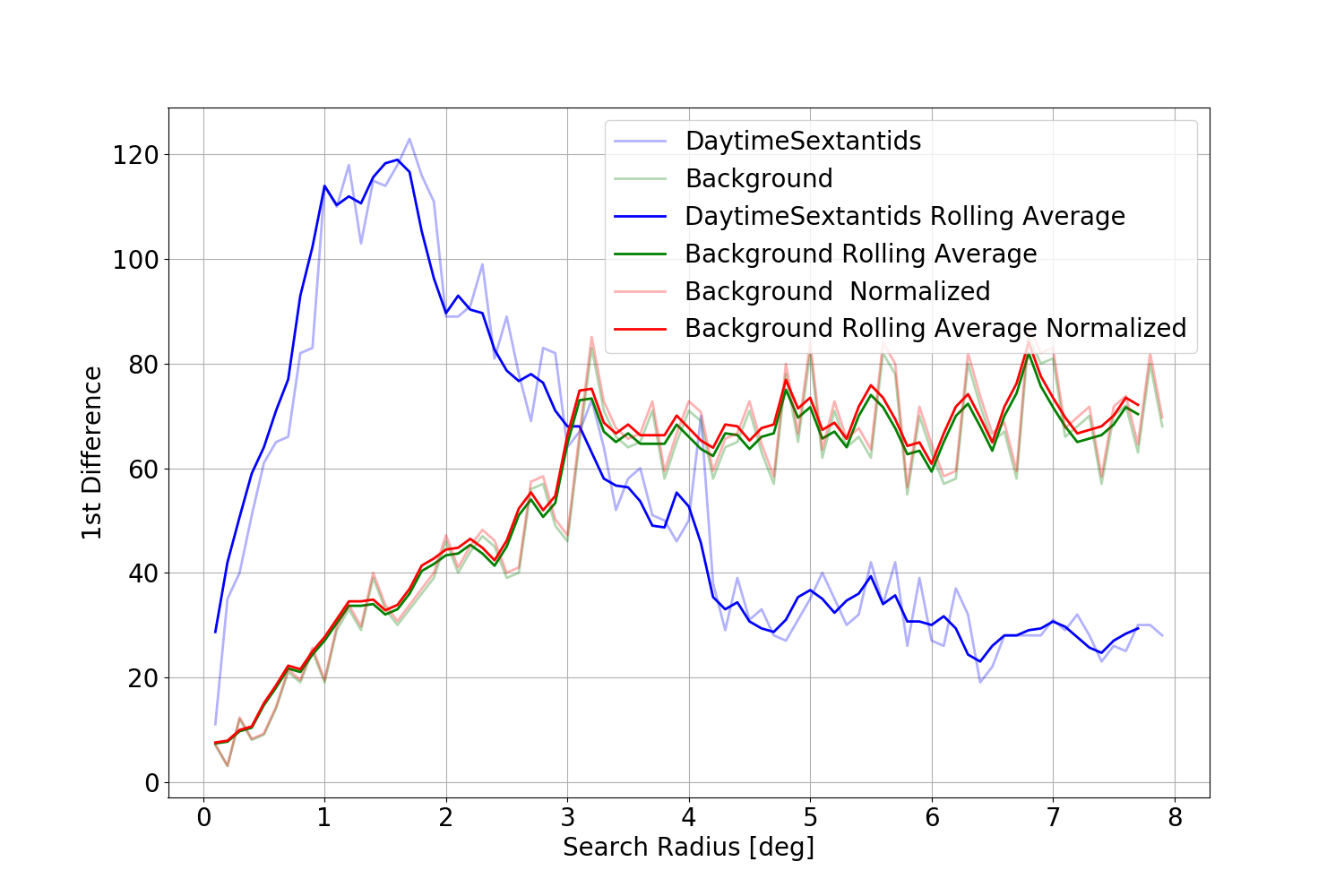}
	
	\end{subfigure}
	\medskip
	
	\begin{subfigure}{0.6\textwidth}
		\includegraphics[trim = 2cm 1cm 3.5cm 1.5cm, clip,width=\linewidth]{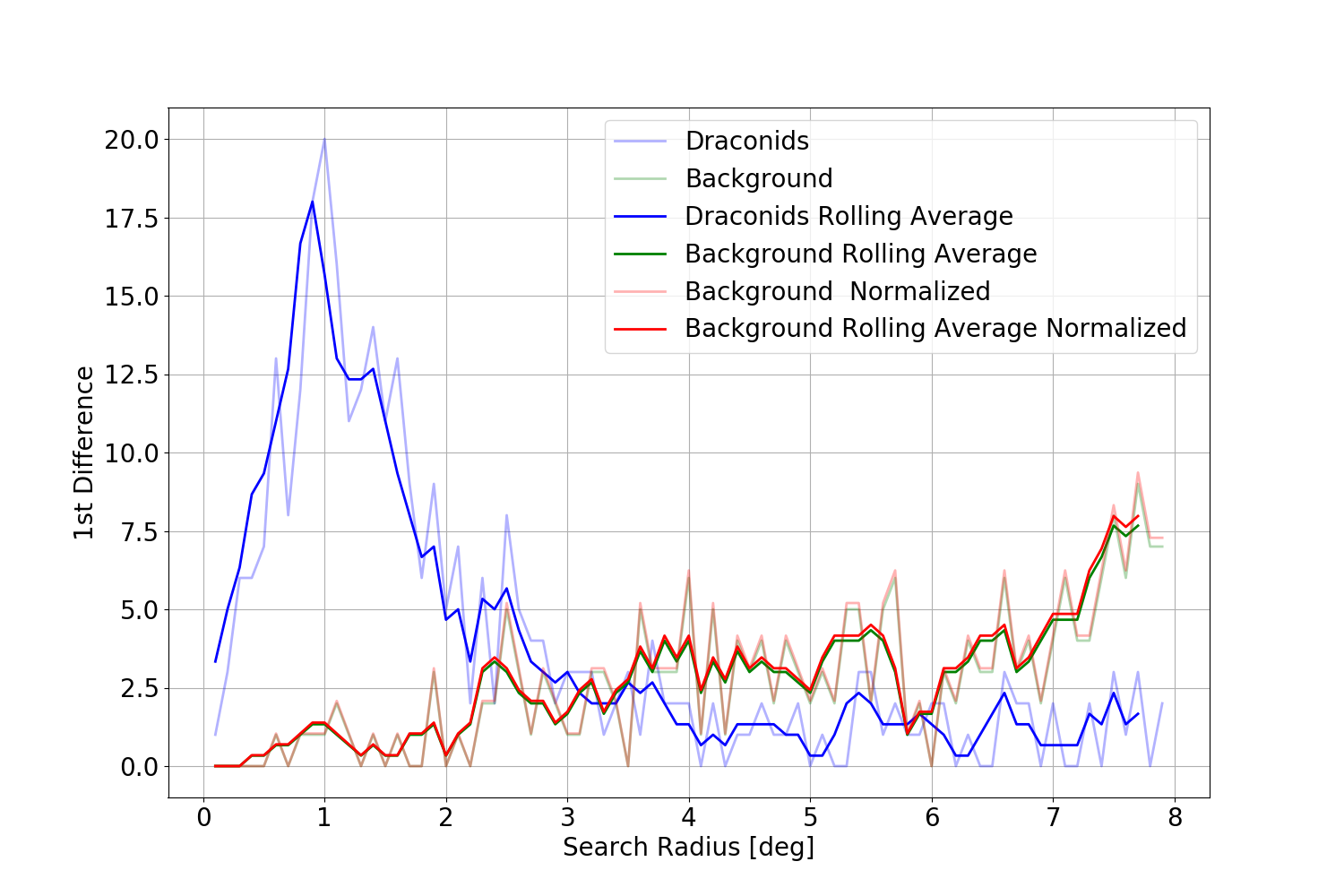}
	\end{subfigure}
		\hspace*{\fill}
	\begin{subfigure}{0.6\textwidth}
		\includegraphics[trim = 2cm 1cm 3.5cm 1.5cm, clip, width=1\linewidth]{ThesisUSE_Geminids1stdiff.png}
	\end{subfigure}
	\medskip
	\begin{subfigure}{0.6\textwidth}
		\includegraphics[trim = 2cm 1cm 3.5cm 1.5cm, clip,width=\linewidth]{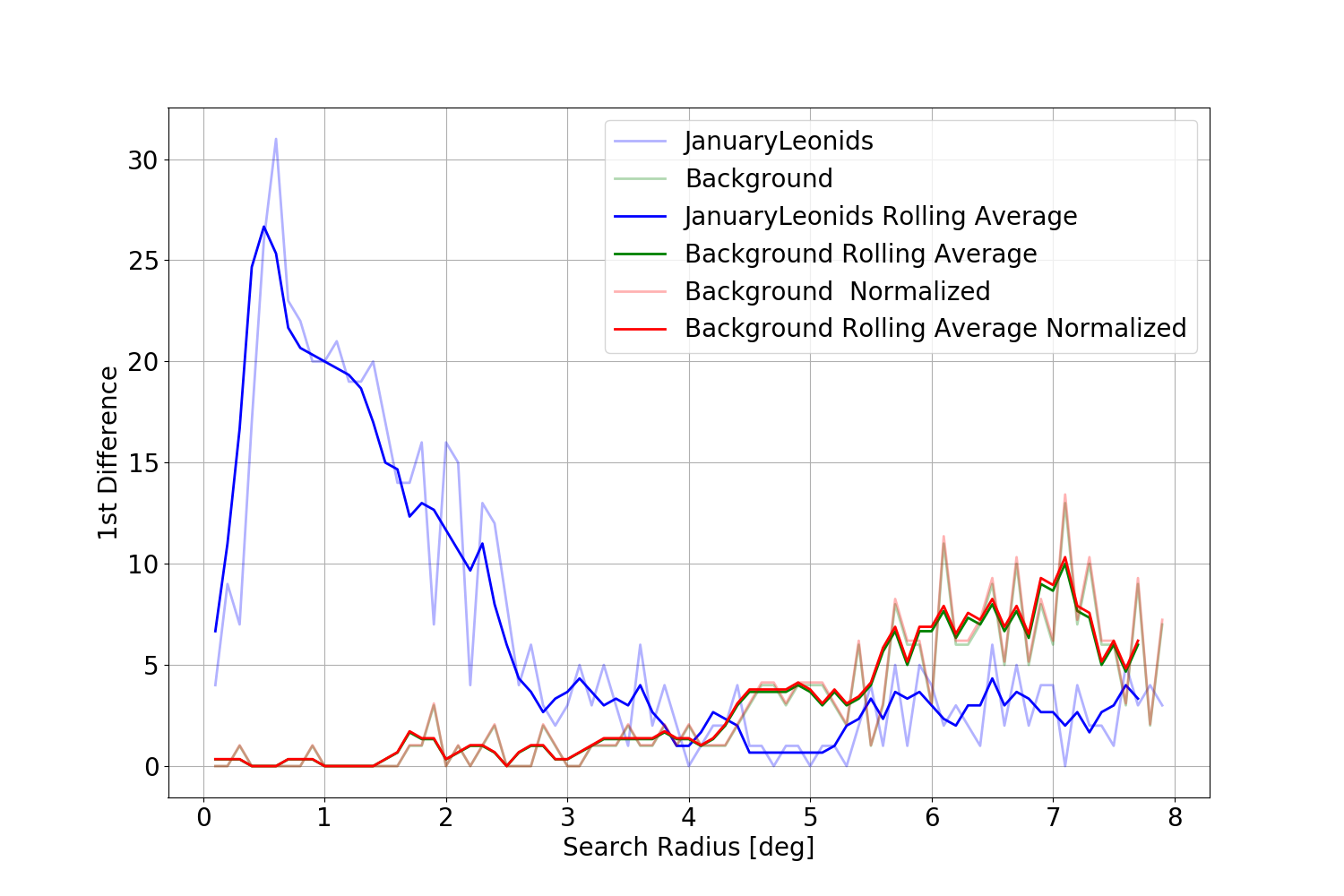}
	\end{subfigure}
	\hspace*{\fill}
	\begin{subfigure}{0.6\textwidth}
		\includegraphics[trim = 2cm 1cm 3.5cm 1.5cm, clip, width=1\linewidth]{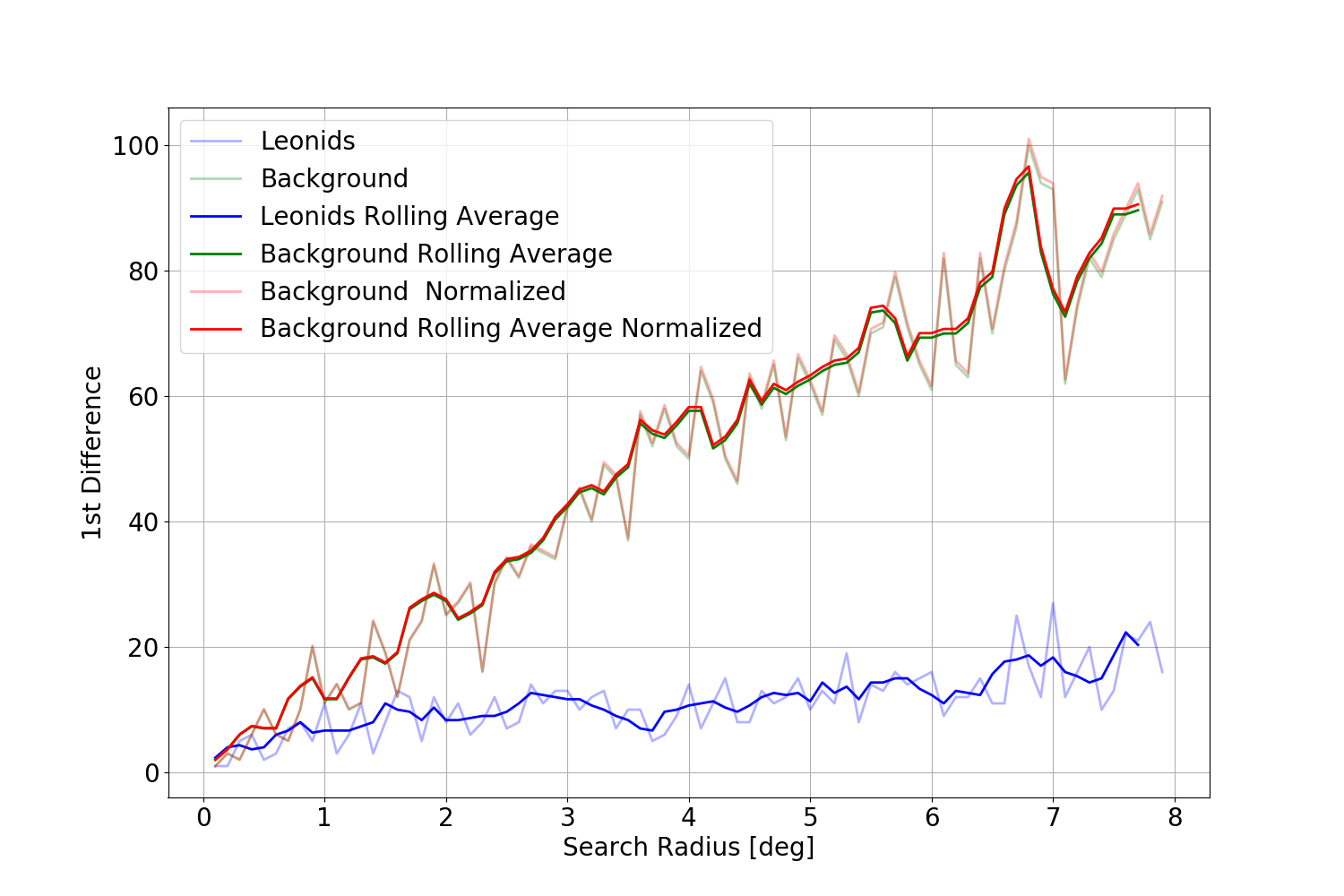}
	\end{subfigure}
	\caption[app1]{First difference in count of shower meteors and sporadic background meteors for all showers used in velocity correction . Note: no cross-over was found for Leonids, therefore 5 degree search radius was used.}
	\label{apx_1stdiff1}
\end{figure}

\begin{figure}[h!]
\vspace*{-0.5cm}
\advance\leftskip-2cm
\advance\rightskip-2cm
\begin{subfigure}{0.6\textwidth}
\includegraphics[trim = 2cm 1cm 3.5cm 1.5cm, clip, clip,width=\linewidth]{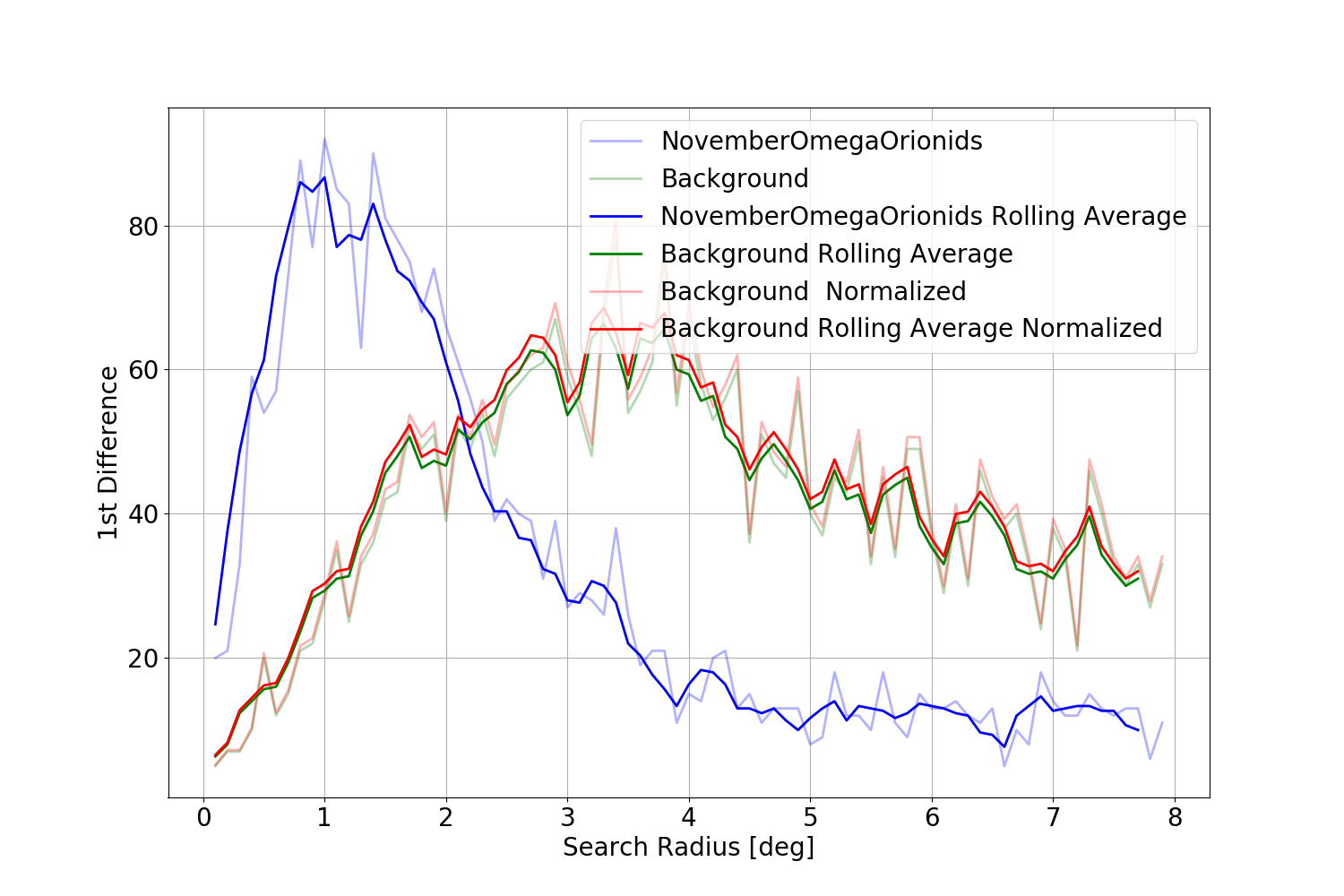}

\end{subfigure}\hspace*{\fill}
\begin{subfigure}{0.6\textwidth}
\includegraphics[trim = 2cm 1cm 3.5cm 1.5cm, clip, clip,width=\linewidth]{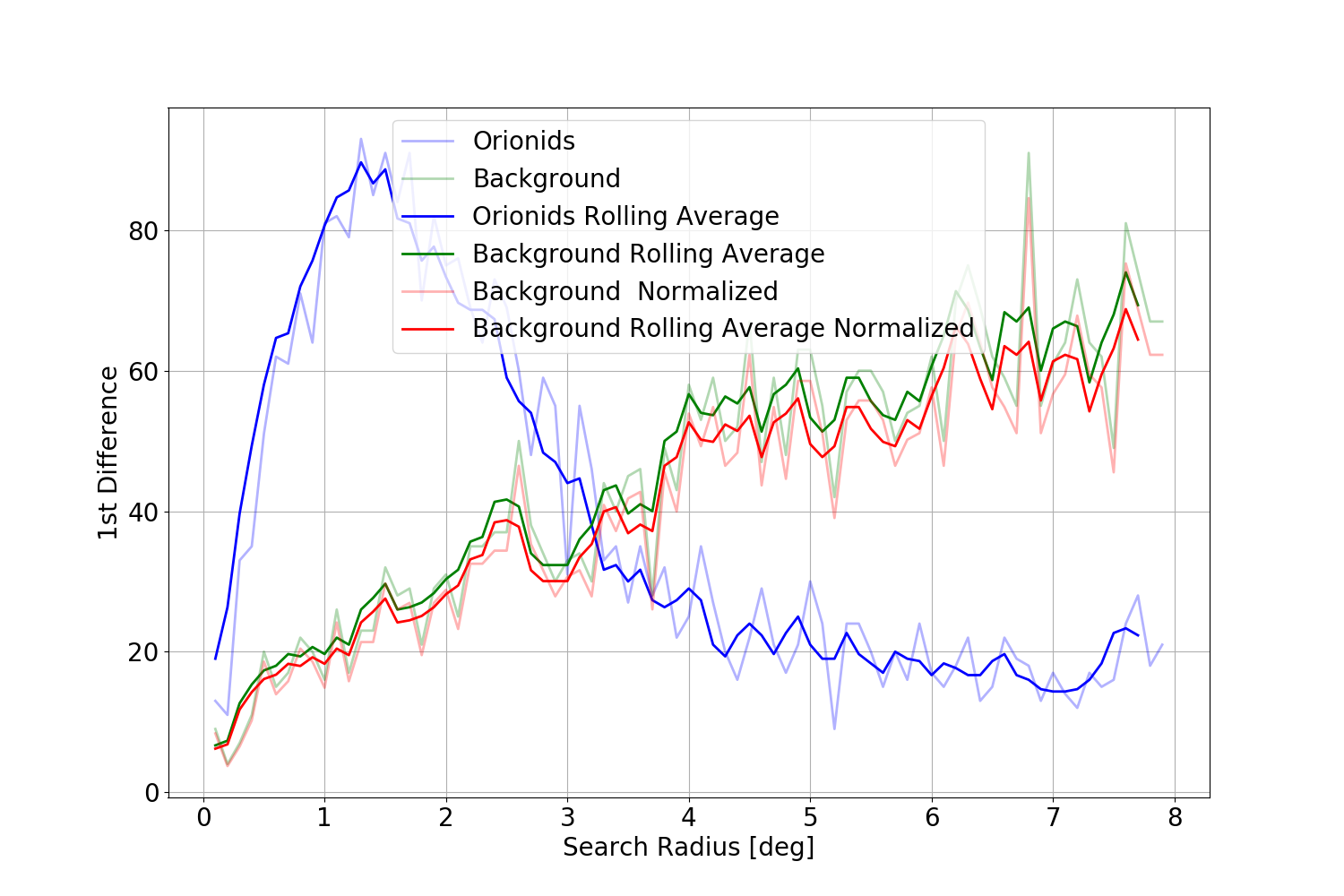}

\end{subfigure}

\medskip
\begin{subfigure}{0.6\textwidth}
\includegraphics[trim = 2cm 1cm 3.5cm 1.5cm, clip,width=\linewidth]{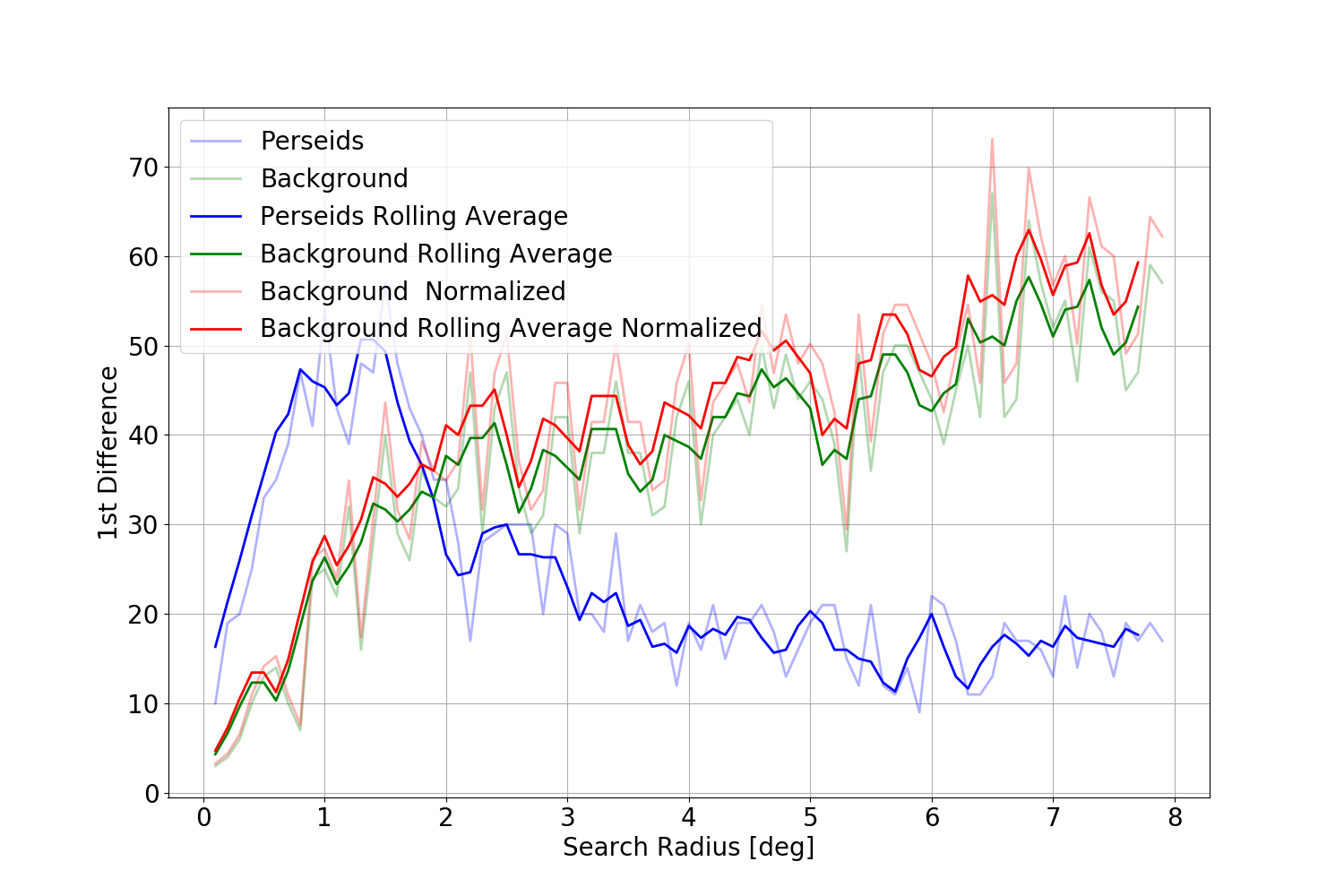}

\end{subfigure}\hspace*{\fill}
\begin{subfigure}{0.6\textwidth}
\includegraphics[trim = 2cm 1cm 3.5cm 1.5cm, clip, clip,width=\linewidth]{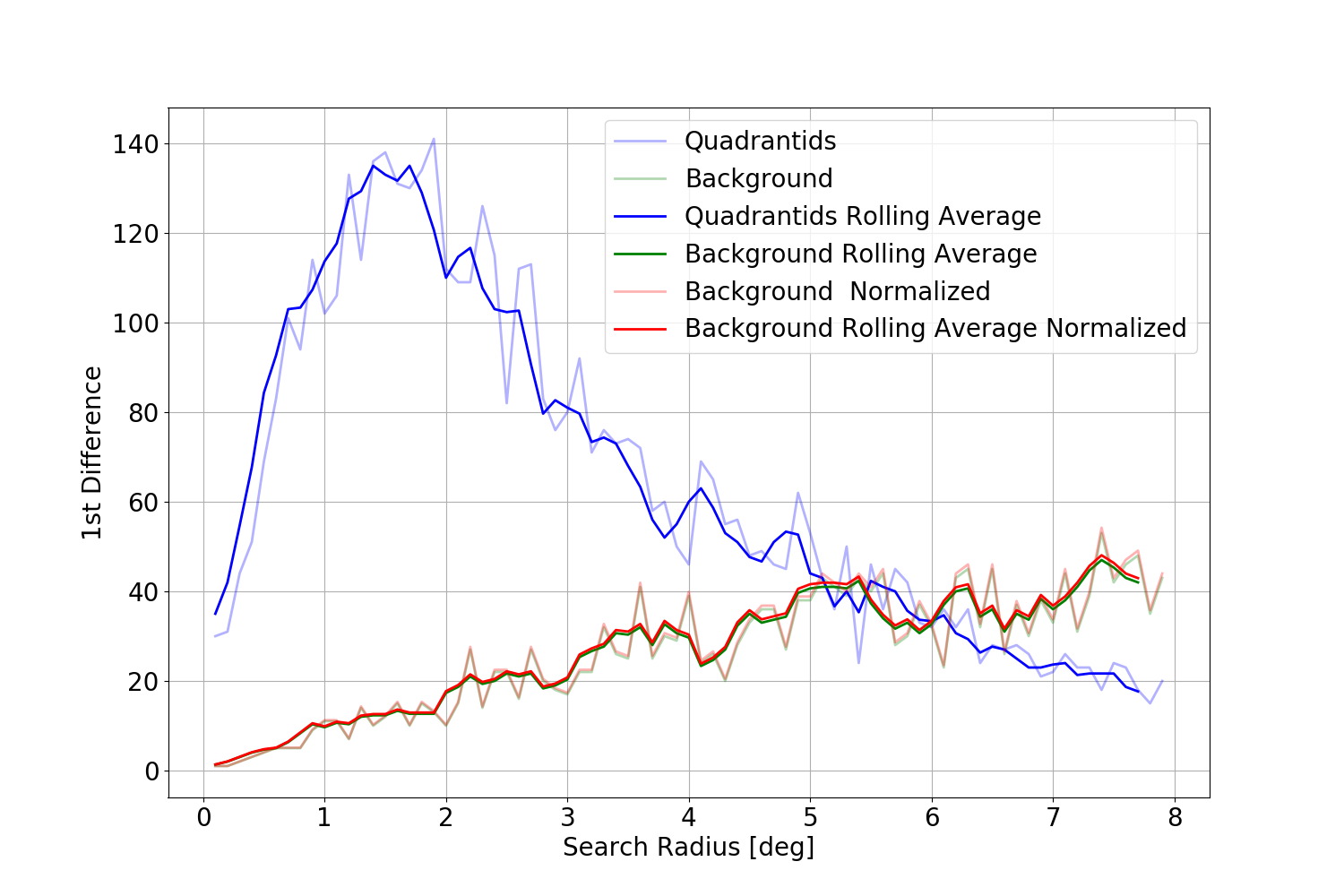}

\end{subfigure}

\medskip
\begin{subfigure}{0.6\textwidth}
\includegraphics[trim = 2cm 1cm 3.5cm 1.5cm, clip, clip,width=\linewidth]{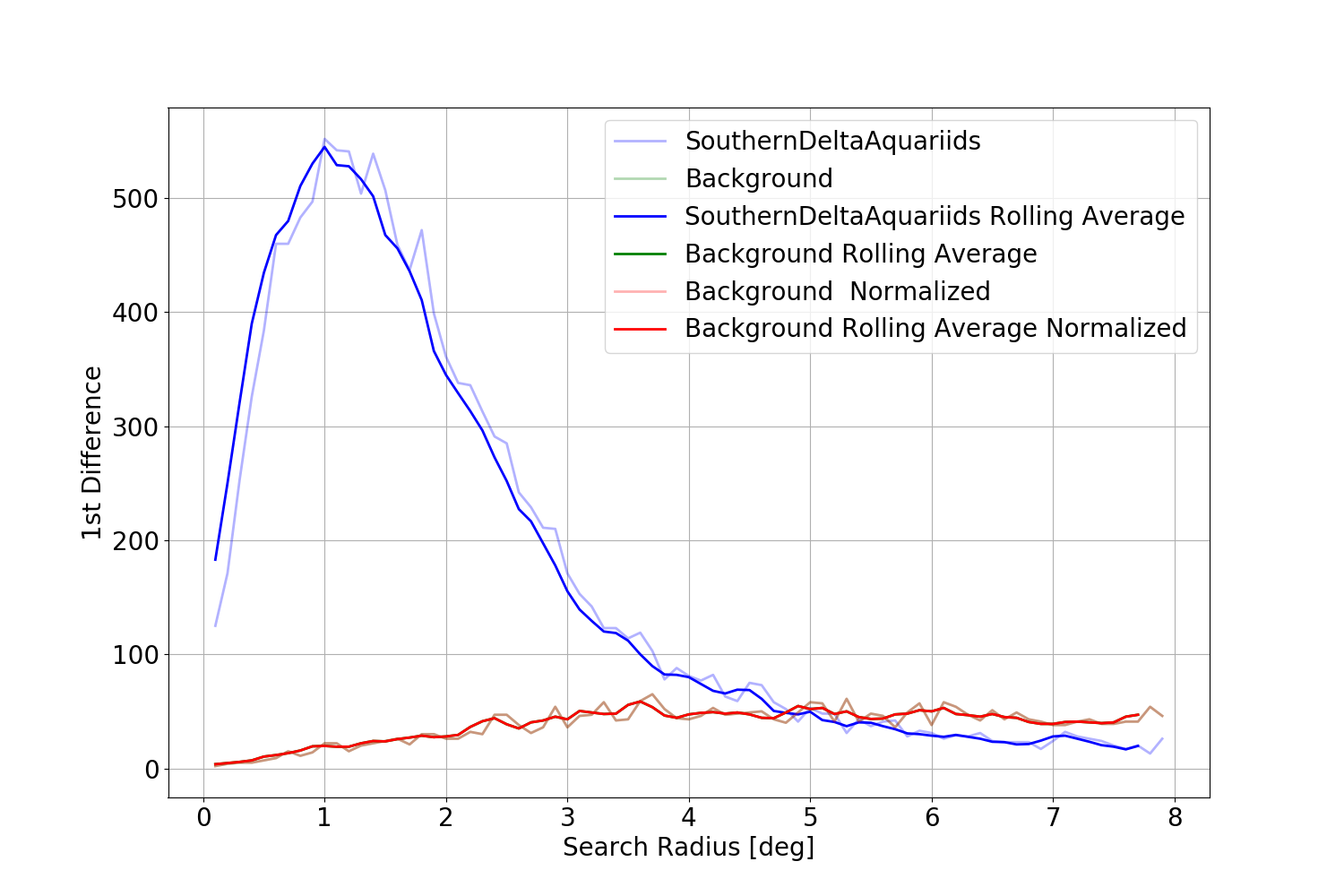}
\end{subfigure}\hspace*{\fill}
\begin{subfigure}{0.6\textwidth}
\includegraphics[trim = 2cm 1cm 3.5cm 1.5cm, clip, width=\linewidth]{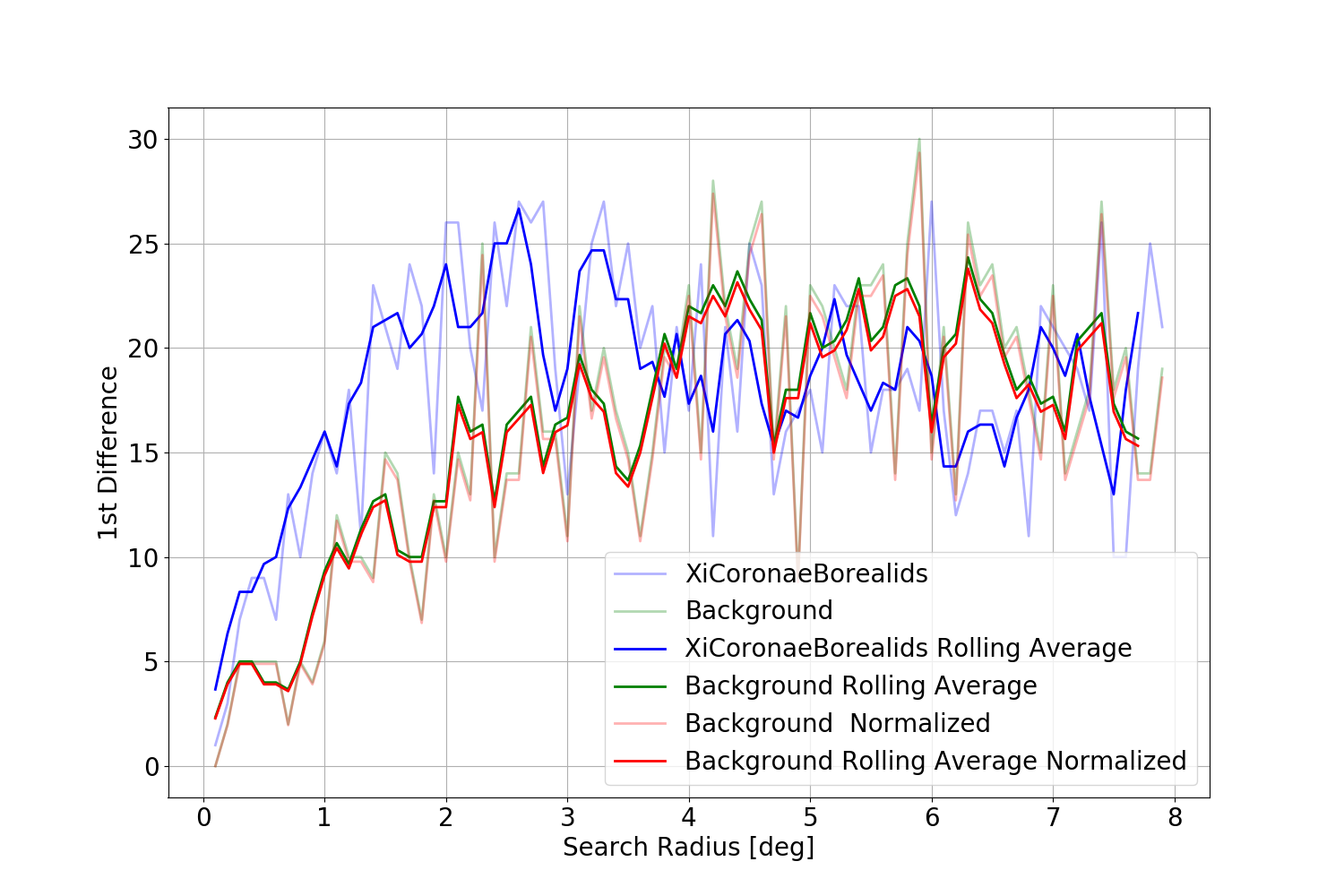}

\end{subfigure}
\caption[1st difference in count of shower and sporadic meteors for showers used in velocity correction (2/2)]{1st difference in count of shower meteors and sporadic background meteors for all showers used in velocity correction (part 2/2)}
\label{apx_1stdiff2}
\end{figure}

\newpage

\begin{figure}[t!]
\vspace*{-0.5cm}
\advance\leftskip-3cm
\advance\rightskip-3cm
\begin{subfigure}{0.7\textwidth}
\includegraphics[trim = 2.5cm 1.25cm 3.5cm 1.5cm,  clip,width=\linewidth]{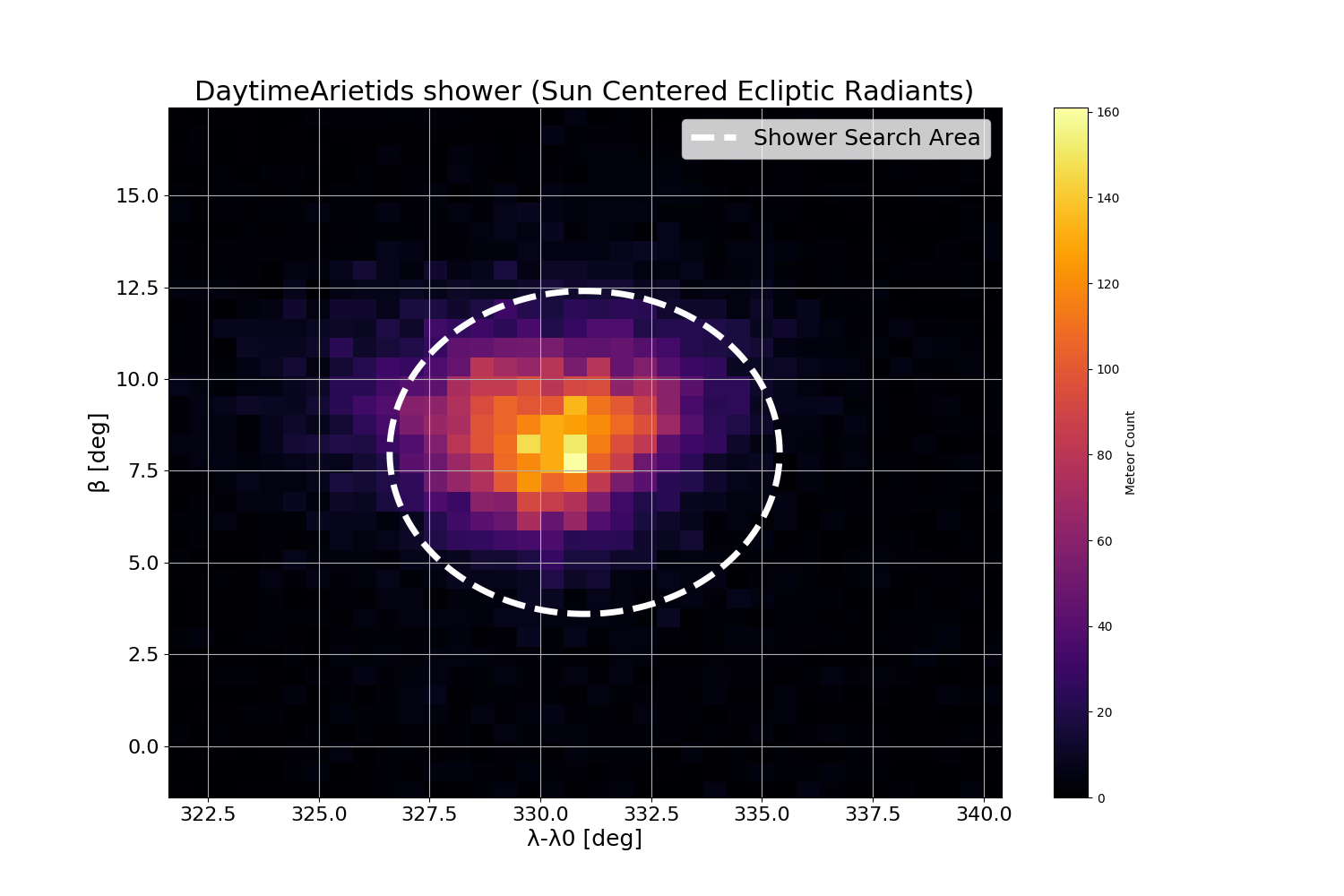}

\end{subfigure}\hspace*{\fill}
\begin{subfigure}{0.7\textwidth}
\includegraphics[trim = 2.5cm 1.25cm 3.5cm 1.5cm,  clip,width=\linewidth]{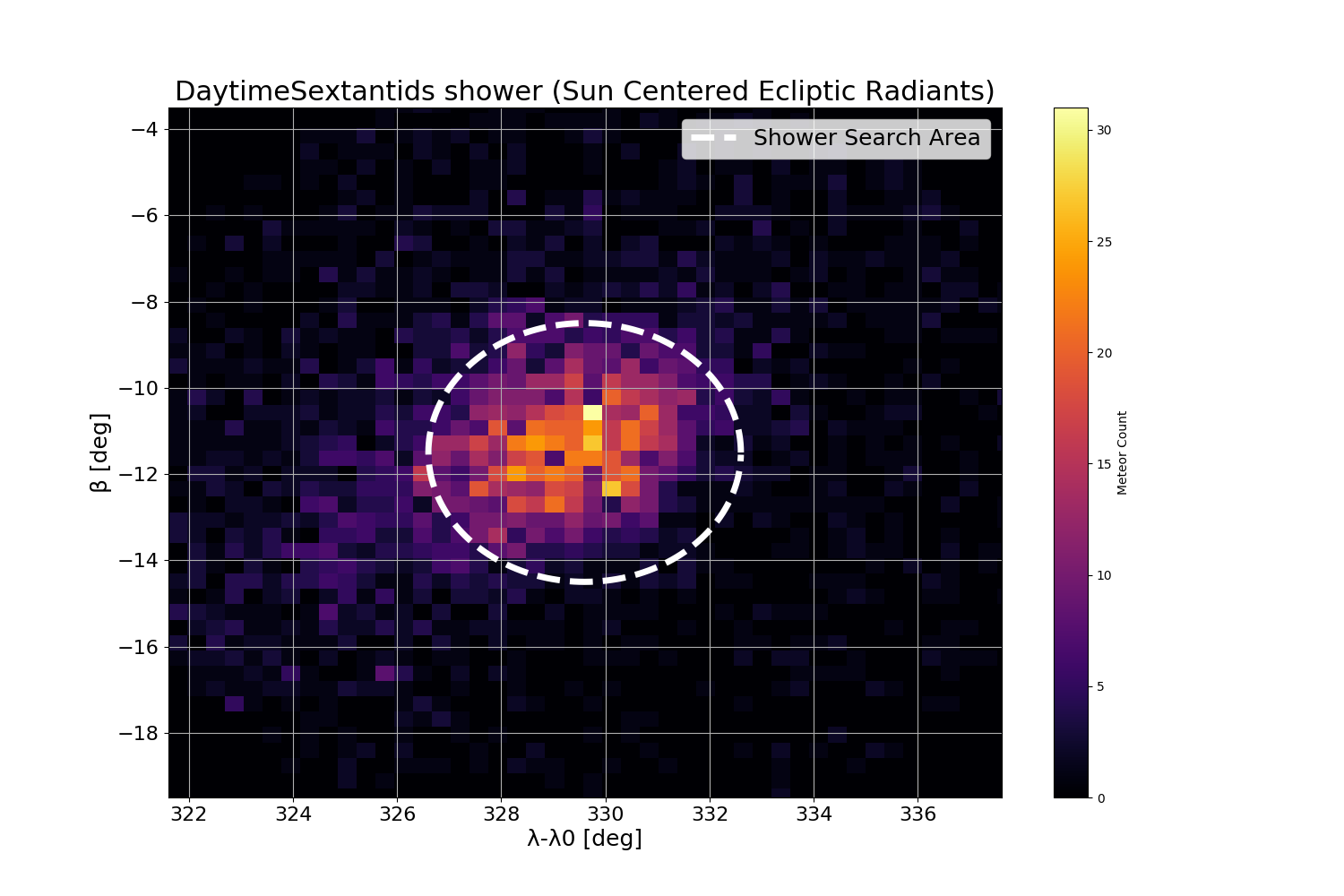}

\end{subfigure}

\medskip
\begin{subfigure}{0.7\textwidth}
\includegraphics[trim = 2.5cm 1.25cm 3.5cm 1.5cm,  clip,width=\linewidth]{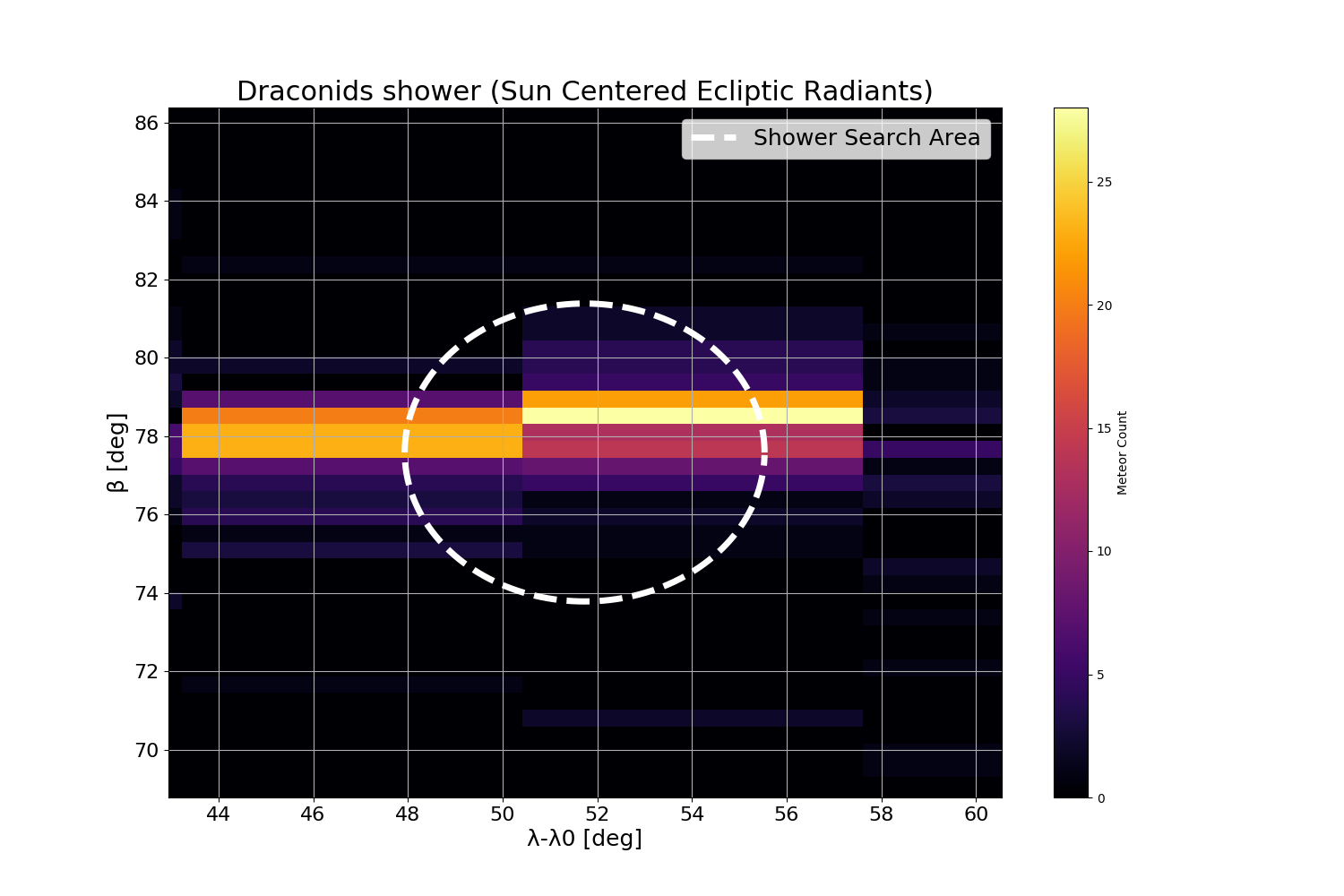}

\end{subfigure}\hspace*{\fill}
\begin{subfigure}{0.7\textwidth}
\includegraphics[trim = 2.5cm 1.25cm 3.5cm 1.5cm,  clip,width=\linewidth]{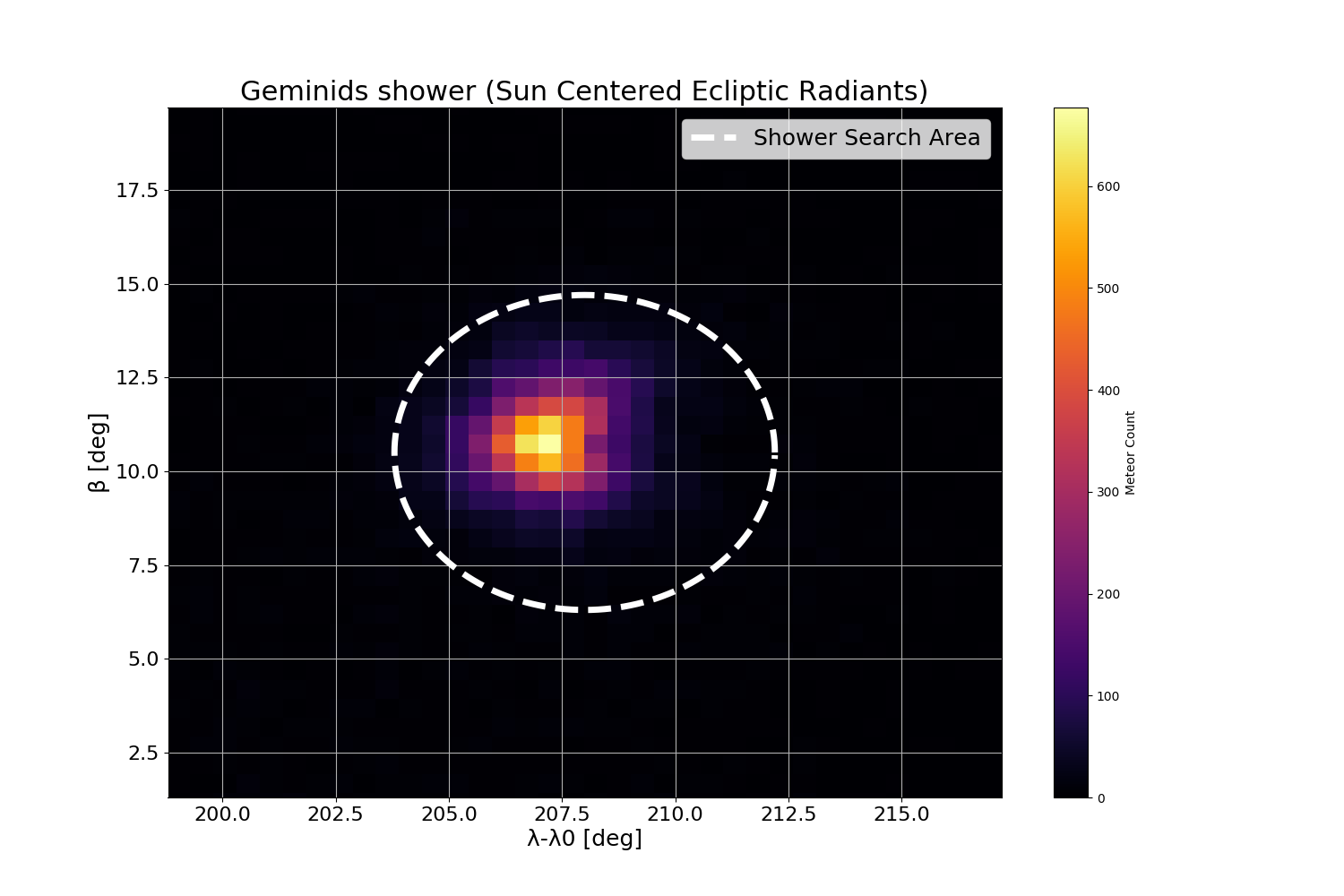}

\end{subfigure}

\medskip
\begin{subfigure}{0.7\textwidth}
\includegraphics[trim = 2.5cm 1.25cm 3.5cm 1.5cm,  clip,width=\linewidth]{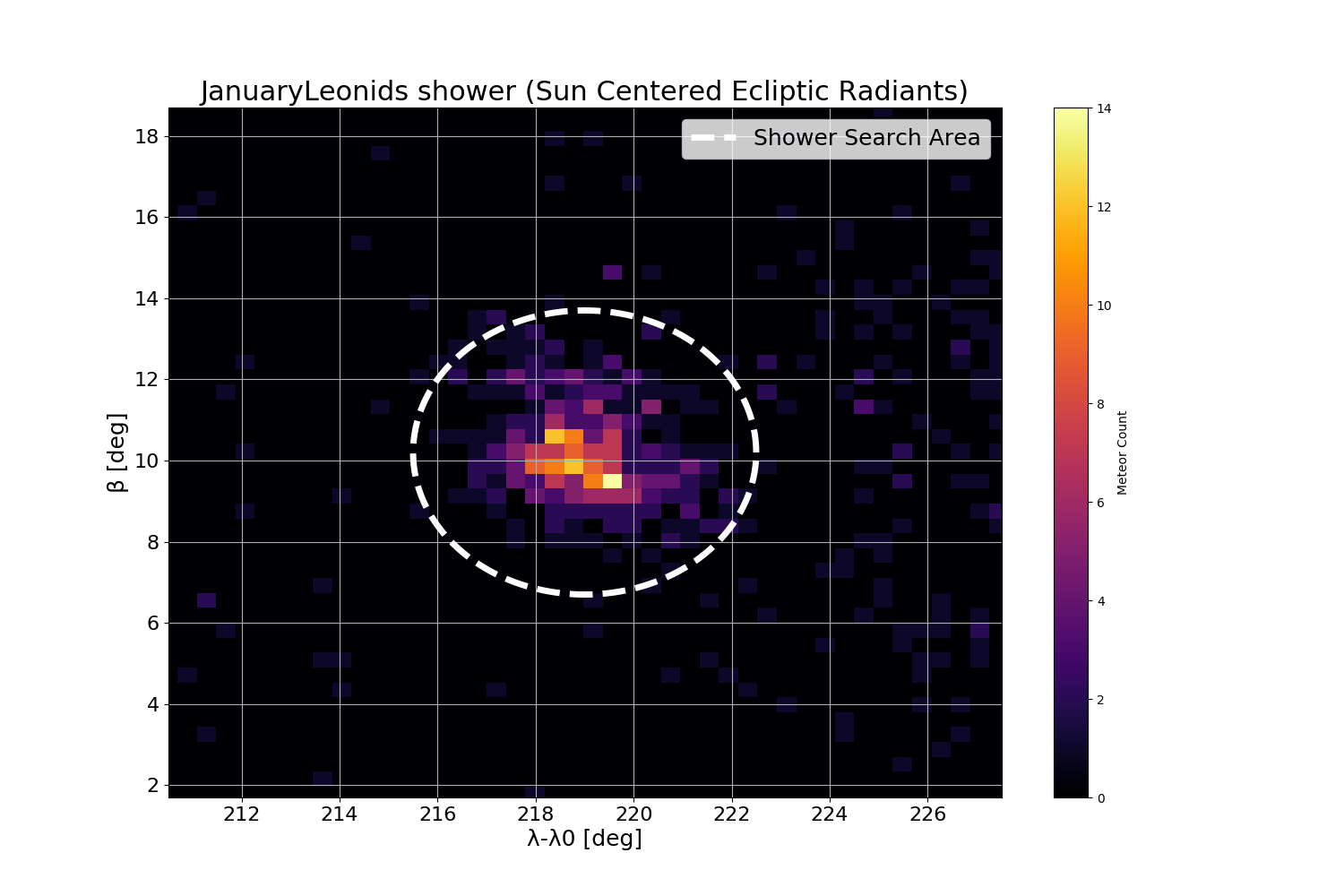}

\end{subfigure}\hspace*{\fill}
\begin{subfigure}{0.7\textwidth}
\includegraphics[trim = 2.5cm 1.25cm 3.5cm 1.5cm,  clip,width=\linewidth]{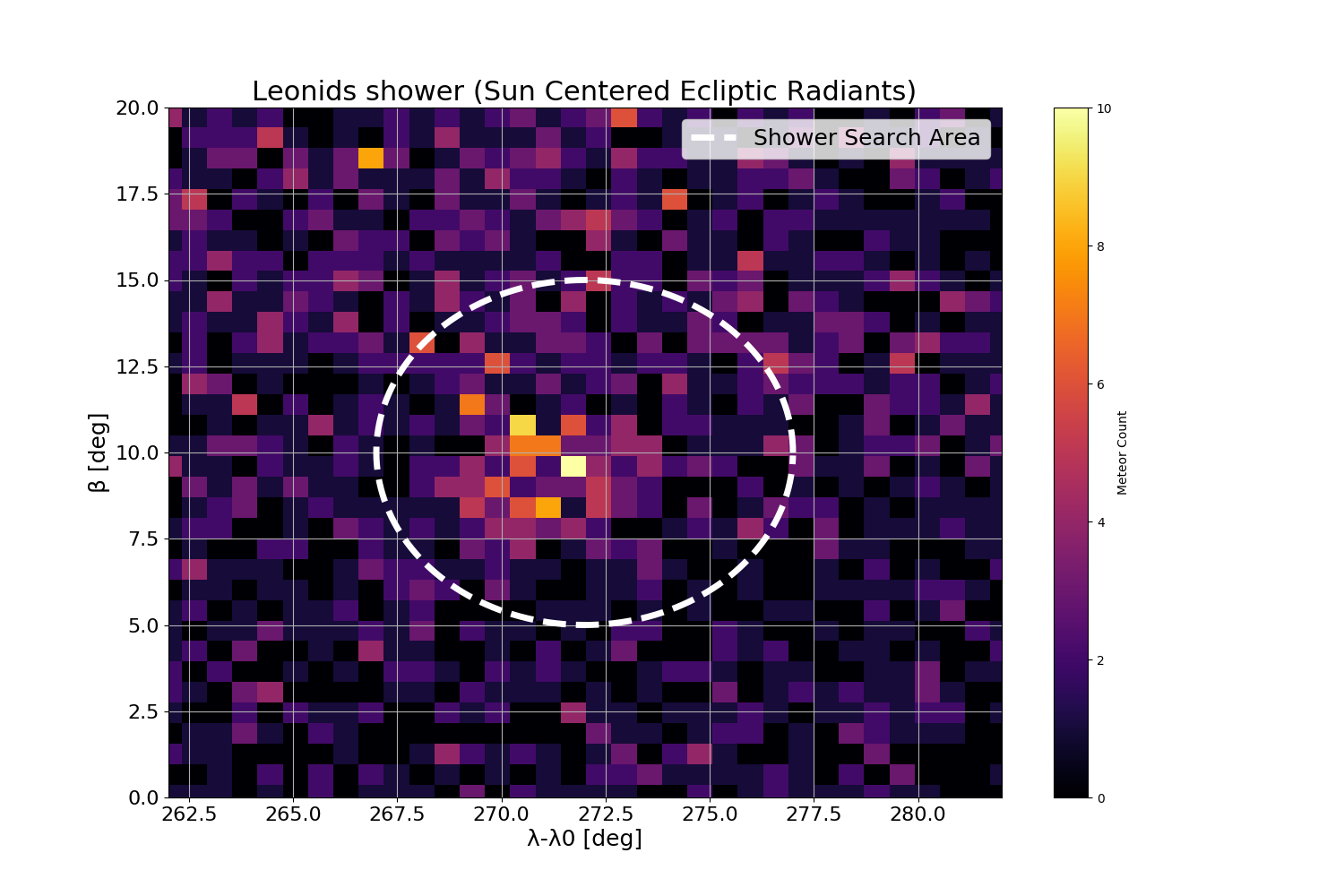}

\end{subfigure}
\caption[$\beta$ and $\lambda -\lambda 0$ positions for meteors used in velocity correction (1/2)]{2D histogram of $\beta$ and $\lambda -\lambda 0$ positions for meteor counts for all showers used in velocity correction (part 1/2)}
\label{apx_showerdens1}
\end{figure}

\clearpage
\newpage

\begin{figure}[t!]
\vspace*{-0.5cm}
\advance\leftskip-3cm
\advance\rightskip-3cm
\begin{subfigure}{0.7\textwidth}
\includegraphics[trim = 2.5cm 1.25cm 3.5cm 1.5cm, clip,width=\linewidth]{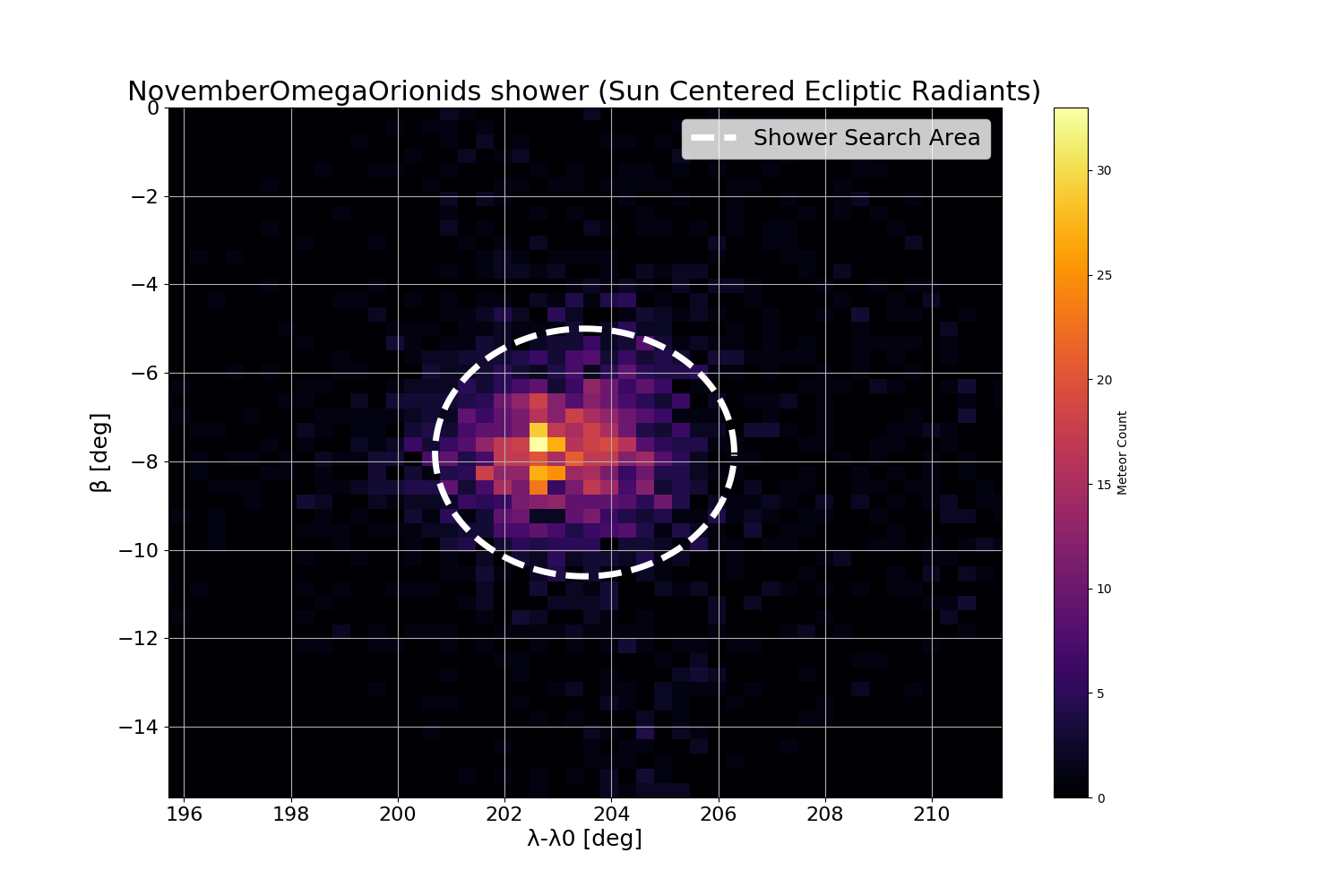}

\end{subfigure}\hspace*{\fill}
\begin{subfigure}{0.7\textwidth}
\includegraphics[trim = 2.5cm 1.25cm 3.5cm 1.5cm, clip,width=\linewidth]{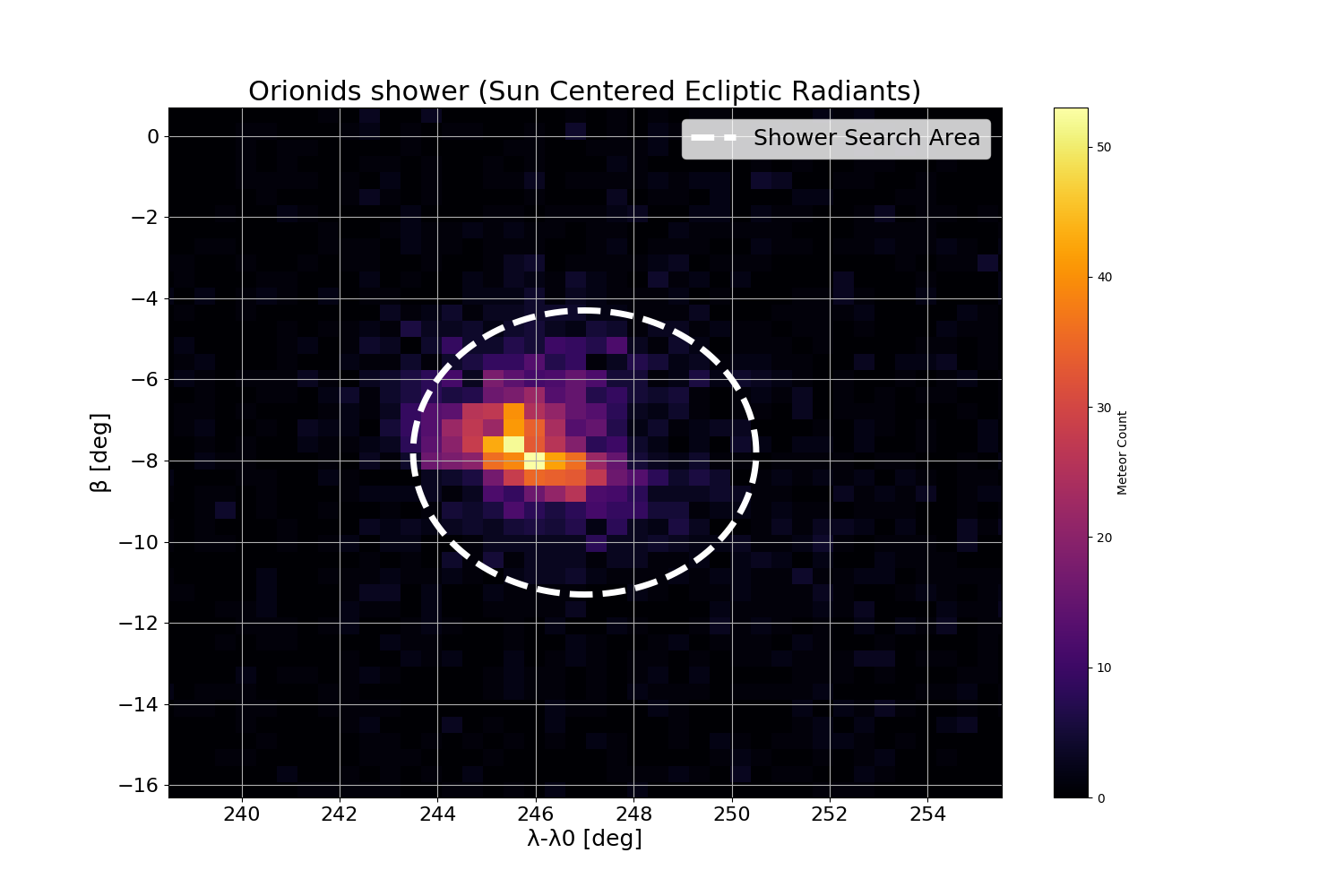}

\end{subfigure}

\medskip
\begin{subfigure}{0.7\textwidth}
\includegraphics[trim = 2.5cm 1.25cm 3.5cm 1.5cm, clip,width=\linewidth]{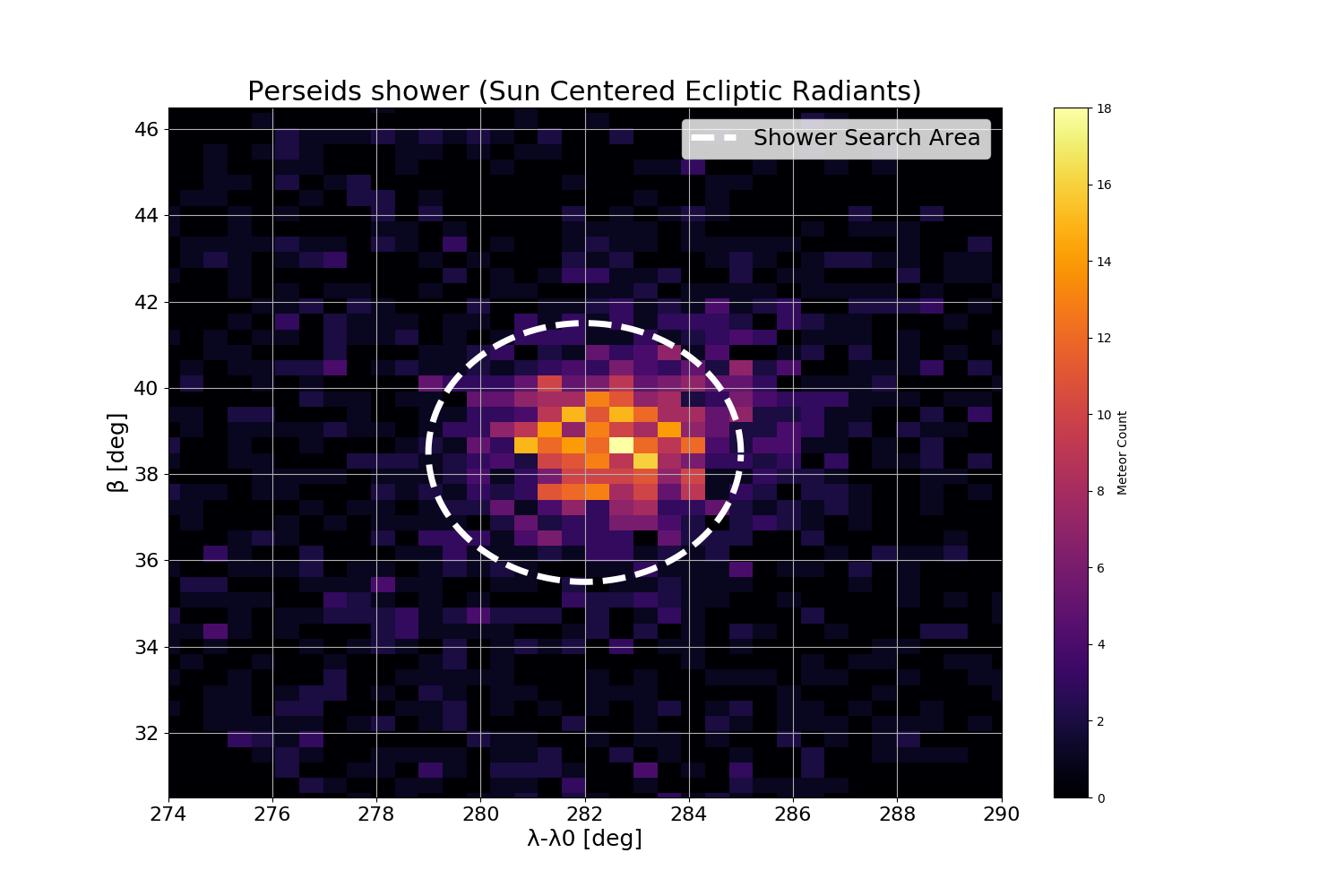}

\end{subfigure}\hspace*{\fill}
\begin{subfigure}{0.7\textwidth}
\includegraphics[trim = 2.5cm 1.25cm 3.5cm 1.5cm, clip,width=\linewidth]{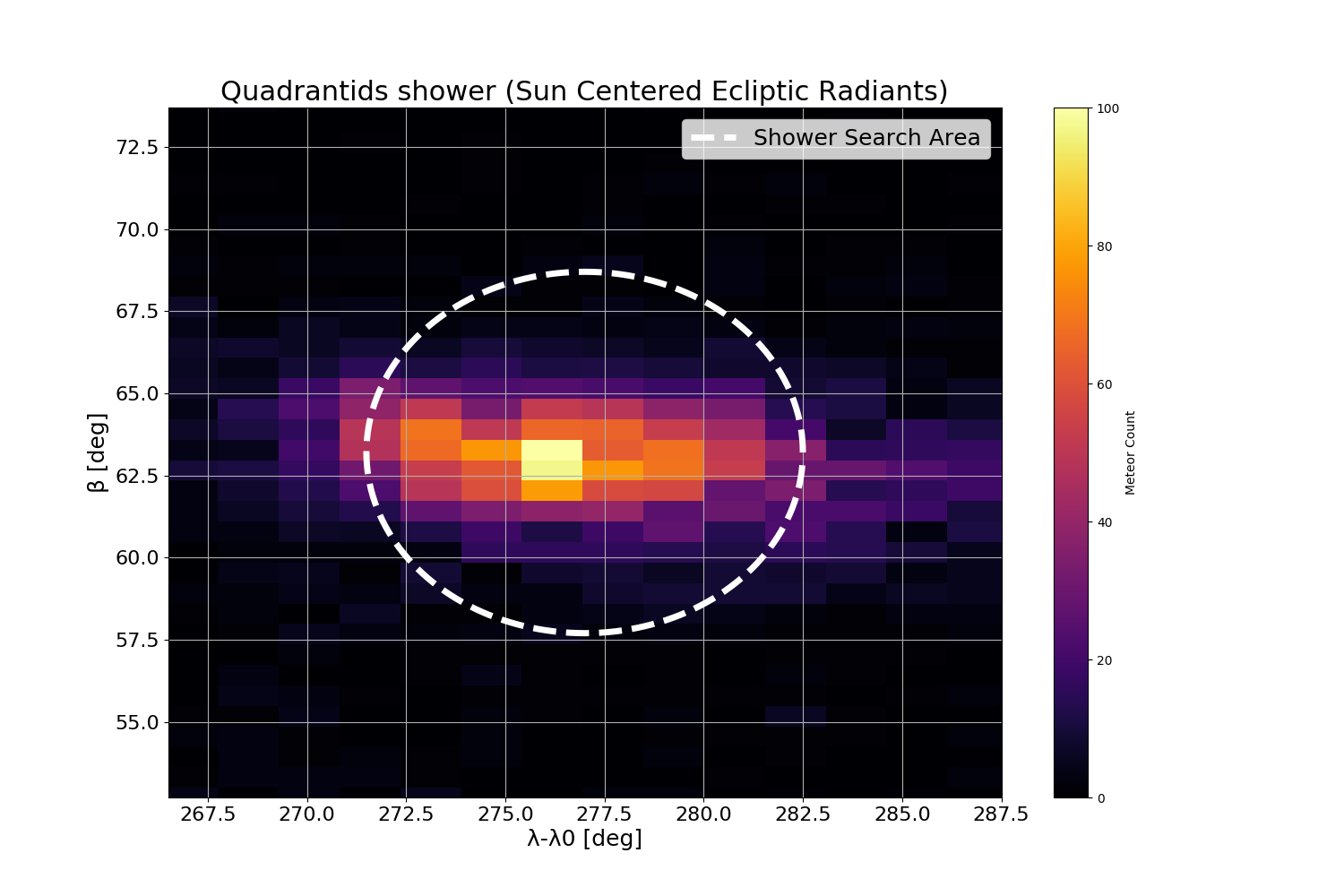}

\end{subfigure}

\medskip
\begin{subfigure}{0.7\textwidth}
\includegraphics[trim = 2.5cm 1.25cm 3.5cm 1.5cm, clip,width=\linewidth]{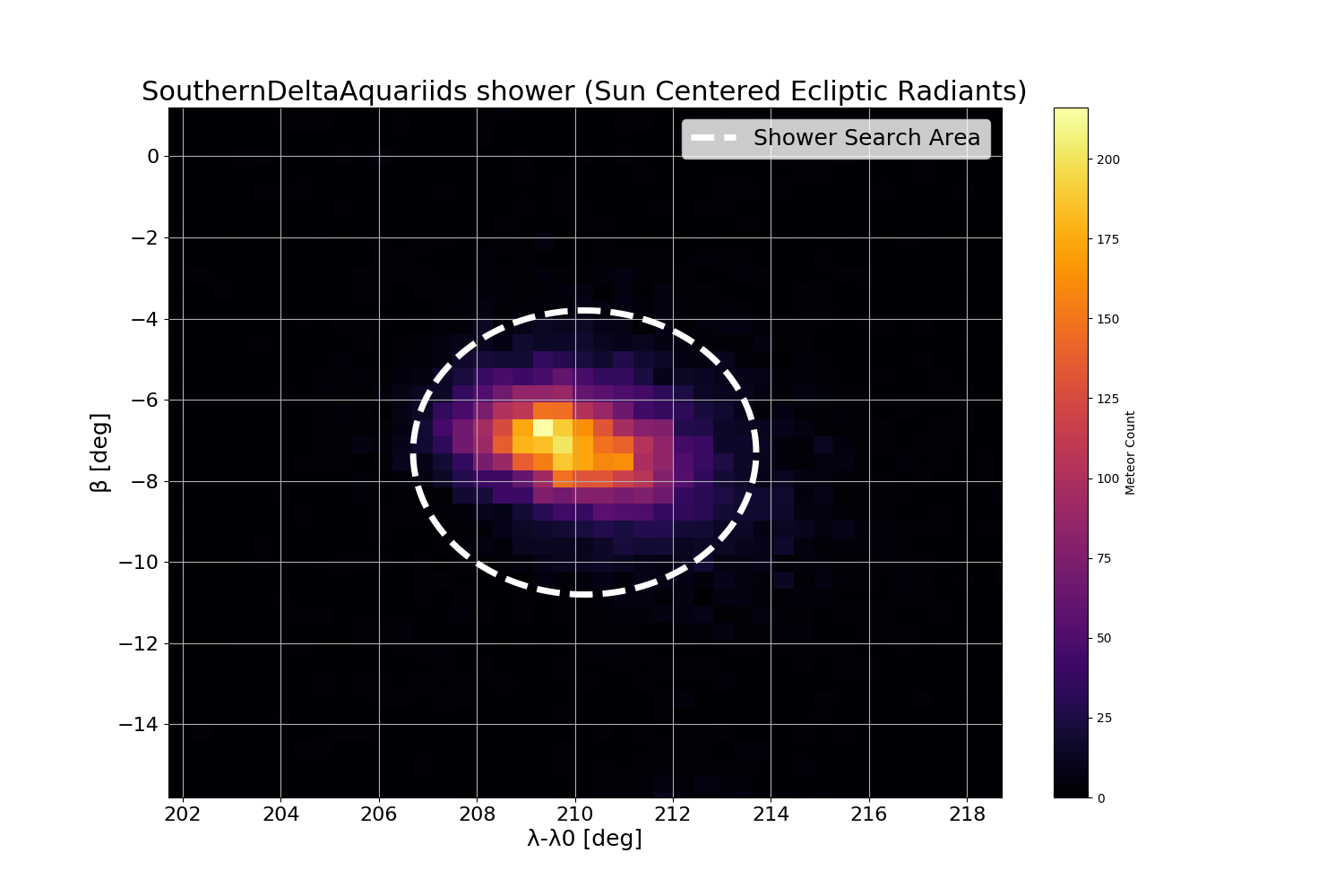}
\end{subfigure}\hspace*{\fill}
\begin{subfigure}{0.7\textwidth}
\includegraphics[trim = 2.5cm 1.25cm 3.5cm 1.5cm, clip, width=\linewidth]{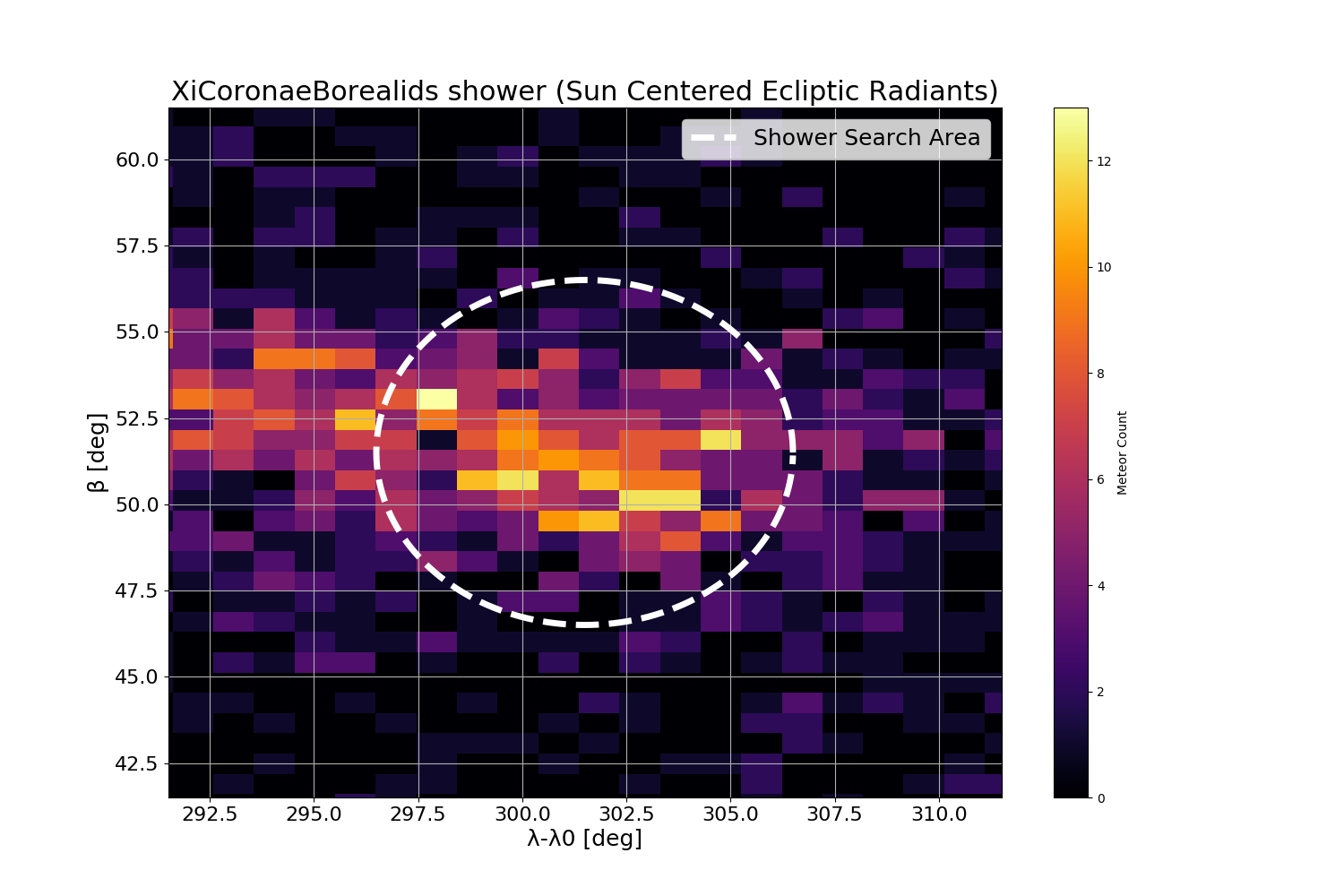}

\end{subfigure}
\caption[$\beta$ and $\lambda -\lambda 0$ positions for meteors used in velocity correction (2/2)]{2D histogram of $\beta$ and $\lambda -\lambda 0$ positions for meteor counts for all showers used in velocity correction (part 2/2)}
\label{apx_showerdens2}
\end{figure}


 \clearpage
 \newpage

\begin{figure}[t!]
\vspace*{-0.5cm}
\advance\leftskip-3cm
\advance\rightskip-3cm
\begin{subfigure}{0.8\textwidth}
\includegraphics[trim = 2cm 1cm 3.5cm 1.5cm, clip, clip,width=\linewidth]{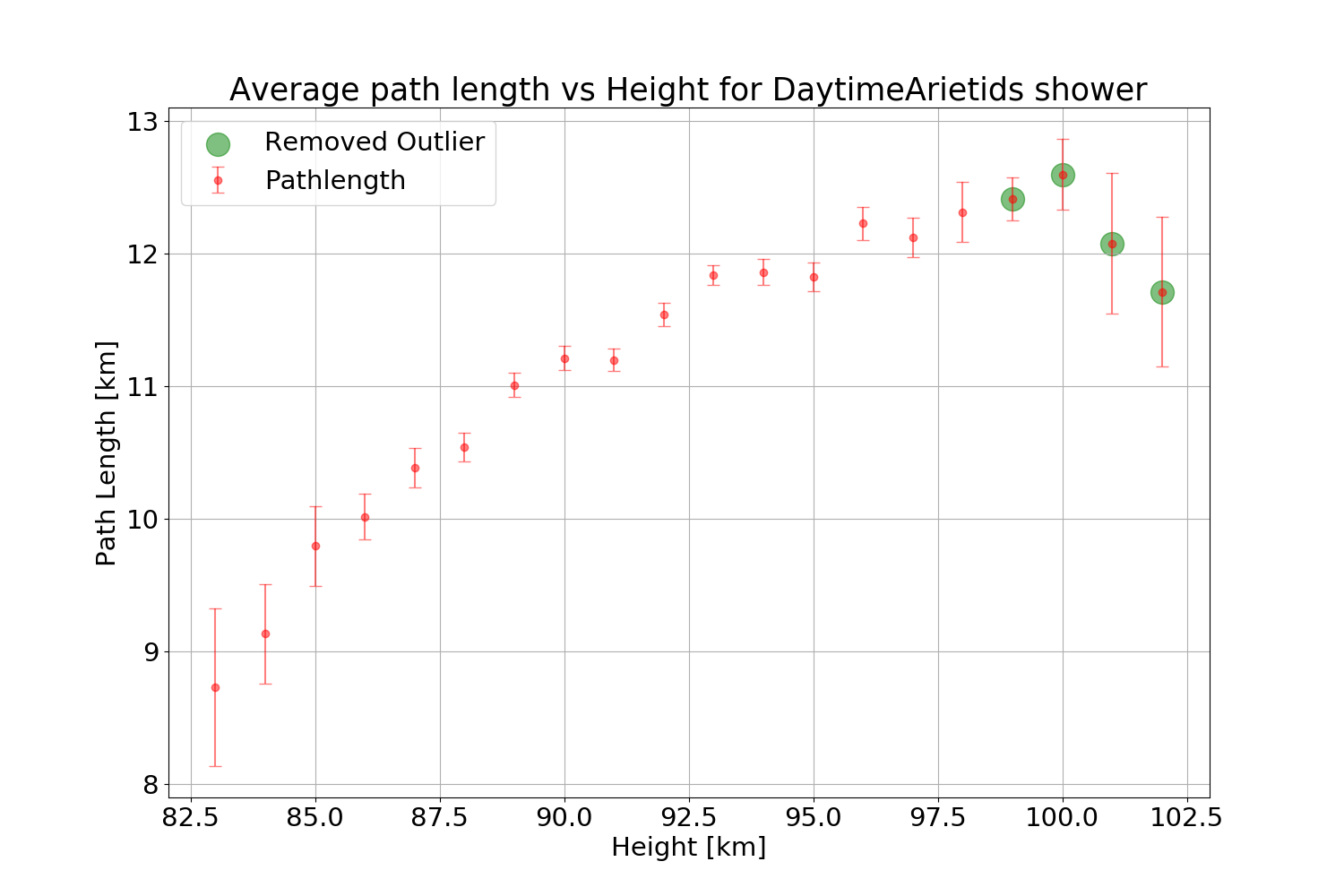}

\end{subfigure}\hspace*{\fill}
\begin{subfigure}{0.8\textwidth}
\includegraphics[trim = 2cm 1cm 3.5cm 1.5cm, clip, clip,width=\linewidth]{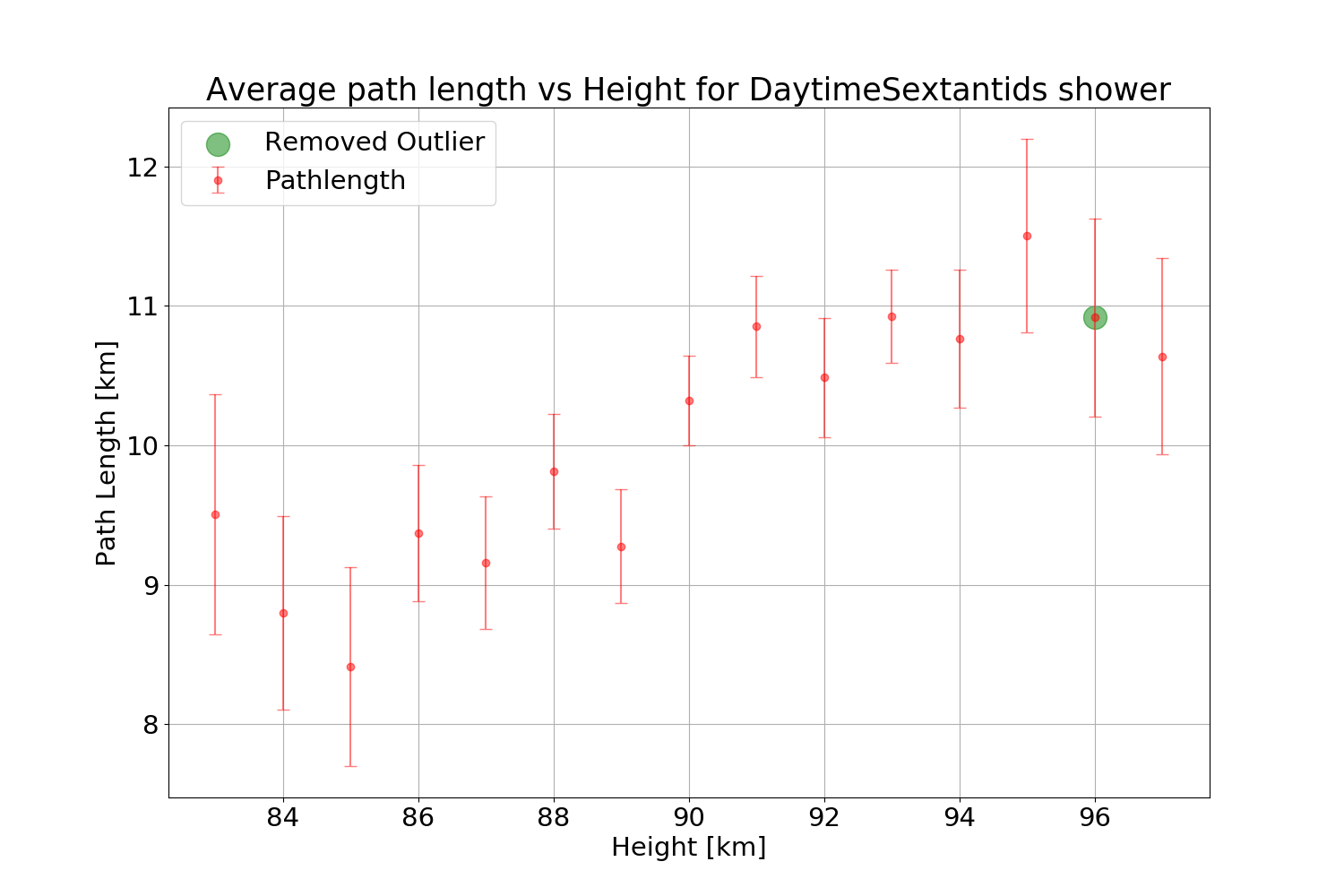}

\end{subfigure}

\medskip
\begin{subfigure}{0.8\textwidth}
\includegraphics[trim = 2cm 1cm 3.5cm 1.5cm, clip, clip,width=\linewidth]{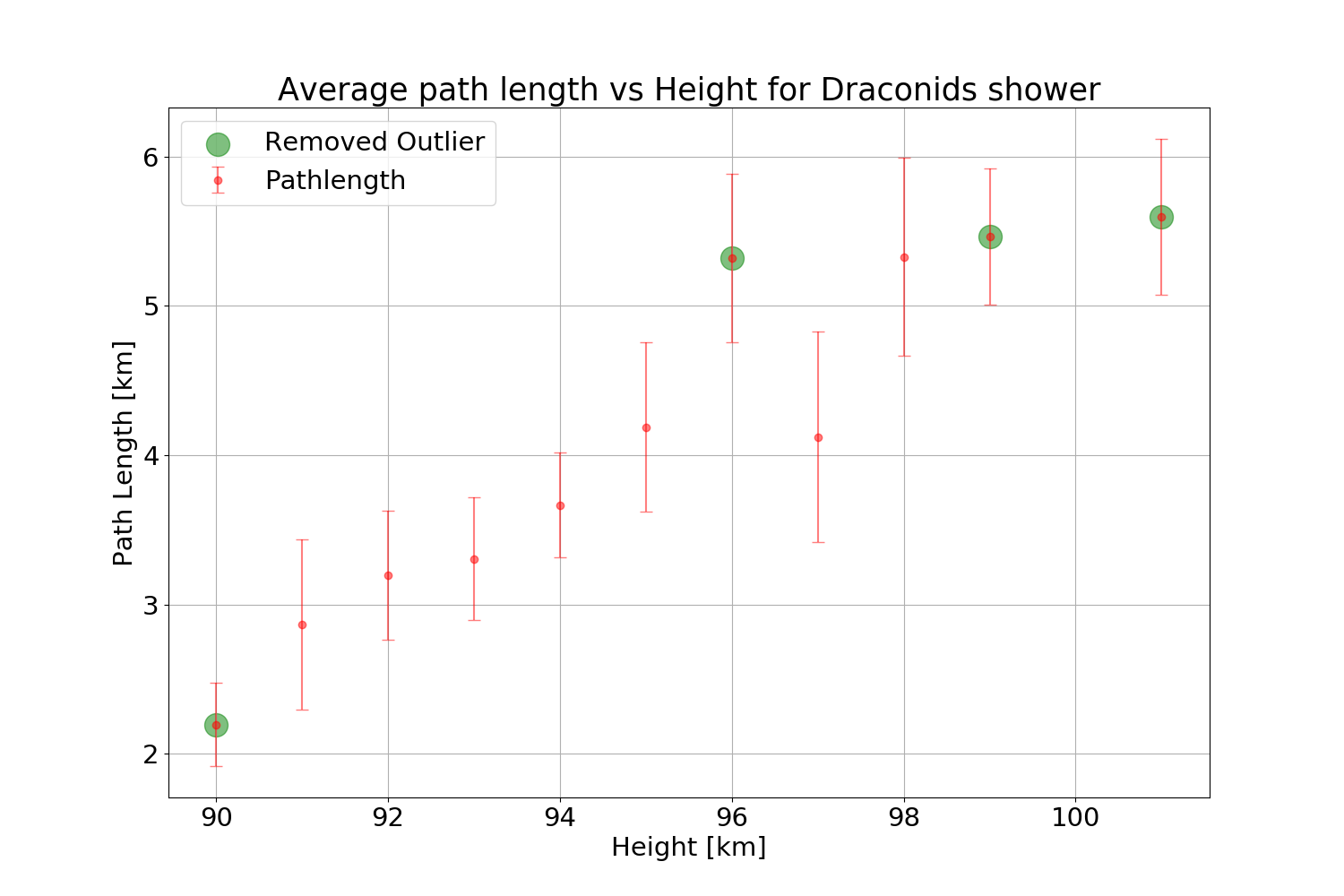}

\end{subfigure}\hspace*{\fill}
\begin{subfigure}{0.8\textwidth}
\includegraphics[trim = 2cm 1cm 3.5cm 1.5cm, clip, clip,width=\linewidth]{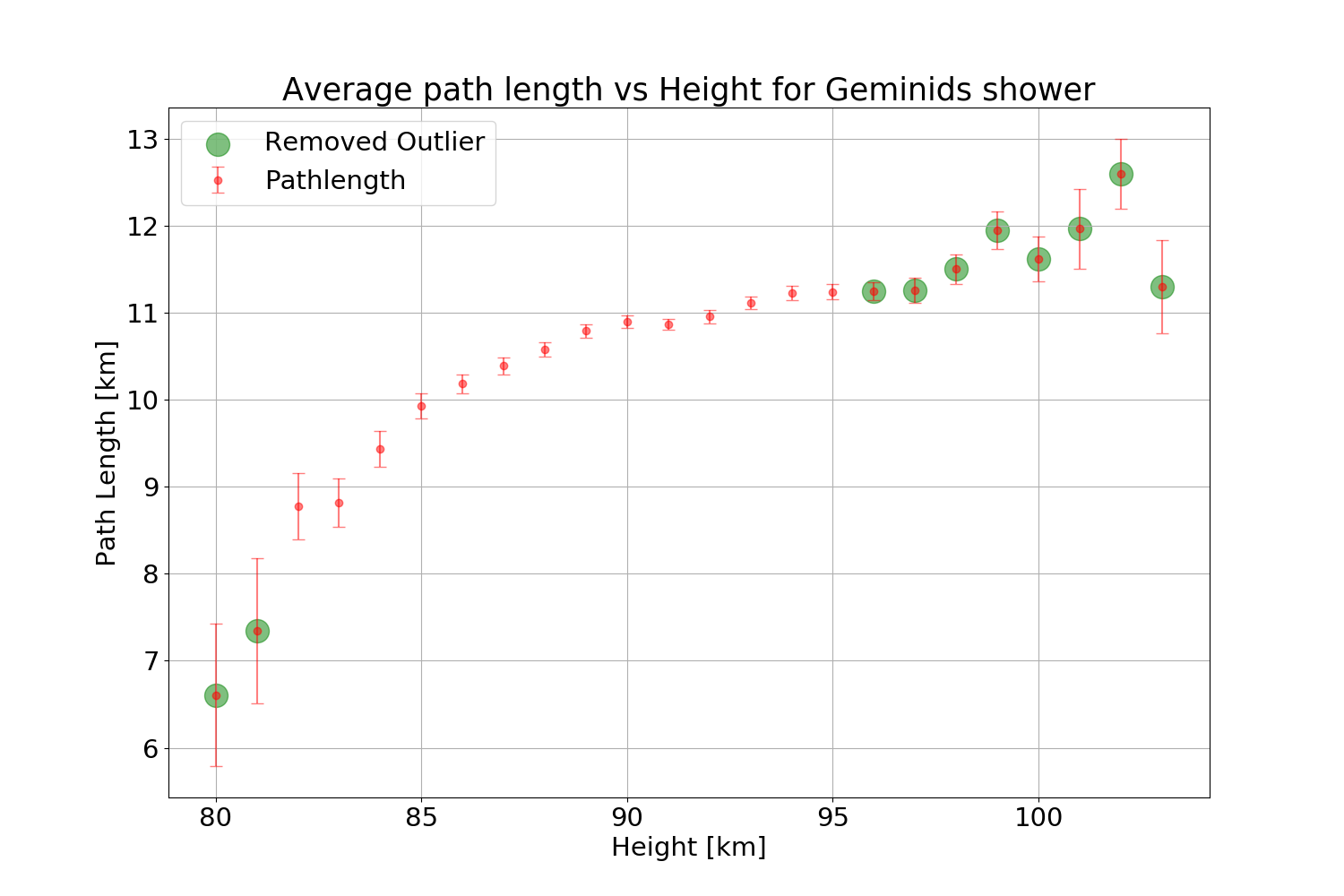}

\end{subfigure}

\medskip
\begin{subfigure}{0.8\textwidth}
\includegraphics[trim = 2cm 1cm 3.5cm 1.5cm, clip, clip,width=\linewidth]{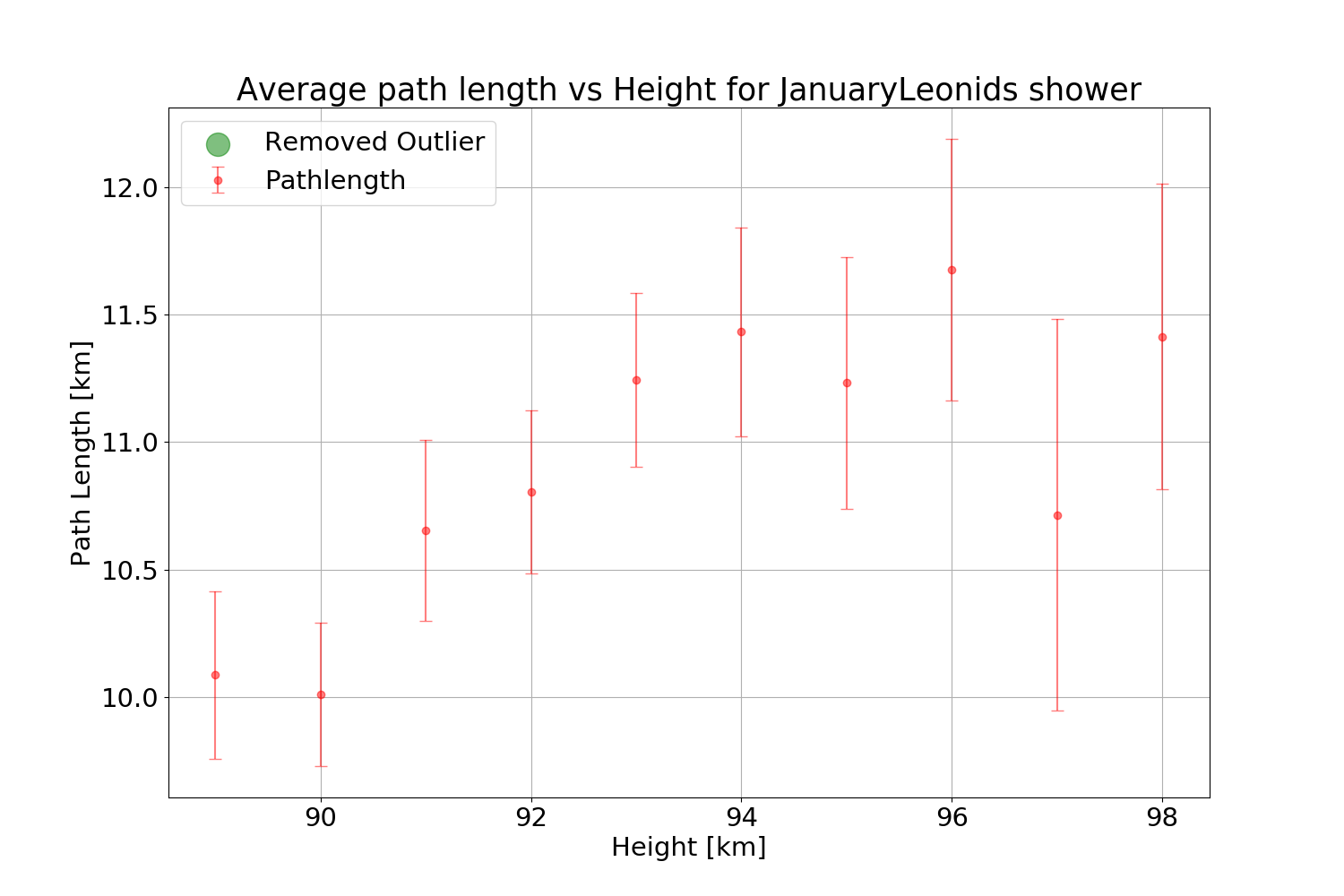}

\end{subfigure}\hspace*{\fill}
\begin{subfigure}{0.8\textwidth}
\includegraphics[trim = 2cm 1cm 3.5cm 1.5cm, clip, clip,width=\linewidth]{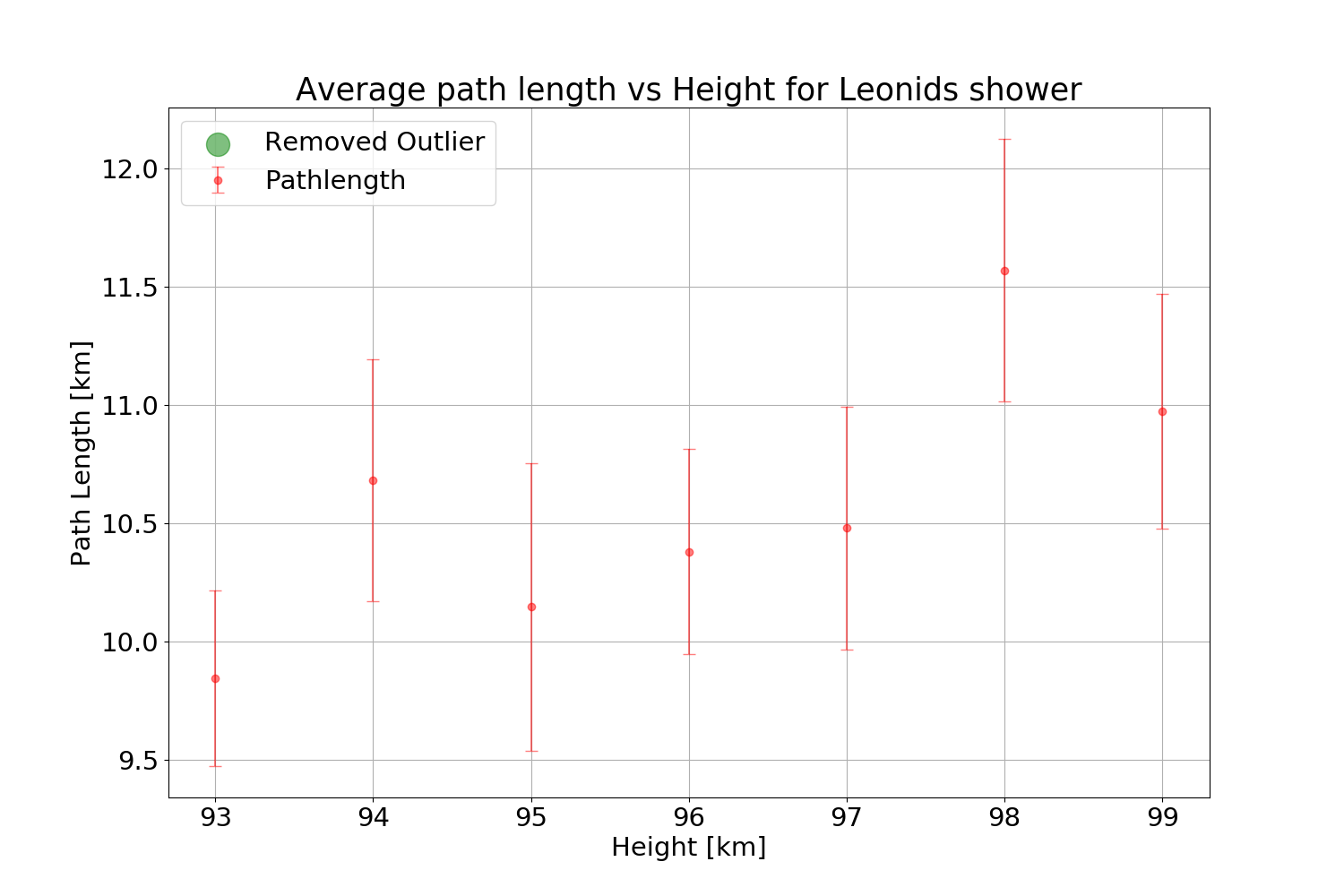}

\end{subfigure}
\caption[Path length vs height for showers used in velocity correction (1/2)]{Average path length versus height plots for all showers used in velocity correction (part 1/2)} \label{apx_plmean1}
\end{figure}

\clearpage
\newpage

\begin{figure}[t!]
\vspace*{-0.5cm}
\advance\leftskip-3cm
\advance\rightskip-3cm
\begin{subfigure}{0.8\textwidth}
\includegraphics[trim = 2cm 1cm 3.5cm 1.5cm, clip, clip,width=\linewidth]{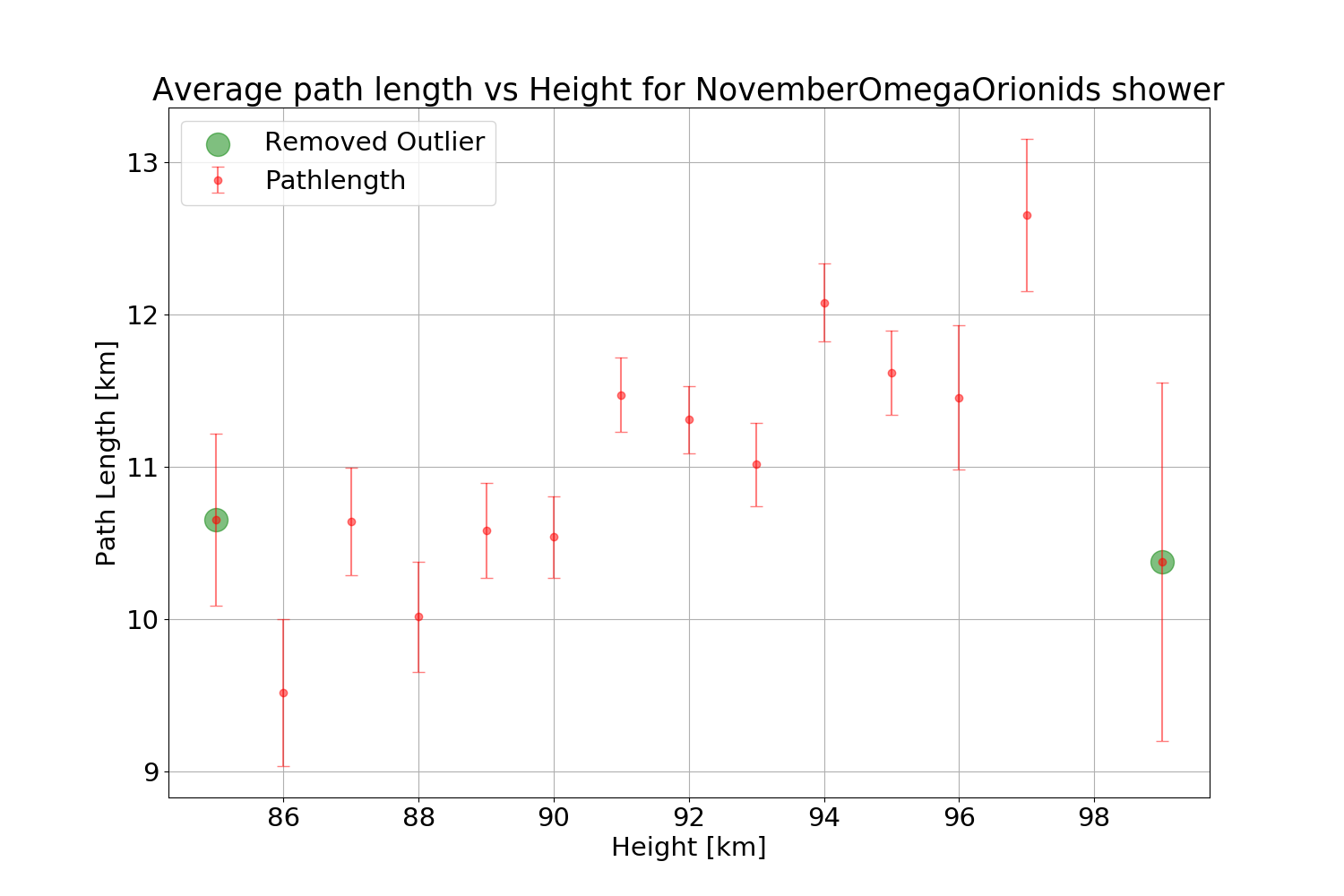}

\end{subfigure}\hspace*{\fill}
\begin{subfigure}{0.8\textwidth}
\includegraphics[trim = 2cm 1cm 3.5cm 1.5cm, clip, clip,width=\linewidth]{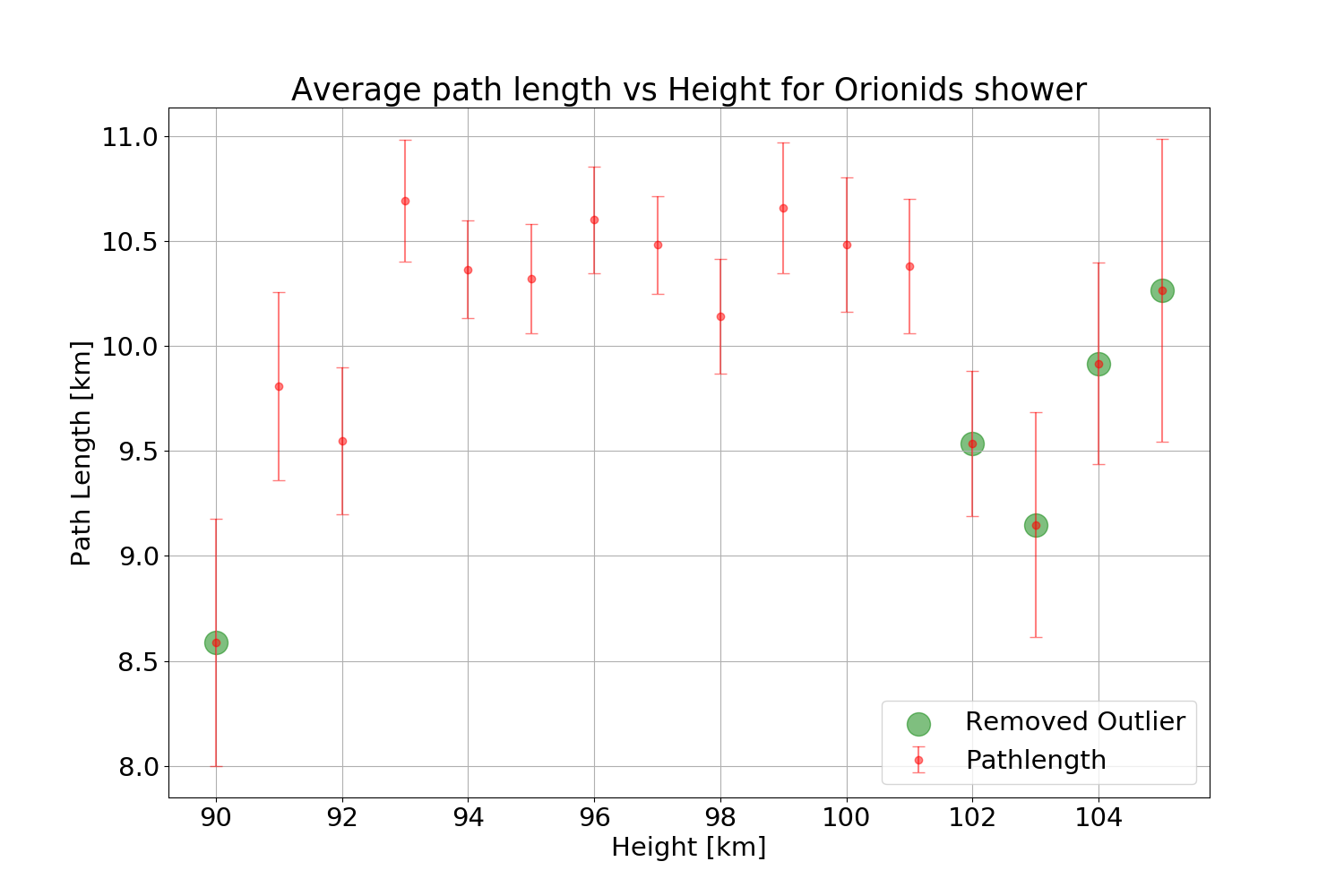}

\end{subfigure}

\medskip
\begin{subfigure}{0.8\textwidth}
\includegraphics[trim = 2cm 1cm 3.5cm 1.5cm, clip, clip,width=\linewidth]{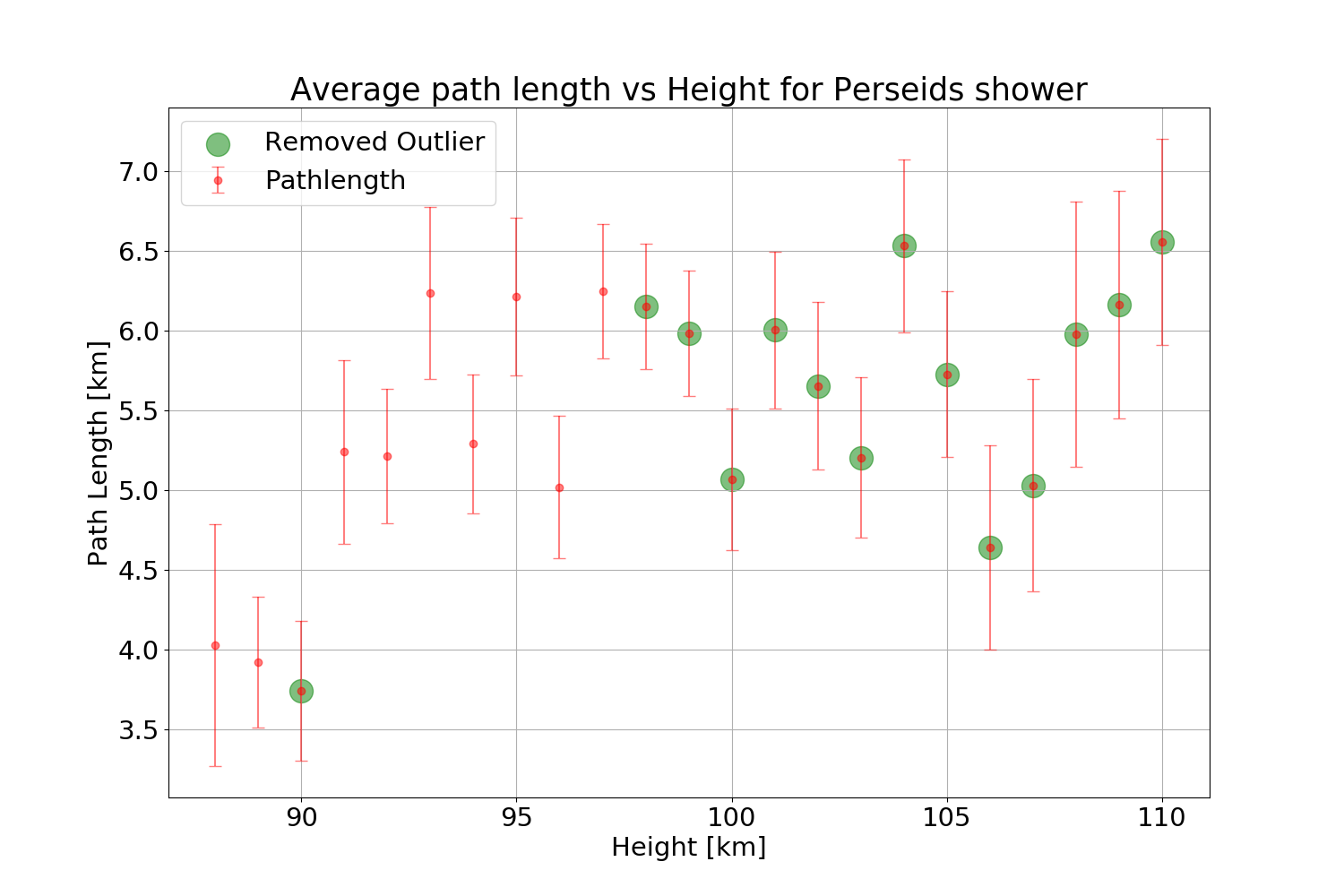}

\end{subfigure}\hspace*{\fill}
\begin{subfigure}{0.8\textwidth}
\includegraphics[trim = 2cm 1cm 3.5cm 1.5cm, clip, clip,width=\linewidth]{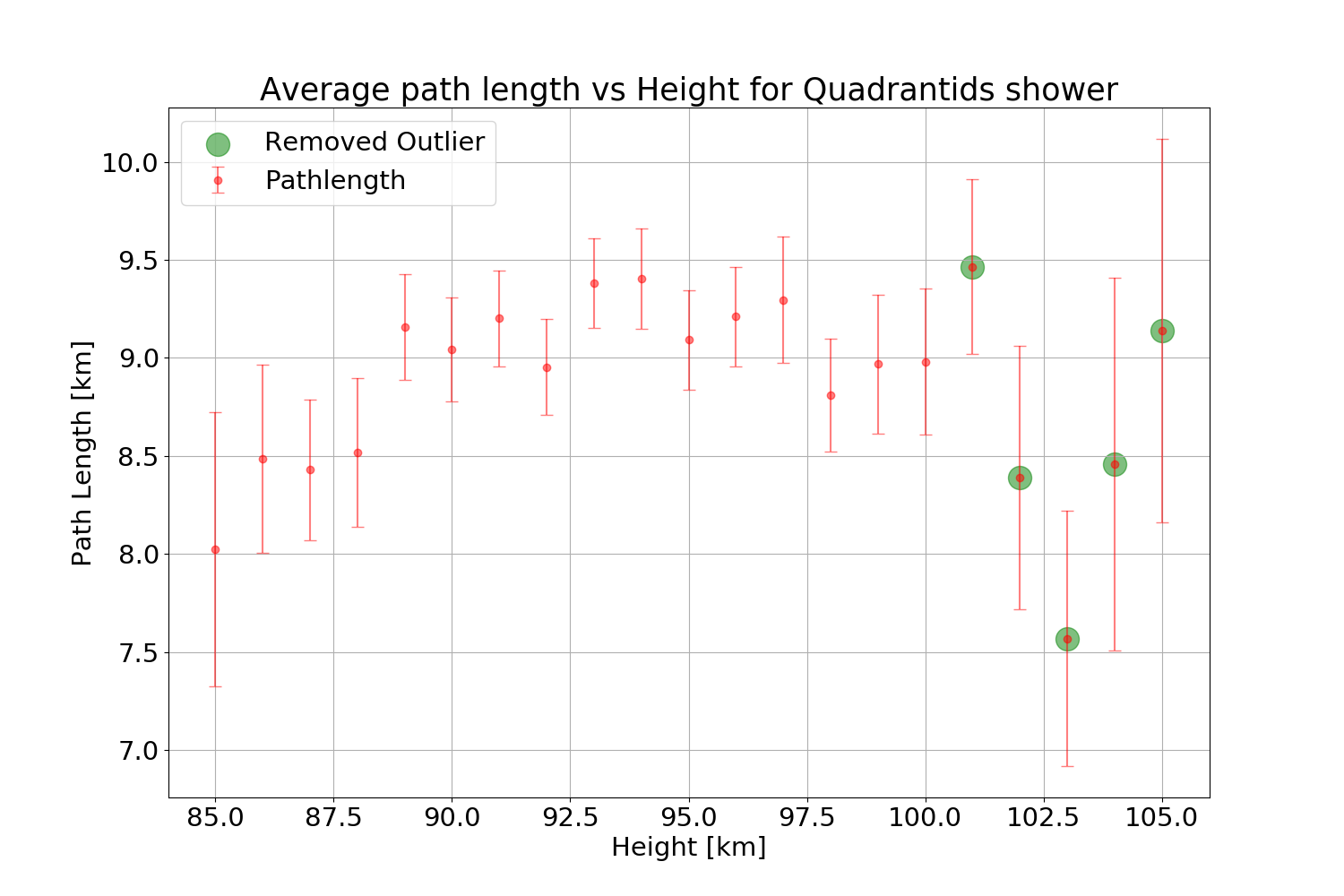}

\end{subfigure}

\medskip
\begin{subfigure}{0.8\textwidth}
\includegraphics[trim = 2cm 1cm 3.5cm 1.5cm, clip, clip,width=\linewidth]{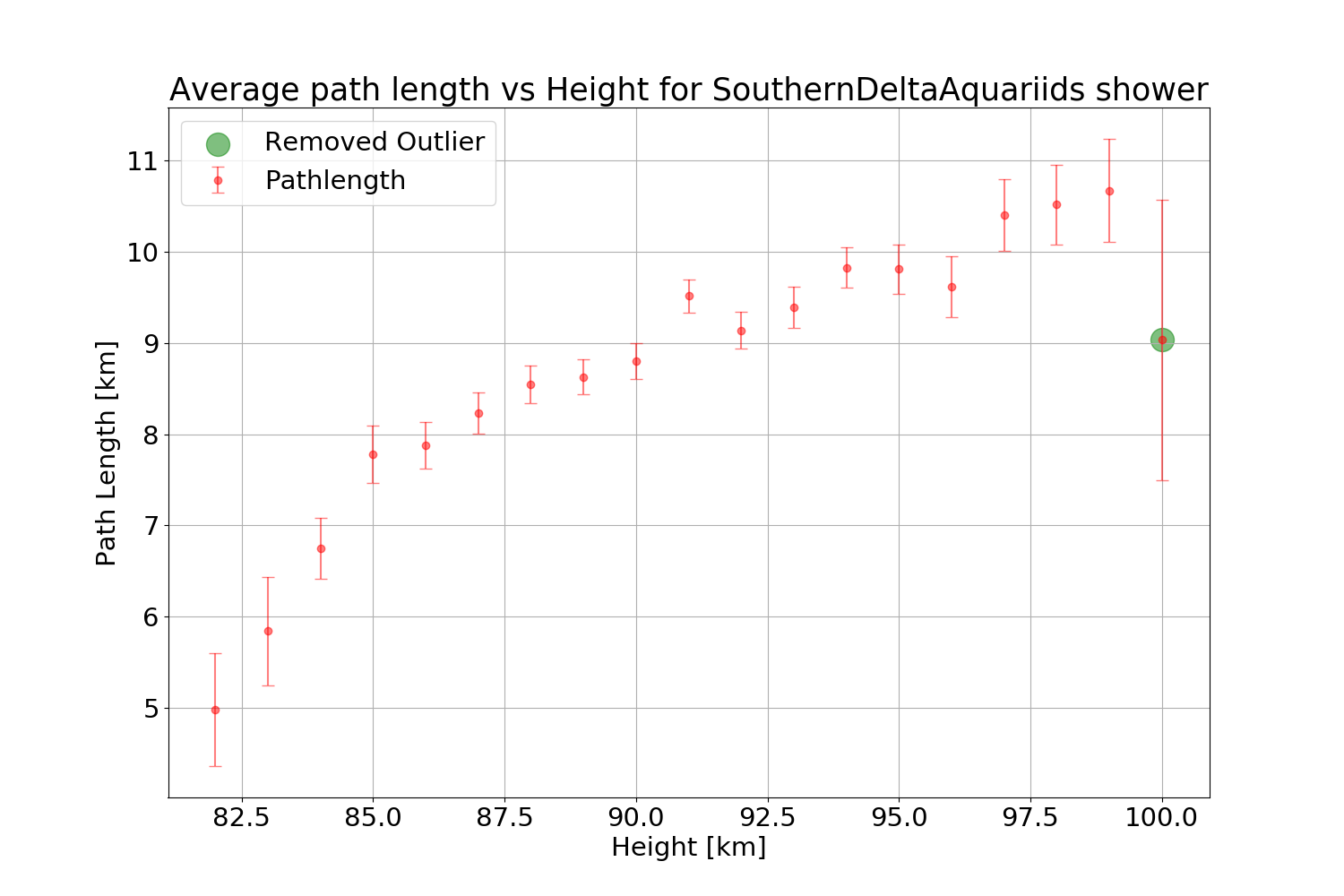}
\end{subfigure}\hspace*{\fill}
\begin{subfigure}{0.8\textwidth}
\includegraphics[trim = 2cm 1cm 3.5cm 1.5cm, clip, width=\linewidth]{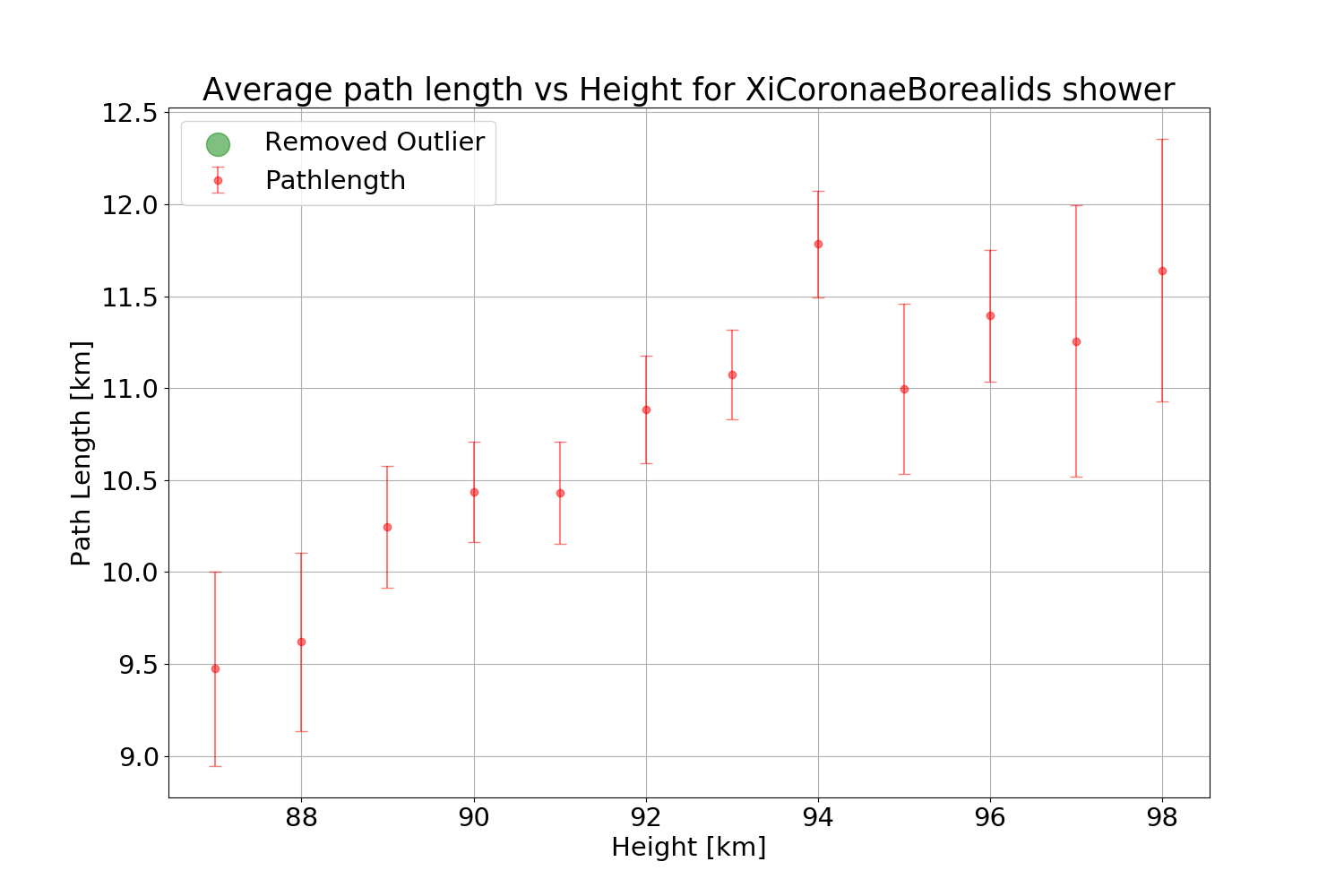}

\end{subfigure}
\caption[Path length vs height for showers used in velocity correction (2/2)]{Average path length versus height plots for all showers used in velocity correction (part 2/2)}
\label{apx_plmean2}
\end{figure}

\clearpage
\newpage

\begin{figure}[t!]
\vspace*{-0.5cm}
\advance\leftskip-3cm
\advance\rightskip-3cm
\begin{subfigure}{0.8\textwidth}
\includegraphics[trim = 2cm 1cm 3.5cm 1.5cm, clip, clip,width=\linewidth]{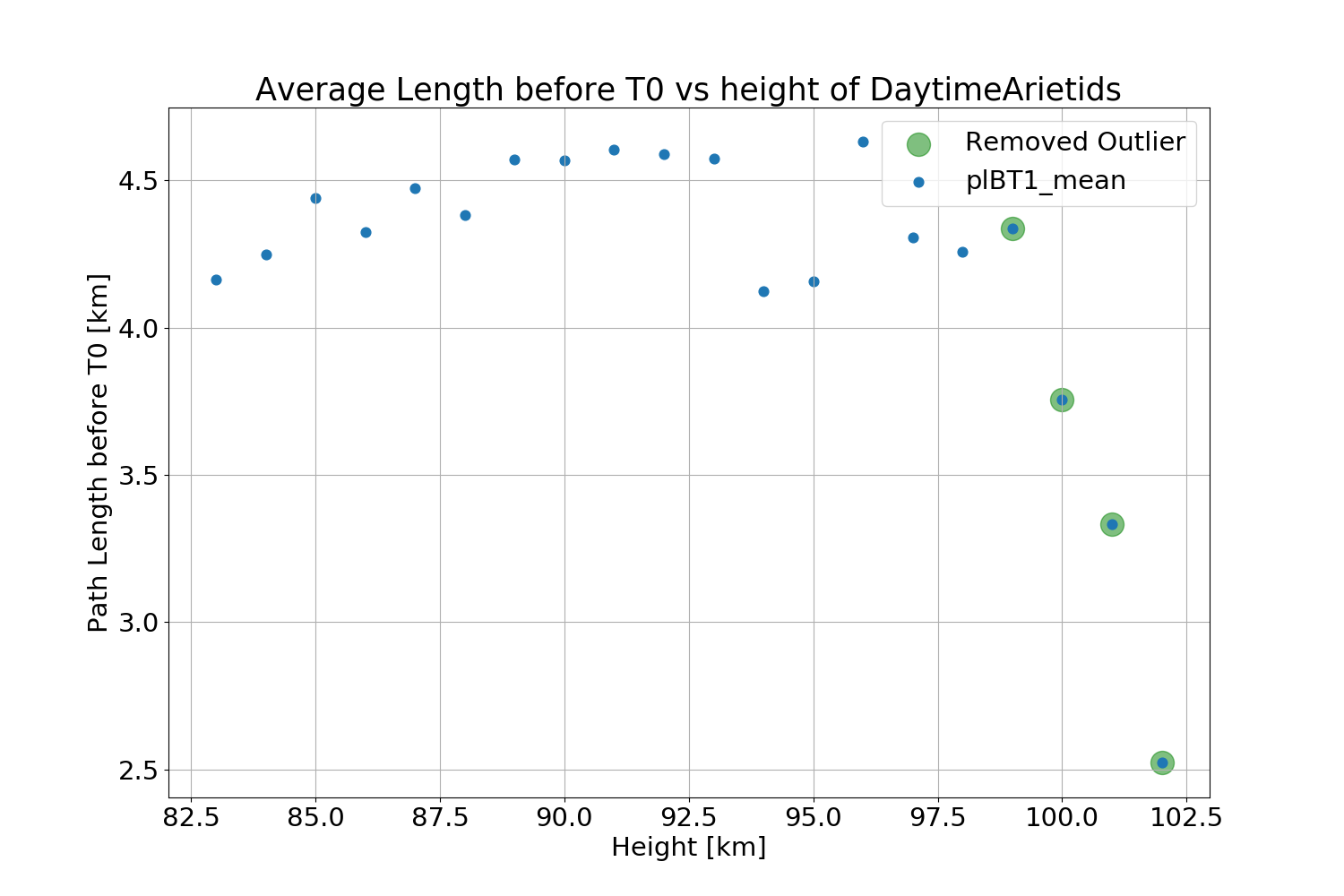}

\end{subfigure}\hspace*{\fill}
\begin{subfigure}{0.8\textwidth}
\includegraphics[trim = 2cm 1cm 3.5cm 1.5cm, clip, clip,width=\linewidth]{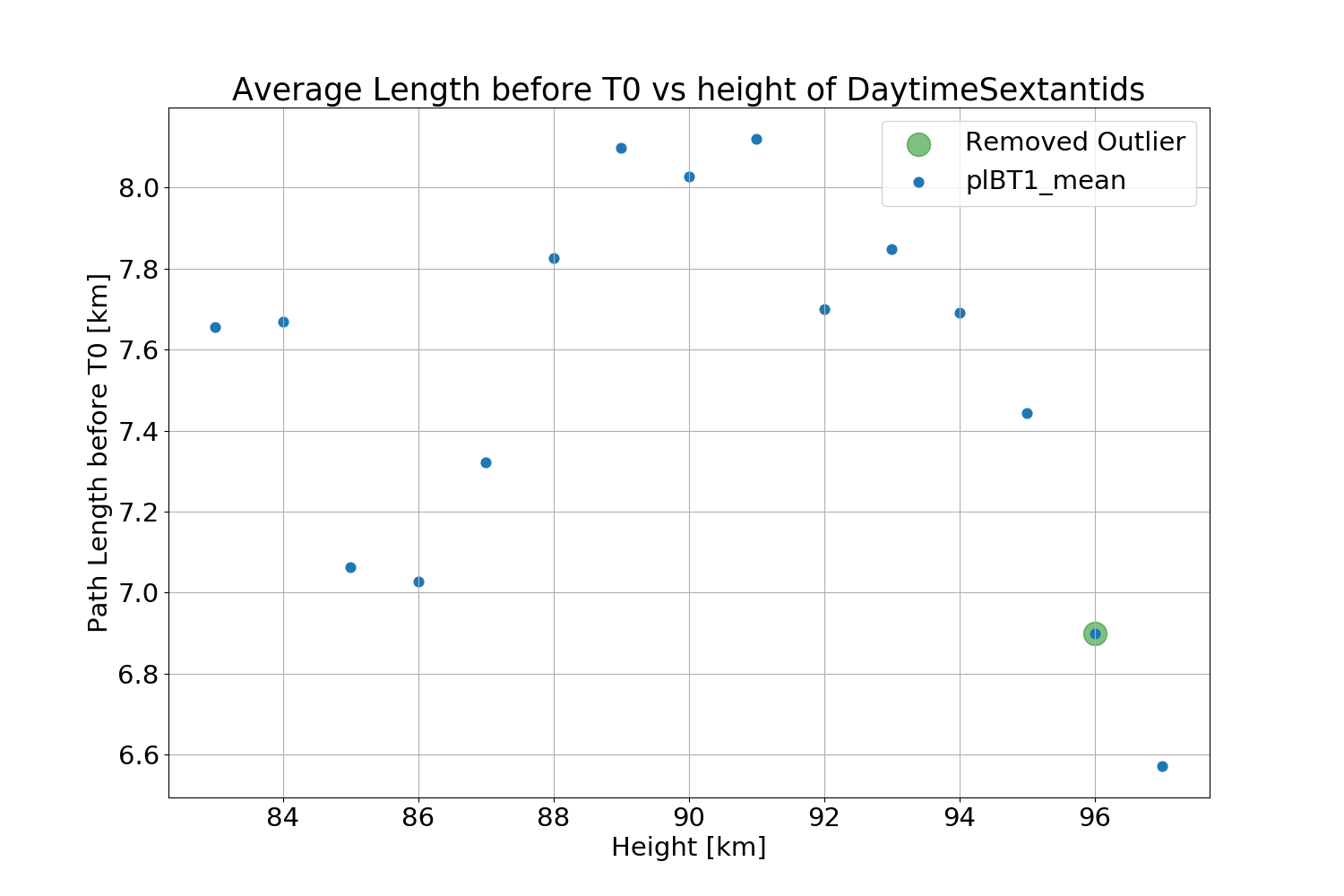}

\end{subfigure}

\medskip
\begin{subfigure}{0.8\textwidth}
\includegraphics[trim = 2cm 1cm 3.5cm 1.5cm, clip, clip,width=\linewidth]{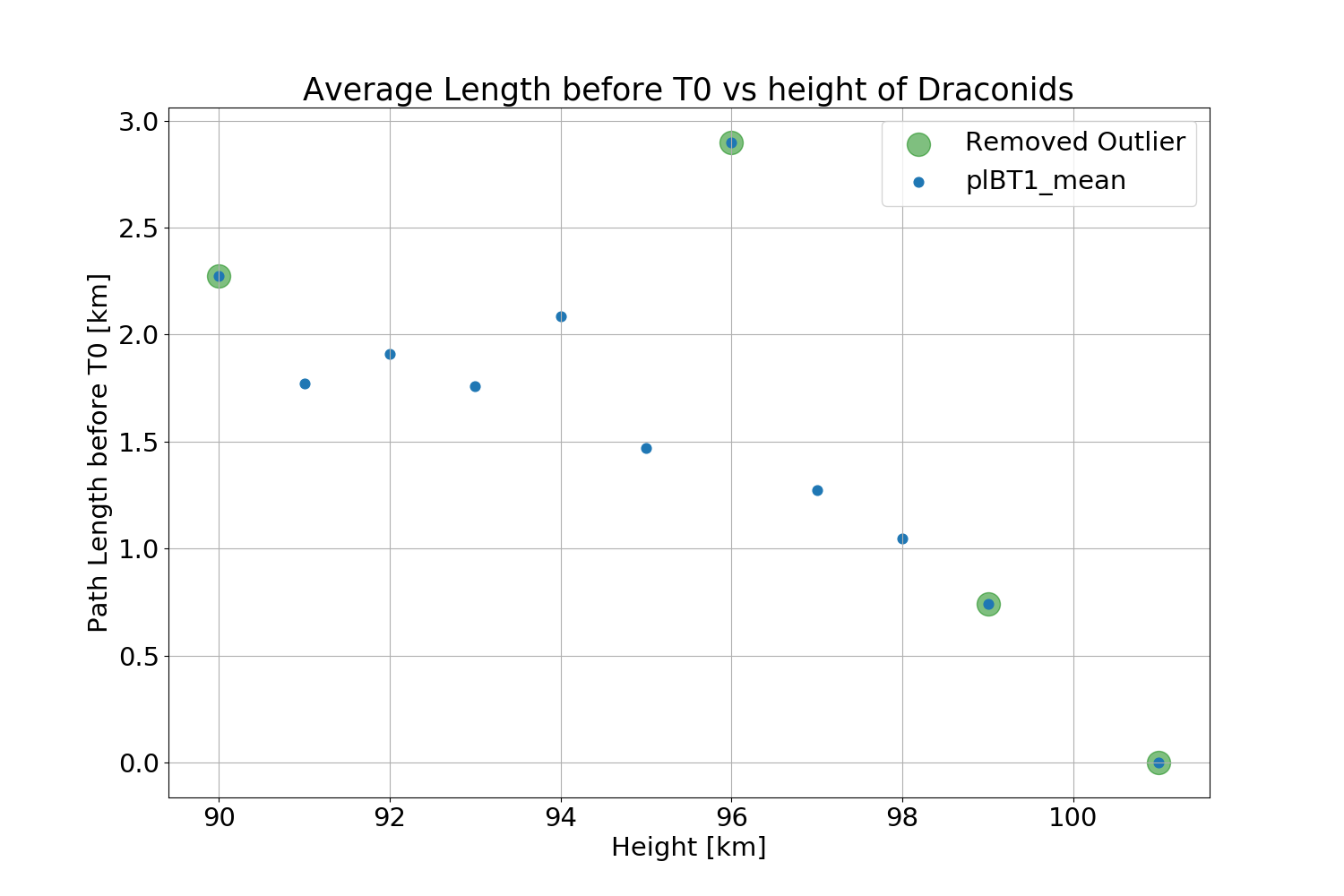}

\end{subfigure}\hspace*{\fill}
\begin{subfigure}{0.8\textwidth}
\includegraphics[trim = 2cm 1cm 3.5cm 1.5cm, clip, clip,width=\linewidth]{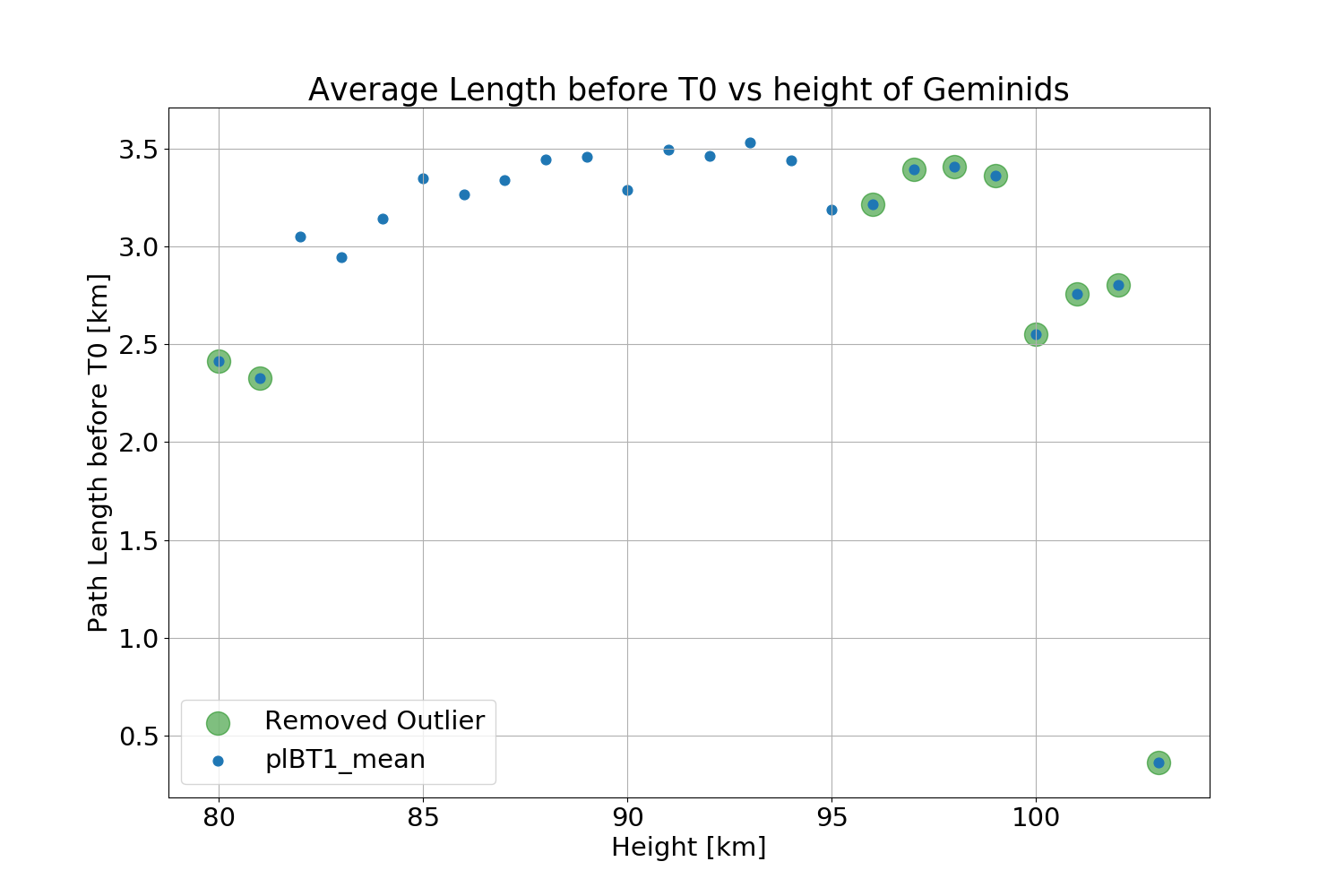}

\end{subfigure}

\medskip
\begin{subfigure}{0.8\textwidth}
\includegraphics[trim = 2cm 1cm 3.5cm 1.5cm, clip, clip,width=\linewidth]{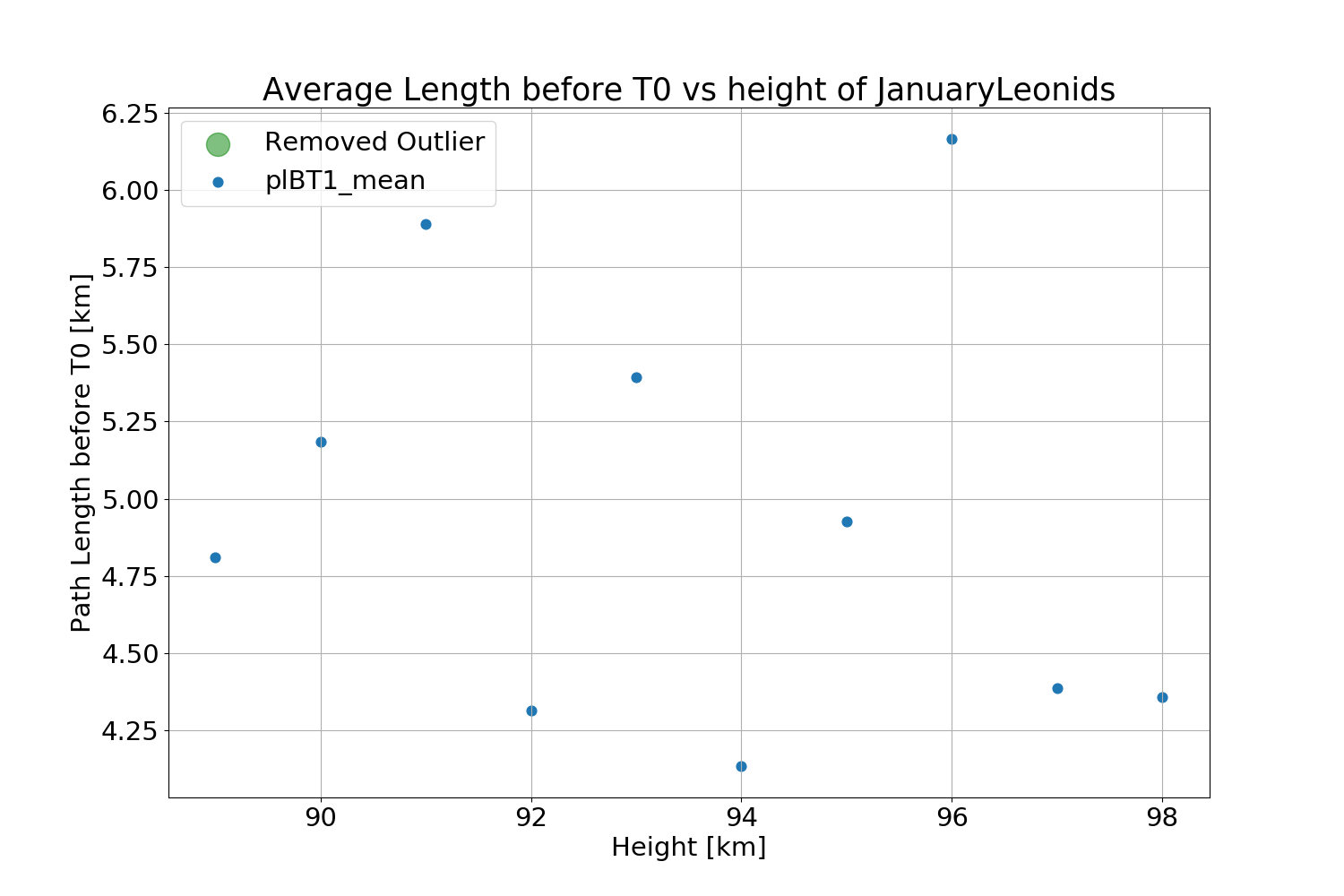}

\end{subfigure}\hspace*{\fill}
\begin{subfigure}{0.8\textwidth}
\includegraphics[trim = 2cm 1cm 3.5cm 1.5cm, clip, clip,width=\linewidth]{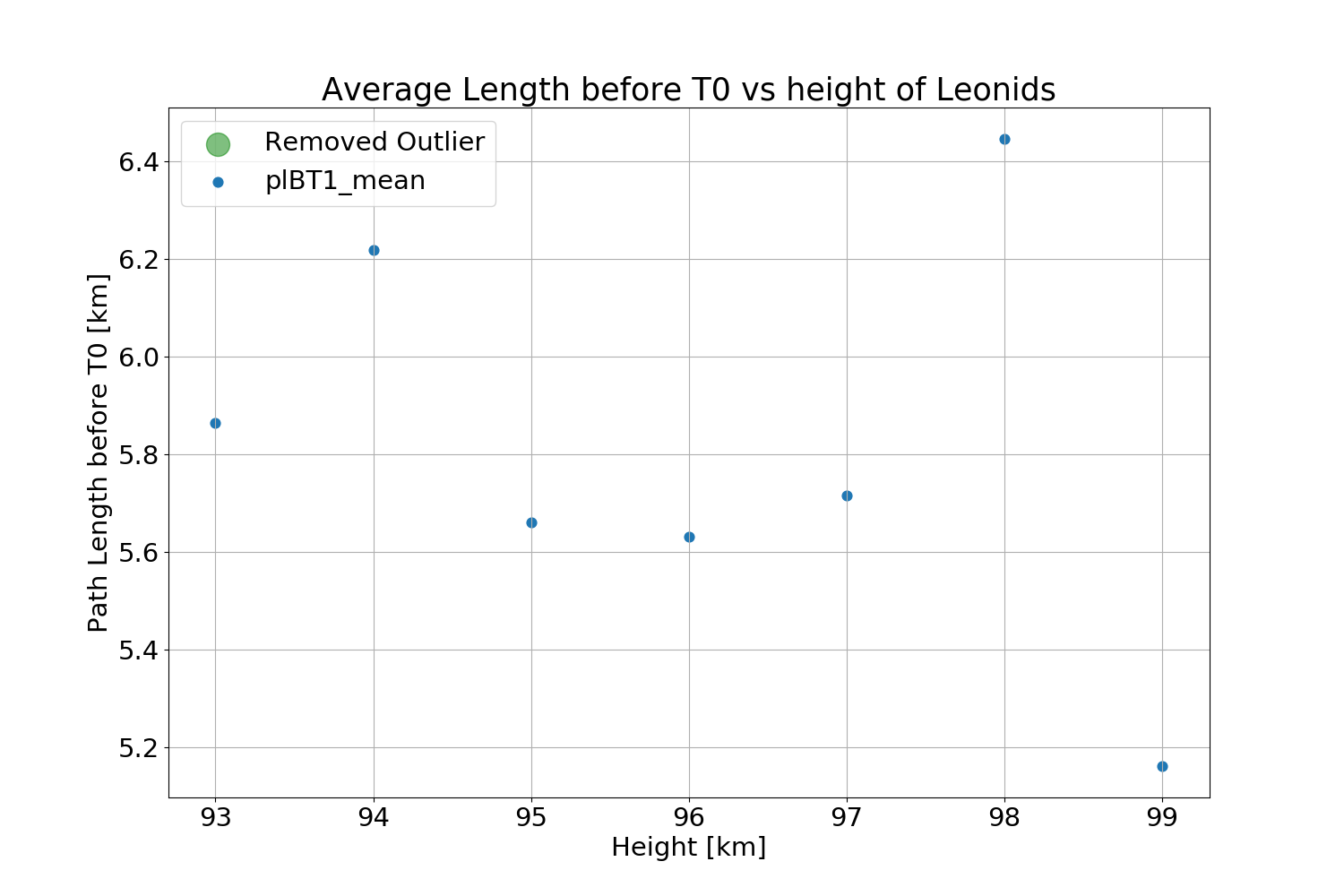}

\end{subfigure}
\caption[Path length before T0 vs height for showers used in velocity correction (1/2)]{Average path length before detection at station T0 versus height plots for all showers used in velocity correction (part 1/2)}
\label{apx_plt11}
\end{figure}

\clearpage
\newpage

\begin{figure}[t!]
\vspace*{-0.5cm}
\advance\leftskip-3cm
\advance\rightskip-3cm
\begin{subfigure}{0.8\textwidth}
\includegraphics[trim = 2cm 1cm 3.5cm 1.5cm, clip, clip,width=\linewidth]{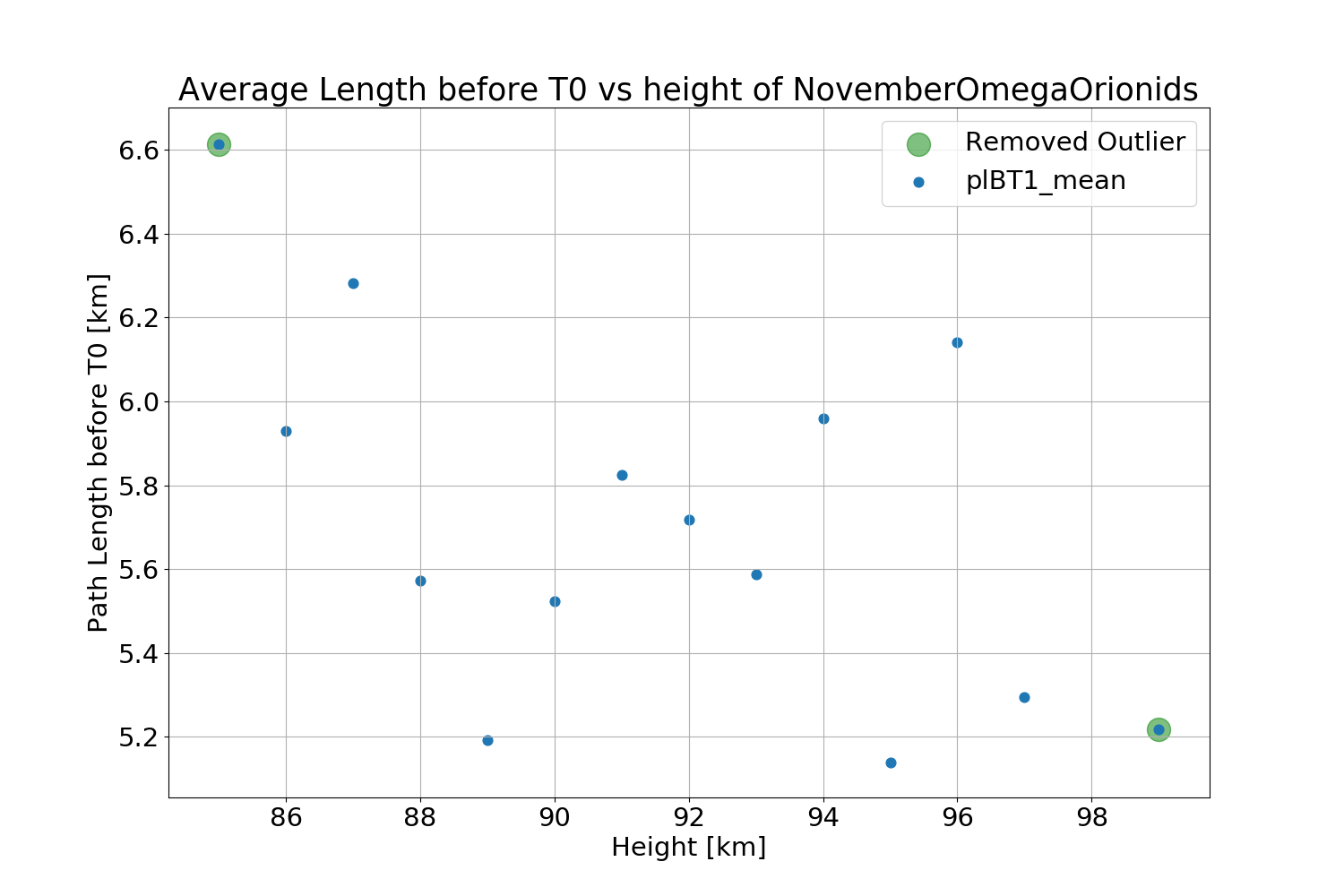}

\end{subfigure}\hspace*{\fill}
\begin{subfigure}{0.8\textwidth}
\includegraphics[trim = 2cm 1cm 3.5cm 1.5cm, clip, clip,width=\linewidth]{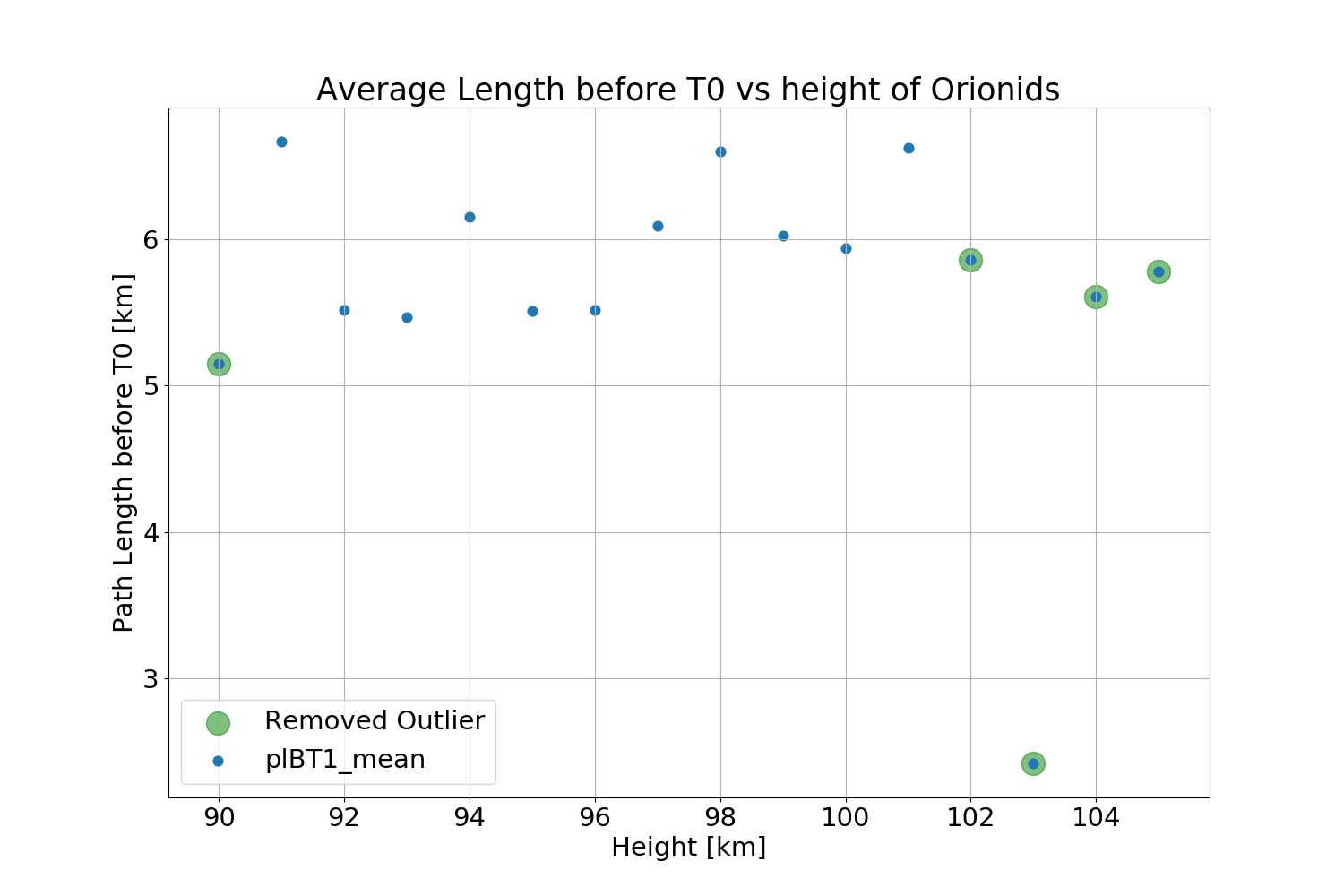}

\end{subfigure}

\medskip
\begin{subfigure}{0.8\textwidth}
\includegraphics[trim = 2cm 1cm 3.5cm 1.5cm, clip, clip,width=\linewidth]{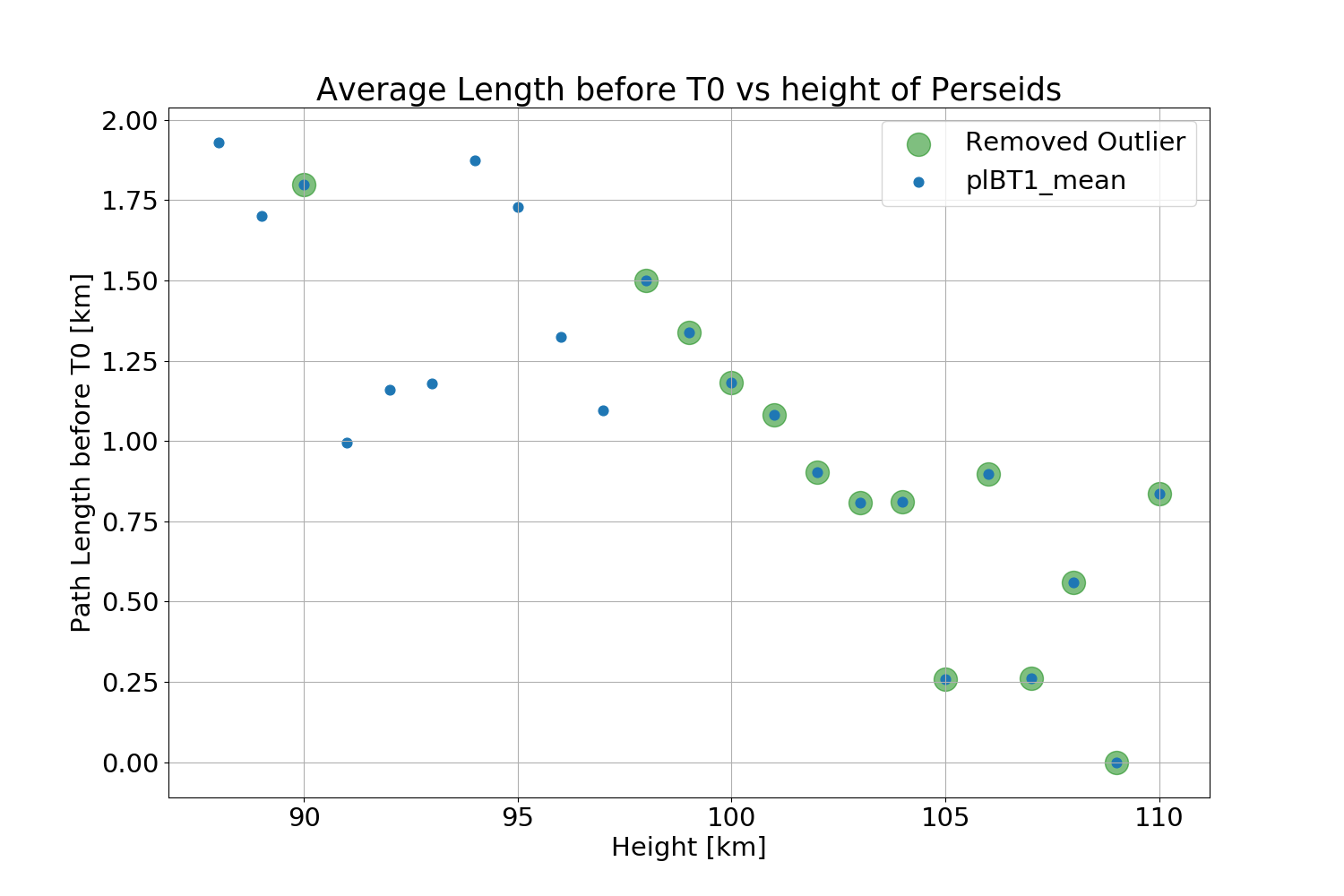}

\end{subfigure}\hspace*{\fill}
\begin{subfigure}{0.8\textwidth}
\includegraphics[trim = 2cm 1cm 3.5cm 1.5cm, clip, clip,width=\linewidth]{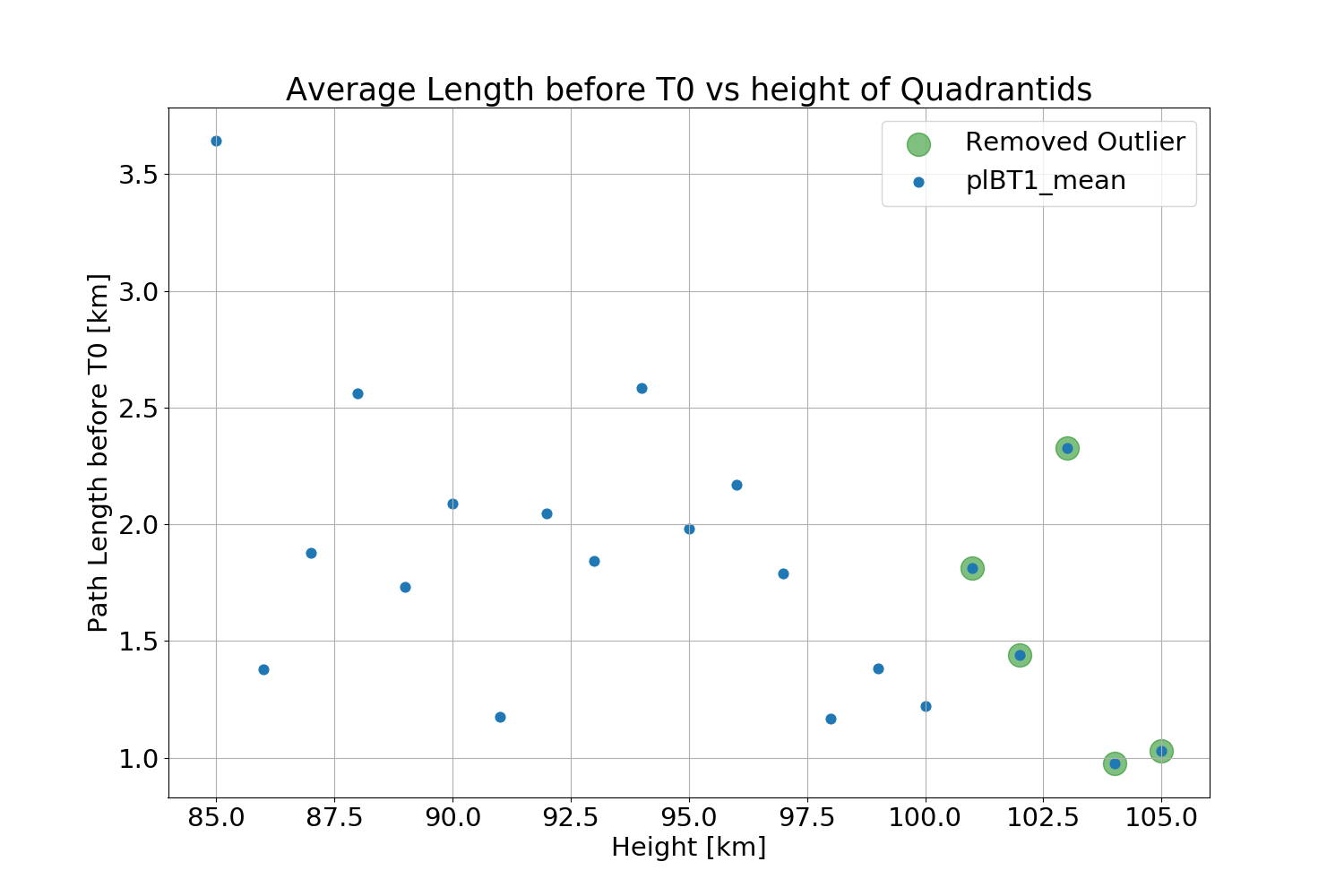}

\end{subfigure}

\medskip
\begin{subfigure}{0.8\textwidth}
\includegraphics[trim = 2cm 1cm 3.5cm 1.5cm, clip, clip,width=\linewidth]{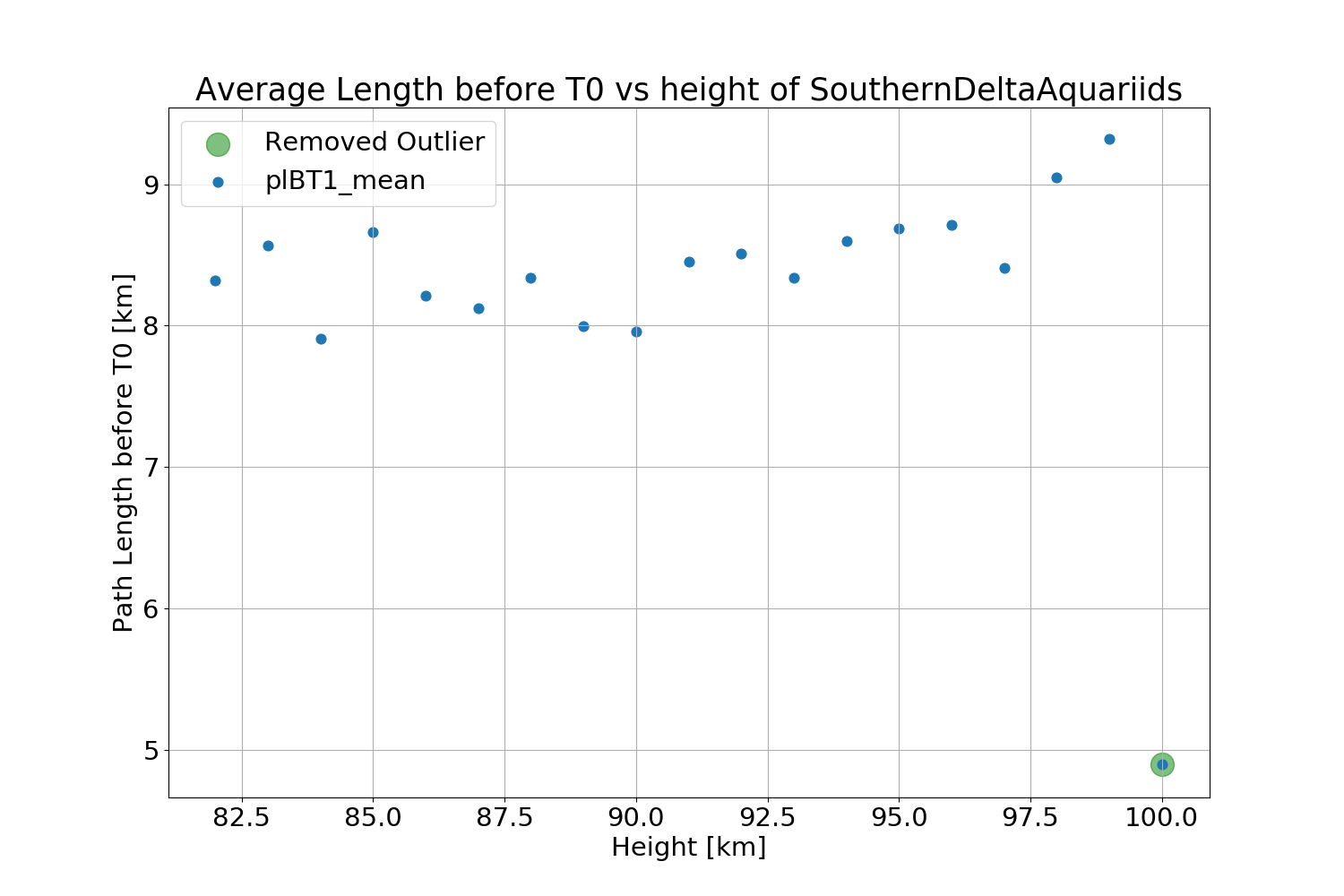}
\end{subfigure}\hspace*{\fill}
\begin{subfigure}{0.8\textwidth}
\includegraphics[trim = 2cm 1cm 3.5cm 1.5cm, clip, width=\linewidth]{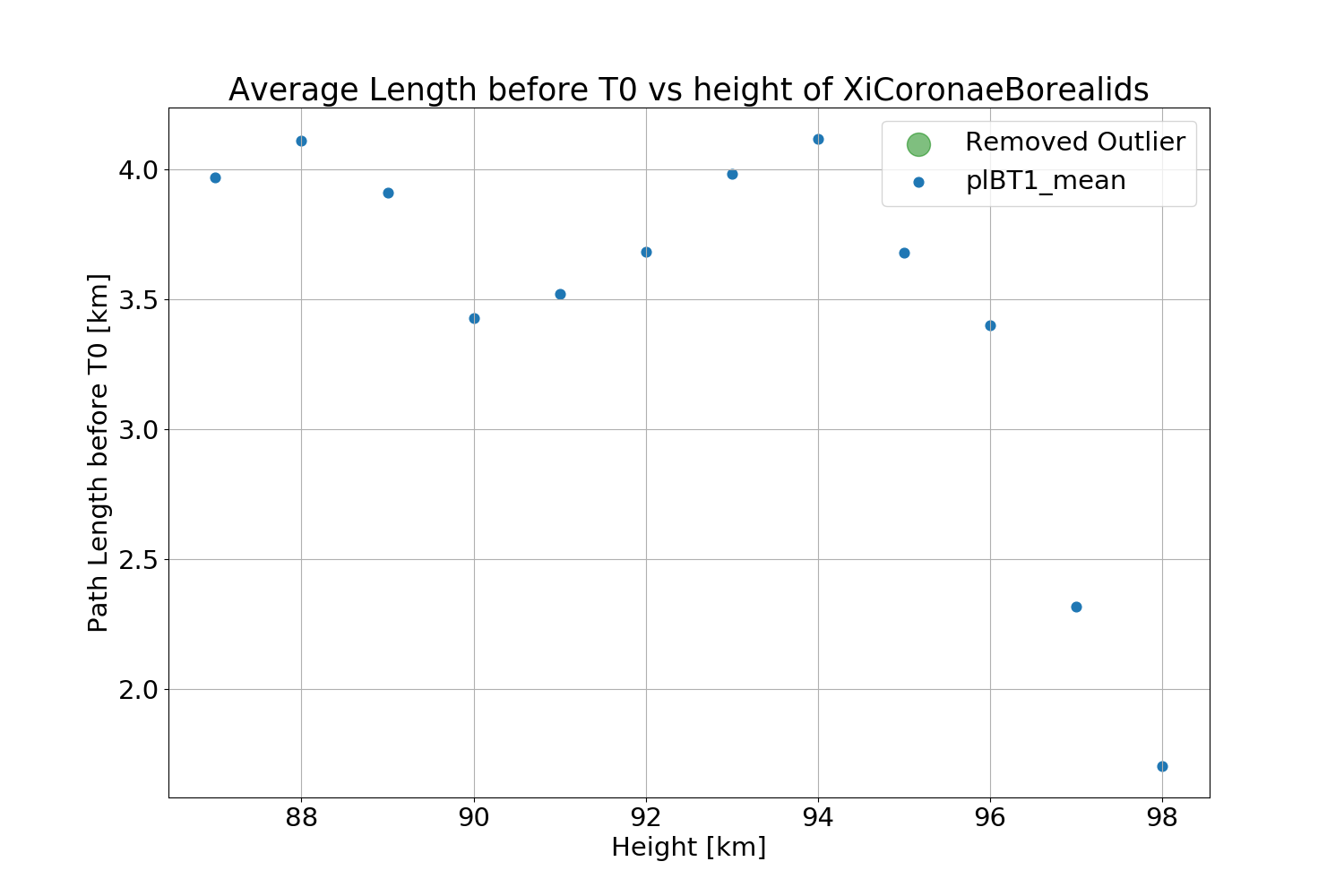}

\end{subfigure}
\caption[Path length before T0 vs height for showers used in velocity correction (2/2)]{Average path length before detection at station T0 versus height plots for all showers used in velocity correction (part 2/2)}
\label{apx_plt12}
\end{figure}

\clearpage
\newpage

\begin{figure}[t!]
\vspace*{-0.5cm}
\advance\leftskip-3cm
\advance\rightskip-3cm
\begin{subfigure}{0.8\textwidth}
\includegraphics[trim = 1.5cm 1cm 3.5cm 1.5cm, clip,width=\linewidth]{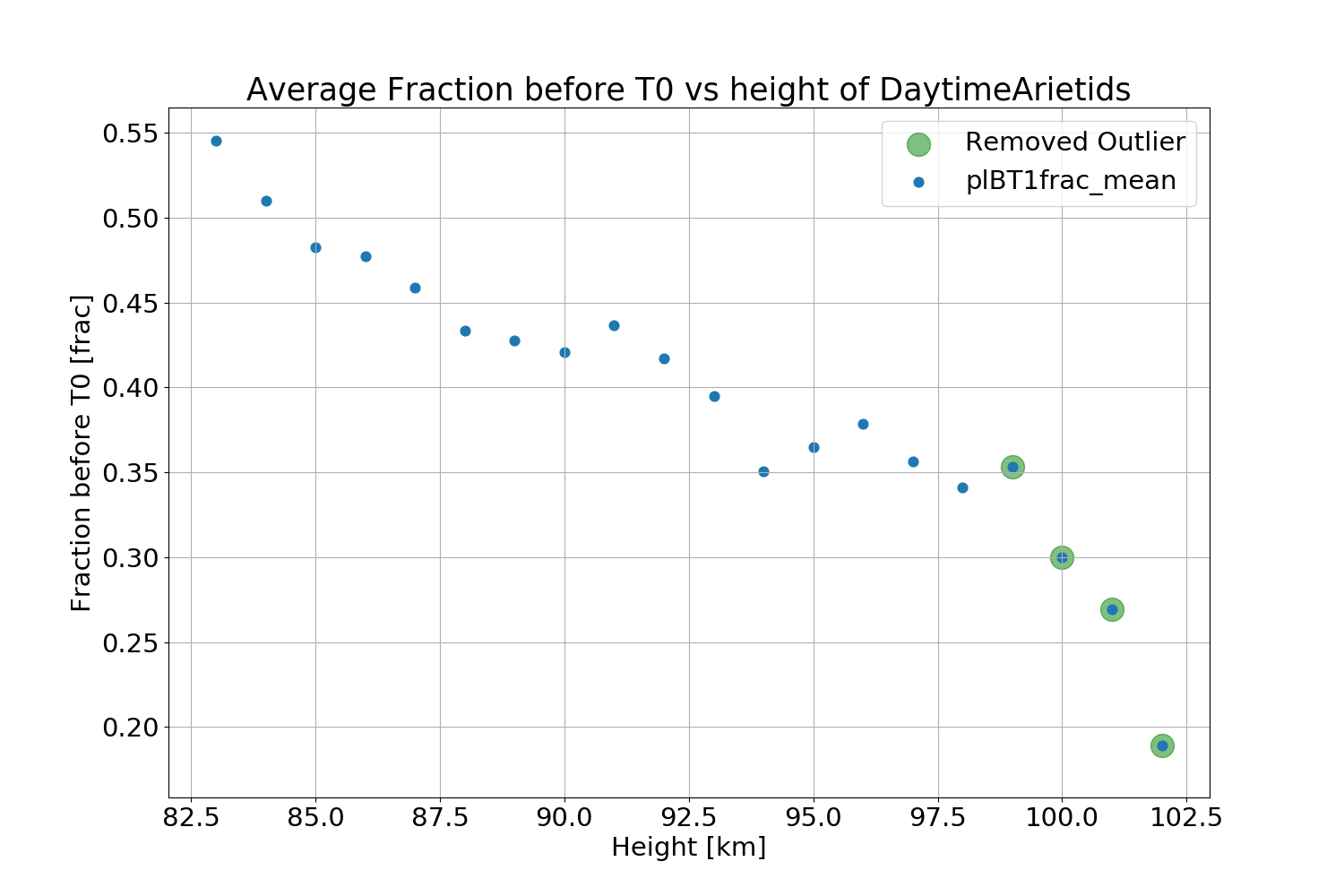}

\end{subfigure}\hspace*{\fill}
\begin{subfigure}{0.8\textwidth}
\includegraphics[trim = 1.5cm 1cm 3.5cm 1.5cm, clip,width=\linewidth]{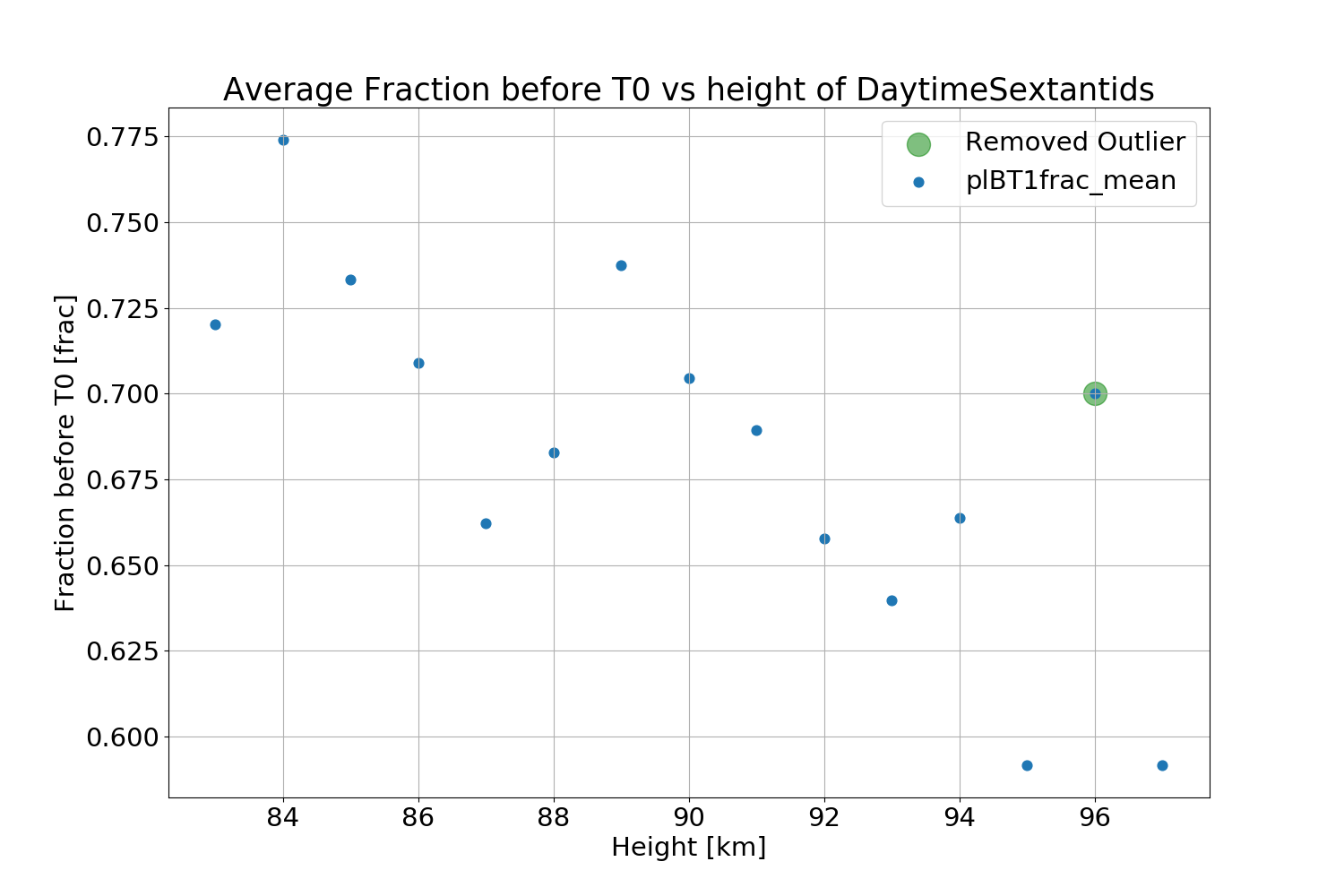}

\end{subfigure}

\medskip
\begin{subfigure}{0.8\textwidth}
\includegraphics[trim = 1.5cm 1cm 3.5cm 1.5cm, clip,width=\linewidth]{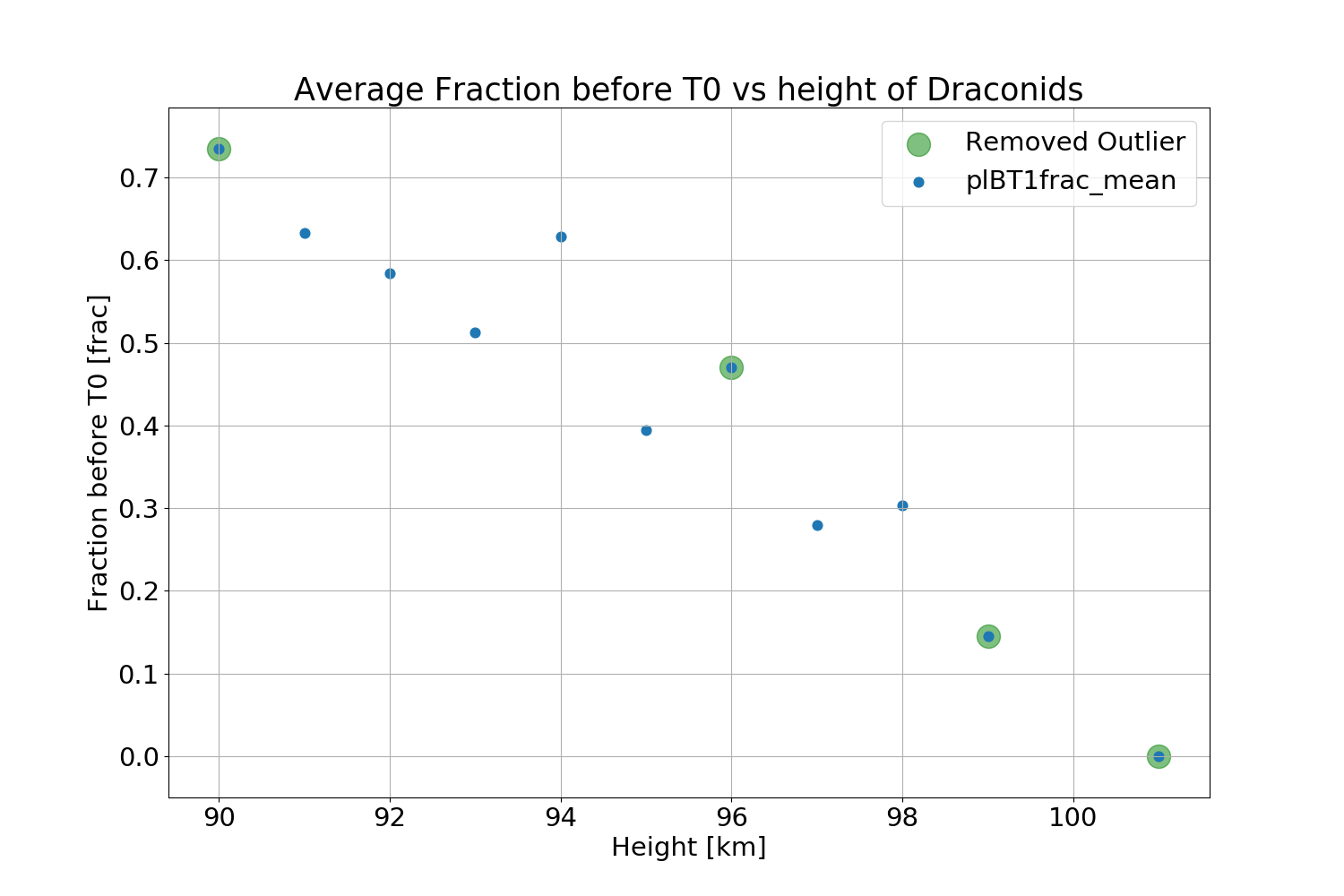}

\end{subfigure}\hspace*{\fill}
\begin{subfigure}{0.8\textwidth}
\includegraphics[trim = 2cm 1cm 3.5cm 1.5cm, clip,width=\linewidth]{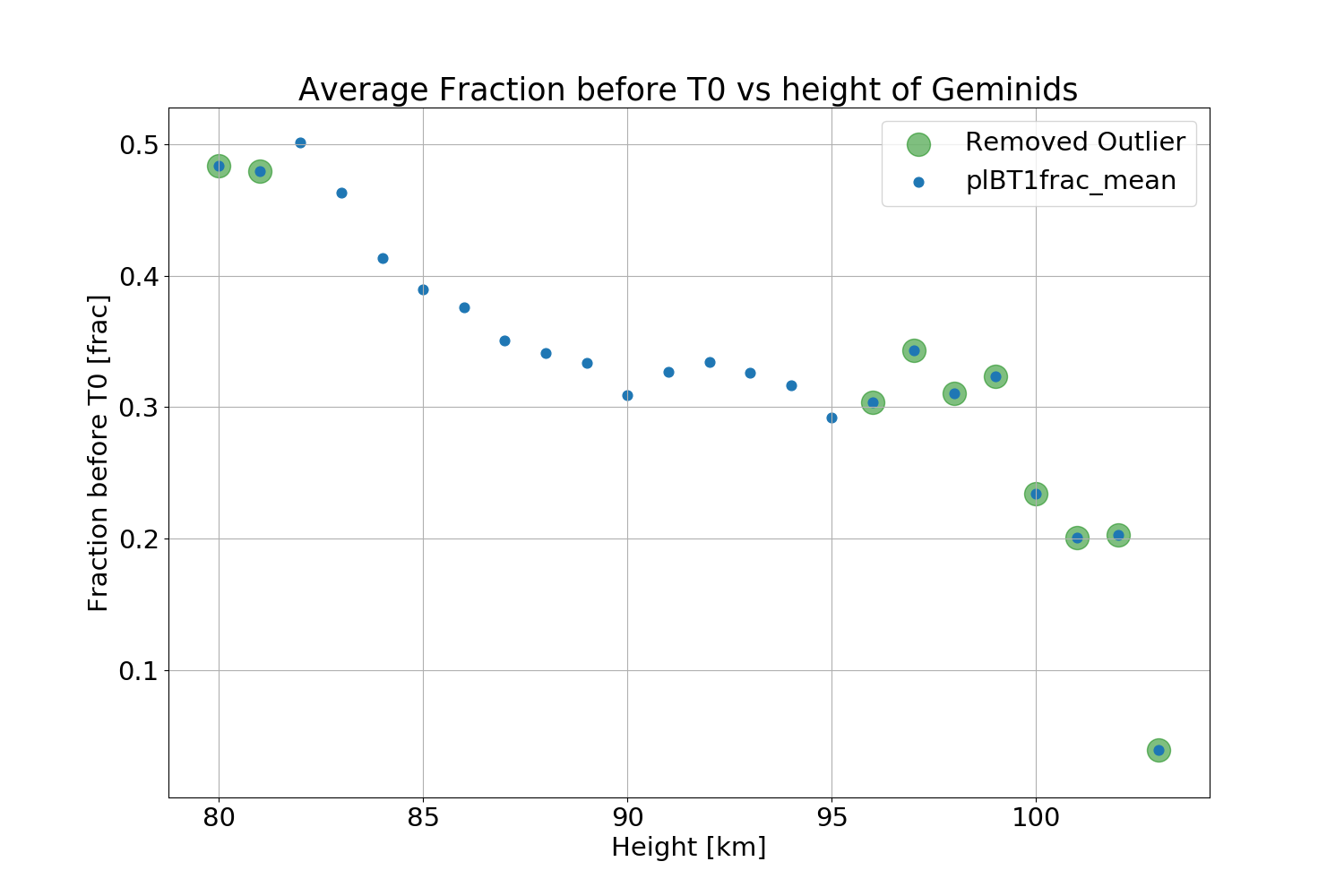}

\end{subfigure}

\medskip
\begin{subfigure}{0.8\textwidth}
\includegraphics[trim = 1.5cm 1cm 3.5cm 1.5cm, clip,width=\linewidth]{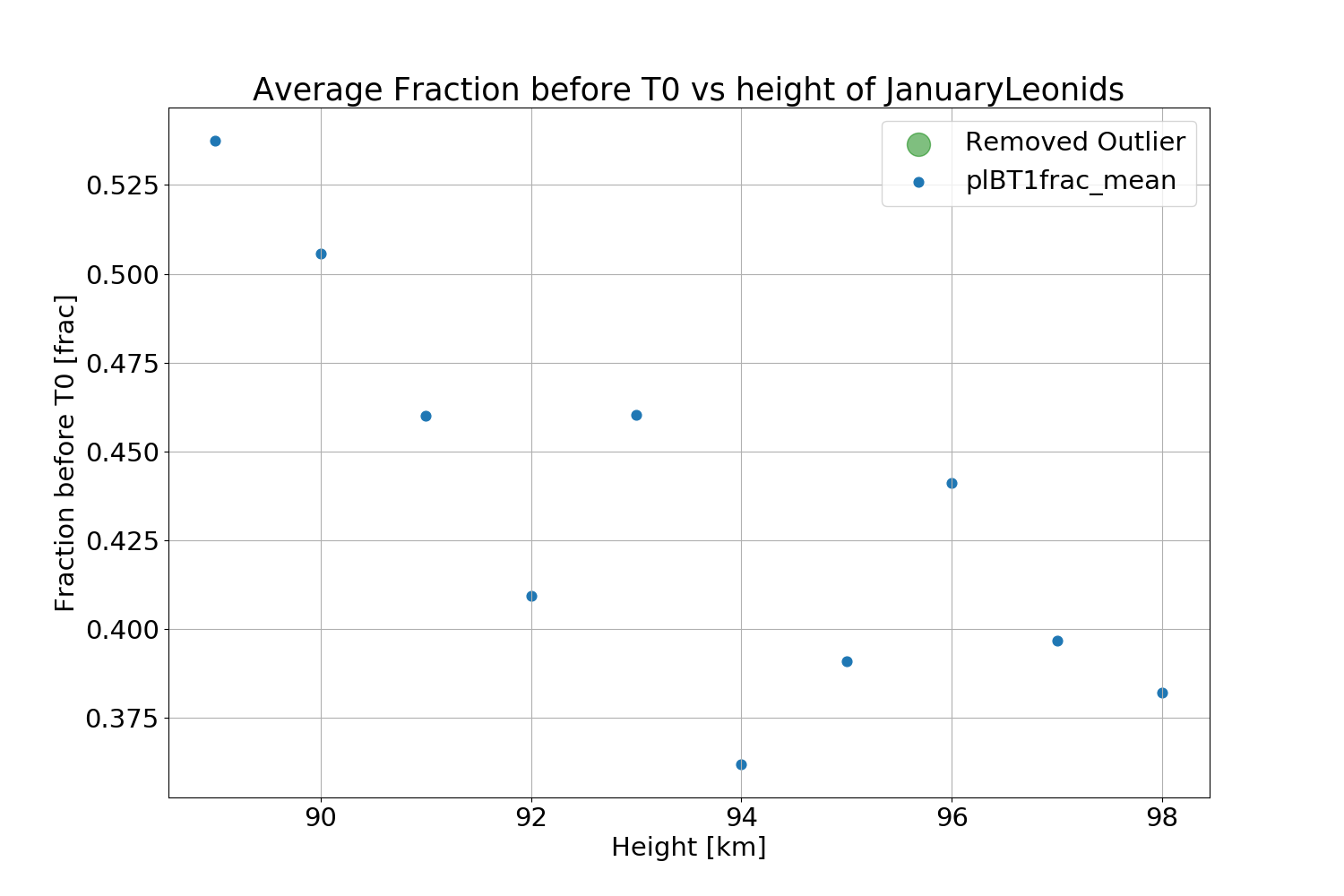}

\end{subfigure}\hspace*{\fill}
\begin{subfigure}{0.8\textwidth}
\includegraphics[trim = 1.5cm 1cm 3.5cm 1.5cm, clip,width=\linewidth]{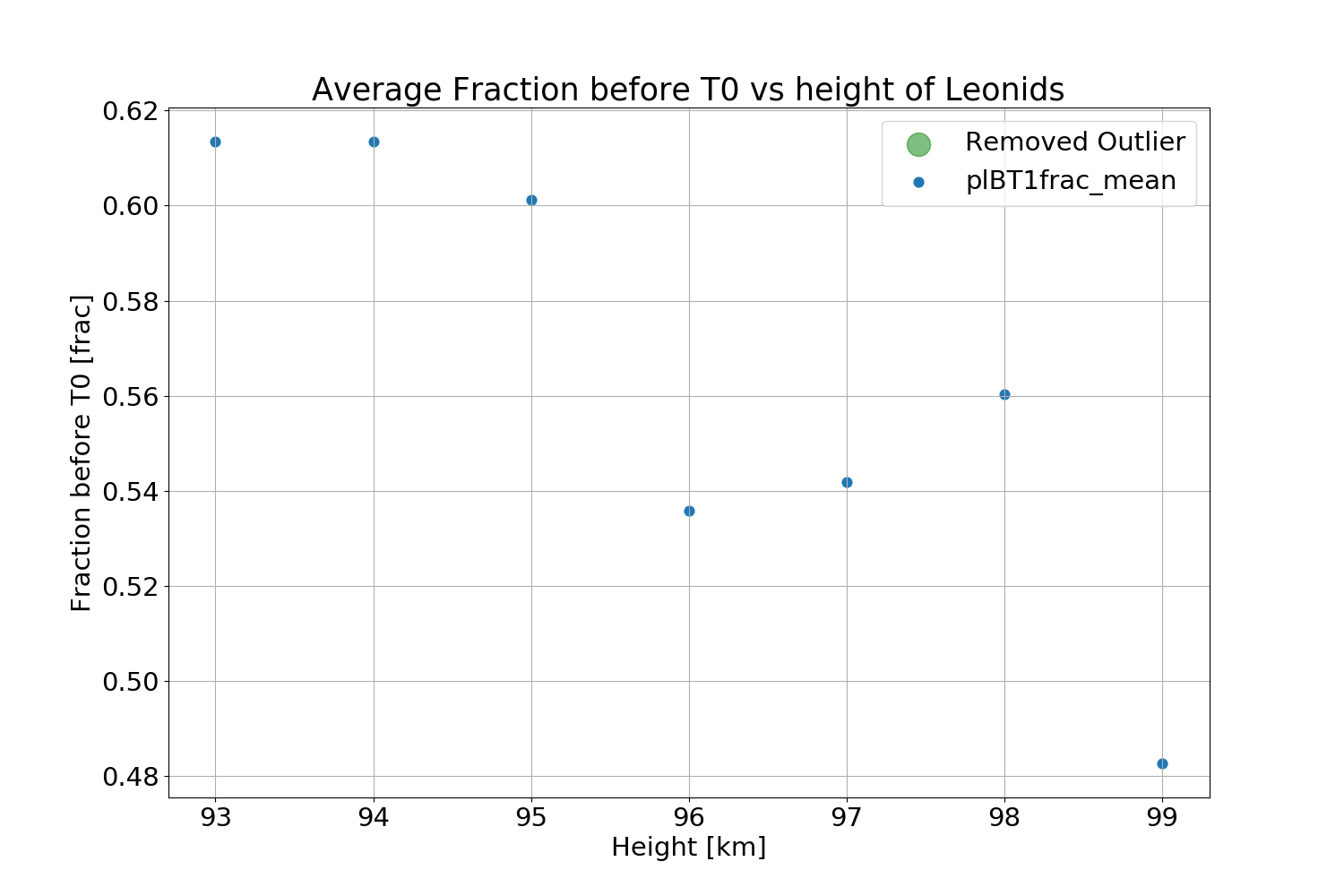}

\end{subfigure}
\caption[Fraction of path length before T0 for showers used in velocity correction (1/2)]{Fraction of meteor path length observed before station T0 for all showers used in velocity correction (part 1/2)} 
\label{apx_t1frac1}
\end{figure}

\clearpage
\newpage

\begin{figure}[t!]
\vspace*{-0.5cm}
\advance\leftskip-3cm
\advance\rightskip-3cm
\begin{subfigure}{0.8\textwidth}
\includegraphics[trim = 1.5cm 1cm 3.5cm 1.5cm, clip,width=\linewidth]{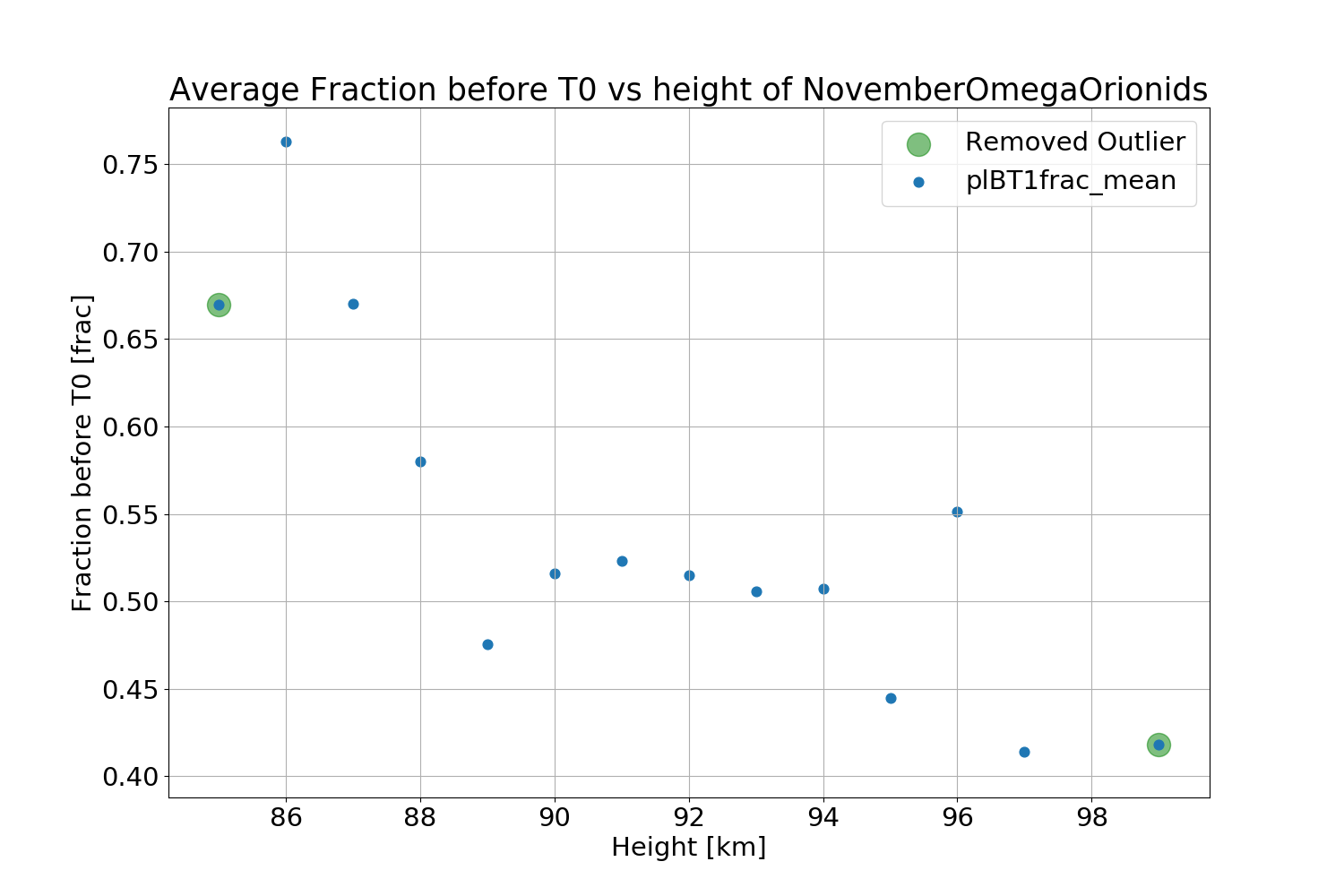}

\end{subfigure}\hspace*{\fill}
\begin{subfigure}{0.8\textwidth}
\includegraphics[trim = 1.5cm 1cm 3.5cm 1.5cm, clip,width=\linewidth]{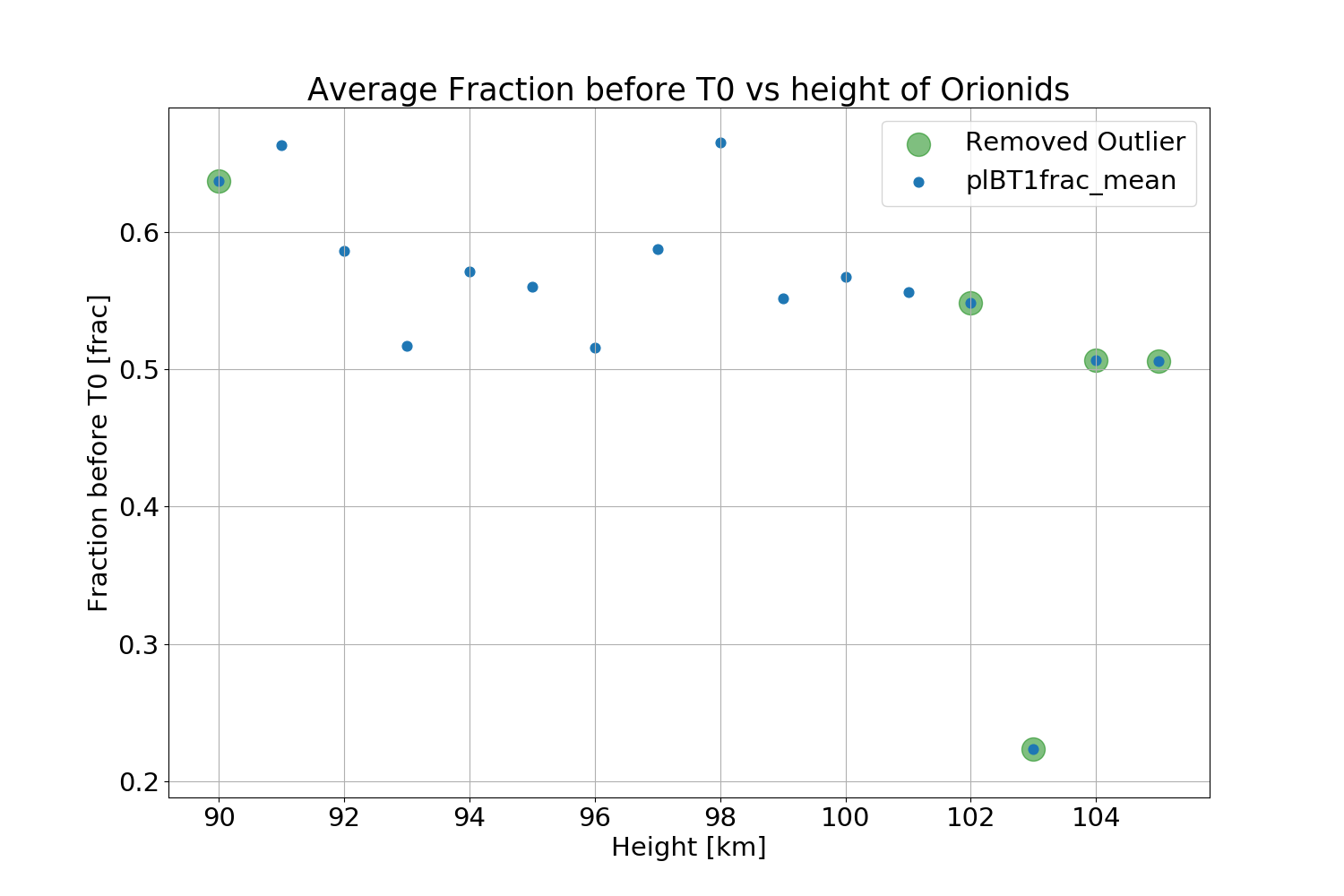}

\end{subfigure}

\medskip
\begin{subfigure}{0.8\textwidth}
\includegraphics[trim = 1.5cm 1cm 3.5cm 1.5cm, clip,width=\linewidth]{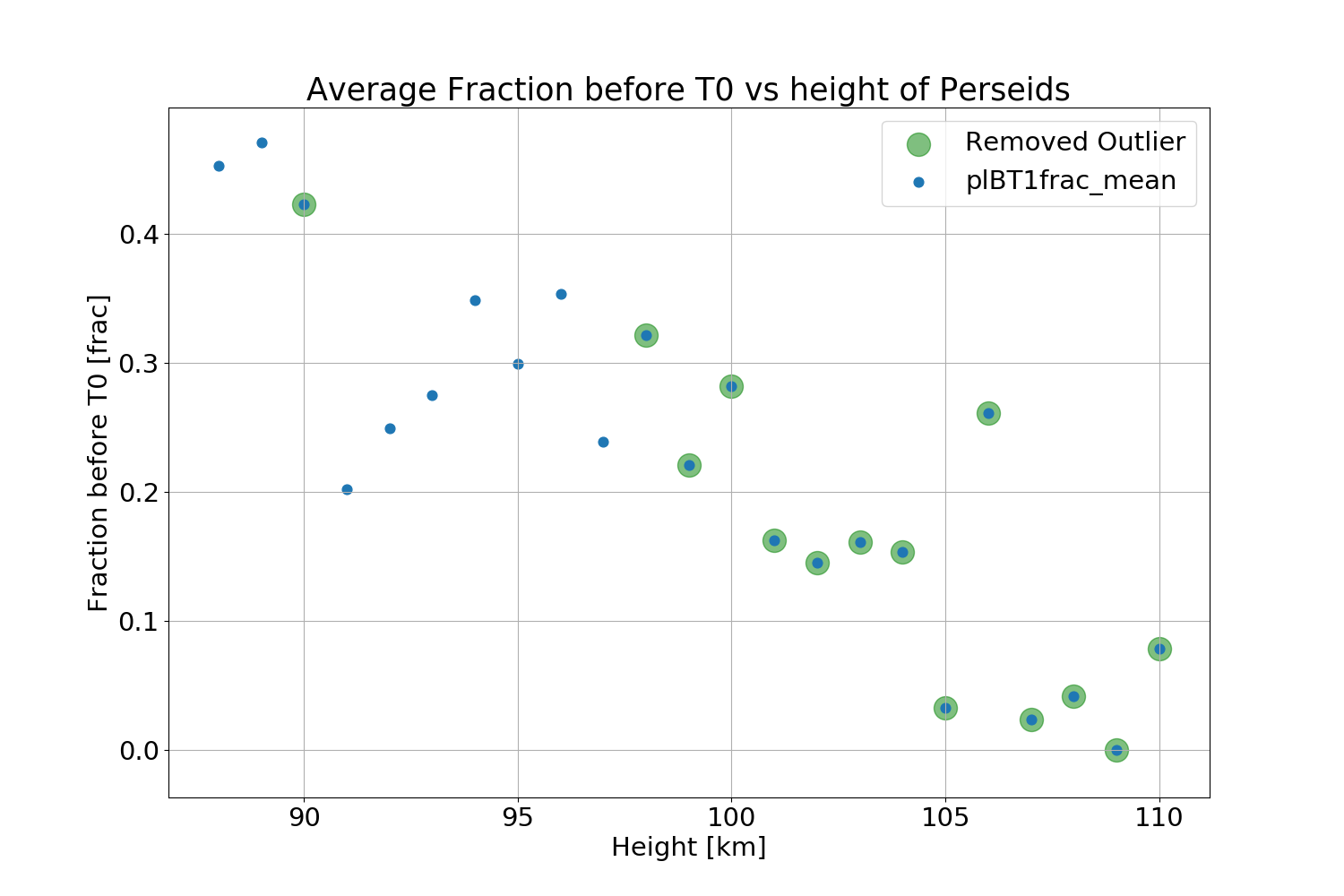}

\end{subfigure}\hspace*{\fill}
\begin{subfigure}{0.8\textwidth}
\includegraphics[trim = 1.5cm 1cm 3.5cm 1.5cm, clip,width=\linewidth]{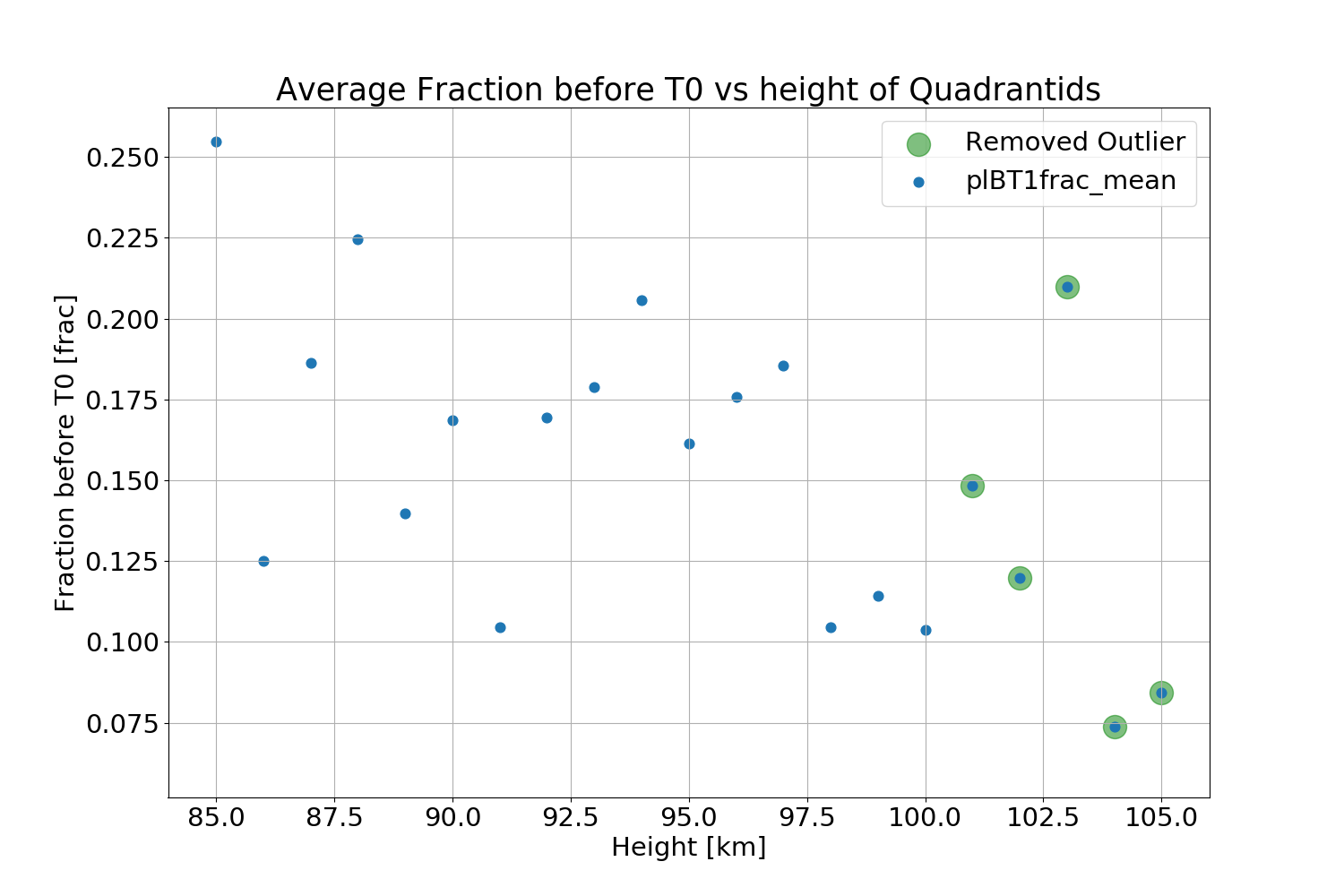}

\end{subfigure}

\medskip
\begin{subfigure}{0.8\textwidth}
\includegraphics[trim = 1.5cm 1cm 2.5cm 1.5cm, clip,width=\linewidth]{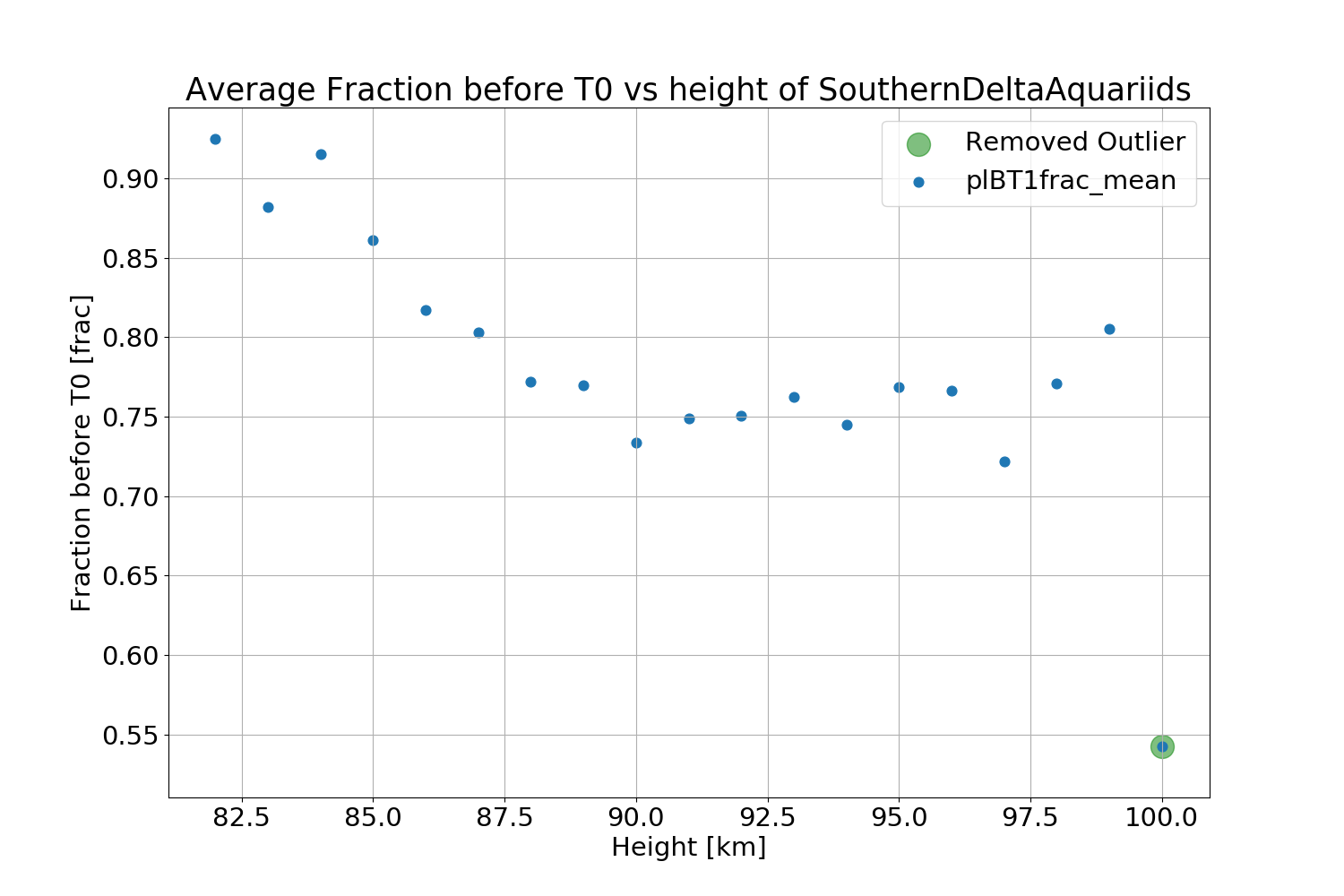}
\end{subfigure}\hspace*{\fill}
\begin{subfigure}{0.8\textwidth}
\includegraphics[trim = 1.5cm 1cm 3.5cm 1.5cm, width=\linewidth]{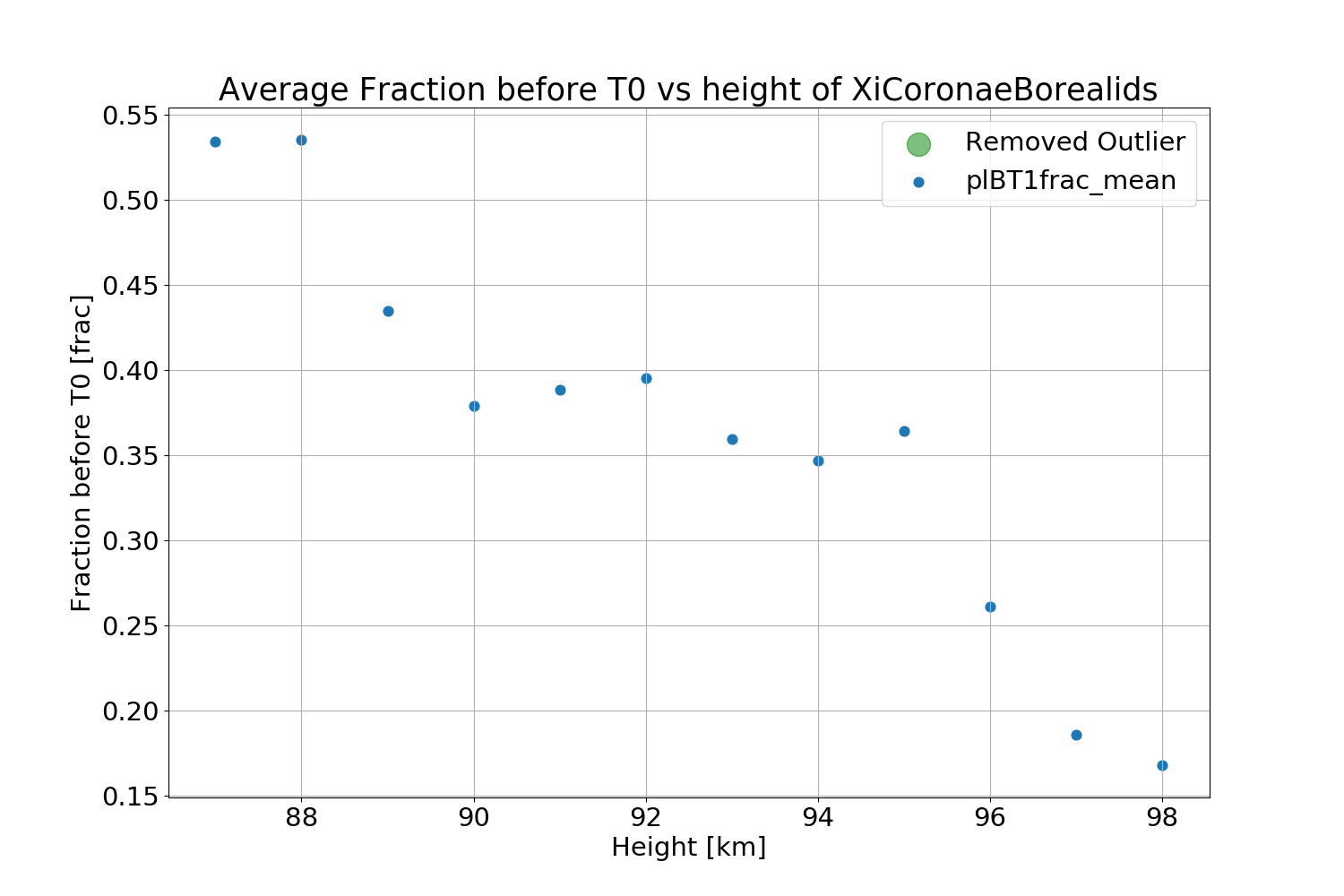}

\end{subfigure}
\caption[Fraction of path length before T0 for showers used in velocity correction (2/2)]{Fraction of meteor path length observed before station T0 for all showers used in velocity correction (part 2/2)}
\label{apx_t1frac2}
\end{figure}

\clearpage
\newpage

\begin{figure}[t!]
\vspace*{-0.5cm}
\advance\leftskip-3cm
\advance\rightskip-3cm
\begin{subfigure}{0.8\textwidth}
\includegraphics[trim = 2cm 1cm 2.5cm 1.25cm, clip,width=\linewidth]{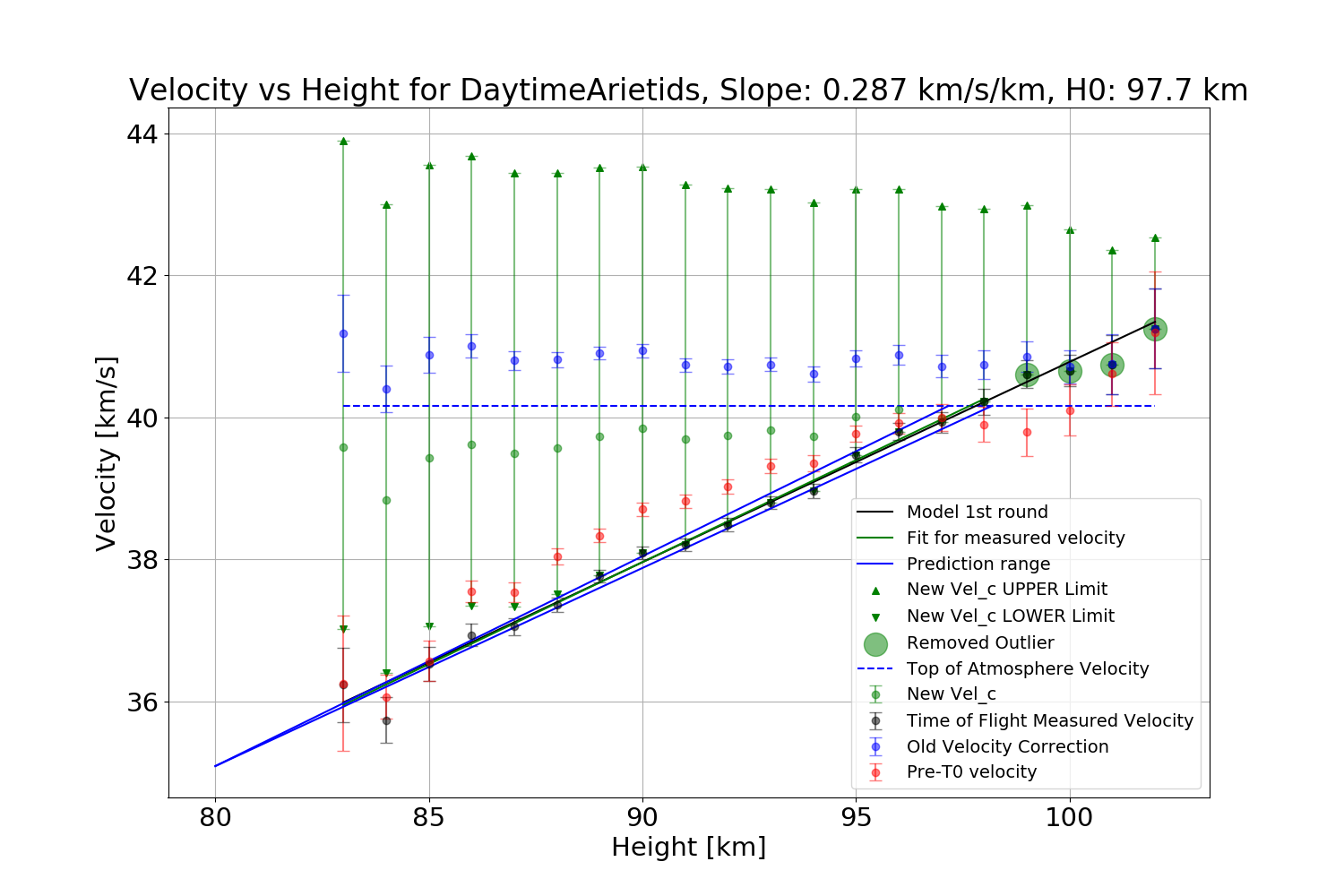}

\end{subfigure}\hspace*{\fill}
\begin{subfigure}{0.8\textwidth}
\includegraphics[trim = 2cm 1cm 2cm 1.25cm, clip,width=\linewidth]{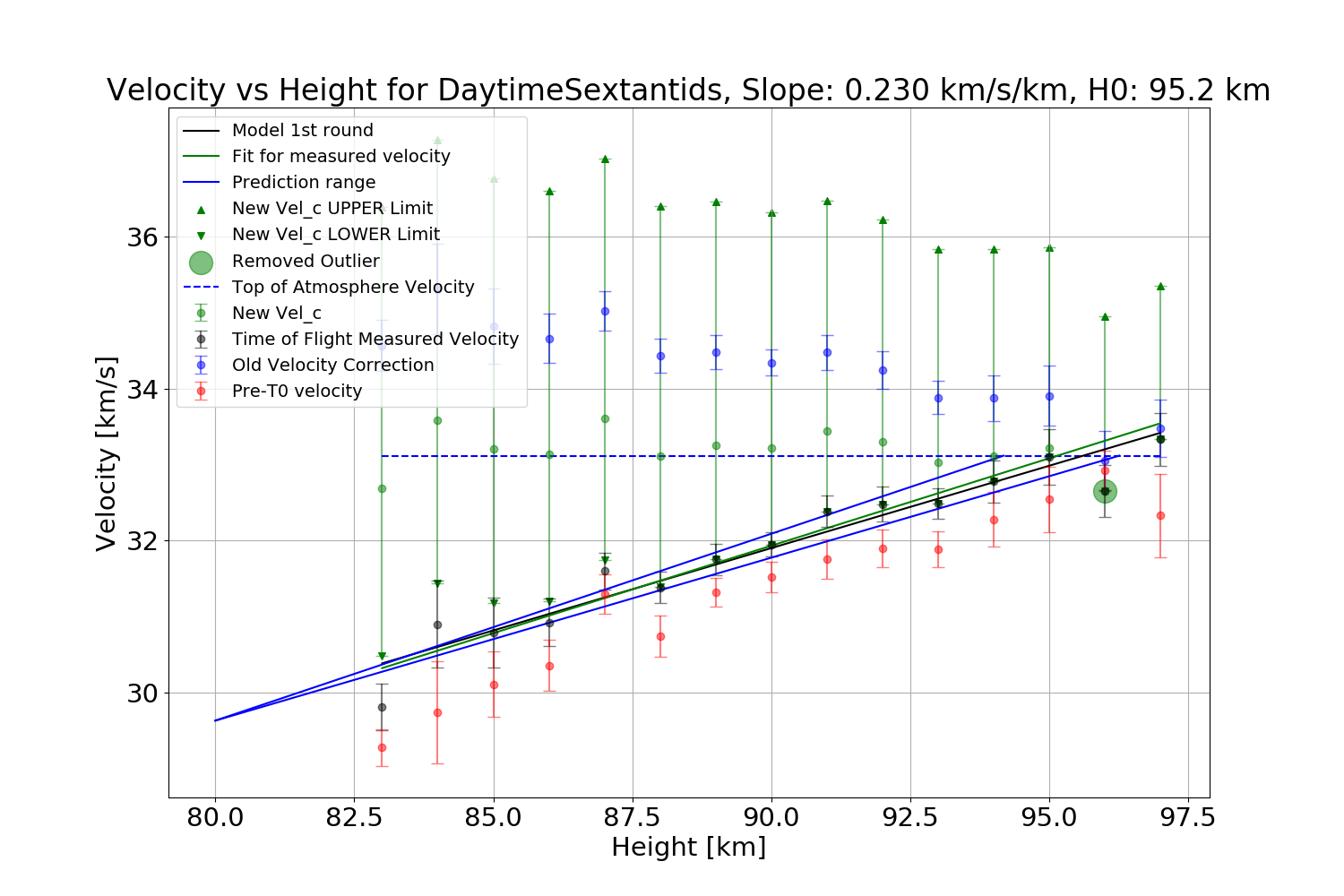}

\end{subfigure}

\medskip
\begin{subfigure}{0.8\textwidth}
\includegraphics[trim = 2cm 1cm 2.5cm 1.25cm, clip,width=\linewidth]{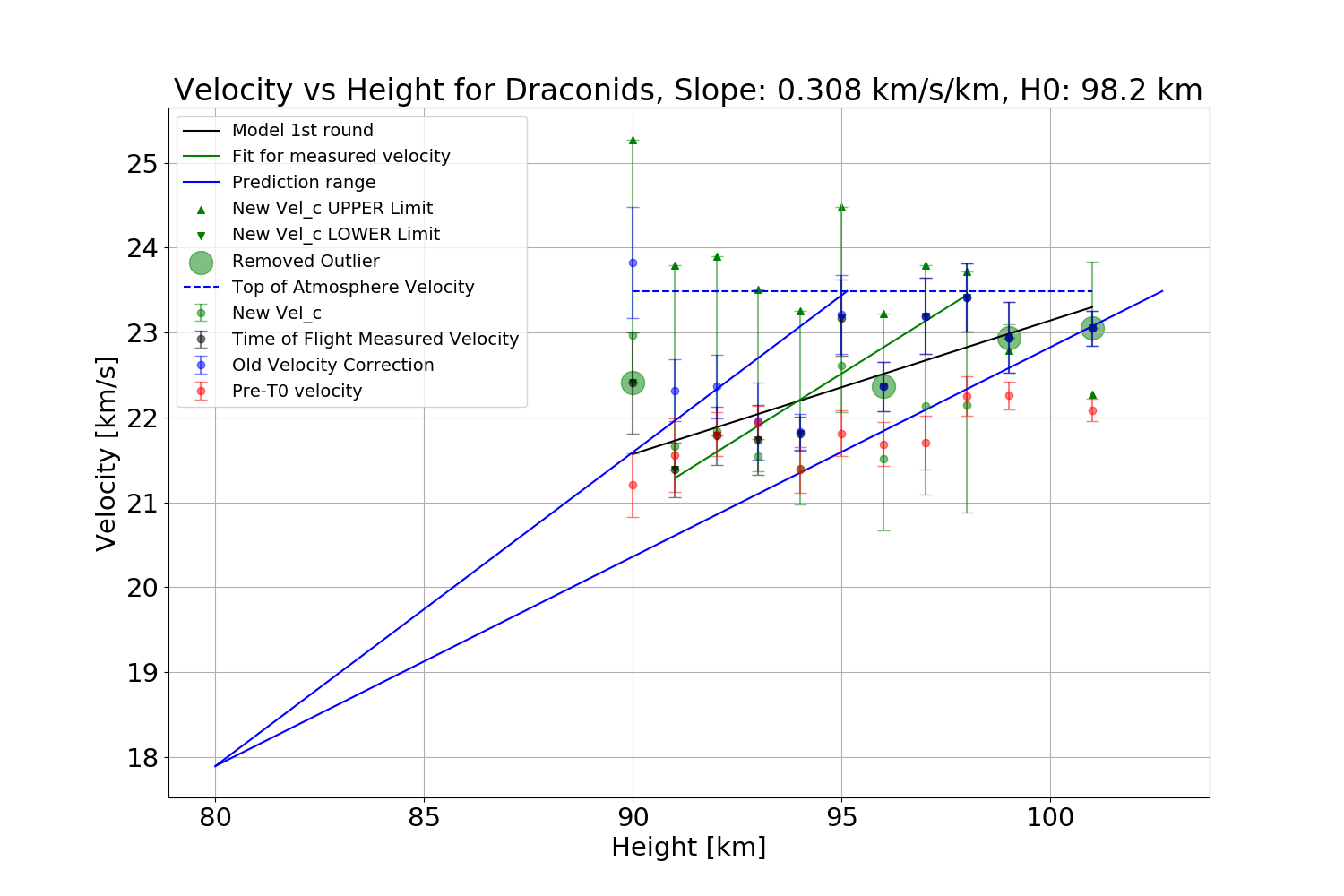}

\end{subfigure}\hspace*{\fill}
\begin{subfigure}{0.8\textwidth}
\includegraphics[trim = 2cm 1cm 2.5cm 1.25cm, clip,width=\linewidth]{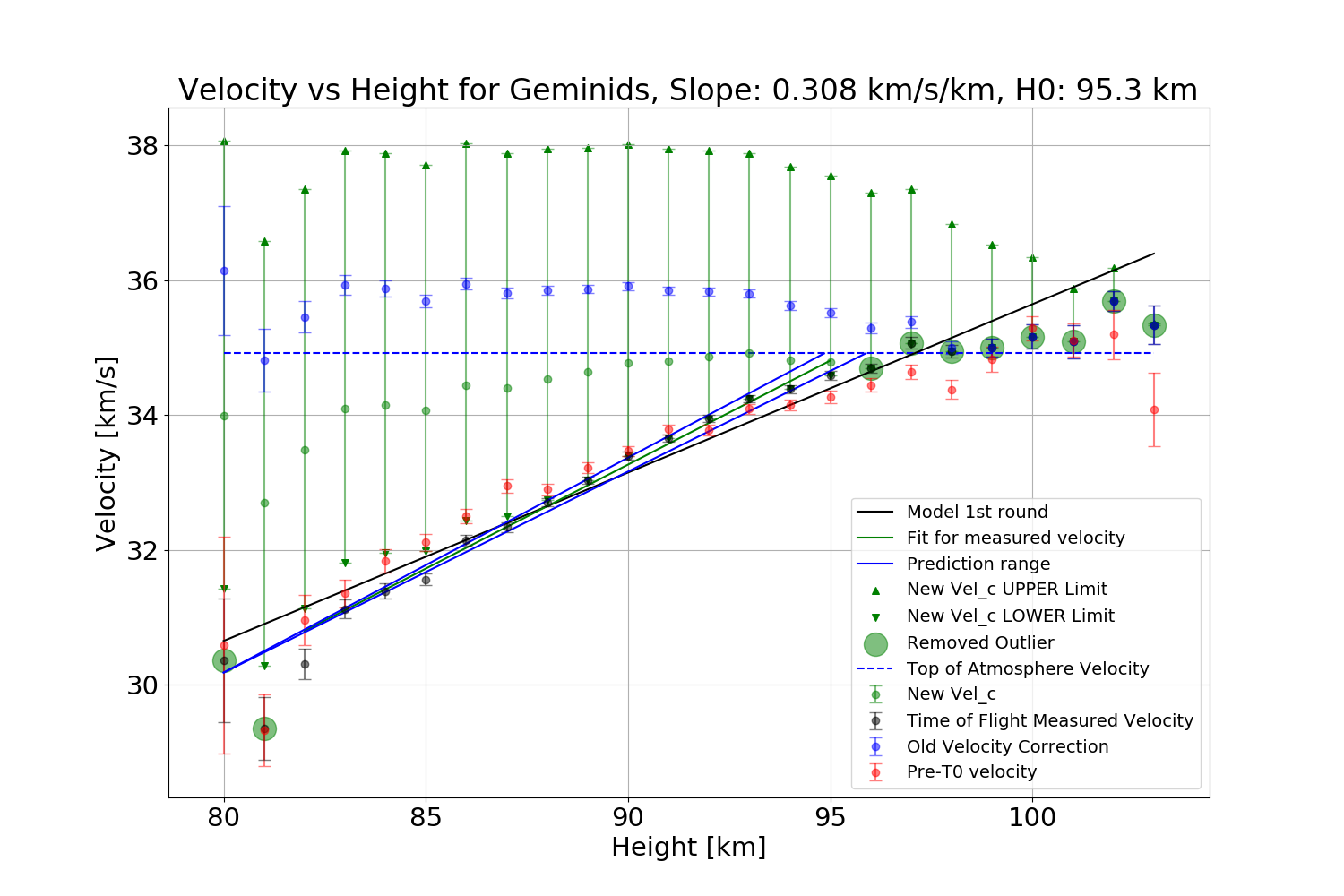}

\end{subfigure}

\medskip
\begin{subfigure}{0.8\textwidth}
\includegraphics[trim = 2cm 1cm 2.5cm 1.25cm, clip,width=\linewidth]{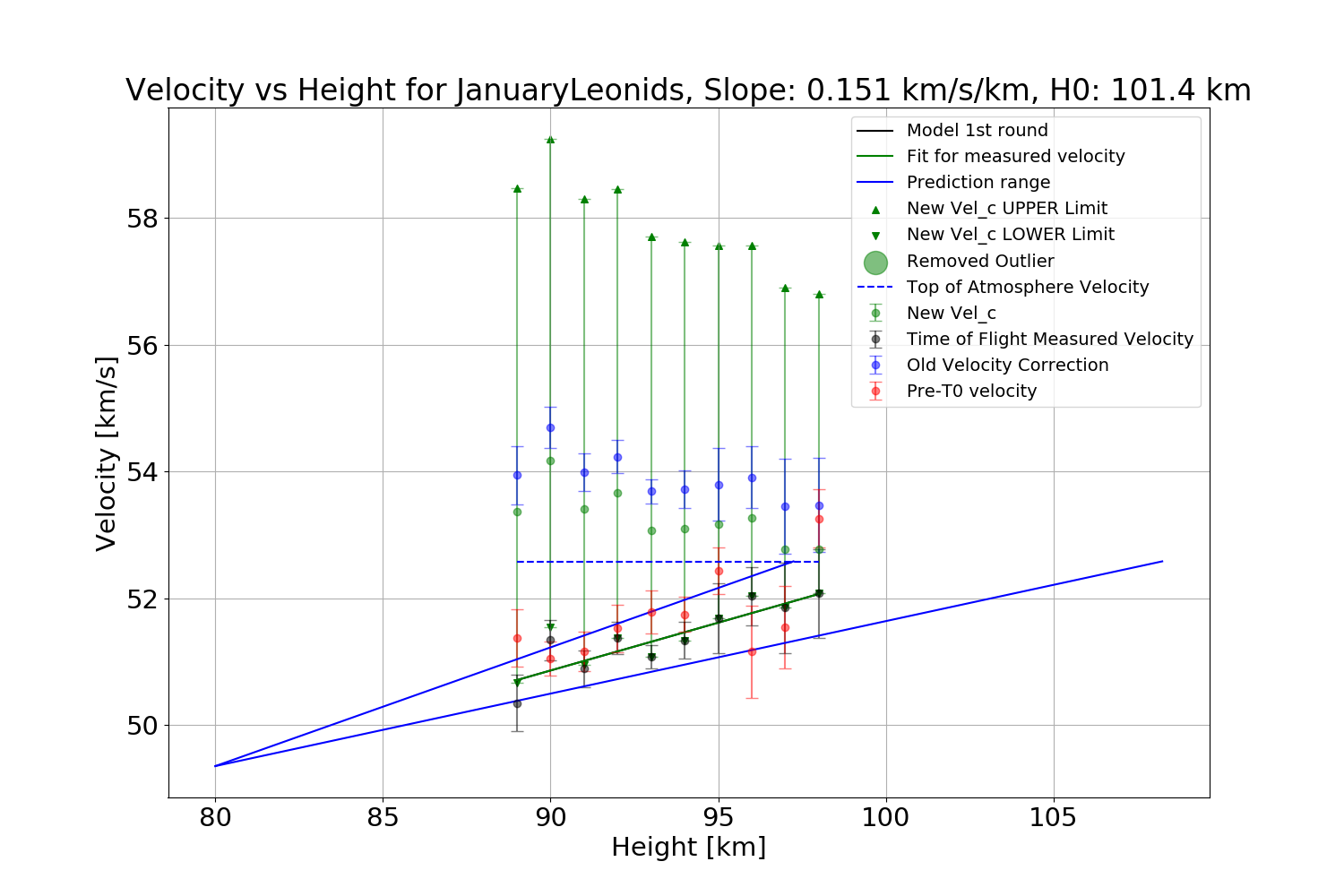}

\end{subfigure}\hspace*{\fill}
\begin{subfigure}{0.8\textwidth}
\includegraphics[trim = 2cm 1cm 2.5cm 1.25cm, clip,width=\linewidth]{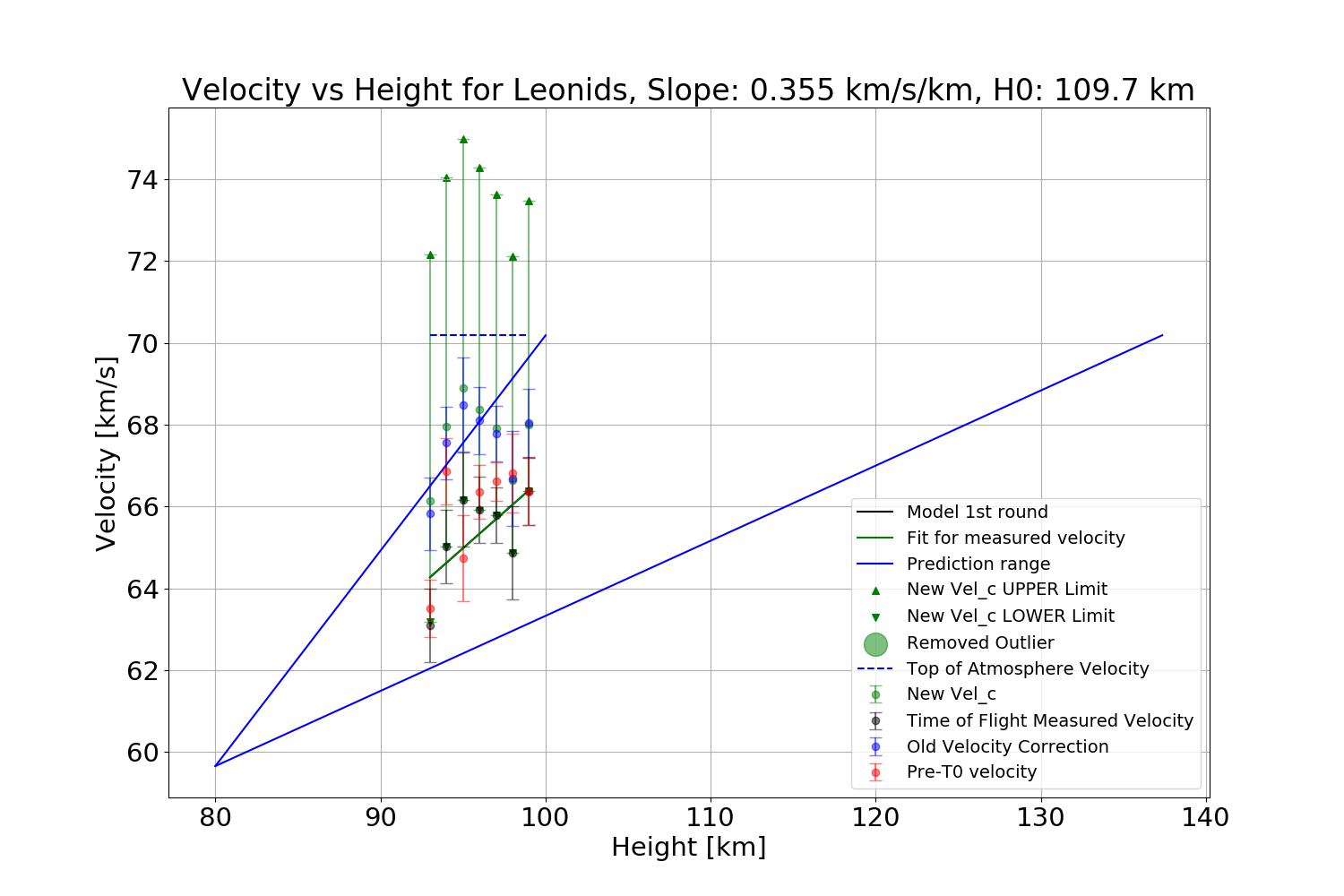}

\end{subfigure}
\caption[Velocities vs height for showers used in velocity correction (1/2)]{Velocities versus height plots for all showers used in velocity correction (part 1/2)} \label{apx_vels1}
\end{figure}

\clearpage
\newpage

\begin{figure}[t!]
\vspace*{-0.5cm}
\advance\leftskip-3cm
\advance\rightskip-3cm
\begin{subfigure}{0.8\textwidth}
\includegraphics[trim = 1cm 1cm 0cm 1cm, clip, width=\linewidth]{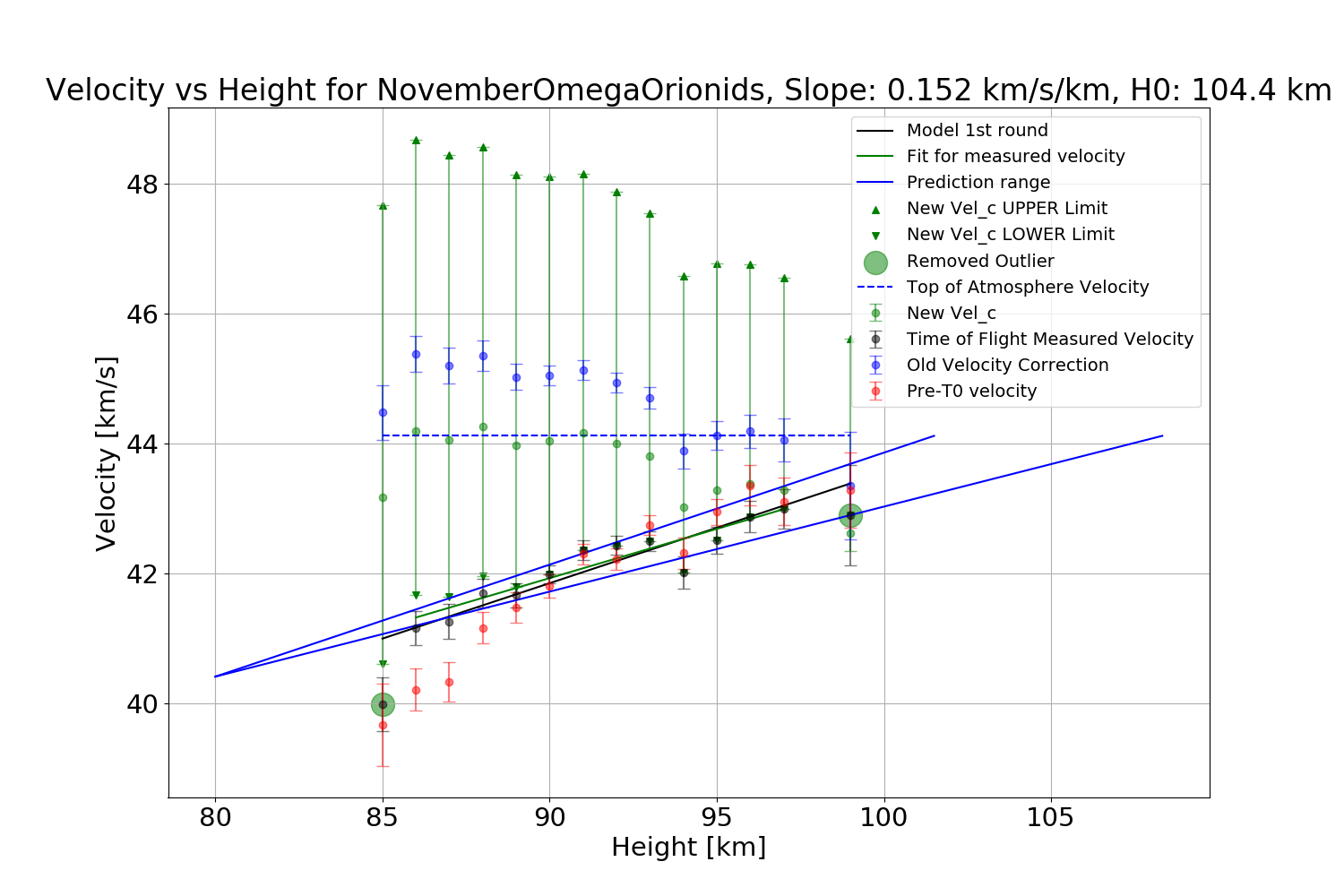}

\end{subfigure}\hspace*{\fill}
\begin{subfigure}{0.8\textwidth}
\includegraphics[trim = 1cm 1cm 0cm 1cm, clip, ,width=\linewidth]{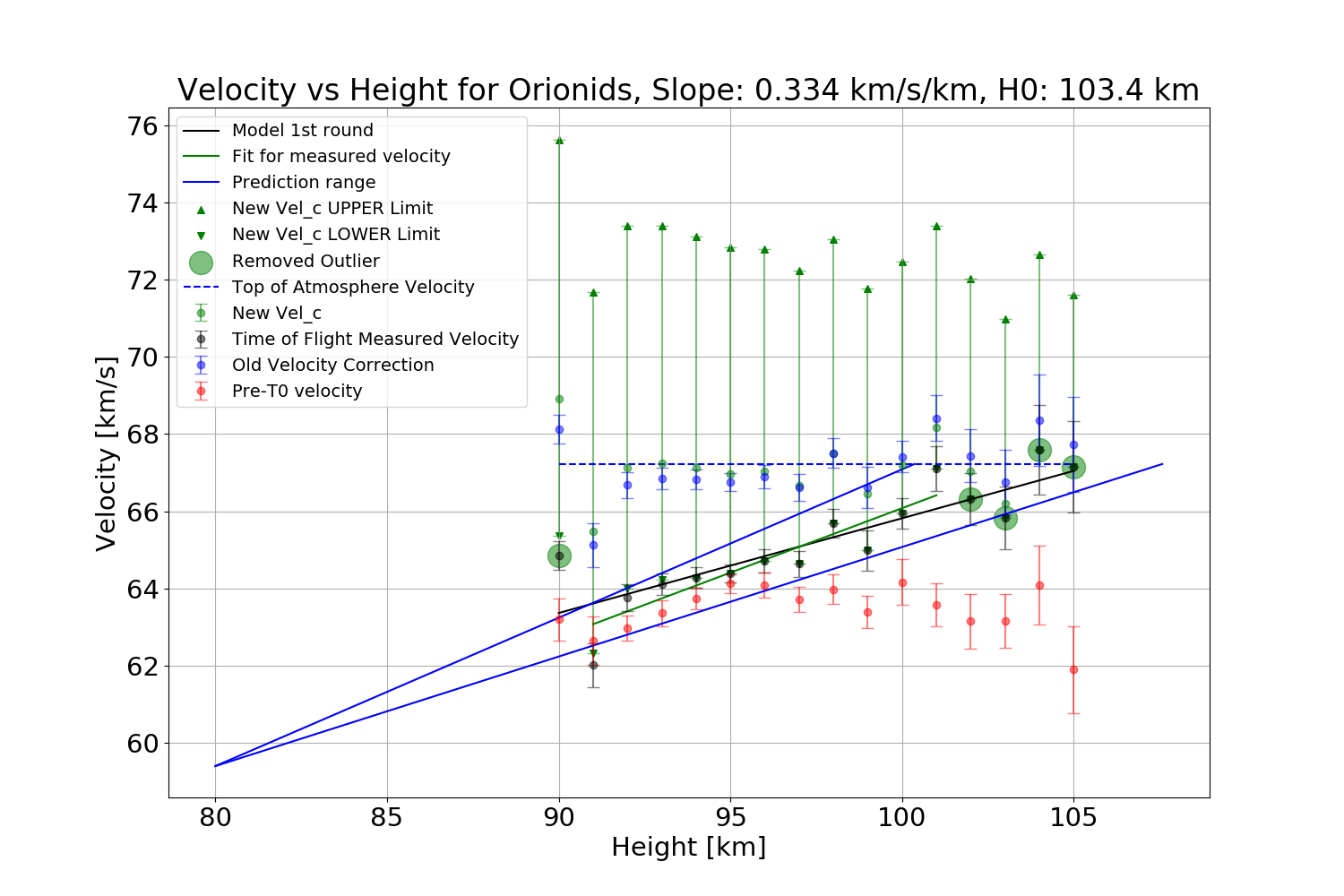}

\end{subfigure}

\medskip
\begin{subfigure}{0.8\textwidth}
\includegraphics[trim = 1cm 1cm 0cm 1cm, clip, width=\linewidth]{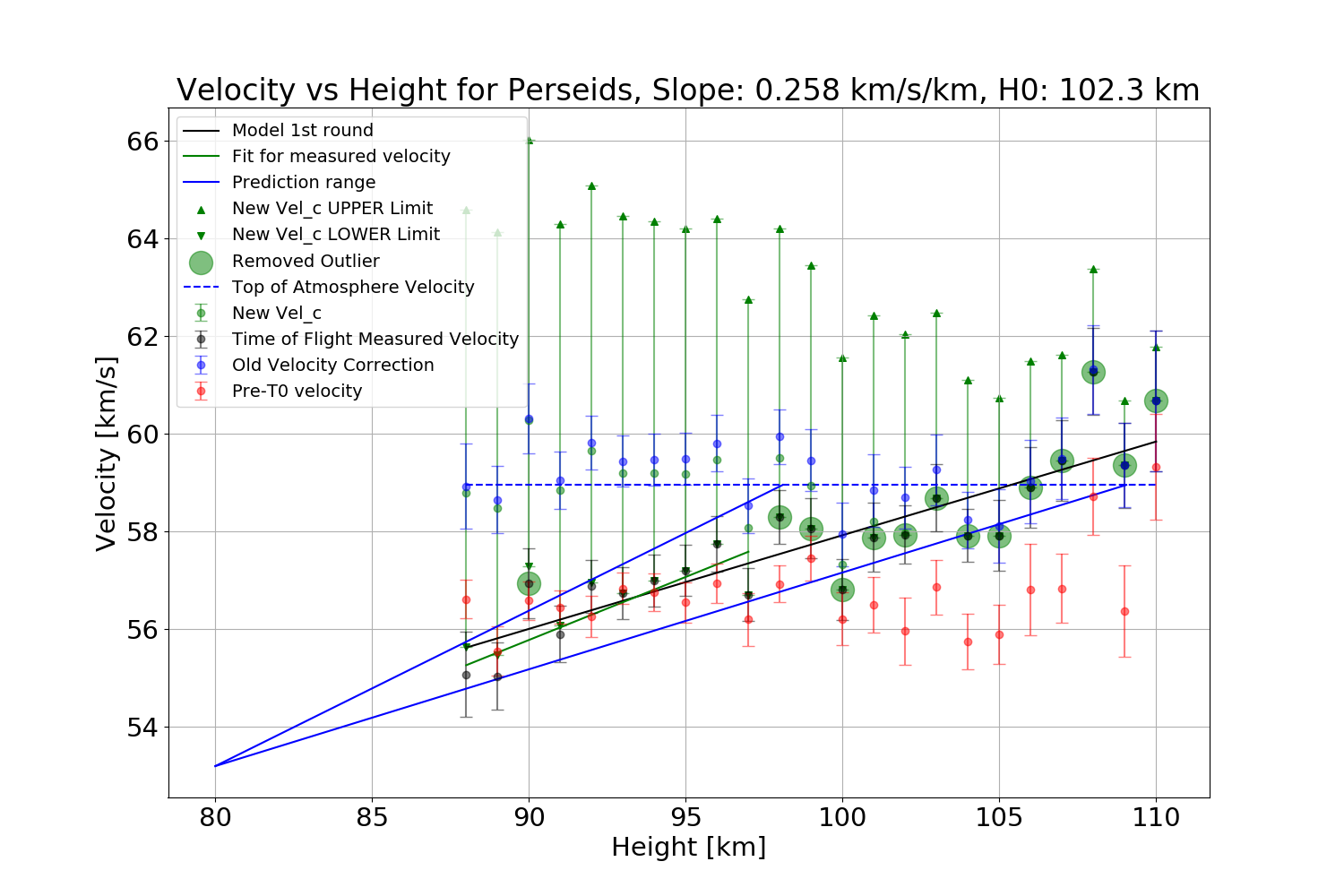}

\end{subfigure}\hspace*{\fill}
\begin{subfigure}{0.8\textwidth}
\includegraphics[trim = 1cm 1cm 0cm 1cm, clip, width=\linewidth]{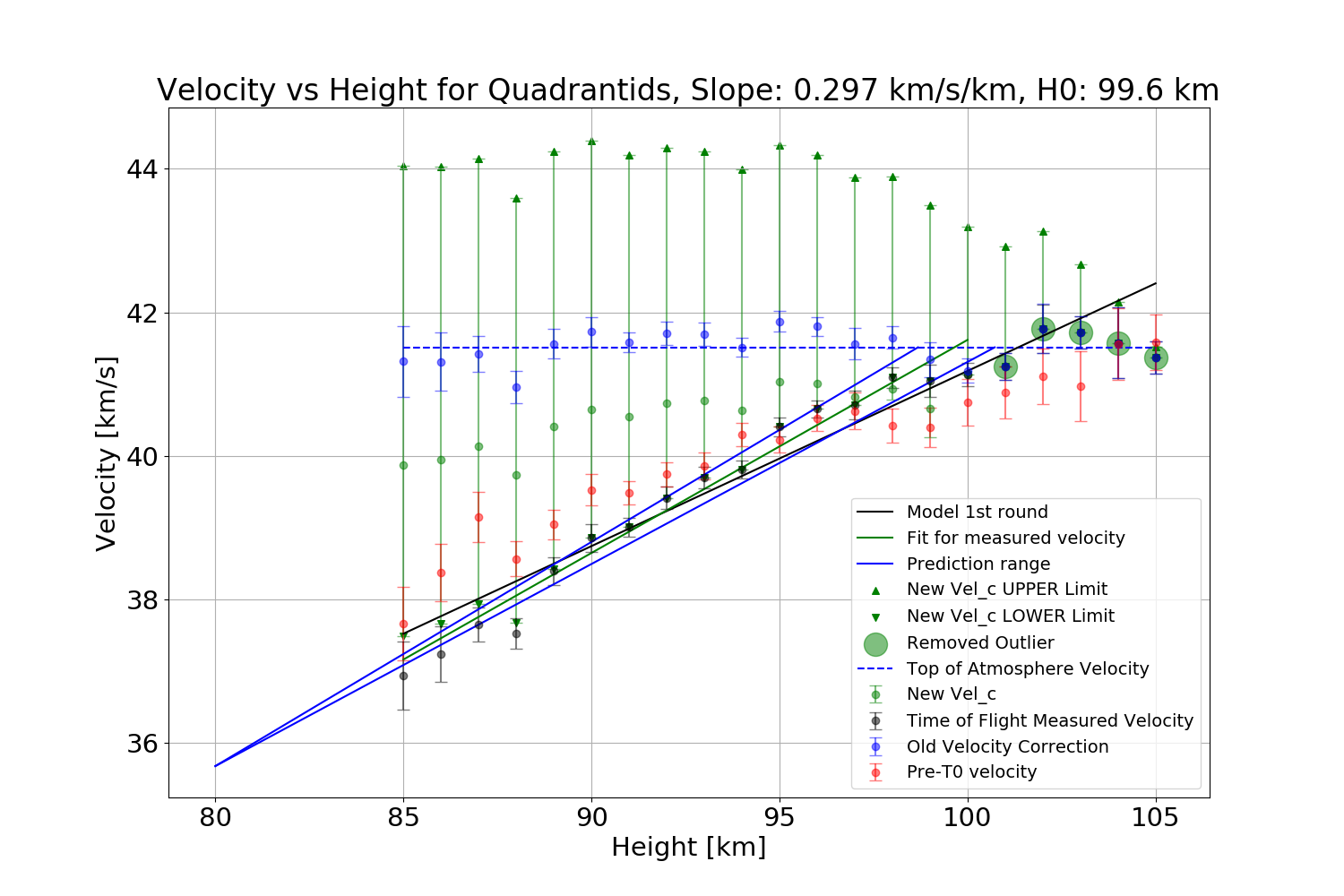}

\end{subfigure}

\medskip
\begin{subfigure}{0.8\textwidth}
\includegraphics[trim = 1cm 1cm 0cm 1cm, clip, width=\linewidth]{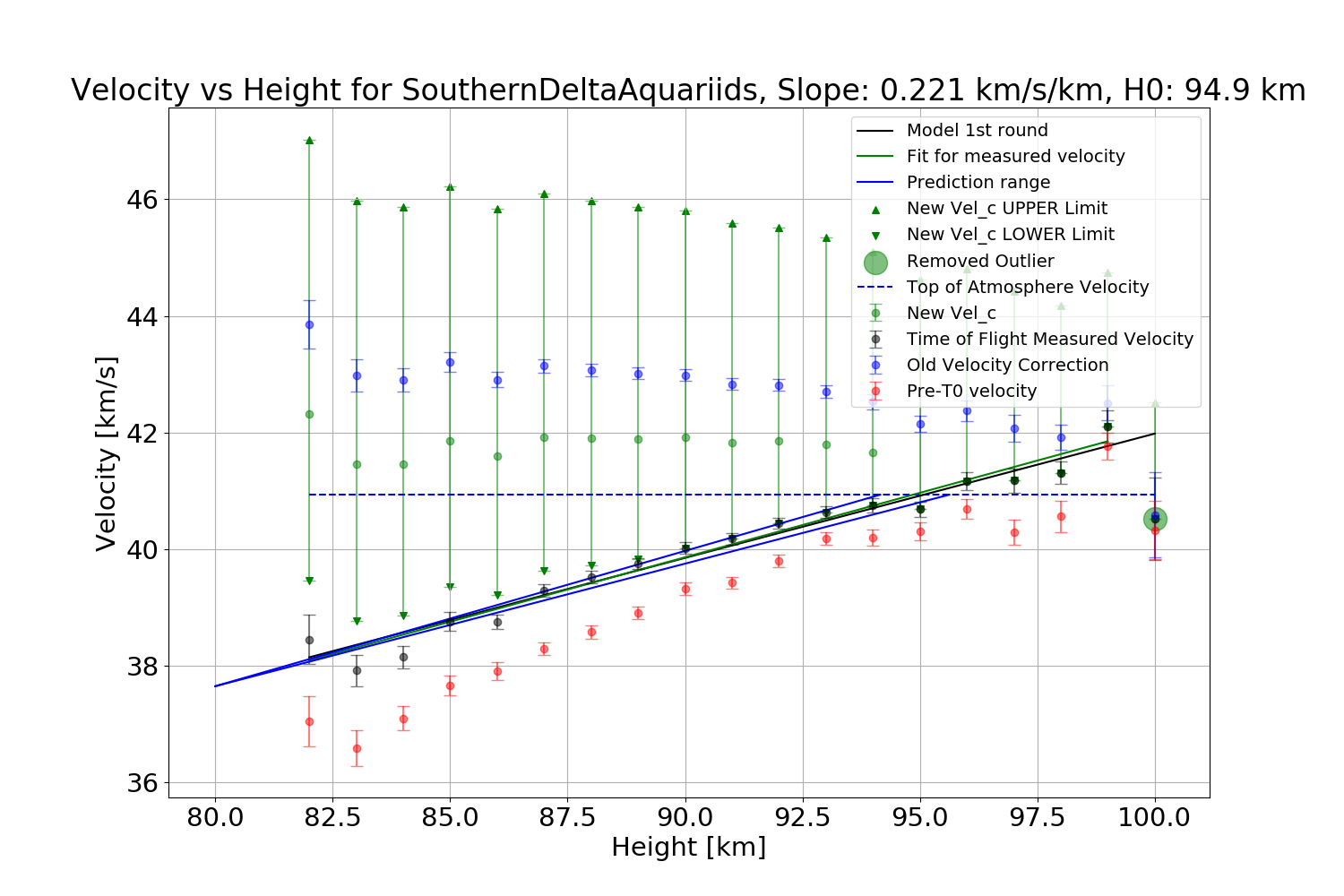}
\end{subfigure}\hspace*{\fill}
\begin{subfigure}{0.8\textwidth}
\includegraphics[trim = 1cm 1cm 0cm 1cm, clip, width=\linewidth]{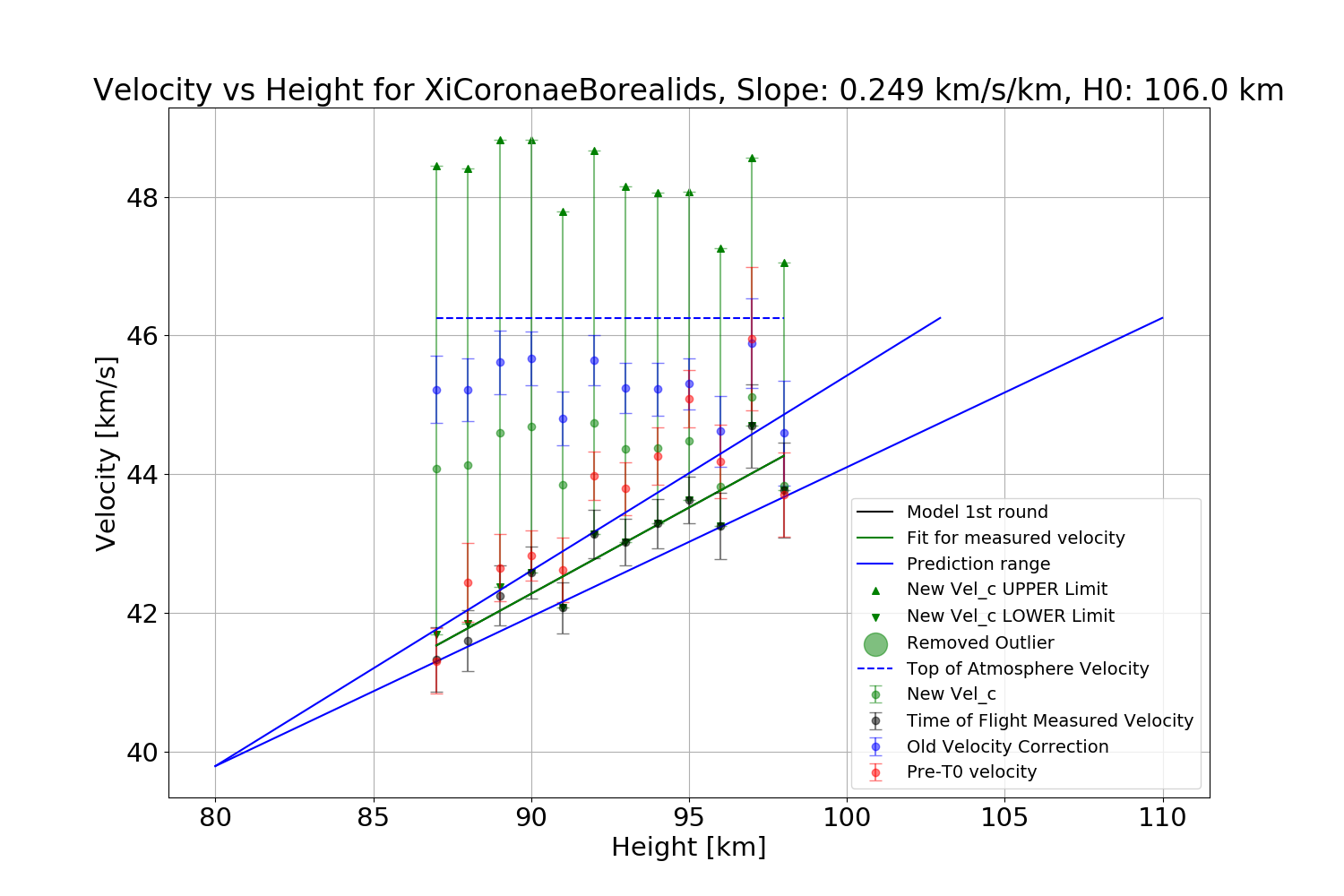}

\end{subfigure}
\caption[Velocities vs height for showers used in velocity correction (2/2)]{Velocities versus height plots for all showers used in velocity correction (part 2/2)}
\label{apx_vels2}
\end{figure}


\clearpage
\newpage

\begin{sidewaystable}

\centering
\advance\rightskip-0.5cm
\advance\leftskip-0.5cm

{\footnotesize
{\linespread{1.2}
{\rmfamily
\begin{tabular}{|p{1.4cm}|p{1.2cm}|p{1.75cm}|c|c|c|c|l|l|l|l|l|l|l|p{0.7cm}|}

\hline
\textbf{Shower}        &  \multicolumn{1}{C{1.5cm}}{\textbf{no. of Meteors}} & \multicolumn{1}{C{1.75cm}}{\textbf{Outliers \newline\tiny{{[}km{]}}}}                                     & \multicolumn{1}{C{1cm}}{\textbf{R1 Slope \tiny{{[}km/s/km{]}}}} & \multicolumn{1}{C{1cm}}{\textbf{SE R1 Slope \tiny{[km/s/km]}}} & \multicolumn{1}{C{1cm}}{\textbf{R1 $H_0$ \tiny{{[}km{]}}}} & \multicolumn{1}{C{1cm}}{\textbf{SE R1 $H_0$ \tiny{{[}km{]}}}} & \multicolumn{1}{C{1cm}}{\textbf{R2 Slope \tiny{{[}km/s/km{]}}}} & \multicolumn{1}{C{1cm}}{\textbf{SE R2 Slope \tiny{{[}km/s/km{]}}}} & \multicolumn{1}{C{1cm}}{\textbf{R2 $H_0$ \tiny{{[}km{]}}}} & \multicolumn{1}{C{1cm}}{\textbf{SE R2 $H_0$ \tiny{{[}km{]}}}} & \multicolumn{1}{C{1cm}}{\textbf{R2 $H_0$ Min. \tiny{{[}km{]}}}} & \multicolumn{1}{C{1cm}}{\textbf{R2 $H_0$ Max. \tiny{{[}km{]}}}} & \multicolumn{1}{C{1cm}}{\textbf{Min.\newline Offset \tiny{{[}s{]}}}} & \multicolumn{1}{C{0.7cm}|}{\textbf{O.R.}} \\ \hline
Geminids               & 7672                                              & {[}80, 81, 96, 97, 98, 99, 100, 101, 102, 103, 104, 105{]}                        & 0.250                                      & 0.014                                                     & 97.069                             & 7.454                                             & 0.308                                      & 0.010                                                     & 95.340                             & 4.371                                             & 94.845                                                        & 95.870                                                        & 0.095                                      & False                                  \\
\hline Quadrantids            & 1629                                            & {[}101, 102, 103, 104, 105, 106, 107{]}                                           & 0.244                                      & 0.017                                                     & 101.349                            & 9.453                                             & 0.297                                      & 0.015                                                     & 99.650                             & 7.072                                             & 98.683                                                        & 100.723                                                       & 0.080                                      & False                                  \\
\hline Leonids                & 112                                             & {[}86, 89, 103{]}                                                                 & 0.355                                      & 0.171                                                     & 109.670                            & 70.384                                            & 0.355                                      & 0.171                                                     & 109.670                            & 70.384                                            & 100.010                                                       & 137.361                                                       & 0.047                                      & False                                  \\
\hline Perseids               & 427                                             & {[}90, 98, 99, 100, 101, 102, 103, 104, 105, 106, 107, 108, 109, 110, 111, 112{]} & 0.192                                      & 0.023                                                     & 105.369                            & 17.464                                            & 0.258                                      & 0.060                                                     & 102.297                            & 32.153                                            & 98.092                                                        & 109.048                                                       & 0.000                                      & True                                   \\
\hline Orionids               & 624                                             & {[}89, 90, 102, 103, 104, 105, 106, 107, 108, 109, 110{]}                         & 0.245                                      & 0.043                                                     & 105.725                            & 25.052                                            & 0.334                                      & 0.050                                                     & 103.425                            & 21.220                                            & 100.360                                                       & 107.578                                                       & 0.049                                      & False                                  \\
\hline January Leonids         & 190                                              & {[}85, 99, 100{]}                                                                 & 0.151                                      & 0.036                                                     & 101.416                            & 33.268                                            & 0.151                                      & 0.036                                                     & 101.416                            & 33.268                                            & 97.245                                                        & 108.248                                                       & 0.063                                      & False                                  \\
\hline Daytime\newline  Arietids        & 5124                                              & {[}80, 99, 100, 101, 102, 103, 104{]}                                             & 0.282                                      & 0.007                                                     & 97.809                             & 3.139                                             & 0.287                                      & 0.008                                                     & 97.663                             & 3.846                                             & 97.169                                                        & 98.187                                                        & 0.082                                      & False                                  \\
\hline Daytime\newline  Sextantids      & 756                                            & {[}80, 81, 82, 96, 98, 99, 100, 101, 102, 103{]}                                  & 0.217                                      & 0.017                                                     & 95.615                             & 10.472                                            & 0.230                                      & 0.016                                                     & 95.152                             & 8.960                                             & 94.181                                                        & 96.265                                                        & 0.100                                      & False                                  \\
\hline Southern \newline Delta \newline Aquariids & 4051                                              & {[}80, 100, 101, 102, 103, 104{]}                                                 & 0.213                                      & 0.013                                                     & 95.087                             & 8.301                                             & 0.221                                      & 0.011                                                     & 94.856                             & 6.488                                             & 94.156                                                        & 95.628                                                        & 0.081                                      & False                                  \\
\hline Draconids              & 159                                               & {[}87, 90, 96, 99, 100, 101{]}                                                    & 0.157                                      & 0.042                                                     & 102.198                            & 37.153                                            & 0.308                                      & 0.062                                                     & 98.152                             & 27.138                                            & 95.132                                                        & 102.679                                                       & 0.000                                      & True                                   \\
\hline Xi Coronae\newline  Borealids     & 397                                              & {[}81, 100{]}                                                                     & 0.249                                      & 0.033                                                     & 106.005                            & 18.677                                            & 0.249                                      & 0.033                                                     & 106.005                            & 18.677                                            & 102.955                                                       & 109.989                                                       & 0.072                                      & False                                  \\
\hline November Omega\newline  Orionids  & 685                                             & {[}82, 83, 85, 99, 101, 102, 103, 104, 105{]}                                     & 0.171                                      & 0.025                                                     & 103.296                            & 20.242                                            & 0.152                                      & 0.021                                                     & 104.446                            & 19.019                                            & 101.501                                                       & 108.327                                                       & 0.075                                      & False \\                                
\hline
\end{tabular}

}
}
}
\caption[Shower velocity correction parameters used in generating fit data]{Shower velocity correction parameters used in generating fit data. R1,R2 represent parameters generated from first and second round fitting, respectively. SE represents standard error. O.R. represents whether minimum time offset override was applied for shower.}
\label{tab:showermodel}
\end{sidewaystable}

\end{changemargin}

\end{document}